\documentclass{article}
\usepackage{epubtk}
\usepackage{amsmath}
\usepackage{amssymb}
\usepackage{epsf}
\usepackage{lscape}
\usepackage{booktabs}
\usepackage{textcomp}

\usepackage{graphicx}

\newcounter{boxcounter}
\newenvironment{mybox}[1]
{\refstepcounter{boxcounter}
 \vspace{2 em}
 \noindent
 \hrulefill
 \begin{center}
   {\bf Box~\theboxcounter. #1}
 \end{center}
 \hrulefill}
{\hrulefill
 \vspace{1 em}}

\newcommand{\cl}[1]{{\centerline{#1}}}

\DeclareMathOperator{\diag}{diag}

\begin{document}

\title{The Confrontation between General Relativity \\ and Experiment}

\author{\epubtkAuthorData{Clifford M.\ Will}
                {Department of Physics \\
        University of Florida \\
        Gainesville FL 32611, U.S.A.}
        {cmw@physics.ufl.edu}
        {http://www.phys.ufl.edu/~cmw/}}
\date{}

\maketitle

%%%%%%%%%%%%%%%%%%%%%%%%%%%%%%%%%%%%%%%%%%%%%%%%%%%%%%%%%%%%%%%%%%%%%%%%%%%%%%%%%%%
%%%%%%%%%%%%%%%%%%%%%%%%%%%%%%%%%%%%%%%%%%%%%%%%%%%%%%%%%%%%%%%%%%%%%%%%%%%%%%%%%%%
%%%%%%%%%%%%%%%%%%%%%%%%%%%%%%%%%%%%%%%%%%%%%%%%%%%%%%%%%%%%%%%%%%%%%%%%%%%%%%%%%%%

\begin{abstract}
  The status of experimental tests of general relativity and of
  theoretical frameworks for analyzing them are reviewed and updated. Einstein's
  equivalence principle (EEP) is well supported by experiments such as
  the E\"otv\"os experiment, tests of local Lorentz invariance and clock experiments. 
  Ongoing tests of EEP and of the
  inverse square law are  searching for new interactions arising from
  unification or quantum gravity. Tests of general relativity at the
  post-Newtonian level have reached high precision, including the
  light deflection, the Shapiro time delay, the perihelion advance of
  Mercury,  the Nordtvedt effect in lunar motion, and frame-dragging.
  Gravitational wave damping has been detected in an amount that
  agrees with general relativity to better than half a percent using
  the Hulse--Taylor binary pulsar, and a growing family of
  other binary pulsar systems is 
  yielding new tests, especially of strong-field effects.  Current and future tests of relativity will center on strong gravity and gravitational waves.
\end{abstract}

%\epubtkKeywords{tests of relativistic gravity, theories of gravity, post-Newtonian
%limit, gravitational radiation}

\newpage

\tableofcontents

%%%%%%%%%%%%%%%%%%%%%%%%%%%%%%%%%%%%%%%%%%%%%%%%%%%%%%%%%%%%%%%%%%%%%%%%%%%%%%%%%%%
%%%%%%%%%%%%%%%%%%%%%%%%%%%%%%%%%%%%%%%%%%%%%%%%%%%%%%%%%%%%%%%%%%%%%%%%%%%%%%%%%%%
%%%%%%%%%%%%%%%%%%%%%%%%%%%%%%%%%%%%%%%%%%%%%%%%%%%%%%%%%%%%%%%%%%%%%%%%%%%%%%%%%%%

\section{Introduction}
\label{S1}

When general relativity was born 100 years ago, experimental
confirmation was almost a side issue.  Admittedly, Einstein did calculate observable
effects of general relativity, such as the perihelion advance of Mercury,
which he knew to be an unsolved problem, and the deflection of light,
which was subsequently verified.  But compared to the inner consistency and
elegance of the theory, he regarded such empirical questions as almost
secondary.   He famously stated that if the measurements of light deflection disagreed with the theory he would ``feel sorry for the dear Lord, for the theory {\em is} correct!''.  

By contrast, today experimental gravitation is a major component of
the field, characterized by continuing efforts to test the theory's
predictions, both in the solar system and in the astronomical world, to detect gravitational waves from astronomical sources, and to search for possible gravitational imprints of phenomena originating in the quantum, high-energy or cosmological realms.

The modern history of experimental relativity can be divided roughly into
four periods: Genesis, Hibernation, a Golden Era, and the Quest for Strong
Gravity. The Genesis (1887\,--\,1919) comprises the period of the two great
experiments which were the foundation of relativistic physics -- the
Michelson--Morley experiment and the E\"otv\"os experiment -- and the two
immediate confirmations of general relativity -- the deflection of light and the
perihelion advance of Mercury.  Following this was a period of Hibernation
(1920\,--\,1960) during which theoretical work temporarily outstripped
technology and experimental possibilities, and, as a consequence, the
field stagnated and was relegated to the backwaters of physics and
astronomy.

But beginning around 1960, astronomical discoveries (quasars, pulsars,
cosmic background radiation) and new experiments pushed general relativity to the
forefront. Experimental gravitation experienced a Golden Era (1960\,--\,1980)
during which a systematic, world-wide effort took place to understand the
observable predictions of general relativity, to compare and contrast them with the
predictions of alternative theories of gravity, and to perform new
experiments to test them.  New technologies -- atomic clocks, radar and laser ranging, space probes, cryogenic capabilities, to mention only a few -- played a central role in this golden era.  The period began with an experiment to confirm
the gravitational frequency shift of light (1960) and ended with the
reported decrease in the orbital period of the Hulse-Taylor binary pulsar
at a rate consistent with the general relativistic prediction of gravitational-wave energy loss (1979). The results all supported general relativity, and most
alternative theories of gravity fell by the wayside (for a popular review,
see~\cite{WER}).

Since that time, the field has entered what might be termed a Quest for Strong
Gravity. 
Much like modern art, the term ``strong'' means different things to different people.  To one steeped in general relativity, the principal figure of merit
that distinguishes strong from weak gravity is the quantity $\epsilon \sim
GM/Rc^2$, where $G$ is the Newtonian gravitational constant, $M$ is the
characteristic mass scale of the phenomenon, $R$ is the characteristic
distance scale, and $c$ is the speed of light.  Near the event horizon of
a non-rotating black hole, or for the expanding observable universe,
$\epsilon \sim 1$; for neutron stars, $\epsilon \sim 0.2$. These are
the regimes of strong gravity. For the solar system, $\epsilon < 10^{-5}$;
this is the regime of weak gravity.   

An alternative view of ``strong'' gravity comes from the world of particle physics.  Here the figure of merit is $GM/R^3c^2 \sim \ell^{-2}$, where the Riemann curvature of spacetime associated with the phenomenon, represented by the left-hand-side, is comparable to the inverse square of a favorite length scale $\ell$.  If $\ell$ is the Planck length, this would correspond to the regime where one expects conventional quantum gravity effects to come into play.   Another possible scale for $\ell$ is the TeV scale associated with many models for unification of the forces, or models with extra spacetime dimensions.   From this viewpoint, strong gravity is where the curvature is comparable to the inverse length squared.  Weak gravity is where the curvature is much smaller than this.  The universe at the Planck time is strong gravity.  Just outside the event horizon of an astrophysical black hole is weak gravity. 

Considerations of the possibilities for new physics from either point of view have led to a wide range of questions that will motivate new tests of general relativity as we move into its second century:

\begin{itemize}

\item
Are the black holes that are in evidence throughout the universe truly the black holes of general relativity?

\item
Do gravitational waves propagate with the speed of light and do they contain more than the two basic polarization states of general relativity?

\item
Does general relativity hold on cosmological distance scales?

\item
Is Lorentz invariance strictly valid, or could it be violated at some detectable level?

\item
Does the principle of equivalence break down at some level?

\item
Are there testable effects arising from the quantization of gravity?

\end{itemize}  

In this update of our {\em Living Review} , we will summarize the current status of
experiments, and attempt to chart the future of the subject.  We will not
provide complete references to early work done in this field but instead
will refer the reader to selected recent papers and to the appropriate review articles and monographs, specifically to \textit{Theory and Experiment in Gravitational
Physics}~\cite{tegp}, hereafter referred to as TEGP;  references to
TEGP will be by chapter or section, e.g., ``TEGP~8.9''.     Additional 
reviews in this subject
are~\cite{PhysRevD.86.010001,shapiro,2008ARNPS..58..207T}. 
The ``Resource Letter'' by the author \cite{2010AmJPh..78.1240W}, contains an annotated list of 100 key references for experimental gravity.

\newpage

%%%%%%%%%%%%%%%%%%%%%%%%%%%%%%%%%%%%%%%%%%%%%%%%%%%%%%%%%%%%%%%%%%%%%%%%%%%%%%%%%%%
%%%%%%%%%%%%%%%%%%%%%%%%%%%%%%%%%%%%%%%%%%%%%%%%%%%%%%%%%%%%%%%%%%%%%%%%%%%%%%%%%%%
%%%%%%%%%%%%%%%%%%%%%%%%%%%%%%%%%%%%%%%%%%%%%%%%%%%%%%%%%%%%%%%%%%%%%%%%%%%%%%%%%%%

\section{Tests of the Foundations of Gravitation Theory}
\label{S2}

%%%%%%%%%%%%%%%%%%%%%%%%%%%%%%%%%%%%%%%%%%%%%%%%%%%%%%%%%%%%%%%%%%%%%%%%%%%%%%%%%%%
%%%%%%%%%%%%%%%%%%%%%%%%%%%%%%%%%%%%%%%%%%%%%%%%%%%%%%%%%%%%%%%%%%%%%%%%%%%%%%%%%%%

\subsection{The Einstein equivalence principle}
\label{eep}

The principle of equivalence has historically played an important role in
the development of gravitation theory. Newton regarded this principle as
such a cornerstone of mechanics that he devoted the opening paragraph of
the \textit{Principia} to it. In 1907, Einstein used the principle as a
basic element in his development of
general relativity (GR). We now regard the principle of
equivalence as the foundation, not of Newtonian gravity or of GR, but of
the broader idea that spacetime is curved.
Much of
this viewpoint can be traced back to Robert Dicke, who contributed crucial
ideas
about the foundations of gravitation theory between 1960 and 1965. These
ideas were summarized in his influential Les Houches lectures of
1964~\cite{dicke64},  and resulted in what has come to be called the
Einstein equivalence principle (EEP).

One elementary equivalence principle is the kind Newton had in mind when
he stated that the property of a body called ``mass'' is proportional to
the ``weight'', and is known as the weak equivalence principle (WEP).
An alternative statement of WEP is that the trajectory of a freely
falling ``test'' 
body (one not acted upon by such forces as electromagnetism and
too small to be affected by tidal gravitational forces) is independent of
its internal structure and composition. In the simplest case of dropping
two different bodies in a gravitational field, WEP states that the bodies
fall with the same acceleration (this is often termed the Universality of
Free Fall, or UFF).

The Einstein equivalence principle (EEP) is a more powerful and
far-reaching concept; it states that:
\begin{enumerate}
\item WEP is valid.
\item The outcome of any local non-gravitational experiment is
  independent of the velocity of the freely-falling reference frame in
  which it is performed.
\item The outcome of any local non-gravitational experiment is
  independent of where and when in the universe it is performed.
\end{enumerate}
The second piece of EEP is called local Lorentz invariance (LLI), and the
third piece is called local position invariance (LPI).

For example, a measurement of the electric force between two charged
bodies is a local non-gravitational experiment; a measurement of the
gravitational force between two bodies (Cavendish experiment) is not.

The Einstein equivalence principle is the heart and soul of gravitational
theory, for it is possible to argue convincingly that if EEP is valid,
then gravitation must be a ``curved spacetime'' phenomenon, in other
words, the effects of gravity must be equivalent to the effects of living
in a curved spacetime. As a consequence of this argument, the only
theories of gravity that can fully embody EEP are those that satisfy the
postulates of ``metric theories of gravity'', which are:
\begin{enumerate}
\item Spacetime is endowed with a symmetric metric.
\item The trajectories of freely falling test bodies are geodesics of
  that metric.
\item In local freely falling reference frames, the non-gravitational
  laws of physics are those written in the language of special
  relativity.
\end{enumerate}

The argument that leads to this conclusion simply notes that, if EEP is
valid, then in local freely falling frames, the laws governing experiments
must be independent of the velocity of the frame (local Lorentz
invariance), with constant values for the various atomic constants (in
order to be independent of location). The only laws we know of that
fulfill this are those that are compatible with special relativity, such
as Maxwell's equations of electromagnetism. Furthermore, in local freely
falling frames, test bodies appear to be unaccelerated, in other words
they move on straight lines; but such ``locally straight'' lines simply
correspond to ``geodesics'' in a curved spacetime (TEGP~2.3~\cite{tegp}).

General relativity is a metric theory of gravity, but then so are
many others, including the Brans--Dicke theory and its generalizations. 
Theories in which varying non-gravitational constants are
associated with dynamical fields that couple to matter directly are
not metric theories. Neither, in
this narrow sense, is
superstring theory (see Section~\ref{newinteractions}), which, while
based fundamentally on a spacetime metric, introduces additional
fields (dilatons, moduli)
that can couple to material stress-energy in a
way that can lead to violations, say, of WEP. It is important to point out,
however, that there is some ambiguity in whether one treats such fields as
EEP-violating gravitational fields, or simply as additional matter fields, like
those that carry electromagnetism or the weak interactions.  
Still,
the notion of curved spacetime is a very general and fundamental
one, and therefore it is important to test the various aspects of
the Einstein equivalence principle thoroughly.
We first survey the experimental tests, and describe some of the theoretical
formalisms that have been developed to interpret them. 
For other reviews of
EEP and its experimental and theoretical significance,
see~\cite{hauganlammer2, lammer03}; for a pedagogical review of the variety of equivalence principles, see~\cite{2013arXiv1310.7426D}.

%%%%%%%%%%%%%%%%%%%%%%%%%%%%%%%%%%%%%%%%%%%%%%%%%%%%%%%%%%%%%%%%%%%%%%%%%%%%%%%%%%%

\subsubsection{Tests of the weak equivalence principle}
\label{wep}

A direct test of WEP is the comparison of the acceleration of two
laborat\-ory-sized bodies of different composition in an external
gravitational field.
If the principle were violated, then the
accelerations of different bodies would differ. The simplest
way to quantify such possible violations of WEP in a form
suitable for comparison with experiment is to suppose that for
a body with inertial mass $m_\mathrm{I}$, the passive gravitational
mass $m_\mathrm{P}$ is no longer equal to $m_\mathrm{I}$, so that in a
gravitational field $g$, the acceleration is given
by $m_\mathrm{I} a= m_\mathrm{P} g$. Now the inertial mass of a typical
laboratory body is made up of several types of mass-energy:  rest
energy, electromagnetic energy, weak-interaction energy, and so
on. If one of these forms of energy contributes to $m_\mathrm{P}$
differently than it does to $m_\mathrm{I}$, a violation of WEP would
result. One could then write
\begin{equation}
  m_\mathrm{P} = m_\mathrm{I} + \sum_A \frac{\eta^A E^A}{c^2},
  \label{E1}
\end{equation}
where $E^A$ is the internal energy of the body generated by
interaction $A$, $\eta^A$ is a dimensionless parameter that
measures the strength of the violation of WEP induced by that
interaction, and $c$ is the speed of light. A measurement or limit
on the fractional difference in acceleration between two bodies
then yields a quantity called the ``E\"otv\"os ratio'' given by
\begin{equation}
  \eta \equiv 2 \frac{|a_1 - a_2|}{|a_1 + a_2|} =
  \sum_A \eta^A \left( \frac{E_1^A}{m_1 c^2} -
  \frac{E_2^A}{m_2 c^2} \right),
  \label{E2}
\end{equation}
where we drop the subscript ``I'' from the inertial masses.
Thus, experimental limits on $\eta$ place limits on the
WEP-violation parameters $\eta^A$.

%%%%%%%%%%%%%%%%%%%%%%%%%%%%

%\epubtkImage{wep.png}{
\begin{figure}[h!t]
\centering
 % \def\epsfsize#1#2{0.5#1}
% \centerline{\epsfbox{livingwep.eps}}
\includegraphics[width = 4in]{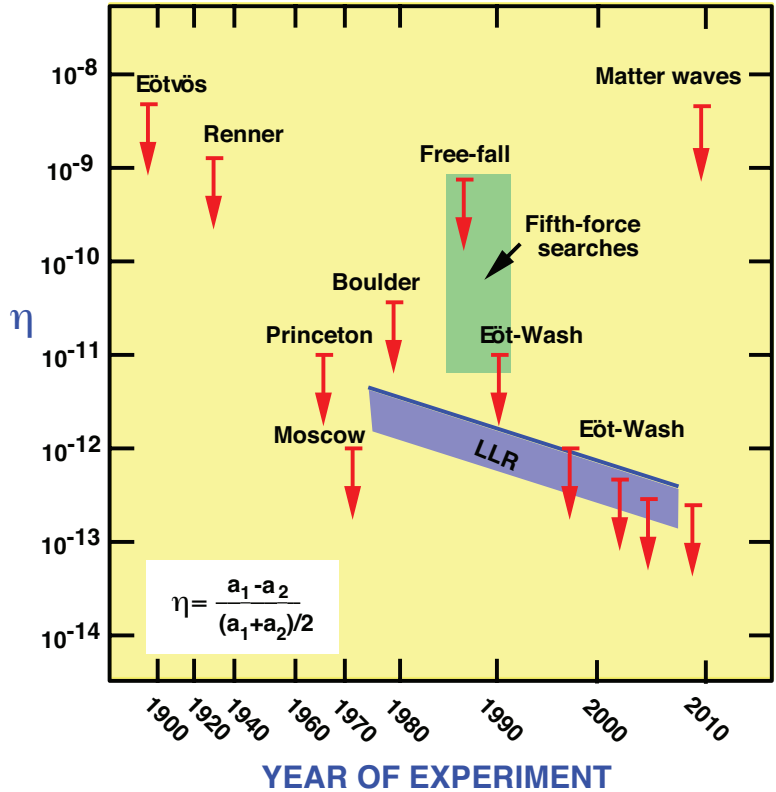}
 \caption{Selected tests of the weak equivalence principle,
    showing bounds on $\eta$, which measures fractional difference in
    acceleration of different materials or bodies. The free-fall and
    E\"ot-Wash experiments were originally performed to search for a
    fifth force (green region, representing many experiments). 
    The blue band shows evolving bounds on $\eta$ for
    gravitating bodies from lunar laser ranging (LLR).}
    \label{wepfig}
    \end{figure}%}
    
%%%%%%%%%%%%%%%%%%%%%%%%%%%

Many high-precision E\"otv\"os-type experiments have been performed, from
the pendulum experiments of Newton, Bessel, and Potter to the classic
torsion-balance measurements of E\"otv\"os~\cite{eotvos},
Dicke~\cite{dicke1}, Braginsky~\cite{braginsky}, and their collaborators.
In the modern torsion-balance experiments, two objects of different
composition are connected by a rod or placed on a tray and suspended in a
horizontal orientation by a fine wire. If the gravitational acceleration
of the bodies differs, and this difference has a component
perpendicular to
the suspension wire, there will be a torque induced on the 
wire, related to the angle between the wire and the direction of the
gravitational acceleration {\boldmath $g$}. If the entire apparatus is
rotated about some direction with angular velocity $\omega$, the torque
will be modulated with period $2 \pi / \omega$. In the experiments of
E\"otv\"os and his collaborators, the wire and {\boldmath $g$} were not
quite parallel because of the centripetal acceleration on the apparatus
due to the Earth's rotation; the apparatus was rotated about the direction
of the wire. In the Dicke and Braginsky experiments, {\boldmath $g$} was
that of the Sun, and the rotation of the Earth provided the modulation of
the torque at a period of 24~hr (TEGP~2.4~(a)~\cite{tegp}). Beginning in
the late 1980s, numerous experiments were carried out primarily to search
for a ``fifth force'' (see Section~\ref{fifthforce}), but their null
results also constituted tests of WEP. In the ``free-fall Galileo
experiment'' performed at the University of Colorado, the relative
free-fall acceleration of two bodies made of uranium and copper was
measured using a laser interferometric technique. The ``E\"ot-Wash''
experiments carried out at the University of Washington used a
sophisticated torsion balance tray to compare the accelerations of various
materials toward local topography on Earth, movable laboratory masses, the
Sun and the galaxy~\cite{Su94, baessler99}, and have reached levels of
$2 \times 10^{-13}$~\cite{adelberger01,2008PhRvL.100d1101S,2012CQGra..29r4002W}. 

The recent development of atom interferometry has yielded tests of WEP, albeit to modest accuracy, comparable to that of the original E\"otv\"os experiment.  In these experiments, one measures the local acceleration of the two separated wavefunctions of an atom such as Cesium by studying the interference pattern when the wavefunctions are combined, and compares that with the acceleration of a nearby macroscopic object of different composition~\cite{2010Metro..47L...9M,2010Natur.463..926M}.
A claim 
that these experiments test the gravitational redshift~\cite{2010Natur.463..926M}  was subsequently shown to be incorrect~\cite{2011CQGra..28n5017W}.

The resulting upper limits on $\eta$ are summarized
in Figure~\ref{wepfig} (TEGP~14.1~\cite{tegp}; for a bibliography of
experiments up to 1991, see~\cite{fischbach92}).

A number of projects are in the development or planning stage to push
the bounds on $\eta$ even lower. The project MICROSCOPE
is
designed to test WEP to $10^{-15}$. It is being developed
by the French space agency CNES for  launch in late 2015,
for a one-year mission~\cite{microscope}.
The drag-compensated 
satellite will be in a Sun-synchronous polar orbit at 700~km
altitude, with a payload consisting of two differential
accelerometers, one with elements made of the same material
(platinum), and another with elements made of different materials
(platinum and titanium).  Other concepts for future improvements include advanced space experiments (Galileo-Galilei, STEP), experiments on sub-orbital rockets, lunar laser ranging (see Sec.\ \ref{Nordtvedteffect}), binary pulsar observations, and experiments with anti-hydrogen.  For a recent focus issue on past and future tests of WEP, see Vol.\ 29, Number 18 of {\em Classical and Quantum Gravity} \cite{0264-9381-29-18-180301}.

%%%%%%%%%%%%%%%%%%%%%%%%%%%%%%%%%%%%%%%%%%%%%%%%%%%%%%%%%%%%%%%%%%%%%%%%%%%%%%%%%%%

\subsubsection{Tests of local Lorentz invariance}
\label{lli}

Although special relativity itself never benefited from the kind of
``crucial'' experiments, such as the perihelion advance of Mercury and
the deflection of light, that contributed so much to the initial acceptance
of
GR and to the fame of Einstein, the steady
accumulation of experimental support, together with the successful
merger of special relativity
with quantum mechanics, led to its acceptance
by mainstream physicists by the late 1920s, ultimately to become part of
the standard toolkit of every working physicist.
This accumulation included
\begin{itemize}
\item the classic Michelson--Morley experiment and its
  descendents~\cite{mm, shankland, jaseja, brillethall},
\item the Ives--Stillwell, Rossi--Hall, and other tests of
  time-dilation~\cite{ives, rossi, farley},
\item tests of the independence of the speed of light of the velocity
  of the source, using both binary X-ray stellar sources and
  high-energy pions~\cite{brecher, alvager},
\item tests of the isotropy of the speed of light~\cite{Champeney,
    riis, krisher90}.
\end{itemize}

In addition to these direct experiments, there was the Dirac equation
of quantum mechanics and its prediction of anti-particles and spin;
later would come the stunningly successful relativistic theory of
quantum electrodynamics.  For a pedagogical review on the occasion of the 2005 centenary of special relativity, see~\cite{2006eins.book...33W}.

In 2015, on the 110th anniversary of the introduction of special relativity,
one might ask ``what is there to test?'' Special relativity has
been so thoroughly
integrated into the fabric of modern physics that its validity is
rarely
challenged, except by cranks and crackpots.
It is ironic then, that during the past several years, a vigorous
theoretical and experimental effort has been launched, on an
international
scale, to find violations of special relativity.
The motivation for this effort is not a desire
to repudiate Einstein, but to look for
evidence of new physics ``beyond'' Einstein, such as apparent, or ``effective''
violations
of Lorentz invariance that might result from certain models of quantum
gravity.
Quantum gravity asserts that there is a fundamental length scale
given by the Planck length, $\ell_\mathrm{Pl} = (\hbar G/c^3)^{1/2} = 1.6 \times
10^{-33} \mathrm{\ cm}$, but since length is not an invariant quantity
(Lorentz--FitzGerald contraction), then there could be a violation of
Lorentz
invariance at some level in quantum gravity. In brane-world
scenarios, while
physics may be locally Lorentz invariant in the higher dimensional
world,
the confinement of the interactions of normal physics to our
four-dimensional ``brane'' could induce apparent Lorentz violating
effects.
And in models such as string theory, the presence of additional
scalar,
vector, and tensor long-range fields that couple to matter of the
standard
model could induce effective violations of Lorentz symmetry.
These and other ideas have motivated
a serious
reconsideration of how to test Lorentz invariance with better
precision and
in new ways.

A simple and useful way of interpreting some of these modern
experiments, called the $c^2$-formalism,  is to suppose that the
electromagnetic interactions suffer a slight violation of Lorentz
invariance, through a change in the speed of electromagnetic radiation $c$
relative to the limiting speed of material test particles ($c_0$, made 
to take the value unity via a choice of units), in other words, $c \ne 1$ (see
Section~\ref{c2formalism}). Such a violation necessarily selects a preferred
universal rest frame, presumably that of the cosmic background radiation,
through which we are moving at about $370 \mathrm{\ km\ s}^{-1}$~\cite{lineweaver96}. Such a
Lorentz-non-invariant electromagnetic interaction would cause shifts in
the energy levels of atoms and nuclei that depend on the orientation of
the quantization axis of the state relative to our universal velocity
vector, and on the quantum numbers of the state. The presence or absence
of such energy shifts can be examined by measuring the energy of one such
state relative to another state that is either unaffected or is affected
differently by the supposed violation. One way is to look for a shifting
of the energy levels of states that are ordinarily equally spaced, such as
the Zeeman-split
$2J+1$ ground states of a nucleus of total spin $J$ in a magnetic field;
another is to compare the levels of a complex nucleus
with the atomic hyperfine levels of a hydrogen maser clock. 
The magnitude of these ``clock
anisotropies'' turns out to be proportional to
$\delta \equiv | c^{-2}-1|$.

\begin{figure}[h!t]
\centering
 % \def\epsfsize#1#2{0.5#1}
 % \centerline{\epsfbox{livinglli.eps}}
 \includegraphics[width = 4in]{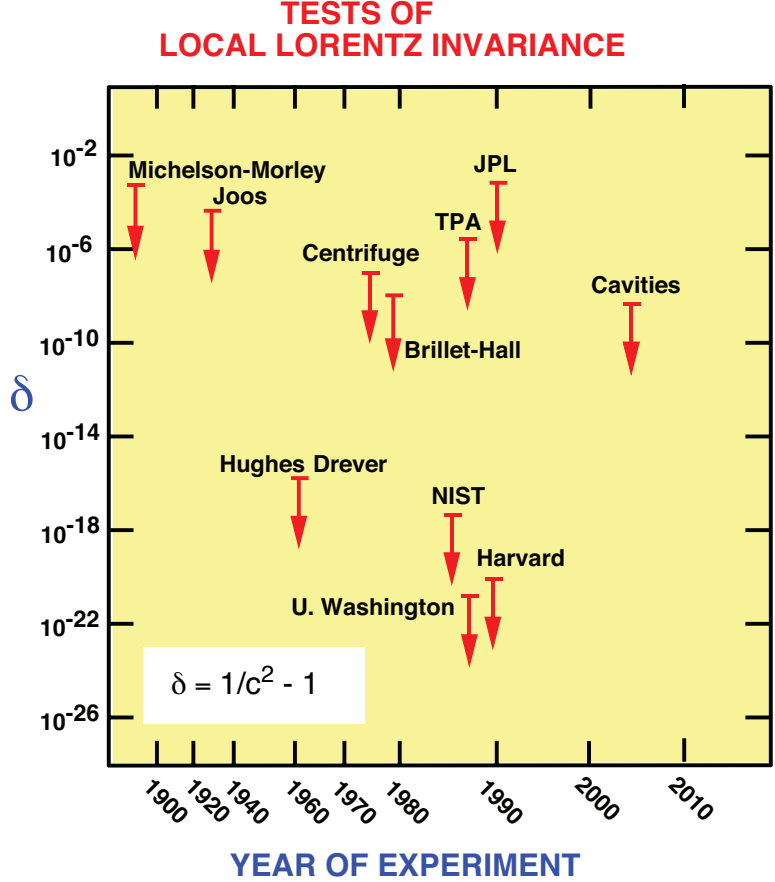}
  \caption{Selected tests of local Lorentz invariance showing the
    bounds on the parameter $\delta$, which measures the degree of
    violation of Lorentz invariance in electromagnetism. The
    Michelson--Morley, Joos, Brillet--Hall and cavity experiments test the
    isotropy of the round-trip speed of light. 
    The centrifuge, two-photon absorption (TPA) and
    JPL experiments test the isotropy of light speed using one-way
    propagation. The most precise experiments test isotropy of
    atomic energy levels. The limits assume a speed of Earth of
    $370 \mathrm{\ km\ s}^{-1}$ relative to the mean rest frame of the universe.}
  \label{llifig}
\end{figure}

The earliest clock anisotropy experiments were the
Hughes--Drever experiments, performed in the period
1959\,--\,60 independently by Hughes and collaborators at Yale University, and
by Drever at Glasgow University, although their original motivation was 
somewhat different~\cite{hughes, drever}. 
The Hughes--Drever 
experiments yielded extremely accurate results, quoted as limits
on the parameter $\delta \equiv c^{-2}-1$ in Figure~\ref{llifig}. 
Dramatic improvements were made in the 1980s using
laser-cooled trapped atoms and ions~\cite{prestage85, lamoreaux86, chupp}.
This technique made
it possible
to reduce the broading of resonance lines caused by collisions,
leading to improved bounds on $\delta$ shown in Figure~\ref{llifig}
(experiments
labelled NIST, U.\ Washington and Harvard, respectively).

Also
included for comparison is the corresponding limit obtained from
Michelson--Morley type experiments (for a review, see~\cite{hauganwill}).
In those experiments, when viewed from the preferred frame, the speed of
light down the two arms of the moving interferometer is $c$, while it can be
shown using the electrodynamics of the $c^2$ formalism,
that the compensating
Lorentz--FitzGerald contraction of the parallel arm is governed by the speed 
$c_0=1$. Thus the Michelson--Morley
experiment and its descendants also measure the coefficient
$c^{-2}-1$. One of these is the Brillet--Hall experiment~\cite{brillethall},
which used a Fabry--Perot laser interferometer. 
In a  recent series of experiments,
the frequencies of electromagnetic cavity oscillators in various
orientations
were compared with each other or
with atomic clocks as a function of
the orientation of the 
laboratory~\cite{wolf03, lipa03, muller03, antonini05, stanwix05}.    
These
placed bounds on $c^{-2}-1$ at the level of better than a part in $10^9$.
Haugan and L\"ammerzahl~\cite{hauganlammer1} have considered
the bounds that Michelson--Morley type experiments could place on a modified
electrodynamics involving a ``vector-valued'' effective photon mass.

%%%%%%%%%%%%%%%%%%%%%%%%%

%\epubtkImage{livinglli.png}{
%}

%%%%%%%%%%%%%%%%%

The $c^2$ framework focusses exclusively on classical electrodynamics.
It has recently been extended to the entire standard model of particle physics
by Kosteleck\'y and
colleagues~\cite{colladay97, colladay98, kosteleckymewes02}. The ``Standard
Model Extension'' (SME) has a large number of Lorentz-violating
parameters, opening up many new opportunities for experimental tests
(see Section~\ref{SME}).
A variety of clock anisotropy experiments have been carried out to
bound the electromagnetic parameters of the SME 
framework~\cite{kosteleckylane99}. 
For example,
the cavity experiments described above~\cite{wolf03, lipa03, muller03}
placed bounds on the coefficients of the
tensors $\tilde{\kappa}_\mathrm{e-}$ and $\tilde{\kappa}_\mathrm{o+}$ 
(see Section~\ref{SME} for definitions) at the levels of
$10^{-14}$ and $10^{-10}$, respectively.
Direct comparisons between atomic clocks based on different nuclear species
place bounds on SME parameters in the neutron and proton sectors, depending
on the nature of the transitions involved. The bounds achieved range from
$10^{-27}$ to $10^{-32} \mathrm{\ GeV}$.   Recent examples include~\cite{2006PhRvL..96f0801W,2011PhRvL.107q1604S}.

Astrophysical observations have also been used to bound Lorentz violations.
For example, if photons satisfy the Lorentz violating dispersion
relation
\begin{equation}
  E^2 = p^2 c^2 + E_\mathrm{Pl} f^{(1)} |p|c + f^{(2)} p^2c^2 +
  \frac{f^{(3)}}{E_\mathrm{Pl}} |p|^3c^3 + \dots,
  \label{dispersion}
\end{equation}
where $E_\mathrm{Pl}= (\hbar c^5/G)^{1/2}$ is the Planck energy, 
then the speed of light 
$v_\gamma=\partial E /\partial p$ would
be given, to linear order in the $f^{(n)}$ by
\begin{equation}
  \frac{v_\gamma}{c} \approx 1 + \sum_{n \ge 1} 
  \frac{(n-1)f_\gamma^{(n)} E^{n-2}}{2E_\mathrm{Pl}^{n-2}}.
\end{equation}
Such a Lorentz-violating dispersion relation could be a relic of
quantum gravity, for instance.
By bounding the difference in arrival time of high-energy photons from a
burst source at large distances, one could bound contributions to the
dispersion for $n >2$. One limit, $|f^{(3)}| < 128$ comes 
from observations of 1 and 2~TeV gamma rays from the blazar
Markarian~421~\cite{biller}.   Another limit comes from birefringence in
photon propagation: In many Lorentz violating models, different photon polarizations  
may propagate with different speeds, causing the plane of polarization
of a wave to rotate. If the frequency dependence of this rotation has
a dispersion relation similar to Eq.~ (\ref{dispersion}), then by
studying ``polarization diffusion'' of light from a polarized source
in a given bandwidth, one can effectively place a bound 
$|f^{(3)}| < 10^{-4}$~\cite{gleiser}.   Measurements of the spectrum of ultra-high-energy cosmic rays using data from the HiRes and Pierre Auger observatories show no evidence for violations of Lorentz invariance~\cite{2009NJPh...11h5003S,2009PhRvD..79h3015B}.
Other testable effects of Lorentz invariance violation include threshold
effects in particle reactions, 
gravitational Cerenkov radiation, and neutrino oscillations.

For thorough and up-to-date surveys of both the theoretical
frameworks and the experimental results for tests of LLI see the reviews by
Mattingly~\cite{mattingly}, Liberati~\cite{Liberati2013} and Kosteleck\'y and Russell~\cite{RevModPhys.83.11}.  The last article gives ``data tables'' showing experimental bounds on all the various parameters of the SME.

%%%%%%%%%%%%%%%%%%%%%%%%%%%%%%%%%%%%%%%%%%%%%%%%%%%%%%%%%%%%%%%%%%%%%%%%%%%%%%%%%%%

\subsubsection{Tests of local position invariance}
\label{lpi}

The principle of local position invariance, the third part of
EEP, can be tested by the gravitational redshift
experiment,
 the first experimental test of gravitation proposed by Einstein. Despite
the fact that Einstein regarded this as a crucial test of GR, we now
realize that it does not distinguish between GR and any other metric
theory of gravity, but is only a test of EEP.  The iconic gravitational
redshift experiment measures the frequency or wavelength shift $Z \equiv
\Delta \nu / \nu = - \Delta \lambda / \lambda$ between two identical
frequency standards (clocks) placed at rest at different heights in a
static gravitational field. If the frequency of a given type of atomic
clock is the same when measured in a local, momentarily comoving freely
falling frame (Lorentz frame), independent of the location or velocity of
that frame, then the comparison of frequencies of two clocks at rest at
different locations boils down to a comparison of the velocities of two
local Lorentz frames, one at rest with respect to one clock at the moment
of emission of its signal, the other at rest with respect to the other
clock at the moment of reception of the signal. The frequency shift is
then a consequence of the first-order Doppler shift between the frames.
The structure of the clock plays no role whatsoever. The result is a
shift
\begin{equation}
  Z = \frac{\Delta U}{c^2},
  \label{E3}
\end{equation}
where $\Delta U$ is the difference in the Newtonian gravitational
potential between the receiver and the emitter. If LPI is not
valid, then it turns out that the shift can be written
\begin{equation}
  Z = (1 + \alpha) \frac{\Delta U}{c^2},
  \label{E4}
\end{equation}
where the parameter $\alpha$ may depend upon the nature of the
clock whose shift is being measured (see TEGP~2.4~(c)~\cite{tegp} for
details).

\begin{figure}[h!t]
\centering
  %\def\epsfsize#1#2{0.5#1}
  %\centerline{\epsfbox{livinglpi.eps}}
  \includegraphics[width=4in]{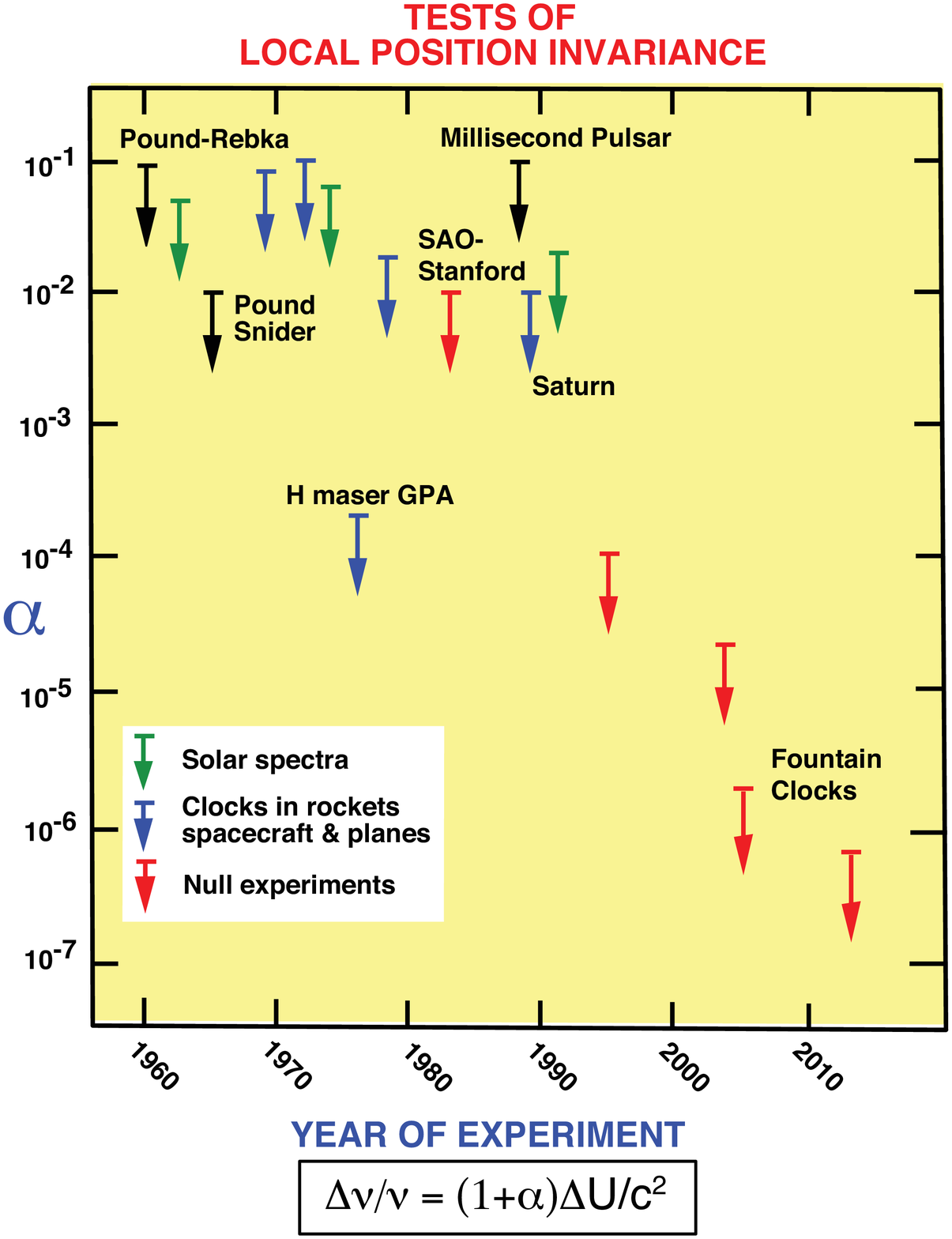}
  \caption{Selected tests of local position invariance via
    gravitational redshift experiments, showing bounds on $\alpha$,
    which measures degree of deviation of redshift from the formula
    $\Delta \nu / \nu = \Delta U/c^2$. In null redshift experiments, the
    bound is on the difference in $\alpha$ between different kinds of
    clocks.}
  \label{lpifig}
\end{figure}

The first successful, high-precision redshift measurement was the
series of Pound--Rebka--Snider experiments of 1960\,--\,1965 that
measured the frequency shift of gamma-ray photons from ${}^{57} \mathrm{Fe}$
as they ascended or descended the Jefferson Physical Laboratory
tower at Harvard University. The high accuracy
achieved -- one percent -- was obtained by making
use of the M\"ossbauer effect to produce a narrow resonance line
whose shift could be accurately determined. Other experiments
since 1960 measured the shift of spectral lines in the Sun's
gravitational field and the change in rate of atomic clocks
transported aloft on aircraft, rockets and satellites. Figure~\ref{lpifig}
summarizes the important redshift experiments that have been
performed since 1960 (TEGP~2.4~(c)~\cite{tegp}).

%%%%%%%%%%%%%%%%%%%%

%\epubtkImage{livinglpi.png}{
%}

%%%%%%%%%%%%%%%%%%%

After almost 50 years of inconclusive or contradictory measurements, the
gravitational redshift of solar spectral lines was finally  measured
reliably. During the early years of GR, the failure to
measure this effect in solar lines
was siezed upon by some as reason to doubt the theory (see~\cite{Crelinsten} for an engaging history of this period).
Unfortunately, the measurement is not simple.
Solar spectral lines are subject to the ``limb effect'', a variation
of spectral line wavelengths between the center of the solar disk and
its edge or ``limb''; this effect is actually a Doppler shift caused by complex
convective and turbulent motions in the photosphere and lower
chromosphere, and is expected to be minimized by observing at the
solar limb, where the motions are predominantly transverse. The
secret is to use strong, symmetrical lines, leading to unambiguous
wavelength measurements. Successful measurements were finally made
in 1962 and 1972 (TEGP~2.4~(c)~\cite{tegp}). In 1991, LoPresto et
al.~\cite{lopresto91} 
measured the solar shift in agreement with LPI to about 2~percent by 
observing the oxygen triplet lines both in
absorption in the limb and in emission just off the limb.

The most precise standard redshift
test to date was the Vessot--Levine rocket
experiment known as Gravity Probe-A (GPA)
 that took place in June 1976~\cite{vessot}. A
hydrogen-maser clock was flown on a rocket to an altitude of
about 10,000~km and its frequency compared to a hydrogen-maser clock on
the ground. The experiment took advantage of the masers'
frequency stability by monitoring the frequency shift as a
function of altitude. A sophisticated data acquisition scheme
accurately eliminated all effects of the first-order Doppler
shift due to the rocket's motion, while tracking data were used
to determine the payload's location and the velocity (to evaluate
the potential difference $\Delta U$, and the special relativistic
time dilation). Analysis of the data yielded a limit
$| \alpha | < 2 \times 10^{-4}$.

A ``null'' redshift experiment performed in 1978 tested whether the {\it
relative} rates of two different clocks depended upon position. Two
hydrogen maser clocks and an ensemble of three superconducting-cavity
stabilized oscillator (SCSO) clocks were compared over a 10-day period.
During the period of the experiment, the solar potential $U/c^2$ within the laboratory was known to change
sinusoidally with a 24-hour period by $3 \times10^{-13}$ because of the
Earth's rotation, and to change linearly at $3 \times 10^{-12}$ per day
because the Earth is 90 degrees from perihelion in April. However,
analysis of the data revealed no variations of either type within
experimental errors, leading to a limit on the LPI violation parameter $|
\alpha^\mathrm{H} - \alpha^\mathrm{SCSO} | < 2 \times 10^{-2}$~\cite{turneaure}.
This bound has been improved using more stable frequency
standards, such as atomic fountain clocks~\cite{Godone, prestage95, bauch02,2008PhRvL.100n0801B}. The best
current bounds, from comparing a Rubidium atomic fountain with a Cesium-133 fountain or with a hydrogen maser~\cite{2012PhRvL.109h0801G,2013PhRvA..87a0102P}, and from comparing transitions of two different isotopes of Dysprosium~\cite{2013PhRvL.111f0801L}, hover around the one part per million mark. 

The Atomic Clock Ensemble in Space (ACES) project will place both a cold trapped atom clock based on Cesium called PHARAO (Projet d'Horloge Atomique par Refroidissement d'Atomes en Orbite), and an advanced hydrogen maser clock on the International Space Station to measure the gravitational redshift to parts in $10^6$, as well as to carry out a number of fundamental physics experiments and to enable improvements in global timekeeping~\cite{2009SSRv..148..233R}.  Launch is currently scheduled for May 2016.  

The varying gravitational redshift of
Earth-bound clocks relative to the highly stable millisecond pulsar PSR
1937+21, caused by the Earth's motion in the solar gravitational field
around the Earth-Moon center of mass (amplitude 4000~km), was 
measured to about 10~percent~\cite{Taylor87}.
Two measurements of the  redshift using stable oscillator clocks
on spacecraft were made at the one percent level: One
used the Voyager spacecraft in Saturn's gravitational 
field~\cite{krisher90a}, while another  
used the Galileo spacecraft in the Sun's field~\cite{krisher93}.

The gravitational redshift could be improved to the $10^{-10}$
level
using an array of laser cooled atomic clocks on board
a spacecraft which would travel to
within four solar radii of the Sun~\cite{maleki01}.  Sadly, the Solar Probe Plus mission, scheduled for launch in 2018, has been formulated as an exclusively heliophysics mission, and thus will not be able to test fundamental gravitational physics.

Modern advances in navigation using Earth-orbiting atomic clocks and
accurate time-transfer must routinely take gravitational redshift and
time-dilation effects into account. For example, the Global Positioning
System (GPS) provides absolute positional
accuracies of around 15~m (even better in
its military mode), and 50 nanoseconds
in time transfer accuracy, anywhere on Earth.
Yet the difference in rate between
satellite and ground clocks as a result of 
relativistic effects is a whopping 39~\emph{microseconds} per day
($46 \mathrm{\ \mu s}$ from the gravitational redshift, and $-7
\mathrm{\ \mu s}$
from time dilation). If these effects were not accurately accounted for,
GPS would fail to function at its stated accuracy. This represents a
welcome practical application of GR! (For the role of GR in GPS,
see~\cite{ashby1, ashby2}; for a popular essay, see~\cite{physicscentral}.)

A final example of the almost ``everyday'' implications of the gravitational redshift is a remarkable measurement using optical clocks based on trapped aluminum ions of the frequency shift over a height of 1/3 of a meter~\cite{2010Sci...329.1630C}.

Local position invariance also refers to position in time. If LPI is
satisfied, the fundamental constants of non-gravitational physics
should be constants in time. Table~\ref{varyconstants} shows current bounds on
cosmological variations in selected dimensionless constants. For
discussion and references to early work, see TEGP~2.4~(c)~\cite{tegp}
or~\cite{dyson72}. For a comprehensive recent review 
both of experiments and of 
theoretical ideas that underly proposals for
varying constants, 
see~\cite{2011LRR....14....2U}.

Experimental bounds on varying constants come in two types: bounds on
the present rate of variation, and bounds on the difference between
today's value and a value in the distant past. The main example of
the former type is the clock comparison test, in which highly stable
atomic clocks of different fundamental type are intercompared over
periods ranging from months to years (variants of the null redshift
experiment). If the frequencies of the
clocks depend differently on the electromagnetic
fine structure constant $\alpha_\mathrm{EM}$, the
electron-proton mass ratio $m_\mathrm{e}/m_\mathrm{p}$, or the
gyromagnetic ratio of the proton $g_\mathrm{p}$, for example,
then a limit on a drift of the fractional frequency
difference translates into a limit on a drift of the constant(s). The
dependence of the frequencies on the constants may be quite complex,
depending on the atomic species involved.  Experiments
have exploited the techniques of laser cooling and trapping, and of
atom fountains, in order to achieve extreme clock stability, and
compared the Rubidium-87 hyperfine transition~\cite{salomon03}, 
the Mercury-199 ion electric quadrupole transition~\cite{bize03}, 
the atomic Hydrogen $1S\mbox{--}2S$ transition~\cite{fischer04}, 
or an optical transition in Ytterbium-171~\cite{peik04},
against the ground-state hyperfine transition in Cesium-133.  More recent experiments have used Strontium-87 atoms trapped in optical lattices~\cite{2008PhRvL.100n0801B} compared with Cesium to obtain $\dot \alpha_\mathrm{EM}/\alpha_\mathrm{EM} < 6 \times 10^{-16}
\mathrm{\ yr}^{-1}$,  compared Rubidium-87 and Cesium-133 fountains~\cite{2012PhRvL.109h0801G} to obtain
$\dot \alpha_\mathrm{EM}/\alpha_\mathrm{EM} < 2.3 \times 10^{-16}
\mathrm{\ yr}^{-1}$, or compared two isotopes of Dysprosium~\cite{2013PhRvL.111f0801L} to obtain
$\dot \alpha_\mathrm{EM}/\alpha_\mathrm{EM} < 1.3 \times 10^{-16}
\mathrm{\ yr}^{-1}$,. 

The second type of bound involves measuring the relics of or signal from
a process that occurred in the distant
past and comparing the inferred value of the constant with the value
measured in the laboratory today. 
One sub-type uses astronomical
measurements of spectral lines at large redshift, while the other uses
fossils of nuclear processes on Earth to infer values of constants
early in geological history. 

\begin{table}[hptb]
  \caption[Bounds on cosmological variation of fundamental
    constants of non-gravitational physics.]{Bounds on cosmological
    variation of fundamental constants of non-gravitational
    physics. For an in-depth review, see~\cite{2011LRR....14....2U}.}
  \label{varyconstants}
  \renewcommand{\arraystretch}{1.2}
  \centering
  \begin{tabular}{p{4.0 cm}|rlp{4.0 cm}}
    \hline \hline
    Constant $k$ &
    \multicolumn{1}{c}{Limit on $\rule{0 em}{1.2 em}\dot{k} / k$} &
    \multicolumn{1}{c}{Redshift} &
    \multicolumn{1}{c}{Method} \\ 
    &
    \multicolumn{1}{c}{($ \mathrm{yr}^{-1} $)} \\
    \hline \hline
    Fine structure constant \newline
    ($ \alpha_\mathrm{EM} = e^2 / \hbar c $) &
    $ < 1.3 \times 10^{-16} $ &
    $ 0 $ &
    Clock comparisons \newline
    \cite{2008PhRvL.100n0801B,2012PhRvL.109h0801G,2013PhRvL.111f0801L} \\
    & $ < 0.5 \times 10^{-16} $ &
    $ 0.15 $ &
    Oklo Natural Reactor \newline
     \cite{damourdyson, fujii04, 2006PhRvC..74f4610P} \\
    & $ < 3.4 \times 10^{-16} $ &
    $ 0.45 $ &
    $ {}^{187}\mathrm{Re} $
    decay in meteorites \newline
     \cite{olive04} \\
    & $ (6.4 \pm 1.4) \times 10^{-16} $ &
    $ 0.2 \mbox{\,--\,} 3.7 $ &
    Spectra in distant quasars \newline
    \cite{webb99, murphy01,2012MNRAS.422.3370K} \\
    & $ < 1.2 \times 10^{-16} $ &
    $ 0.4 \mbox{\,--\,} 2.3 $ &
    Spectra in distant quasars \newline
     \cite{petitjean1, petitjean2,quast04,2012ApJ...746L..16K,2013MNRAS.430.2454L} \\
    \hline
    Weak interaction constant \newline
    ($ \alpha_\mathrm{W} = G_\mathrm{f} m_\mathrm{p}^2 c / \hbar^3 $) &
    $ < 1 \times 10^{-11} $ &
    $ 0.15 $ &
    Oklo Natural Reactor \newline
    \cite{damourdyson} \\
    & $ < 5 \times 10^{-12} $ & $ 10^9 $ &
    Big Bang nucleosynthesis \newline
     \cite{malaney, reeves} \\
    \hline
    e-p mass ratio &
    $<  3.3 \times 10^{-15}$& 0&
    Clock comparisons \newline
    \cite{2008PhRvL.100n0801B} \\
    &$ < 3 \times 10^{-15} $ &
    $ 2.6 \mbox{\,--\,} 3.0 $ &
    Spectra in distant quasars \newline
     \cite{ivanchik05} \\
    \hline \hline
  \end{tabular}
  \renewcommand{\arraystretch}{1.0}
\end{table}

Earlier comparisons of
spectral lines of different atoms or transitions in distant galaxies
and quasars produced bounds $\alpha_\mathrm{EM}$ or
$g_\mathrm{p}(m_\mathrm{e}/m_\mathrm{p})$ on the order of
a part in 10 per Hubble time~\cite{wolfe76}. Dramatic improvements in
the precision of astronomical and laboratory spectroscopy, in the
ability to model the complex astronomical environments where emission
and absorption lines are produced, and in the ability to reach large
redshift have made it possible to improve the bounds significantly.
In fact, in 1999, Webb et al.~\cite{webb99, murphy01} announced that
measurements of 
absorption lines in Mg, Al, Si, Cr, Fe, Ni, and Zn in quasars in the
redshift range $0.5 < Z < 3.5$ indicated a smaller value of
$\alpha_\mathrm{EM}$
in earlier epochs, namely 
$\Delta \alpha_\mathrm{EM}/\alpha_\mathrm{EM} = (-0.72 \pm 0.18) \times 10^{-5}$, 
corresponding to $\dot \alpha_\mathrm{EM}/\alpha_\mathrm{EM} = (6.4 \pm 1.4)
\times 10^{-16} \mathrm{\ yr}^{-1}$ (assuming a linear drift with time).   The Webb group continues to report changes in $\alpha$ over large redshifts~\cite{2012MNRAS.422.3370K}.
Measurements by other groups have so far 
failed to confirm 
this non-zero  effect~\cite{petitjean1, petitjean2, quast04};
An analysis of Mg absorption systems in quasars at $0.4 < Z < 2.3$
gave 
$\dot \alpha_\mathrm{EM}/\alpha_\mathrm{EM} = 
(-0.6 \pm 0.6) \times 10^{-16} \mathrm{\ yr}^{-1}$~\cite{petitjean1}.  Recent studies have also yielded no evidence for a variation in $\alpha_\mathrm{EM}$~\cite{2012ApJ...746L..16K,2013MNRAS.430.2454L}

Another important set of bounds arises from studies of the ``Oklo''
phenomenon, a group of natural, sustained ${}^{235}\mathrm{U}$  fission reactors that
occurred in the Oklo region of Gabon, Africa, around 1.8~billion years ago.
Measurements of ore samples yielded an abnormally low value for the ratio
of two isotopes of Samarium,
${}^{149}\mathrm{Sm}/^{147}\mathrm{Sm}$. Neither of
these isotopes is a fission product, but ${}^{149}\mathrm{Sm}$ can be depleted by
a flux of neutrons. Estimates of the neutron fluence (integrated dose)
during the reactors' ``on'' phase, combined with the measured abundance
anomaly, yield a value for the neutron cross-section for  ${}^{149}\mathrm{Sm}$
1.8 billion years ago that agrees with the modern value. However, the
capture cross-section is extremely sensitive to the energy of a low-lying
level ($E \sim 0.1 \mathrm{\ eV}$), so that a variation in the energy of this
level of only 20~meV over a billion years would change the capture
cross-section from its present value by more than the observed amount. This
was first analyzed in 1976 by Shlyakter~\cite{shlyakter}. 
Recent reanalyses of the
Oklo data~\cite{damourdyson, fujii04, 2006PhRvC..74f4610P} lead to a bound on $\dot
\alpha_\mathrm{EM}$ at the level of around $5 \times 10^{-17} \mathrm{\ yr}^{-1}$.

In a similar manner, recent 
reanalyses of decay rates of ${}^{187}\mathrm{Re}$ in ancient
meteorites (4.5~billion years old) gave the 
bound $\dot \alpha_\mathrm{EM}/\alpha_\mathrm{EM}
< 3.4 \times 10^{-16} \mathrm{\ yr}^{-1}$~\cite{olive04}.

%%%%%%%%%%%%%%%%%%%%%%%%%%%%%%%%%%%%%%%%%%%%%%%%%%%%%%%%%%%%%%%%%%%%%%%%%%%%%%%%%%%
%%%%%%%%%%%%%%%%%%%%%%%%%%%%%%%%%%%%%%%%%%%%%%%%%%%%%%%%%%%%%%%%%%%%%%%%%%%%%%%%%%%

\subsection{Theoretical frameworks for analyzing EEP}
\label{EEPframeworks}

%%%%%%%%%%%%%%%%%%%%%%%%%%%%%%%%%%%%%%%%%%%%%%%%%%%%%%%%%%%%%%%%%%%%%%%%%%%%%%%%%%%

\subsubsection{Schiff's conjecture}
\label{Schiff}

Because the three parts of the Einstein equivalence principle
discussed above are so very different in their empirical
consequences, it is tempting to regard them as independent
theoretical principles. On the other hand, any complete and
self-consistent gravitation theory must possess sufficient
mathematical machinery to make predictions for the outcomes of
experiments that test each principle, and because there are
limits to the number of ways that gravitation can be meshed with
the special relativistic laws of physics, one might not be
surprised if there were theoretical connections between the three
sub-principles. For instance, the same mathematical formalism
that produces equations describing the free fall of a hydrogen
atom must also produce equations that determine the energy levels
of hydrogen in a gravitational field, and thereby the ticking
rate of a hydrogen maser clock. Hence a violation of EEP in the
fundamental machinery of a theory that manifests itself as a
violation of WEP might also be expected to show up as a violation
of local position invariance. Around 1960, Leonard Schiff conjectured
that this kind of connection was a necessary feature of any
self-consistent theory of gravity. More precisely,
Schiff's conjecture states that \emph{any complete, self-consistent
theory of gravity that embodies WEP necessarily embodies EEP}.
In other words, the validity of WEP alone guarantees the validity
of local Lorentz and position invariance, and thereby of EEP.

If Schiff's conjecture is correct, then E\"otv\"os experiments may
be seen as the direct empirical foundation for EEP, hence for the
interpretation of gravity as a curved-spacetime phenomenon. Of course,
a rigorous proof of such a conjecture is impossible (indeed, some
special counter-examples are known~\cite{Ohanian74, Ni77, Coley82}),
yet a number
of powerful ``plausibility'' arguments can be formulated.

The most general and elegant of these arguments is based upon the
assumption of energy conservation. This assumption allows one to
perform very simple cyclic gedanken experiments in which the
energy at the end of the cycle must equal that at the beginning
of the cycle. This approach was pioneered by Dicke, Nordtvedt, and
Haugan (see, e.g., \cite{haugan}).
A system in a quantum state $A$ decays to state $B$,
emitting a quantum of frequency $\nu$. The quantum falls a height
$H$ in an external gravitational field and is shifted to frequency
$\nu'$, while the system in state $B$ falls with acceleration
$g_B$. At the bottom, state $A$ is rebuilt out of state $B$, the
quantum of frequency $ \nu'$, and the kinetic energy $m_B g_B H$
that state $B$ has gained during its fall. The energy left over
must be exactly enough, $m_A g_A H$, to raise state $A$ to its original
location. (Here an assumption of local Lorentz invariance
permits the inertial masses $m_A$  and $m_B$ to be identified with
the total energies of the bodies.)  If $g_A$ and $g_B$ depend on
that portion of the internal energy of the states that was
involved in the quantum transition from $A$ to $B$ according to
\begin{equation}
  g_A = g \left( 1 + \frac{\alpha E_A}{m_A c^2} \right),
  \qquad
  g_B = g \left( 1 + \frac{\alpha E_B}{m_B c^2} \right),
  \qquad
  E_A - E_B \equiv h \nu
  \label{E5}
\end{equation}
(violation of WEP), then by conservation of energy, there must be
a corresponding violation of LPI in the frequency shift of the
form (to lowest order in $h \nu /mc^2$)
\begin{equation}
  Z = \frac{\nu' - \nu}{\nu'} = (1 + \alpha) \frac{gH}{c^2} =
  (1 + \alpha) \frac{\Delta U}{c^2}.
  \label{E6}
\end{equation}
Haugan generalized this approach to include violations of
LLI~\cite{haugan} (TEGP~2.5~\cite{tegp}).

\begin{mybox}{The \boldmath $ TH\epsilon\mu $ formalism}
  \begin{description}
  \item[Coordinate system and conventions:]~\\ [0.5 em]
    $x^0 = t$: time coordinate associated with the static nature of the
    static spherically symmetric (SSS) gravitational field;
    ${\bf x} = (x, y, z) $: isotropic quasi-Cartesian spatial
    coordinates; spatial vector and gradient operations as in
    Cartesian space.
  \end{description}
  
  \begin{description}
  \item[Matter and field variables:]~\\ [-1.5 em]
    \begin{itemize}
    \item
      $m_{0a}$: rest mass of particle $a$.
    \item
      $e_a$: charge of particle $a$.
    \item
      $x_a^\mu (t)$: world line of particle $a$.
    \item
      $v_a^\mu = dx_a^\mu/dt$: coordinate velocity of particle $a$.
    \item
      $A_\mu = $: electromagnetic vector potential;
      ${\bf E}=\nabla A_0-\partial {\bf A}/\partial t$,
      ${\bf B}=\nabla \times {\bf A}$.
    \end{itemize}
  \end{description}
  
  \begin{description}
  \item[Gravitational potential:]~\\ [0.5 em]
    $U ({\bf x})$.
  \end{description}
  
  \begin{description}
  \item[Arbitrary functions:]~\\ [0.5 em]
    $T(U)$, $H(U)$, $\epsilon (U)$, $\mu (U)$;
    EEP is satisfied if $\epsilon = \mu = (H/T)^{1/2}$ for all $U$.
  \end{description}
  
  \begin{description}
  \item[Action:]~\\ [-0.5 em]
    \begin{displaymath}
      I = - \sum_a m_{0a} \int (T-Hv_a^2)^{1/2} \, dt +
      \sum_a e_a \int A_\mu (x_a^\nu ) v_a^\mu \, dt +
      (8 \pi)^{-1} \int (\epsilon E^2 - \mu^{-1} B^2) \, d^4 x.
    \end{displaymath}
  \end{description}
  
  \begin{description}
  \item[Non-metric parameters:]~\\ [-0.5 em]
    \begin{displaymath}
      \Gamma_0 = - c_0^2 \, \frac{\partial}{\partial U}
      \ln [\epsilon (T/H)^{1/2}]_0,
      \qquad
      \Lambda_0 = - c_0^2 \, \frac{\partial}{\partial U}
      \ln [\mu (T/H)^{1/2}]_0,
      \qquad
      \Upsilon_0 = 1 - (TH^{-1} \epsilon\mu)_0,
    \end{displaymath}
    where $c_0 = (T_0 /H_0 )^{1/2}$ and subscript ``0'' refers to a
    chosen point in space. If EEP is satisfied,
    $\Gamma_0 \equiv \Lambda_0 \equiv \Upsilon_0 \equiv 0$.
  \end{description}
  \label{box1}
\end{mybox}

%%%%%%%%%%%%%%%%%%%%%%%%%%%%%%%%%%%%%%%%%%%%%%%%%%%%%%%%%%%%%%%%%%%%%%%%%%%%%%%%%%%

\subsubsection[The $TH\epsilon\mu$ formalism]{The \boldmath $TH\epsilon\mu$ formalism}
\label{theuformalism}

The first successful attempt to prove Schiff's conjecture more
formally was made by Lightman and Lee~\cite{lightmanlee}. They developed a
framework called the $TH\epsilon\mu$ formalism that encompasses
all metric theories of gravity and many non-metric theories
(see Box~\ref{box1}). It restricts attention to the behavior of charged
particles (electromagnetic interactions only) in an external
static spherically symmetric (SSS) gravitational field, described
by a potential $U$. It characterizes the motion of the charged
particles in the external potential by two arbitrary functions
$T(U)$ and $H(U)$, and characterizes the response of
electromagnetic fields to the external potential (gravitationally
modified Maxwell equations) by two functions $\epsilon (U)$ and
$\mu (U)$. The forms of $T$, $H$, $\epsilon$, and $\mu$ vary from theory
to theory, but every metric theory satisfies
\begin{equation}
  \epsilon = \mu = \left( \frac{H}{T} \right)^{1/2}\!\!\!\!\!\!\!\!,
  \label{E7}
\end{equation}
for all $U$. This consequence follows from the action of
electrodynamics with a ``minimal'' or metric coupling:
\begin{equation}
  I = - \sum_a m_{0a} \int (-g_{\mu\nu} v_a^\mu v_a^\nu )^{1/2} \, dt +
  \sum_a e_a \int A_\mu (x_a^\nu ) v_a^\mu \, dt -
  \frac{1}{16 \pi} \int \sqrt{-g} \,
  g^{\mu \alpha} g^{\nu \beta} F_{\mu \nu} F_{\alpha \beta} \, d^4 x,
  \label{E8}
\end{equation}
where the variables are defined in Box~\ref{box1}, and where
$F_{\mu\nu} \equiv A_{\nu, \mu} - A_{\mu, \nu}$. By identifying
$g_{00} = T$ and $g_{ij} = H \delta_{ij}$ in a SSS field,
$F_{i0} = E_i$ and $F_{ij} = \epsilon_{ijk} B_k$,
one obtains Eq.~ (\ref{E7}).
Conversely, every theory within this class that satisfies Eq.~ (\ref{E7})
can have its electrodynamic equations cast into ``metric'' form.
In a given non-metric theory, the functions $T$, $H$, $\epsilon$, and
$\mu$ will depend in general on the full gravitational environment,
including the potential of the Earth, Sun, and Galaxy, as well as on
cosmological boundary conditions. Which of these factors has the most
influence on a given experiment will depend on the nature of the
experiment.

Lightman and Lee then calculated explicitly the rate of
fall of a ``test'' body made up of interacting charged particles,
and found that the rate was independent of the internal
electromagnetic structure of the body (WEP) if and only if Eq.~ (\ref{E7})
was satisfied. In other words, WEP $\Rightarrow$ EEP and Schiff's
conjecture was verified, at least within the restrictions built
into the formalism.

Certain combinations of the functions $T$, $H$, $\epsilon$, and $\mu$
reflect different aspects of EEP. For instance, position or
$U$-dependence of either of the combinations $\epsilon (T/H)^{1/2}$
and $\mu (T/H)^{1/2}$ signals violations of LPI, the first
combination playing the role of the locally measured electric
charge or fine structure constant. The ``non-metric parameters''
$\Gamma_0$ and $\Lambda_0$ (see Box~\ref{box1}) are measures of such
violations of EEP. Similarly, if the parameter
$\Upsilon_0 \equiv 1-(TH^{-1} \epsilon \mu )_0$ is non-zero anywhere,
then violations of LLI will occur. This parameter is related to the
difference between the speed of light $c$, and the
limiting speed of material test particles $c_0$, given by
\begin{equation}
  c = ( \epsilon_0 \mu_0 )^{- 1/2},
  \qquad
  c_0 = \left( \frac{T_0}{H_0} \right)^{1/2}\!\!\!\!\!\!\!.
  \label{E9}
\end{equation}
In many applications,
by suitable definition of units, $c_0$ can be set equal to unity.
If EEP is valid, $\Gamma_0 \equiv \Lambda_0 \equiv \Upsilon_0 = 0$
everywhere.

The rate of fall of a composite spherical test body of
electromagnetically interacting particles then has the form
\begin{eqnarray}
  {\bf a} & = & \frac{m_\mathrm{P}}{m} \nabla U,
  \label{E10}
  \\
  \frac{m_\mathrm{P}}{m} & = &
  1 + \frac{E_\mathrm{B}^\mathrm{ES}}{Mc_0^2}
  \left[ 2 \Gamma_0 - \frac{8}{3} \Upsilon_0 \right] +
  \frac{E_\mathrm{B}^\mathrm{MS}}{Mc_0^2}
  \left[ 2 \Lambda_0 - \frac{4}{3} \Upsilon_0 \right] + \dots,
  \label{E11}
\end{eqnarray}%
where $E_\mathrm{B}^\mathrm{ES}$ and $E_\mathrm{B}^\mathrm{MS}$ are the electrostatic
and magnetostatic binding energies of the body, given by
\begin{eqnarray}
  E_\mathrm{B}^\mathrm{ES} & = &
  - \frac{1}{4} T_0^{1/2} H_0^{-1} \epsilon_0^{-1}
  \left< \sum_{ab} \frac{e_a e_b}{r_{ab}} \right>,
  \label{E12}
  \\
  E_\mathrm{B}^\mathrm{MS} & = &
  - \frac{1}{8} T_0^{1/2} H_0^{-1} \mu_0
  \left< \sum_{ab} \frac{e_a e_b}{r_{ab}}
  [{\bf v}_a \cdot {\bf v}_b + ( {\bf v}_a \cdot {\bf n}_{ab})
  ({\bf v}_b \cdot {\bf n}_{ab})] \right>,
  \label{E13}
\end{eqnarray}%
where $r_{ab} = | {\bf x}_a  -  {\bf x}_b |$,
${\bf n}_{ab}  =  ( {\bf x}_a  -  {\bf x}_b )/r_{ab}$, and the
angle brackets denote an expectation value of the enclosed
operator for the system's internal state. E\"otv\"os experiments
place limits on the WEP-violating terms in Eq.~ (\ref{E11}), and
ultimately place limits on the non-metric parameters
$| \Gamma_0 | < 2 \times 10^{-10}$ and
$| \Lambda_0 |  <  3  \times  10^{-6}$.
(We set
$\Upsilon_0 = 0$ because of very tight constraints on it from
tests of LLI; see Figure~\ref{llifig}, where $\delta = -\Upsilon_0$.)  
These limits are
sufficiently tight to rule out a number of non-metric theories of
gravity thought previously to be viable (TEGP~2.6~(f)~\cite{tegp}).

The $TH \epsilon\mu$ formalism also yields a
gravitationally modified Dirac equation that can be used to
determine the gravitational redshift experienced by a variety of
atomic clocks. For the redshift parameter $\alpha$
(see Eq.~ (\ref{E4})), the results are (TEGP~2.6~(c)~\cite{tegp}):
\begin{equation}
  \alpha = \left\{
    \begin{array}{ll}
      - 3 \Gamma_0 + \Lambda_0 &
      \mbox{hydrogen hyperfine transition, H-Maser clock,}
      \\ [0.3 em]
      \displaystyle - \frac{1}{2} (3 \Gamma_0 + \Lambda_0) &
      \mbox{electromagnetic mode in cavity, SCSO clock,}
      \\ [0.8 em]
      - 2 \Gamma_0 &
      \mbox{phonon mode in solid, principal transition in hydrogen.}
    \end{array}
  \right.
  \label{E14}
\end{equation}

The redshift is the standard one $( \alpha = 0)$, independently
of the nature of the clock if and only if
$\Gamma_0 \equiv \Lambda_0 \equiv 0$. Thus the Vessot--Levine
rocket redshift experiment sets a limit on the parameter
combination $| 3 \Gamma_0 - \Lambda_0 |$ (see Figure~\ref{lpifig}); the
null-redshift experiment comparing hydrogen-maser and SCSO clocks
sets a limit on
$| \alpha_\mathrm{H} - \alpha_\mathrm{SCSO} | = \frac{3}{2} | \Gamma_0 - \Lambda_0 |$.
Alvarez and Mann~\cite{AlvarezMann96b, AlvarezMann96a, AlvarezMann97a,
AlvarezMann97b, AlvarezMann97c} extended the $TH\epsilon\mu$ formalism to
permit analysis of such effects as the Lamb shift, anomalous magnetic moments
and non-baryonic effects, and placed interesting bounds on EEP violations.

%%%%%%%%%%%%%%%%%%%%%%%%%%%%%%%%%%%%%%%%%%%%%%%%%%%%%%%%%%%%%%%%%%%%%%%%%%%%%%%%%%%

\subsubsection[The ${c^2}$ formalism]{The \boldmath ${c^2}$ formalism}
\label{c2formalism}

The $TH \epsilon \mu$ formalism can also be applied to tests of
local Lorentz invariance, but in this context it can be
simplified. Since most such tests do not concern themselves with
the spatial variation of the functions $T$, $H$, $\epsilon$, and $\mu$,
but rather with observations made in moving frames, we can treat
them as spatial constants. Then by rescaling the time and space
coordinates, the charges and the electromagnetic fields, we can
put the action in Box~\ref{box1} into the form
(TEGP~2.6~(a)~\cite{tegp})
\begin{equation}
  I = - \sum_a m_{0a} \int (1 - v_a^2)^{1/2} \, dt +
  \sum_a e_a \int A_\mu (x_a^\nu ) v_a^\mu \, dt +
  (8 \pi)^{-1} \int (E^2 - c^2 B^2) \, d^4 x,
  \label{E15}
\end{equation}
where $c^2 \equiv H_0 / (T_0 \epsilon_0 \mu_0)=(1-\Upsilon_0)^{-1}$.
This amounts to using units in which the limiting speed $c_0$
of massive test particles is unity, and the speed of light is $c$.
If $c \ne 1$, LLI is violated; furthermore, the form of the
action above must be assumed to be valid only in some preferred
universal rest frame. The natural candidate for such a frame is
the rest frame of the microwave background.

The electrodynamical equations which follow from Eq.~ (\ref{E15})
yield the behavior of rods and clocks, just as in the full $TH
\epsilon \mu$
formalism. For example, the length of a rod 
which moves with velocity $ \bf V$ relative to the rest frame
in a direction parallel to its length
will be observed by a rest observer to be contracted relative to
an identical rod perpendicular to the motion by a factor
$1 - V^2 / 2 + {\cal O} (V^4)$. Notice that $c$ does not appear in this
expression, because only electrostatic interactions are involved, and $c$
appears only in the magnetic sector of the action~(\ref{E15}). 
The energy and momentum of an electromagnetically
bound body 
moving with velocity $ \bf V$ relative to the rest frame
are given by
\begin{equation}
  \begin{array}{rcl}
    E & = & \displaystyle
    M_\mathrm{R} + \frac{1}{2} M_\mathrm{R} V^2 +
    \frac{1}{2} \delta M_\mathrm{I}^{ij} V^i \, V^j +
    {\cal O} (M V^4),
    \\ [1 em]
    P^i & = & M_\mathrm{R} V^i + \delta M_\mathrm{I}^{ij} V^j +
    {\cal O} (M V^3),
  \end{array}
  \label{E16}
\end{equation}
where $M_\mathrm{R} = M_0 - E_\mathrm{B}^\mathrm{ES}$, $M_0$ is the
sum of the particle rest masses, $E_\mathrm{B}^\mathrm{ES}$ is the
electrostatic binding energy of the system (see Eq.~ (\ref{E12}) with
$T_0^{1/2}H_0 \epsilon_0^{-1}=1$), and
\begin{equation}
  \delta M_\mathrm{I}^{ij} = - 2 \left( \frac{1}{c^2} - 1 \right)
  \left[ \frac{4}{3} E_\mathrm{B}^\mathrm{ES} \delta^{ij} +
  \tilde{E}_\mathrm{B}^{\mathrm{ES}\,ij} \right],
  \label{E17}
\end{equation}
where
\begin{equation}
  \tilde E_\mathrm{B}^{\mathrm{ES}\,ij} =
  - \frac{1}{4} \left< \sum_{ab} \frac{e_a e_b}{r_{ab}}
  \left( n_{ab}^i n_{ab}^j - \frac{1}{3} \delta^{ij} \right) \right>.
  \label{E18}
\end{equation}
Note that $(c^{-2} - 1)$ corresponds to the parameter $\delta$
plotted in Figure~\ref{llifig}.

The electrodynamics given by Eq.~ (\ref{E15}) can also be
quantized, so that we may treat the interaction of photons with
atoms via perturbation theory. The energy of a photon is $\hbar$
times its frequency $\omega$, while its momentum is $\hbar \omega /c$.
Using this approach, one finds that the difference in round trip
travel times of light along the two arms of the interferometer in the
Michelson--Morley experiment is given by
$L_0 (v^2 /c)(c^{-2}  - 1)$.
The experimental null result then leads to the bound on
$(c^{-2} - 1)$ shown on Figure~\ref{llifig}. Similarly the anisotropy in
energy levels is clearly illustrated by the tensorial terms
in Eqs.~ (\ref{E16}, \ref{E18}); by
evaluating $\tilde E_\mathrm{B}^{\mathrm{ES}\,ij}$ for each nucleus in the
various Hughes--Drever-type experiments and comparing with the
experimental limits on energy differences, one obtains the extremely
tight bounds also shown on Figure~\ref{llifig}.

The behavior of moving
atomic clocks can also be analyzed in detail, and bounds
on $(c^{-2} - 1)$ can be placed using results from tests of
time dilation and of the propagation of light. In some cases, it is
advantageous to combine the $c^2$ framework with a ``kinematical''
viewpoint that treats a general class of boost transformations between
moving frames. Such kinematical approaches have been discussed by
Robertson, Mansouri and Sexl, and Will (see~\cite{Will92b}).

For example,
in the ``JPL'' experiment, in which
the phases of two hydrogen masers connected by a fiberoptic link were
compared as a function of the Earth's orientation,
the predicted phase difference as a function of
direction is, to first order in $\bf V$, the velocity of the Earth
through the cosmic background,
\begin{equation}
  \frac{\Delta \phi}{\tilde \phi} \approx - \frac{4}{3} (1 - c^2)
  ({\bf V} \cdot {\bf n} - {\bf V} \cdot {\bf n}_0),
  \label{E19}
\end{equation}
where $\tilde \phi  = 2 \pi \nu L$, $\nu$ is the maser frequency,
$L=21 \mathrm{\ km}$ is the baseline, and where ${\bf n}$ and ${\bf n}_0$ are
unit vectors along the direction of propagation of the light at
a given time and at the initial time of the experiment, respectively. The
observed limit on a diurnal variation in the relative phase
resulted in the bound
$| c^{-2}-1 | < 3 \times 10^{-4}$.
Tighter bounds were obtained from a ``two-photon absorption'' (TPA)
experiment, and a 1960s series of
``M\"ossbauer-rotor'' experiments, which tested the isotropy of
time dilation between a gamma ray emitter on the rim of a rotating
disk and an absorber placed at the center~\cite{Will92b}.

%%%%%%%%%%%%%%%%%%%%%%%%%%%%%%%%%%%%%%%%%%%%%%%%%%%%%%%%%%%%%%%%%%%%%%%%%%%%%%%%%%%

\subsubsection{The Standard Model Extension (SME)}
\label{SME}

Kosteleck\'y and collaborators developed a useful and elegant framework for
discussing violations of Lorentz symmetry in the context of the Standard
Model of particle physics~\cite{colladay97, colladay98, kosteleckymewes02}.
Called the Standard Model Extension (SME), it takes the standard
$\mathrm{SU(3)} \times \mathrm{SU(2)} \times \mathrm{U(1)}$ field theory of particle physics, and
modifies the terms in the action by inserting a variety of tensorial
quantities in the quark, lepton, Higgs, and gauge boson sectors 
that could explicitly violate LLI. SME extends the earlier classical
$TH\epsilon\mu$ and  $c^2$ frameworks, and the $\chi-g$ framework of
Ni~\cite{Ni77}
to quantum field theory and
particle physics.
The modified terms split naturally 
into those that are
odd under CPT (i.e.\ that violate CPT) and terms that are even under CPT.
The result is a rich and complex framework, with many parameters to be
analyzed and tested by experiment. 
Such details are  beyond the scope of this review; for a review of SME and
other frameworks, the reader is referred to the Living Review by
Mattingly~\cite{mattingly} or the review by Liberati~\cite{Liberati2013}.   The review of the SME by Kosteleck\'y and Russell~\cite{RevModPhys.83.11} provides ``data tables'' showing experimental bounds on all the various parameters of the SME.

Here we confine our attention to the electromagnetic sector, in order to
link the SME with the $c^2$ framework discussed above. In the SME, the
Lagrangian for a scalar particle $\phi$ with charge $e$ interacting 
with electrodynamics takes the form
\begin{equation}
  {\cal L} = \left[ \eta^{\mu\nu} + (k_\phi)^{\mu\nu} \right]
  (D_\mu \phi)^\dag D_\nu \phi - m^2 \phi^\dag \phi -
  \frac{1}{4} \left[ \eta^{\mu\alpha} \eta^{\nu\beta} +
  (k_F)^{\mu\nu\alpha\beta} \right] F_{\mu \nu} F_{\alpha \beta},
  \label{ESMaction}
\end{equation}
where $D_\mu \phi = \partial_\mu \phi + ieA_\mu \phi$, where 
$(k_\phi)^{\mu\nu}$ is a real symmetric trace-free tensor, and where
$(k_F)^{\mu \nu\alpha \beta}$ is a tensor with the symmetries of the
Riemann tensor, and with vanishing double trace. It has 19 independent
components. There could also be a CPT-odd term in $\cal L$ of the form 
$(k_A)^\mu \epsilon_{\mu\nu\alpha\beta} A^\nu F^{\alpha \beta}$, but because
of a
variety of pre-existing theoretical and experimental constraints,
it is generally set to zero.

The  tensor $(k_F)^{\mu \alpha\nu \beta}$ can be decomposed into
``electric'', ``magnetic'', and ``odd-parity'' components, by defining
\begin{equation}
  \begin{array}{rcl}
    (\kappa_{DE})^{jk} & = & - 2 (k_F)^{0j0k},
    \\ [0.7 em]
    (\kappa_{HB})^{jk} & = & \displaystyle \frac{1}{2}
    \epsilon^{jpq} \epsilon^{krs} (k_F)^{pqrs},
    \\ [1.0 em]
    (\kappa_{DB})^{kj} & = & - (k_{HE})^{jk} =
    \epsilon^{jpq} (k_F)^{0kpq}.
  \end{array}
\end{equation}
In many applications it is useful to use the further decomposition
\begin{equation}
  \begin{array}{rcl}
    \tilde{\kappa}_\mathrm{tr} & = & \displaystyle
    \frac{1}{3} (\kappa_{DE})^{jj},
    \\ [1.0 em]
    (\tilde{\kappa}_\mathrm{e+})^{jk} & = & \displaystyle
    \frac{1}{2} (\kappa_{DE} + \kappa_{HB})^{jk},
    \\ [1.0 em]
    (\tilde{\kappa}_\mathrm{e-})^{jk} & = & \displaystyle
    \frac{1}{2} (\kappa_{DE} - \kappa_{HB})^{jk} -
    \frac{1}{3} \delta^{jk} (\kappa_{DE})^{ii},
    \\ [1.0 em]
    (\tilde{\kappa}_\mathrm{o+})^{jk} & = & \displaystyle
    \frac{1}{2} (\kappa_{DB} + \kappa_{HE})^{jk},
    \\ [1.0 em]
    (\tilde{\kappa}_\mathrm{o-})^{jk} & = & \displaystyle
    \frac{1}{2} (\kappa_{DB} - \kappa_{HE})^{jk}.
  \end{array}
  \label{kappatensors}
\end{equation}
The first expression is a single number, the next three are symmetric
trace-free matrices, and the final is an antisymmetric matrix, accounting
thereby for the 19 components of the original tensor $(k_F)^{\mu \alpha\nu
\beta}$. 

In the rest frame of the universe, these tensors have some form that is
established by the global nature of the solutions of the overarching theory
being used. In a frame that is moving relative to the universe, the tensors
will have components that depend on the velocity of the frame, and on the
orientation of the frame relative to that velocity.

In the case where the theory is rotationally symmetric in the preferred
frame, the tensors $(k_\phi)^{\mu\nu}$ and 
$(k_F)^{\mu \nu \alpha\beta}$ can be expressed in the form
\begin{eqnarray}
  (k_\phi)^{\mu\nu} & = & \tilde{\kappa}_\phi
  \left( u^\mu \, u^\nu + \frac{1}{4} \eta^{\mu\nu} \right),
  \\
  (k_F)^{\mu\nu\alpha\beta} & = & \tilde{\kappa}_\mathrm{tr}
  \left( 4 u^{[\mu} \eta^{\nu][\alpha} u^{\beta]} -
  \eta^{\mu[\alpha} \eta^{\beta]\nu} \right),
\end{eqnarray}%
where $[~]$ around indices denote antisymmetrization, and where $u^\mu$ is
the four-velocity of an observer at rest in the preferred frame. 
With this assumption, all the tensorial quantities in
Eq.~ (\ref{kappatensors}) vanish in the preferred frame, 
and, after suitable rescalings of coordinates and fields,
the action~(\ref{ESMaction}) can be put into
the form of the $c^2$ framework, with
\begin{equation}
  c = \left( \frac{1 - \frac{3}{4} \tilde{\kappa}_\phi}
  {1 + \frac{1}{4} \tilde{\kappa}_\phi} \right)^{1/2}
  {\left( \frac{1 - \tilde{\kappa}_\mathrm{tr}}{1 +
  \tilde{\kappa}_\mathrm{tr}} \right)^{1/2}} \,.
\end{equation}

%%%%%%%%%%%%%%%%%%%%%%%%%%%%%%%%%%%%%%%%%%%%%%%%%%%%%%%%%%%%%%%%%%%%%%%%%%%%%%%%%%%
%%%%%%%%%%%%%%%%%%%%%%%%%%%%%%%%%%%%%%%%%%%%%%%%%%%%%%%%%%%%%%%%%%%%%%%%%%%%%%%%%%%

\subsection{EEP, particle physics, and the search for new interactions}
\label{newinteractions}

Thus far, we have discussed EEP as a principle that strictly divides the
world into metric and non-metric theories, and have implied
that a failure of EEP might invalidate metric theories (and thus general
relativity). 
On the other hand, there is mounting 
theoretical evidence to suggest that EEP is {\it
likely} to be violated at some level, whether by quantum gravity
effects, by effects arising from string theory, or by hitherto
undetected interactions. 
Roughly speaking, in addition to the pure Einsteinian
gravitational interaction, which respects EEP, theories such as string
theory predict
other interactions which do not. In string theory, for example, the
existence of such EEP-violating fields is assured, but the theory is not
yet mature enough to enable a robust calculation of their strength relative to
gravity, or a determination of whether they are long range, like gravity, or
short range, like the nuclear and weak interactions,
and thus too short-range to be detectable.

In one simple example~\cite{dick98}, one can write the Lagrangian for the low-energy
limit of a string-inspired theory in the so-called ``Einstein frame'', in which
the gravitational Lagrangian is purely general relativistic:
\begin{eqnarray}
  \tilde {\cal L} = \sqrt{- \tilde{g}} \biggl( \!\!\! & &
  \tilde{g}^{\mu\nu} \left[ \frac{1}{2\kappa} \tilde{R}_{\mu\nu} -
  \frac{1}{2} \tilde{G} (\varphi) \partial_\mu \varphi \,
  \partial_\nu \varphi \right] - U (\varphi) \, \tilde{g}^{\mu\nu} \,
  \tilde{g}^{\alpha\beta} F_{\mu\alpha} \, F_{\nu\beta}
  \nonumber
  \\
  & & + \overline{\tilde{\psi}} \left[ i \tilde{e}^\mu_a \gamma^a
  \left( \partial_\mu + \tilde{\Omega}_\mu + q A_\mu \right) -
  \tilde{M} (\varphi) \right] \tilde{\psi} \biggr),
  \label{stringlagrangian1}
\end{eqnarray}%
where ${\tilde g}_{\mu\nu}$ is the non-physical metric,
${\tilde R}_{\mu\nu}$ is the Ricci tensor derived from
it,
$\varphi$ is a
dilaton field, and $\tilde G$, $U$ and $\tilde M$ are functions of
$\varphi$. The Lagrangian includes that for the electromagnetic field
$F_{\mu\nu}$, and that for
particles, written in terms of Dirac spinors $\tilde \psi$. This is
not a metric representation because of the coupling of $\varphi$ to
matter via $\tilde M (\varphi)$ and $U(\varphi)$.
A conformal transformation ${\tilde g}_{\mu\nu} =
F(\varphi) g_{\mu\nu}$, $\tilde \psi = F(\varphi)^{-3/4} \psi$,
puts the Lagrangian in the form (``Jordan'' frame)
\begin{eqnarray}
  {\cal L} = \sqrt{- g} \biggl( \!\!\! & & g^{\mu\nu}
  \left[ \frac{1}{2 \kappa} F (\varphi) R_{\mu\nu} -
  \frac{1}{2} F (\varphi) \tilde{G} (\varphi)
  \partial_\mu \varphi \, \partial_\nu \varphi +
  \frac{3}{4 \kappa F (\varphi)} \partial_\mu F \, \partial_\nu F \right]
  \nonumber \\
  & & - U (\varphi) g^{\mu\nu} \, g^{\alpha\beta} F_{\mu\alpha}
  F_{\nu\beta} + \overline{\psi} \left[ i e^\mu_a \gamma^a
  (\partial_\mu + \Omega_\mu + q A_\mu) -
  \tilde M (\varphi) F^{1/2} \right] \psi \biggr).
  \label{stringlagrangian2}
\end{eqnarray}%
One may choose $F(\varphi)= \mathrm{const.}/\tilde M (\varphi)^2$
so that the particle Lagrangian takes the
metric form (no explicit
coupling to $\varphi$), but the electromagnetic Lagrangian
will still couple non-metrically to $U(\varphi)$. The gravitational
Lagrangian here takes the form of a scalar-tensor theory (see Section~\ref{scalartensor}). But the non-metric electromagnetic term will, in
general, produce violations of EEP.  For examples of specific models,
see~\cite{TaylorVeneziano, DamourPolyakov}. Another class of non-metric
theories are included in the ``varying speed of light (VSL)'' theories; for
a detailed review, see~\cite{magueijo03}.

On the other hand, whether one views such effects as a violation of EEP or
as effects arising from additional ``matter'' fields whose interactions,
like those of the electromagnetic field, do not fully embody EEP, is to some
degree a matter of semantics. Unlike the fields of the standard model of
electromagnetic, weak and strong interactions, which couple to properties
other than mass-energy and are either short range or are strongly screened,
the fields inspired by string theory \emph{could} be long range (if they
remain massless by virtue of a symmetry, or at best, acquire a very small
mass), and \emph{can} couple to mass-energy, and thus can mimic gravitational
fields. Still, there appears to be no way to make this precise.

As a result, EEP and related tests are now viewed as ways to discover or place
constraints on new physical interactions, or as a branch of
``non-accelerator particle physics'', searching for the possible imprints
of high-energy particle effects in the low-energy realm of gravity.
Whether current or proposed experiments
can actually probe these phenomena meaningfully is an open
question at the moment, largely because of a dearth of firm
theoretical predictions.

%%%%%%%%%%%%%%%%%%%%%%%%%%%%%%%%%%%%%%%%%%%%%%%%%%%%%%%%%%%%%%%%%%%%%%%%%%%%%%%%%%%

\subsubsection{The ``fifth'' force}
\label{fifthforce}

On the phenomenological side, the idea of
using EEP tests in this way may have originated in the middle 1980s,
with the search for a ``fifth'' force.
In 1986, as a result
of a detailed reanalysis of E\"otv\"os' original data,
Fischbach et al.~\cite{fischbach5} suggested the existence of a fifth
force of nature, with a strength of about a percent that of
gravity, but with a range (as defined by the range $\lambda$ of a Yukawa
potential, $e^{-r/\lambda} /r$) of a few hundred meters.
This proposal dovetailed with earlier hints of a deviation from the
inverse-square law of Newtonian gravitation derived from measurements of
the gravity profile down deep mines in Australia,
and with emerging
ideas from particle physics suggesting the possible
presence of very low-mass particles with gravitational-strength couplings.
During the next four years
numerous experiments looked for evidence of the fifth force by searching
for composition-dependent differences in acceleration, with variants of
the E\"otv\"os experiment or with free-fall Galileo-type experiments.
Although two early experiments reported positive evidence, the others
all yielded null results. Over the range between one and $10^4$ meters,
the null experiments produced upper limits on the strength of a postulated
fifth force between $10^{-3}$ and $10^{-6}$ of the strength of gravity.
Interpreted as tests of WEP (corresponding to the limit of
infinite-range forces), the results of two representative experiments from
this period, the free-fall Galileo experiment
and the early E\"ot-Wash experiment, are shown in
Figure~\ref{wepfig}. At the same time, tests of the inverse-square
law of gravity were carried out by comparing variations in gravity
measurements up tall towers or down mines or boreholes with gravity
variations predicted using the inverse square law together with Earth
models and surface gravity data mathematically ``continued'' up
the tower or down the hole. Despite early reports of anomalies,
independent tower, borehole, and seawater measurements 
ultimately showed no evidence of a
deviation. Analyses of orbital data from planetary range
measurements, lunar laser ranging (LLR), and laser tracking of the LAGEOS
satellite verified the inverse-square law to parts in $10^8$ over
scales of $10^3$ to $10^5 \mathrm{\ km}$, and to parts in $10^9$ over planetary
scales of several astronomical units~\cite{talmadge}.
A consensus emerged that there was no
credible experimental evidence for a fifth force of nature, of a type
and range proposed by Fischbach et al. For reviews and bibliographies
of this episode, see~\cite{fischbach92, FischbachTalmadge,
FischbachTalmadge2, Adelberger91, WillSky}.

%%%%%%%%%%%%%%%%%%%%%%%%%%%%%%%%%%%%%%%%%%%%%%%%%%%%%%%%%%%%%%%%%%%%%%%%%%%%%%%%%%%

\subsubsection{Short-range modifications of Newtonian gravity}
\label{shortrange}

Although the idea of an intermediate-range violation of Newton's
gravitational law was dropped,  new ideas emerged to suggest the possibility
that the inverse-square law could be violated at very short ranges, below
the centimeter range of existing laboratory verifications of the $1/r^2$
behavior. One set of ideas~\cite{antoniadis98, add98, randall1, randall2} posited 
that some of the extra spatial dimensions that come with string theory
could extend over macroscopic scales, rather than being rolled
up at the Planck scale of $10^{-33} \mathrm{\ cm}$, which was then
the conventional
viewpoint. On laboratory distances large
compared to the relevant scale 
of the extra dimension, gravity would fall off as the
inverse square, whereas on short scales, gravity would fall off as
$1/R^{2+n}$,  where $n$ is the number of large extra dimensions. Many
models favored $n=1$ or $n=2$.
Other possibilities for effective modifications of gravity at short range
involved the exchange of light scalar particles.

Following these proposals,
many of the high-precision, low-noise methods that were
developed for tests of WEP were adapted to carry out
laboratory tests of the inverse square law of
Newtonian gravitation at millimeter scales and below. 
The challenge of these experiments has been to
distinguish gravitation-like interactions from electromagnetic and
quantum
mechanical (Casimir) effects. No deviations from
the inverse square law have been found to date at distances between tens of nanometers and $10 \mathrm{\ mm}$~\cite{long99, hoyle01, hoyle04, kapitulnik, long03,2007PhRvL..98b1101K,2007PhRvL..98m1104A,2007PhRvL..98t1101T,2008PhRvD..78b2002G,2011PhRvL.107q1101S,2011PhRvD..83g5004B,2012PhRvL.108h1101Y,2013PhRvD..87l5031K}.
For a comprehensive review of both the theory and the experiments circa 2002,
see~\cite{adelberger03}.

\subsubsection{The Pioneer anomaly}
\label{pioneer}

In 1998, Anderson et al.~\cite{1998PhRvL..81.2858A} reported the presence of an anomalous deceleration in the motion of the Pioneer 10 and 11 spacecraft at distances between 20 and 70 astronomical units from the Sun.  Although the anomaly was the result of a rigorous analysis of Doppler data taken over many years, it might have been dismissed as having no real significance for new physics, where it not for the fact that the acceleration, of order $10^{-9} \, {\rm m/s}^2$, when divided by the speed of light, was strangely close to the inverse of the Hubble time.   The Pioneer anomaly prompted an outpouring of hundreds of papers, most attempting to explain it via modifications of gravity or via new physical interactions, with a small subset trying to explain it by conventional means.  

Soon after the publication of the initial Pioneer anomaly paper~\cite{1998PhRvL..81.2858A}, Katz pointed out that the anomaly could be accounted for as the result of the anisotropic emission of radiation from the radioactive thermal generators (RTG)
that continued to power the spacecraft decades after their launch~\cite{1999PhRvL..83.1892K}.   At the time, there was insufficient data on the performance of the RTG over time or on the thermal characteristics of the spacecraft to justify more than an order-of-magnitude estimate.  However, the recovery of an extended set of Doppler data covering a longer stretch of the orbits of both spacecraft, together with the fortuitous discovery of project documentation and of telemetry data giving on-board temperature information, made it possible both to improve the orbit analysis and to develop detailed thermal models of the spacecraft in order to quantify the effect of thermal emission anisotropies.  Several independent analyses now confirm that the anomaly is almost entirely due to the recoil of the spacecraft from the anisotropic emission of residual thermal radiation~\cite{2011AnP...523..439R,2012PhRvL.108x1101T,2013arXiv1311.4978M}.   For a thorough review of the Pioneer anomaly published just as the new analyses were underway, see the {\em Living Review} by Turyshev and Toth~\cite{2010LRR....13....4T}.

\newpage

%%%%%%%%%%%%%%%%%%%%%%%%%%%%%%%%%%%%%%%%%%%%%%%%%%%%%%%%%%%%%%%%%%%%%%%%%%%%%%%%%%%
%%%%%%%%%%%%%%%%%%%%%%%%%%%%%%%%%%%%%%%%%%%%%%%%%%%%%%%%%%%%%%%%%%%%%%%%%%%%%%%%%%%
%%%%%%%%%%%%%%%%%%%%%%%%%%%%%%%%%%%%%%%%%%%%%%%%%%%%%%%%%%%%%%%%%%%%%%%%%%%%%%%%%%%

\section{Metric Theories of Gravity and the PPN Formalism}
\label{S3_1}

%%%%%%%%%%%%%%%%%%%%%%%%%%%%%%%%%%%%%%%%%%%%%%%%%%%%%%%%%%%%%%%%%%%%%%%%%%%%%%%%%%%
%%%%%%%%%%%%%%%%%%%%%%%%%%%%%%%%%%%%%%%%%%%%%%%%%%%%%%%%%%%%%%%%%%%%%%%%%%%%%%%%%%%

\subsection{Metric theories of gravity and the strong equivalence principle}
\label{metrictheories}

%%%%%%%%%%%%%%%%%%%%%%%%%%%%%%%%%%%%%%%%%%%%%%%%%%%%%%%%%%%%%%%%%%%%%%%%%%%%%%%%%%%

\subsubsection{Universal coupling and the metric postulates}
\label{universal}

The empirical evidence supporting the Einstein
equivalence principle, discussed in Section~\ref{S2},
supports the conclusion that the only theories of
gravity that have a hope of being viable are metric
theories, or possibly theories that are metric apart from very weak
or short-range non-metric couplings (as in string theory). Therefore for
the remainder of this review, we shall turn our attention
exclusively to metric theories of gravity, which assume that
\begin{enumerate}
\item there exists a symmetric metric,
\item test bodies follow geodesics of the metric, and
\item in local Lorentz frames, the non-gravitational laws of physics
  are those of special relativity.
\end{enumerate}

The property that all non-gravitational fields should couple in
the same manner to a single gravitational field is sometimes
called ``universal coupling''. Because of it, one can discuss the
metric as a property of spacetime itself rather than as a field
over spacetime. This is because its properties may be measured
and studied using a variety of different experimental devices,
composed of different non-gravitational fields and particles,
and, because of universal coupling, the results will be
independent of the device. Thus, for instance, the proper time
between two events is a characteristic of spacetime and of the
location of the events, not of the clocks used to measure it.

Consequently, if EEP is valid, the non-gravitational laws of
physics may be formulated by taking their special relativistic
forms in terms of the Min\-kowski metric {\boldmath $\eta$} and simply
``going over'' to new forms in terms of the curved spacetime
metric {\boldmath $g$}, using the mathematics of differential geometry.
The details of this ``going over'' can be found in standard
textbooks (see~\cite{MTW, Weinberg,2009fcgr.book.....S,PW2014}, TEGP~3.2.~\cite{tegp}).

%%%%%%%%%%%%%%%%%%%%%%%%%%%%%%%%%%%%%%%%%%%%%%%%%%%%%%%%%%%%%%%%%%%%%%%%%%%%%%%%%%%

\subsubsection{The strong equivalence principle}
\label{sep}

In any metric theory of gravity, matter and non-gravitational
fields respond only to the spacetime metric {\boldmath $g$}. In
principle, however, there could exist other gravitational fields
besides the metric, such as scalar fields, vector fields, and so
on. If, by our strict definition of metric theory,
matter does not couple to these fields, what can their role
in gravitation theory be?  Their role must be that of mediating
the manner in which matter and non-gravitational fields generate
gravitational fields and produce the metric; once determined,
however, the metric alone acts back on the matter in the manner
prescribed by EEP.

What distinguishes one metric theory from another, therefore, is
the number and kind of gravitational fields it contains in
addition to the metric, and the equations that determine the
structure and evolution of these fields. From this viewpoint,
one can divide all metric theories of gravity into two fundamental
classes:  ``purely dynamical'' and ``prior-geometric''.

By ``purely dynamical metric theory''
 we mean any metric theory
whose gravitational fields have their structure and evolution
determined by coupled partial differential field equations. In
other words, the behavior of each field is influenced to some
extent by a coupling to at least one of the other fields in the
theory. By ``prior geometric''
 theory, we mean any metric theory
that contains ``absolute elements'', fields or equations whose
structure and evolution are given \emph{a priori}, and are
independent of the structure and evolution of the other fields of
the theory. These ``absolute elements'' typically include flat
background metrics {\boldmath $\eta$} or cosmic time coordinates
$t$.

General relativity is a purely dynamical theory since it contains
only one gravitational field, the metric itself, and its
structure and evolution are governed by partial differential
equations (Einstein's equations). Brans--Dicke theory and its
generalizations are purely
dynamical theories; the field equation for the metric involves the
scalar field (as well as the matter as source), and that for the
scalar field involves the metric. Visser's bimetric massive gravity theory~\cite{visser} is a
prior-geometric theory:  It has a non-dynamical, Riemann-flat
background metric
{\boldmath $\eta$}, and the field equations for the physical metric
{\boldmath $g$} involve {\boldmath $\eta$}.

By discussing metric theories of gravity from this broad point of
view, it is possible to draw some general conclusions about the
nature of gravity in different metric theories, conclusions that
are reminiscent of the Einstein equivalence principle, but that
are subsumed under the name ``strong equivalence principle''.

Consider a local, freely falling frame in any metric theory of
gravity. Let this frame be small enough that inhomogeneities in
the external gravitational fields can be neglected throughout its
volume. On the other hand, let the frame be large enough to
encompass a system of gravitating matter and its associated
gravitational fields. The system could be a star, a black hole,
the solar system, or a Cavendish experiment. Call this frame a
``quasi-local Lorentz frame''. To determine the behavior
of the system we must calculate the metric. The computation
proceeds in two stages. First we determine the external
behavior of the metric and gravitational fields, thereby
establishing boundary values for the fields generated by the
local system, at a boundary of the quasi-local frame ``far'' from
the local system. Second, we solve for the fields generated by
the local system. But because the metric is coupled directly or
indirectly to the other fields of the theory, its structure and
evolution will be influenced by those fields, and in particular
by the boundary values taken on by those fields far from the
local system. This will be true even if we work in a coordinate
system in which the asymptotic form of $g_{\mu\nu}$ in the
boundary region between the local system and the external world
is that of the Minkowski metric. Thus the gravitational
environment in which the local gravitating system resides can
influence the metric generated by the local system via the
boundary values of the auxiliary fields. Consequently, the
results of local gravitational experiments may depend on the
location and velocity of the frame relative to the external
environment. Of course, local \emph{non}-gravitational
experiments are unaffected since the gravitational fields they
generate are assumed to be negligible, and since those
experiments couple only to the metric, whose form can always be
made locally Minkowskian at a given spacetime event.
Local gravitational experiments might
include Cavendish experiments, measurement of the acceleration of
massive self-gravitating bodies,
studies of the structure of stars and planets, or
analyses of the periods of ``gravitational clocks''. We
can now make several statements about different kinds of metric
theories.
\begin{itemize}
\item A theory which contains only the metric {\boldmath $g$} yields
  local gravitational physics which is independent of the location and
  velocity of the local system. This follows from the fact that the
  only field coupling the local system to the environment is
  {\boldmath $g$}, and it is always possible to find a coordinate
  system in which {\boldmath $g$} takes the Minkowski form at the
  boundary between the local system and the external environment
  (neglecting inhomogeneities in the external gravitational
  field). Thus the asymptotic values of $g_{\mu\nu}$ are constants
  independent of location, and are asymptotically Lorentz invariant,
  thus independent of velocity. GR is an example of
  such a theory.
\item A theory which contains the metric {\boldmath $g$} and dynamical
  scalar fields $\varphi_A$ yields local gravitational physics which
  may depend on the location of the frame but which is independent of
  the velocity of the frame. This follows from the asymptotic Lorentz
  invariance of the Minkowski metric and of the scalar fields, but now
  the asymptotic values of the scalar fields may depend on the
  location of the frame. An example is Brans--Dicke theory, where the
  asymptotic scalar field determines the effective value of the
  gravitational constant, which can thus vary as $\varphi$ varies. On
  the other hand, a form of velocity dependence in local physics can
  enter indirectly if the asymptotic values of the scalar field vary
  with time cosmologically. Then the \emph{rate} of variation of the
  gravitational constant could depend on the velocity of the frame.
\item A theory which contains the metric {\boldmath $g$} and
  additional dynamical vector or tensor fields or prior-geometric
  fields yields local gravitational physics which may have both
  location and velocity-dependent effects.
\end{itemize}
These ideas can be summarized in the strong equivalence principle
(SEP), which states that:
\begin{enumerate}
\item WEP is valid for self-gravitating bodies as well as for test
  bodies.
\item The outcome of any local test experiment is independent of the
  velocity of the (freely falling) apparatus.
\item The outcome of any local test experiment is independent of where
  and when in the universe it is performed.
\end{enumerate}
The distinction between SEP and EEP is the inclusion of bodies
with self-gravitational interactions (planets, stars) and of
experiments involving gravitational forces (Cavendish
experiments, gravimeter measurements). Note that SEP contains
EEP as the special case in which local gravitational forces are
ignored.  For further discussion of SEP and EEP, see~\cite{2013arXiv1310.7426D}.

The above discussion of the coupling of auxiliary fields to local
gravitating systems indicates that if SEP is strictly valid, there must be
one and only one gravitational field in the universe, the metric
{\boldmath $g$}. These arguments are only suggestive however, and no
rigorous proof of this statement is available at present.
Empirically it has been found that almost every metric theory other than
GR introduces auxiliary gravitational fields,
either dynamical or prior geometric, and thus predicts violations
of SEP at some level (here we ignore quantum-theory inspired
modifications to GR involving ``$R^2$'' terms). The one exception is
Nordstr\"om's 1913 conformally-flat scalar theory~\cite{nordstrom13}, which
can be written purely in terms of the metric; the theory satisfies SEP, but
unfortunately violates experiment by predicting no deflection of light.
General relativity seems to be the only viable
metric theory that embodies SEP completely. In
Section~\ref{septests}, we shall discuss experimental evidence for the
validity of SEP.

%%%%%%%%%%%%%%%%%%%%%%%%%%%%%%%%%%%%%%%%%%%%%%%%%%%%%%%%%%%%%%%%%%%%%%%%%%%%%%%%%%%
%%%%%%%%%%%%%%%%%%%%%%%%%%%%%%%%%%%%%%%%%%%%%%%%%%%%%%%%%%%%%%%%%%%%%%%%%%%%%%%%%%%

\subsection{The parametrized post-Newtonian formalism}
\label{ppn}

Despite the possible existence of long-range gravitational fields
in addition to the metric in various metric theories of gravity,
the postulates of those theories demand that matter and
non-gravitational fields be completely oblivious to them. The
only gravitational field that enters the equations of motion is
the metric {\boldmath $g$}. The role of the other fields that a theory may
contain can only be that of helping to generate the spacetime
curvature associated with the metric. Matter may create these
fields, and they plus the matter may generate the metric, but they
cannot act back directly on the matter. Matter responds only to
the metric.

Thus the metric and the equations of motion for matter become the
primary entities for calculating observable effects,
and all that distinguishes one
metric theory from another is the particular way in which matter
and possibly other gravitational fields generate the metric.

The comparison of metric theories of gravity with each other and
with experiment becomes particularly simple when one takes the
slow-motion, weak-field limit. This approximation, known as the
post-Newtonian limit, is sufficiently accurate to encompass most
solar-system tests that can be performed in the foreseeable
future. It turns out that, in this limit, the spacetime metric
{\boldmath $g$} predicted by nearly every metric theory of gravity has the
same structure. It can be written as an expansion about the
Minkowski metric ($ \eta_{\mu\nu} = \diag (-1,1,1,1)$) in terms
of dimensionless gravitational potentials of varying degrees of
smallness.
These potentials are constructed from the matter
variables (see Box~\ref{box2}) in imitation of the Newtonian gravitational
potential
\begin{equation}
  U ({\bf x}, t) \equiv \int \rho ({\bf x}', t)
  |{\bf x} - {\bf x}'|^{-1} \, d^3 x'.
  \label{E20}
\end{equation}
The ``order of smallness'' is determined according to the rules
$U \sim~v^2 \sim \Pi \sim p/ \rho \sim \epsilon$,
$v^i \sim | d/dt | / | d/dx | \sim \epsilon^{1/2}$, and so on (we use units
in which $G=c=1$; see Box~\ref{box2} for definitions and conventions).

A consistent post-Newtonian limit requires determination of $g_{00}$
correct through ${\cal O} (\epsilon^2)$,
$g_{0i}$ through ${\cal O} (\epsilon^{3/2})$, and $g_{ij}$
through ${\cal O} (\epsilon)$ (for details see TEGP~4.1~\cite{tegp}). The only way that
one metric theory differs from another is in the numerical values
of the coefficients that appear in front of the metric
potentials. The parametrized post-Newtonian (PPN) formalism
inserts parameters in place of these coefficients, parameters
whose values depend on the theory under study. In the current
version of the PPN  formalism, summarized in Box~\ref{box2}, ten
parameters are used, chosen in such a manner that they measure or
indicate general properties of metric theories of gravity
(see Table~\ref{ppnmeaning}). Under reasonable assumptions about the
kinds of potentials that can be present at post-Newtonian order
(basically only Poisson-like potentials), one finds that ten PPN
parameters exhaust the possibilities.

\begin{table}[hptb]
  \caption{The PPN Parameters and their significance (note that
    $\alpha_3$ has been shown twice to indicate that it is a measure
    of two effects).}
  \label{ppnmeaning}
  \renewcommand{\arraystretch}{1.2}
  \centering
  \begin{tabular}{p{1.5 cm}|p{4.0 cm}ccc}
    \hline \hline
    Parameter &
    \centering What it measures relative to GR &
    \multicolumn{1}{p{1.2 cm}}{\centering Value in GR} &
    \multicolumn{1}{p{2.2 cm}}{\centering Value in semiconservative
      theories} &
    \multicolumn{1}{p{2.2 cm}}{\centering Value in fully conservative
      theories} \\
    \hline \hline
    $\gamma$ &
    How much space-curva\-ture produced by unit rest mass? &
    $1$ & $\gamma$ & $\gamma$ \\
    \hline
    $\beta$ &
    How much ``nonlinearity'' in the superposition law for gravity? &
    $1$ & $\beta$ & $\beta$ \\
    \hline
    $\xi$ &
    Preferred-location effects? &
    $0$ & $\xi$ & $\xi$ \\
    \hline
    $\alpha_1$ &
    Preferred-frame effects? &
    $0$ & $\alpha_1$ & $0$ \\
    $\alpha_2$ & &
    $0$ & $\alpha_2$ & $0$ \\
    $\alpha_3$ & &
    $0$ & $0$ & $0$ \\
    \hline
    $\alpha_3$ &
    Violation of conservation &
    $0$ & $0$ & $0$ \\
    $\zeta_1$ &
    of total momentum? &
    $0$ & $0$ & $0$ \\
    $\zeta_2$ & &
    $0$ & $0$ & $0$ \\
    $\zeta_3$ & &
    $0$ & $0$ & $0$ \\
    $\zeta_4$ & &
    $0$ & $0$ & $0$ \\
    \hline \hline
  \end{tabular}
  \renewcommand{\arraystretch}{1.0}
\end{table}

The parameters $\gamma$ and $\beta$ are the usual
Eddington--Robertson--Schiff parameters used to describe the
``classical'' tests of GR, and are in some sense the most important; they
are the only non-zero parameters in GR and scalar-tensor gravity.
The parameter $\xi$ is non-zero in
any theory of gravity that predicts preferred-location effects
such as a galaxy-induced anisotropy in the local gravitational
constant $G_\mathrm{L}$ (also called ``Whitehead'' effects);
$\alpha_1$, $\alpha_2$, $\alpha_3$ measure whether or
not the theory predicts post-Newtonian preferred-frame
effects; $\alpha_3$, $\zeta_1$, $\zeta_2$,
$\zeta_3$, $\zeta_4$ measure whether or not the theory
predicts violations of global conservation laws for total
momentum. 
In Table~\ref{ppnmeaning} we show the values these
parameters take%
\begin{enumerate}
\item in GR,
\item in any theory of gravity that possesses conservation laws for
  total momentum, called ``semi-conservative'' (any theory that is
  based on an invariant action principle is semi-conservative), and
\item in any theory that in addition possesses six global conservation
  laws for angular momentum, called ``fully conservative'' (such
  theories automatically predict no post-Newtonian preferred-frame
  effects).
\end{enumerate}
Semi-conservative theories have five free PPN  parameters ($\gamma$,
$\beta$, $\xi$, $\alpha_1$, $\alpha_2$) while fully conservative
theories have three ($\gamma$, $\beta$, $\xi$).

The PPN  formalism was pioneered by
Kenneth Nordtvedt~\cite{nordtvedt2}, who studied the post-Newtonian
metric of a system of gravitating point masses, extending earlier
work by Eddington, Robertson and Schiff (TEGP~4.2~\cite{tegp}). 
Will~\cite{Will71a} generalized the framework to perfect fluids. A
general and unified version of the PPN  formalism was developed by
Will and Nordtvedt~\cite{willnordtvedt72}. The canonical version, with
conventions altered to be more in accord with standard textbooks
such as~\cite{MTW}, is discussed in detail in TEGP~4~\cite{tegp}. 
Other versions
of the PPN  formalism have been developed to deal with point
masses with charge, fluid with anisotropic stresses,
bodies with strong internal gravity, and
post-post-Newtonian effects (TEGP~4.2, 14.2~\cite{tegp}).

\begin{mybox}{The Parametrized Post-Newtonian formalism}
  \begin{description}
  \item[Coordinate system:]~\\ [0.5 em]
    The framework uses a nearly globally Lorentz coordinate system in
    which the coordinates are $(t, x^1, x^2, x^3)$. Three-dimensional,
    Euclidean vector notation is used throughout. All coordinate
    arbitrariness (``gauge freedom'') has been removed by
    specialization of the coordinates to the standard PPN  gauge
    (TEGP~4.2~\cite{tegp}). Units are chosen so that $G = c = 1$,
    where $G$ is the physically measured Newtonian constant far from
    the solar system.
  \end{description}

  \begin{description}
  \item[Matter variables:]~\\ [-1.5 em]
    \begin{itemize}
    \item $\rho$: density of rest mass as measured in a local freely
      falling frame momentarily comoving with the gravitating matter.
    \item $v^i=(dx^i/dt)$: coordinate velocity of the matter.
    \item $w^i$: coordinate velocity of the PPN coordinate system
      relative to the mean rest-frame of the universe.
    \item $p$: pressure as measured in a local freely falling frame
      momentarily comoving with the matter.
    \item $\Pi$: internal energy per unit rest mass (it includes all
      forms of non-rest-mass, non-gravitational energy, e.g., energy
      of compression and thermal energy).
    \end{itemize}
  \end{description}

  \begin{description}
  \item[PPN parameters:]~\\ [0.5 em]
    $\gamma$, $\beta$, $\xi$, $\alpha_1$, $\alpha_2$, $\alpha_3$,
    $\zeta_1$, $\zeta_2$, $\zeta_3$, $\zeta_4$.
  \end{description}

  \begin{description}
  \item[Metric:]~\\ [-1 em]
    \begin{eqnarray*}
      g_{00} & = & - 1 + 2 U - 2 \beta U^2 - 2 \xi \Phi_W +
      (2 \gamma + 2 + \alpha_3 + \zeta_1 - 2 \xi ) \Phi_1 +
      2 (3 \gamma - 2 \beta + 1 + \zeta_2 + \xi) \Phi_2 \\
      & & + 2 (1 + \zeta_3) \Phi_3 + 2 (3 \gamma + 3 \zeta_4 - 2 \xi)
      \Phi_4 - (\zeta_1 - 2 \xi ) {\cal A} -
      (\alpha_1 - \alpha_2 - \alpha_3) w^2 U -
      \alpha_2 w^i w^j U_{ij} \\
      & & + (2 \alpha_3 - \alpha_1 ) w^i V_i + {\cal O} (\epsilon^3),
      \\ [0.5 em]
      g_{0i} & = & - \frac{1}{2} (4 \gamma + 3 + \alpha_1 -
      \alpha_2 + \zeta_1 - 2 \xi ) V_i - \frac{1}{2}
      (1 + \alpha_2 - \zeta_1 + 2 \xi) W_i -
      \frac{1}{2} ( \alpha_1 - 2 \alpha_2 ) w^i U \\
      & & - \alpha_2 w^j U_{ij} + {\cal O} (\epsilon^{5/2}),
      \\ [0.5 em]
      g_{ij} & = & (1 + 2 \gamma U) \delta_{ij} + {\cal O} (\epsilon^2).
    \end{eqnarray*}%
  \end{description}

  \begin{description}
  \item[Metric potentials:]~\\ [-1 em]
      \begin{eqnarray*}
        U & = &
        \int \frac{\rho'}{|{\bf x}-{\bf x}'|} \, d^3x',
        \\ [0.5 em]
        U_{ij} & =
        & \int \frac{\rho' (x-x')_i (x-x')_j}
        {|{\bf x}-{\bf x}'|^3} \, d^3x',
        \\ [0.5 em]
        \Phi_W & = &
        \int \frac{\rho' \rho'' ({\bf x}-{\bf x}')}
        {|{\bf x}-{\bf x}'|^3} \cdot
        \left( \frac{{\bf x}'-{\bf x}''}{|{\bf x}-{\bf x}''|} -
        \frac{{\bf x}-{\bf x}''}{| {\bf x}'-{\bf x}''|} \right)
        \, d^3x' \, d^3x'',
        \\ [0.5 em]
        {\cal A} & = &
        \int \frac{\rho' [{\bf v}' \cdot ({\bf x}-{\bf x}')]^2}
        {|{\bf x}-{\bf x}'|^3} \, d^3x',
        \\ [0.5 em]
        \Phi_1 & = &
        \int \frac{\rho' v'^2}{|{\bf x}-{\bf x}'|} \, d^3x',
        \\ [0.5 em]
        \Phi_2 & = &
        \int \frac{\rho' U'}{|{\bf x} - {\bf x}'|} \, d^3x',
        \\ [0.5 em]
        \Phi_3 & = &
        \int \frac{\rho' \Pi'}{|{\bf x}-{\bf x}'|} \, d^3x',
        \\ [0.5 em]
        \Phi_4 & = &
        \int \frac{p'}{|{\bf x}-{\bf x}'|} \, d^3x',
        \\ [0.5 em]
        V_i & = &
        \int \frac{\rho' v_i'}{|{\bf x}-{\bf x}'|} \, d^3x',
        \\ [0.5 em]
        W_i & = &
        \int \frac{\rho' [{\bf v}' \cdot ({\bf x}-{\bf x}')](x-x')_i}
        {|{\bf x}-{\bf x}'|^3} \, d^3x'.
      \end{eqnarray*}%
  \end{description}

  \begin{description}
  \item[Stress--energy tensor (perfect fluid):]~\\ [-1.5 em]
    \begin{eqnarray*}
      T^{00} &=& \rho (1+ \Pi + v^2 +2U), \\
      T^{0i} &=& \rho v^i \left( 1 + \Pi + v^2 +2U + \frac{p}{\rho} \right), \\
      T^{ij} &=& \rho v^iv^j \left( 1 + \Pi + v^2 +2U + \frac{p}{\rho} \right) +
      p \delta^{ij} (1 - 2\gamma U).
    \end{eqnarray*}%
  \end{description}

  \begin{description}
  \item[Equations of motion:]~\\ [-1.5 em]
    \begin{itemize}
    \item Stressed matter:
      ${T^{\mu\nu}}_{;\nu}= 0$.
    \item Test bodies:
      $\displaystyle \frac{d^2 x^\mu}{d \lambda^2} + {\Gamma^\mu}_{\nu \lambda}
      \frac{dx^\nu}{d \lambda}\frac{dx^\lambda}{d \lambda} = 0$.
    \item Maxwell's equations:
      ${F^{\mu \nu}}_{; \nu} = 4 \pi J^\mu,
      \qquad
      F_{\mu \nu} = A_{\nu ; \mu} - A_{\mu; \nu }$.
    \end{itemize}
  \end{description}
  \label{box2}
\end{mybox}

\newpage

%%%%%%%%%%%%%%%%%%%%%%%%%%%%%%%%%%%%%%%%%%%%%%%%%%%%%%%%%%%%%%%%%%%%%%%%%%%%%%%%%%%
%%%%%%%%%%%%%%%%%%%%%%%%%%%%%%%%%%%%%%%%%%%%%%%%%%%%%%%%%%%%%%%%%%%%%%%%%%%%%%%%%%%

\subsection{Competing theories of gravity}
\label{theories}

One of the important applications of the PPN formalism is the
comparison and classification of alternative metric theories of
gravity. The population of viable theories has fluctuated over
the years as new effects and tests have been discovered, largely
through the use of the PPN  framework, which eliminated many
theories thought previously to be viable. The theory population
has also fluctuated as new, potentially viable theories have been
invented.

In this review, we shall focus on GR, the
general class of scalar-tensor modifications of it, of which the
Jordan--Fierz--Brans--Dicke theory (Brans--Dicke, for short)
is the classic example, and vector-tensor theories. 
The reasons are several-fold:
\begin{itemize}
\item A full compendium of alternative theories circa 1981 is given in
  TEGP~5~\cite{tegp}.
\item Many alternative metric theories developed during the 1970s and
  1980s could be viewed as ``straw-man'' theories, invented to prove
  that such theories exist or to illustrate particular properties. Few
  of these could be regarded as well-motivated theories from the point
  of view, say, of field theory or particle physics.
\item A number of theories fall into the class of ``prior-geometric''
  theories, with absolute elements such as a flat background metric in
  addition to the physical metric. Most of these theories predict
  ``preferred-frame'' effects, that have been tightly constrained by
  observations (see Section~\ref{preferred}). An example is Rosen's
  bimetric theory.
\item A large number of alternative theories of gravity predict
  gravitational wave emission substantially different from that of
  GR, in strong disagreement with observations of the
  binary pulsar (see Section~\ref{S5}).
\item Scalar-tensor modifications of GR have become very popular in
  unification schemes such as string theory, and in cosmological model
  building. Because the scalar fields could be massive, the potentials
  in the post-Newtonian limit could be modified by Yukawa-like terms.
\item Theories that also incorporate vector fields have attracted recent attention, in the
  spirit of the SME (see Section~\ref{SME}), as models for violations
  of Lorentz invariance in the gravitational sector, and as potential candidates to account for phenomena such as galaxy rotation curves without resorting to dark matter.
\end{itemize}

%%%%%%%%%%%%%%%%%%%%%%%%%%%%%%%%%%%%%%%%%%%%%%%%%%%%%%%%%%%%%%%%%%%%%%%%%%%%%%%%%%%

\subsubsection{General relativity}
\label{generalrelativity}

The metric {\boldmath $g$} is the sole
dynamical field, and the theory contains no arbitrary functions or
parameters, apart from the value of the Newtonian coupling constant
$G$, which is measurable in laboratory experiments. Throughout this
article, we ignore the cosmological constant $\Lambda_\mathrm{C}$. 
We do this despite recent evidence, from supernova data, of an accelerating
universe, which would indicate either a non-zero
cosmological constant or a dynamical
``dark energy'' contributing about 70~percent of the critical density.
Although
$\Lambda_\mathrm{C}$ has significance for quantum field theory, quantum
gravity, and cosmology, on the scale of the solar-system or of stellar
systems its effects are negligible, for the values of $\Lambda_\mathrm{C}$
inferred from supernova observations. 

The field equations of GR are derivable from an invariant action
principle $\delta I=0$, where
\begin{equation}
  I = (16 \pi G)^{-1} \int R (-g)^{1/2} \, d^4 x +
  I_\mathrm{m} (\psi_\mathrm{m}, g_{\mu\nu}),
  \label{E21}
\end{equation}
where $R$ is the Ricci scalar, and $I_\mathrm{m}$ is the matter action, which
depends on matter fields $\psi_\mathrm{m}$ universally coupled to the metric
{\boldmath $g$}. By varying the action with respect to $g_{\mu\nu}$, we
obtain the field equations
\begin{equation}
  G_{\mu\nu} \equiv R_{\mu\nu} - \frac{1}{2} g_{\mu\nu} R = 8 \pi G T_{\mu\nu},
  \label{E22}
\end{equation}
where $T_{\mu\nu}$ is the matter energy-momentum tensor. General
covariance of the matter action implies the equations of motion
${T^{\mu\nu}}_{;\nu}=0$; varying $I_\mathrm{m}$ with respect to
$\psi_\mathrm{m}$
yields the matter field equations of the Standard Model. 
By virtue of the \emph{absence} of
prior-geometric elements, the equations of motion are also a
consequence of the field equations via the Bianchi identities
${G^{\mu\nu}}_{;\nu}=0$.  According to our choice of units, we set $G=1$.

The general procedure for deriving the post-Newtonian limit of metric
theories is spelled out in TEGP~5.1~\cite{tegp}, and is described in
detail for GR in TEGP~5.2~\cite{tegp}. (see also Chapters 6 -- 8 of~\cite{PW2014}).  The PPN parameter values are
listed in Table~\ref{ppnvalues}.

\begin{table}[t]
  \caption[Metric theories and their PPN parameter values.]{Metric
    theories and their PPN parameter values ($\alpha_3 = \zeta_i=0$
    for all cases). The parameters $\gamma^\prime$, $\beta^\prime$,
    $\alpha_1^\prime$, and $\alpha_2^\prime$ denote complicated
    functions of the arbitrary constants and matching parameters.}
  \label{ppnvalues}
  \centering
  \begin{tabular}{lccccccc}
    \hline \hline
    Theory &Arbitrary&Cosmic&\multicolumn{5}{p{5 cm}}{\cl{PPN parameters}}\\[-1 em]
    &functions&matching&\multicolumn{5}{p{5 cm}}{\cl{\hrulefill}}\\[-1 em]
    &or constants&parameters&$\gamma$ & $\beta$ & $\xi$ & 
           $\alpha_1$ & $\alpha_2$ \\
    %\multicolumn{1}{p{3.0 cm}}{\cl{Arbitrary} \cl{functions} \cl{or constants}} &
    %\multicolumn{1}{p{2.8 cm}}{\cl{Cosmic} \cl{matching} \cl{parameters}} &
    %\multicolumn{5}{p{6.8 cm}}{\cl{PPN parameters} \cl{\hrulefill}} \\ [-1 em]
    %& & & $\gamma$ & $\beta$ & $\xi$ & $\alpha_1$ & $\alpha_2$ \\
    \hline \hline \\
    General relativity
    & none &
    none &
    $1$ &
    $1$ &
    $0$ &
    $0$ &
    $0$ \\
    \hline \\
    Scalar-tensor \\
    \quad Brans--Dicke &
    $\omega_\mathrm{BD}$ &
    $\phi_0$ &
    $ \displaystyle \frac{1+\omega_\mathrm{BD}}{2+\omega_\mathrm{BD}}$ &
    $1$ &
    $0$ &
    $0$ &
    $0$ \\
    \quad General, $f(R)$ &
    $A(\varphi)$,
    $V(\varphi)$ &
    $\varphi_0$ &
    $ \rule{0 cm}{0.6cm} \displaystyle \frac{1+\omega}{2+\omega}$ &
    $\displaystyle 1+\frac{\lambda}{4+2\omega}$ &
    $0$ &
    $0$ &
    $0$ \\ [0.6 em]
    \hline \\
    Vector-tensor \\
    \quad Unconstrained &
    $\omega, c_1, c_2, c_3, c_4$ &
    $u$ &
    $\gamma^\prime$ &
    $\beta^\prime$ &
    $0$ &
    $\alpha_1^\prime$ &
    $\alpha_2^\prime$ \\
    \quad Einstein-{\AE}ther &
    $c_1, c_2, c_3, c_4$ &
    none &
    $1$ &
    $1$ &
    $0$ &
    $\alpha_1^\prime$ &
    $\alpha_2^\prime$ \\ [0.6 em]
    \hline \\
    Tensor-Vector-Scalar
    &$k,c_1, c_2, c_3, c_4$&
    $\phi_0$&
    $1$ &
    $1$ &
    $0$ &
    $\alpha_1^\prime$ &
    $\alpha_2^\prime$ \\
       \hline \hline
  \end{tabular}
  \renewcommand{\arraystretch}{1.2}
\end{table}

%%%%%%%%%%%%%%%%%%%%%%%%%%%%%%%%%%%%%%%%%%%%%%%%%%%%%%%%%%%%%%%%%%%%%%%%%%%%%%%%%%%

\subsubsection{Scalar-tensor theories}
\label{scalartensor}

These theories contain the metric {\boldmath $g$}, a
scalar field $\varphi$, a potential  function $V(\varphi)$, and a
coupling function $A(\varphi)$ (generalizations to more than one scalar
field have also been carried out~\cite{DamourEspo92}).
For some purposes, the action is conveniently written in a non-metric
representation, sometimes denoted the ``Einstein frame'', in which the
gravitational action looks exactly like that of GR and the scalar action looks like a minimally coupled scalar field with a potential:
\begin{equation}
  \tilde I = (16 \pi G)^{-1} \int \left[ \tilde{R} -
  2 \tilde{g}^{\mu\nu} \partial_\mu \varphi \, \partial_\nu \varphi -
  V (\varphi) \right] (- \tilde{g})^{1/2} \, d^4 x + I_\mathrm{m}
  \left( \psi_\mathrm{m}, A^2 (\varphi) \tilde{g}_{\mu\nu} \right),
  \label{E23}
\end{equation}
where $\tilde R \equiv \tilde g^{\mu\nu} \tilde R_{\mu\nu}$ is the
Ricci scalar of the
``Einstein'' metric $\tilde g_{\mu\nu}$. (Apart from the scalar potential term
$V(\varphi)$, this corresponds to Eq.~ (\ref{stringlagrangian1})
with $\tilde G(\varphi) \equiv (4\pi G)^{-1}$, $U(\varphi) \equiv 1$,
and $\tilde M(\varphi) \propto A(\varphi)$.)  This representation is a
``non-metric'' one because the matter fields $\psi_\mathrm{m}$ couple to a
combination of $\varphi$ and $\tilde g_{\mu\nu}$.
Despite appearances, however,
it is a metric theory, because it can be put
into a metric representation by identifying the ``physical metric''
\begin{equation}
  g_{\mu\nu} \equiv A^2 (\varphi) \tilde g_{\mu\nu}.
  \label{E24}
\end{equation}
The action can then be rewritten in the metric form
\begin{equation}
  I = (16 \pi G)^{-1} \int \left[ \phi R - \phi^{-1} \omega(\phi)
  g^{\mu\nu} \partial_\mu \phi \partial_\nu \phi - \phi^2 V \right]
  (-g)^{1/2} \, d^4 x + I_\mathrm{m} (\psi_\mathrm{m}, g_{\mu\nu}),
  \label{E25}
\end{equation}
where
\begin{equation}
  \begin{array}{rcl}
    \phi & \equiv & A (\varphi)^{-2},
    \\ [0.5 em]
    3 + 2 \omega (\phi) & \equiv & \alpha (\varphi)^{-2},
    \\ [0.5 em]
    \alpha (\varphi) & \equiv & \displaystyle
    \frac{d(\ln A(\varphi))}{d\varphi}.
  \end{array}
  \label{E26}
\end{equation}
The Einstein frame is useful for discussing general characteristics of
such theories, and for some cosmological applications, while the metric
representation is most useful for calculating observable effects.
The field equations, post-Newtonian limit and PPN  parameters are
discussed in TEGP~5.3~\cite{tegp}, and the values of the PPN  parameters are
listed in Table~\ref{ppnvalues}.

The parameters that enter the post-Newtonian limit are
\begin{equation}
  \omega \equiv \omega(\phi_0),
  \qquad
  \lambda \equiv \left[ \frac{\phi \, d\omega/d\phi}
  {(3 + 2 \omega) (4 + 2 \omega)} \right]_{\phi_0}\!\!\!,
  \label{E27}
\end{equation}
where $\phi_0$ is the value of $\phi$ today far from the
system being studied, as determined by appropriate cosmological boundary
conditions.   The Newtonian gravitational constant $G_N$, which is set equal to unity by our choice of units, is related to the coupling constant $G$, $\phi_0$ and $\omega$ by
\begin{equation}
G_N \equiv 1 = \frac{G}{\phi_0} \left (\frac{4 + 2\omega}{3 + 2\omega} \right )_0 \,.
\end{equation}
In Brans--Dicke theory ($\omega(\phi) \equiv \omega_\mathrm{BD}= \mathrm{const.}$),
the larger the value of
$\omega_\mathrm{BD}$, the smaller the effects of the scalar field, and in the
limit $\omega_\mathrm{BD} \to \infty$ ($\alpha_0 \to 0$),
the theory becomes indistinguishable from
GR in all its predictions. In more general
theories, the function $\omega ( \phi )$ could have the
property that, at the present epoch, and in weak-field situations,
the value of the scalar field $\phi_0$ is such that
$\omega$ is very large and $\lambda$ is very small (theory almost
identical to GR today), but that for past or
future values of $\phi$, or in strong-field regions such as the
interiors of neutron stars, $\omega$ and $\lambda$ could take on values
that would lead to significant differences from GR. It is useful to point
out that all versions of scalar-tensor gravity predict that $\gamma \le 1$
(see Table~\ref{ppnvalues}).

Damour and Esposito-Far\`ese~\cite{DamourEspo92} have adopted an alternative
parametrization of scalar-tensor theories, in which  one
expands $\ln A(\varphi)$
about a cosmological background field value $\varphi_0$:
\begin{equation}
  \ln A (\varphi) = \alpha_0 (\varphi -\varphi_0) +
  \frac{1}{2} \beta_0 (\varphi - \varphi_0)^2 + \dots
  \label{alphaexpand}
\end{equation}
A precisely linear coupling function produces Brans--Dicke theory, with
$\alpha_0^2 = 1/(2 \omega_\mathrm{BD} +3)$, or  
$1/(2+\omega_\mathrm{BD})=2\alpha_0^2/(1+\alpha_0^2)$.
The function $\ln A(\varphi)$ acts
as a potential for the scalar field $\varphi$ within matter, 
and, if $\beta_0
>0$, then during cosmological evolution,
the scalar field naturally evolves toward the minimum of the
potential, i.e.\ toward $\alpha \approx 0$, 
$\omega \to \infty$, or toward
a theory close to, though not precisely GR~\cite{DamourNord93a, DamourNord93b}.
Estimates of the
expected relic deviations from GR today in such theories depend on the
cosmological model, but range from $10^{-5}$ to a few times $10^{-7}$
for $|\gamma-1|$.

Negative
values of $\beta_0$ correspond to a ``locally unstable'' 
scalar potential (the overall theory is still stable in the sense of having
no tachyons or ghosts).
In this case, objects such as neutron stars can experience a
``spontaneous scalarization'', whereby the interior values of $\varphi$
can take on values very different from the exterior values, through
non-linear interactions between strong gravity and the scalar field,
dramatically affecting the stars' internal structure and leading to
strong violations of SEP. On the other hand, in the case 
$\beta_0 <0$, one must confront that fact that, 
with an unstable $\varphi$ potential,
cosmological evolution would presumably drive the system away from the
peak where $\alpha \approx 0$,
toward parameter values that could be excluded
by solar system experiments. 

Scalar fields coupled to gravity or matter are also
ubiquitous in particle-physics-inspired models of unification, such as
string theory~\cite{TaylorVeneziano, maeda88, DamourPolyakov,
damourpiazza02a, damourpiazza02b}.
In some models, the coupling to matter may lead to
violations of EEP, which could be  tested or bounded
by the experiments described in Section~\ref{eep}. In
many models the scalar field could be massive; if the Compton wavelength is
of macroscopic scale, its effects are those of a ``fifth force''.
Only if the theory can be cast as a metric theory with a
scalar field of infinite range or of range long compared to the scale
of the system in question (solar system) can the PPN  framework be
strictly
applied. If the mass of the scalar field is sufficiently large that its
range is microscopic, then, on solar-system scales, the scalar field is
suppressed, and the theory is essentially equivalent to general
relativity. 

For a detailed review of scalar-tensor theories see~\cite{2007sttg.book.....F}.

\subsubsection{$f(R)$ theories}
\label{f(R)}

These are theories whose action has the form
\begin{equation}
  I = \frac{c^3}{16 \pi G} \int f(R) (-g)^{1/2} \, d^4 x +
  I_\mathrm{m} (\psi_\mathrm{m}, g_{\mu\nu}),
  \label{fRaction}
\end{equation}
where $f$ is a function chosen so that at cosmological scales, the universe will experience accelerated expansion without  resorting to either a cosmological constant or dark energy.  However, it turns out that such theories are equivalent to scalar-tensor theories:  replace $f(R)$ by $f(\chi) - f_{,\chi} (\chi) (R-\chi)$, where $\chi$ is a dynamical field.  Varying the action with respect to $\chi$ yields $f_{,\chi\chi}(R-\chi) = 0$, which implies that $\chi = R$ as long as $f_{,\chi\chi} \ne 0$.  Then defining a scalar field $\phi \equiv f_{,\chi} (\chi) $ one puts the action into the form of a scalar-tensor theory given by Eq.\ (\ref{E25}), with $\omega(\phi) =0$ and $\phi^2 V = \phi \chi(\phi) - f(\chi(\phi))$.  As we will see, this value of $\omega$ would ordinarily strongly violate solar-system experiments, but it turns out that in many models, the potential $V(\phi)$ has the effect of giving the scalar field a large effective mass in the presence of matter (the so-called ``chameleon mechanism''~\cite{2004PhRvL..93q1104K}), so that the scalar field is suppressed at distances that extend outside bodies like the Sun and Earth.  In this way, with only modest fine tuning, $f(R)$ theories can claim to obey standard tests, while providing interesting, non general-relativistic behavior on cosmic scales.  For detailed reviews of this class of theories, see \cite{2010RvMP...82..451S} and 
\cite{lrr-2010-3}.

%%%%%%%%%%%%%%%%%%%%%%%%%%%%%%%%%%%%%%%%%%%%%%%%%%%%%%%%%%%%%%%%%%%%%%%%%%%%%%%%%%%

\subsubsection{Vector-tensor theories}
\label{vectortensor}

These theories contain the metric {\boldmath $g$} and a dynamical, 
typically timelike, four-vector
field $u^\mu$. In some models, the four-vector is unconstrained, while
in others, called Einstein-{\AE}ther theories it is constrained to be timelike
with unit norm. The most general action for such theories
that is quadratic in derivatives of the vector is given by
\begin{equation}
  I = (16 \pi G)^{-1} \int \left[ (1 + \omega u_\mu u^\mu) R -
  K^{\mu\nu}_{\alpha\beta} \nabla_\mu u^\alpha \nabla_\nu u^\beta +
  \lambda (u_\mu u^\mu + 1) \right] (-g)^{1/2} \, d^4x +
  I_\mathrm{m} (\psi_\mathrm{m}, g_{\mu\nu}),
  \label{aetheraction}
\end{equation}
where
\begin{equation}
  K^{\mu\nu}_{\alpha\beta} = c_1 g^{\mu\nu} g_{\alpha\beta} +
  c_2 \delta^\mu_\alpha \delta^\nu_\beta +
  c_3 \delta^\mu_\beta\delta^\nu_\alpha -
  c_4 u^\mu u^\nu g_{\alpha\beta}.
  \label{ktensor}
\end{equation}
The coefficients $c_i$ are arbitrary. In the unconstrained theories,
$\lambda \equiv 0$ and  $\omega$ is arbitrary. In the constrained theories,
$\lambda$ is a Lagrange multiplier, and by virtue of the constraint
$u_\mu u^\mu = -1$, the factor $\omega u_\mu u^\mu$
in front of the Ricci scalar can be
absorbed into a rescaling of $G$; equivalently, in the constrained theories,
we can set $\omega=0$. Note that the possible term $u^\mu u^\nu
R_{\mu\nu}$ can be shown under integration by parts to be equivalent to a
linear combination of the terms involving $c_2$ and $c_3$. 

Unconstrained theories were studied during the 1970s as ``straw-man''
alternatives to GR. In addition to having up to four
arbitrary parameters, they also left the magnitude of the vector field 
arbitrary, since it satisfies a linear homogenous vacuum field equation
of the form ${\cal L} u^\mu =0$ ($c_4=0$ in all such cases studied). 
Indeed, this latter fact was one of most serious defects of these theories.
Each unconstrained theory studied corresponds to a special case of the
action~(\ref{aetheraction}), all with $\lambda \equiv 0$:

\begin{description}
\item[General vector-tensor theory; \boldmath $\omega$, $\tau$,
  $\epsilon$, $\eta$ {\rm (see TEGP~5.4~\cite{tegp})}]~\\
  The gravitational Lagrangian for this class of theories had the form
  $R + \omega u_\mu u^\mu R + \eta u^\mu u^\nu R_{\mu\nu}-\epsilon
  F_{\mu\nu}F^{\mu\nu} + \tau \nabla_\mu u_\nu \nabla^\mu u^\nu$,
  where $F_{\mu\nu} = \nabla_\mu u_\nu -\nabla_\nu u_\mu$,
  corresponding to the values $c_1 = 2\epsilon - \tau$, $c_2 = -\eta$,
  $c_1+c_2+c_3= -\tau$, $c_4=0$. In these theories $\gamma$, $\beta$,
  $\alpha_1$, and $\alpha_2$ are complicated functions of the
  parameters and of $u^2 = -u^\mu u_\mu$, while the rest vanish.
\end{description}

\begin{description}
\item[Will--Nordtvedt theory {\rm (see~\cite{willnordtvedt72})}]~\\
  This is the special case $c_1 = -1$, $c_2 = c_3 = c_4 =0$. In this
  theory, the PPN parameters are given by $\gamma = \beta = 1$,
  $\alpha_2 = u^2/(1+u^2/2)$, and zero for the rest.
\end{description}

\begin{description}
\item[Hellings--Nordtvedt theory; \boldmath $\omega$
  {\rm (see~\cite{hellings73})}]~\\
  This is the special case $c_1 =2$, $c_2 =2\omega$,
  $c_1 + c_2 +c_3 = 0 = c_4$. Here $\gamma$, $\beta$, $\alpha_1$ and
  $\alpha_2$ are complicated functions of the parameters and of $u^2$,
  while the rest vanish.

\item[Einstein-{\AE}ther theory; \boldmath $c_1, \, c_2, \,c_3, \,c_4$]~\\
The Einstein-{\AE}ther theories were motivated in part by a desire to explore
possibilities for violations of Lorentz invariance in gravity, in parallel
with similar studies in matter interactions, such as the SME. The general
class of theories was analyzed by Jacobson and
collaborators~\cite{jacobson01, mattingly02, jacobson04, eling04, foster05}, 
motivated
in part by~\cite{kosteleckysamuel}. 
Analyzing the post-Newtonian limit, they were able
to infer values of the PPN parameters $\gamma$ and $\beta$ 
as follows~\cite{foster05}:
\begin{eqnarray}
  \gamma & = & 1,
  \\
  \beta & = & 1,
  \\
  \xi & = & \alpha_3 = \zeta_1 = \zeta_2 = \zeta_3 = \zeta_4 = 0,
  \\
  \alpha_1 & = & -\frac{8 (c_3^2 + c_1 c_4)}{2 c_1 - c_1^2 + c_3^2} \,,
 \label{alpha1AE} \\
  \alpha_2 & = & -\frac{4 (c_3^2 + c_1 c_4)}{2 c_1 - c_1^2 + c_3^2} - \frac{(2c_{13}-c_{14})(c_{13}+c_{14}+3c_2)}{c_{123} (2-c_{14})}\,,
   \label{alpha2AE}
\end{eqnarray}%
where $c_{123}=c_1+c_2+c_3$,  $c_{13}=c_1+c_3$, $c_{14}=c_1+c_4$, subject to
the constraints $c_{123} \ne 0$, $c_{14} \ne 2$, $2c_1-c_1^2+c_3^2 \ne 0$.
By requiring that gravitational wave modes have real (as
opposed to imaginary) frequencies, one can impose the bounds
$c_1/c_{14} \ge 0$ and $c_{123}/c_{14} \ge 0$. 
Considerations of positivity of energy impose the constraints $c_1 > 0$,
$c_{14}>0$ and $c_{123} >0$. 

\end{description}

\subsubsection{Tensor--vector--scalar (TeVeS) theories}
\label{sec:TeVeS}

This class of theories was invented to provide a fully relativistic theory of gravity that could mimic the phenomenological behavior of so-called Modified Newtonian Dynamics (MOND), whereby in a weak-field regime, Newton's laws hold, namely $a = Gm/r^2$ where $m$  is the mass of a central object, as long as $a$ is large compared to some fundamental scale $a_0$, but in a regime where $a < a_0$, the equations of motion would take the form $a^2/a_0 = Gm/r^2$~\cite{milgrom83}.  With such a behavior, the rotational velocity of a particle far from a central mass would have the form $v \sim \sqrt{ar} \sim (Gma_0)^{1/4}$, thus reproducing the flat rotation curves observed for spiral galaxies, without invoking a distribution of dark matter.

Devising such a theory turned out to be no simple matter, and the final result, TeVeS was rather complicated~\cite{PhysRevD.70.083509}.  Furthermore, it was shown to have unexpected singular behavior that was most simply cured by incorporating features of the Einstein-{\AE}ther theory~\cite{PhysRevD.77.123502}.   The extended theory is based on an ``Einstein'' metric $\tilde{g}_{\mu\nu}$, related to the physical metric ${g}_{\mu\nu}$ by
\begin{equation}
{g}_{\mu\nu} \equiv e^{-2\phi} \tilde{g}_{\mu\nu} - 2 u_\mu u_\nu \sinh (2\phi) \,,
\end{equation}
where $u^\mu$ is a vector field, and $\phi$ is a scalar field.   The action for gravity is the standard GR action of Eq.\ (\ref{E21}), but defined using the Einstein metric $\tilde{g}_{\mu\nu}$, while the matter action is that of a standard metric theory, using ${g}_{\mu\nu}$.  These are supplemented by the vector action, given by that of Einstein-{\AE}ther theory, Eq.\ (\ref{aetheraction}), and a scalar action, given by 
\begin{eqnarray}
I_S &=& -\frac{c^3}{2k^2 \ell^2 G} \int {\cal F} (k\ell^2 h^{\mu\nu} \phi_{,\mu} \phi_{,\nu} ) 
 (-g)^{1/2} \, d^4x \,,
\end{eqnarray}
where $k$ is a constant, $\ell$ is a distance, and $h^{\mu\nu} \equiv \tilde{g}^{\mu\nu} - u^\mu u^\nu$, indices being raised and lowered using the Einstein metric. The function ${\cal F}(y)$ is chosen so that $\mu(y) \equiv d{\cal F}/dy$ is unity in the high-acceleration, or normal Newtonian and post-Newtonian regimes, and nearly zero in the MOND regime.  

The PPN parameters of the theory~\cite{PhysRevD.80.044032}
 have the values $\gamma = \beta =1$ and $\xi = \alpha_3 = \zeta_i = 0$, while the parameters $\alpha_1$ and $\alpha_2$ are given by
\begin{eqnarray}
  \alpha_1 & = & (\alpha_1)_{\AE} -16G \, \frac{ \kappa c_1 (2-c_{14}) - c_3 \sinh 4\phi_0  +2 (1-c_1) \sinh^2 2\phi_0 }{2c_1 -c_1^2+c_3^2} \,,
  \\
  \alpha_2 & = & (\alpha_2)_{\AE} -  2 G \, \left ( A_1 \kappa -2 A_2 \sinh 4\phi_0 - A_3 \sinh^2 2\phi_0 \right ) \,,
\end{eqnarray}%
where $(\alpha_1)_{\AE}$ and $(\alpha_2)_{\AE}$ are given by Eqs.~(\ref{alpha1AE}) and (\ref{alpha2AE}), where
\begin{eqnarray}
A_1 &\equiv& \frac{(2c_{13}-c_{14})^2}{c_{123} (2-c_{14})} + \frac{4c_1 (2-c_{14})}{2c_1 -c_1^2+c_3^2} -  \frac{6(1+c_{13}-c_{14})}{ 2-c_{14}} \,,
\\
A_2 &\equiv& \frac{(2c_{13}-c_{14})^2}{c_{123} (2-c_{14})^2} -\frac{4(1-c_1) }{2c_1 -c_1^2+c_3^2}
+ \frac{2(1-c_{13})}{2-c_{14}} \left ( \frac{2}{c_{123}}+ \frac{3}{2-c_{14}} \right ) \,,
\\
A_3 &\equiv& \frac{(2c_{13}-c_{14})^2}{c_{123} (2-c_{14})^2} + \frac{4c_3}{2c_1 -c_1^2+c_3^2} +\frac{2}{(2-c_{14})} \left (\frac{3(1-c_{13})}{c_{123}}- \frac{2c_{13}-c_{14}}{2-c_{14}}  \right ) \,,
\end{eqnarray}
where $\kappa \equiv k/8\pi$, 
\begin{equation}
G \equiv \frac{1}{2} \left ( \frac{2-c_{14}}{1+ \kappa (2-c_{14})} \right ) \,,
\end{equation}
and $\phi_0$ is the asymptotic value of the scalar field.  In the limit $\kappa \to 0$ and $\phi_0 \to 0$, $\alpha_1$ and $\alpha_2$ reduce to their Einstein-{\AE}ther forms.

However, these PPN parameter values are computed in the limit where the function ${\cal F}(y)$ is a linear function of its argument $y = k\ell^2 h^{\mu\nu} \phi_{,\mu} \phi_{,\nu}$.  When one takes into account the fact that the function $\mu (y) = d{\cal F}/dy$ must interpolate between unity and zero to reach the MOND regime, it has been found that the dynamics of local systems is more strongly affected by the fields of surrounding matter than was anticipated.  This ``external field effect'' (EFE)~\cite{2009MNRAS.399..474M,2011MNRAS.412.2530B,2011arXiv1105.5815B} produces a quadrupolar contribution to the local Newtonian gravitational potential that depends on the external distribution of matter and on the shape of the function $\mu(y)$, and that can be significantly larger than the galactic tidal contribution.  Although the calculations of EFE have been carried out using phenomenological MOND equations, it should be a generic phenomenon, applicable to TeVeS as well.   Analysis of the orbit of Jupiter using Cassini data has placed interesting constraints on the MOND interpolating function $\mu(y)$~\cite{2014arXiv1402.6950H}.
 
For thorough reviews of MOND and TeVeS, and their confrontation with the dark-matter paradigm, see~\cite{2009CQGra..26n3001S,2012LRR....15...10F}.

\subsubsection{Quadratic gravity and Chern-Simons theories}
\label{sec:Chern}

Quadratic gravity is a recent incarnation of an old idea of adding to the action of GR terms quadratic in the Riemann and Ricci tensors or the Ricci scalar, as ``effective field theory'' models for more fundamental string or quantum gravity theories.  The general action for such theories can be written as
\begin{eqnarray}
I &=& \int \biggl [ \kappa  R + \alpha_1 f_1(\phi) R^2 + \alpha_2 f_2(\phi) R_{\alpha\beta}R^{\alpha\beta}
+ \alpha_3 f_3(\phi) R_{\alpha\beta\gamma\delta}R^{\alpha\beta\gamma\delta}
+ \alpha_4  f_4(\phi) \,^*RR
\nonumber \\
&& \qquad  - \frac{\beta}{2} \biggl (g^{\mu\nu} \partial_\mu \phi \partial_\nu \phi  + 2 V(\phi) \biggr ) \biggr ] (-g)^{1/2} d^4x
+ I_\mathrm{m} (\psi_\mathrm{m}, g_{\mu\nu}) \,,
\label{eq:quadraticaction}
\end{eqnarray}
where $\kappa = (16\pi G)^{-1}$,  $\phi$ is a scalar field, $\alpha_i$ are dimensionless coupling constants (if the functions $f_i(\phi)$ are dimensionless), and $\beta$ is a constant whose dimension depends on that of $\phi$, and where   
$^*RR \equiv {{^*R^\alpha}_\beta}^{\gamma\delta}{R^\beta}_{\alpha\gamma\delta}$, where $^*{{R^\alpha}_\beta}^{\gamma\delta} \equiv \frac{1}{2} \epsilon^{\gamma\delta\rho\sigma} {R^\alpha}_{\beta\rho\sigma}$ is the dual Riemann tensor.   

One challenge inherent in these theories is to find an argument or a mechanism that evades making the natural choice for each of the $\alpha$ parameters to be of order unity.  Such a choice makes the effects of the additional terms essentially unobservable in most laboratory or astrophysical situations because of the enormous scale of $\kappa \propto 1/\ell_\mathrm{Planck}^2$ in the leading term.   This class of theories is too vast and diffuse to cover in this review, and no good review is available, to our knowledge.   

Chern-Simons gravity is the special case of this class of theories in which only the parity-violating term $^*RR$ is present ($\alpha_1 = \alpha_2=\alpha_3 = 0$).   It can arise in various anomaly cancellation schemes in the standard model of particle physics, or in cancelling the Green-Schwarz anomaly in string theory.  It can also arise in loop quantum gravity.  The action in this case is given by 
\begin{equation}
I = \int \left [ \kappa R + \frac{\alpha}{4} \phi \,^*RR - \frac{\beta}{2} \biggl (g^{\mu\nu} \partial_\mu \phi \partial_\nu \phi  + 2 V(\phi) \biggr ) \right ] (-g)^{1/2} d^4x
+ I_\mathrm{m} (\psi_\mathrm{m}, g_{\mu\nu}) \,,
\label{eq:chernsimonsaction}
\end{equation}
where  $\alpha$ and $\beta$ are coupling constants with dimensions $\ell^A$, and $\ell^{2A-2}$, assuming that the scalar field has dimensions $\ell^{-A}$.   

There are two different versions of Chern-Simons theory, a non-dynamical version in which $\beta =0$, so that $\phi$, given {\em a priori} as some specified function of spacetime, plays the role of a Lagrange multiplier enforcing the constraint $^*RR =0$, and a dynamical version, in which $\beta \ne 0$. 

The PPN parameters for a non-dynamical version of the theory with $\alpha = \kappa$ and $\beta =0$ are identical to those of GR; however, there is an additional, parity-even potential in the $g_{0i}$ component of the metric that does not appear in the standard PPN framework, given by
\begin{equation}
\delta g_{0i} = 2 \frac{d\phi}{dt} \left ( {\bf \nabla} \times {\bf V} \right )_i \,.
\end{equation}
Alexander and Yunes~\cite{2009PhR...480....1A} give a thorough review of Chern-Simons gravity.

\subsubsection{Massive gravity}
\label{sec:massive}  

Massive gravity theories attempt to give the putative ``graviton'' a mass.
The simplest attempt to implement this in a ghost-free manner 
suffers from the so-called van Dam--Veltman--Zakharov
(vDVZ) discontinuity~\cite{vdv70, zakharov70}. Because of the 3 additional
helicity states available to the massive spin-2 graviton, the limit of small
graviton mass does not coincide with pure GR, and the predicted
perihelion advance,
for example, violates experiment.
A model theory by Visser~\cite{visser} attempts to circumvent the vDVZ
problem by
introducing a non-dynamical flat-background metric. This theory is truly
continuous with GR in the limit of vanishing graviton mass; on the other
hand, its observational implications have been only partially explored.
Braneworld scenarios predict a tower or a continuum of massive gravitons,
and may avoid the vDVZ discontinuity, although the full details are
still a work in progress~\cite{deffayet02, creminelli05}.
Attempts to avert the vDVZ problem involve treating non-linear aspects of the theory at the fundamental level; many models incorporate a second tensor field in addition to the metric. For  recent reviews, see~\cite{2012RvMP...84..671H,2014arXiv1401.4173D}, and a focus issue in Vol. 30, Number 18 of  {\em Classical and Quantum Gravity}.

%%%%%%%%%%%%%%%%%%%%%%%%%%%%%%%%%%%%%%%%%%%%%%%%%%%%%%%%%%%%%%%%%%%%%%%%%%%%%%%%%%
%%%%%%%%%%%%%%%%%%%%%%%%%%%%%%%%%%%%%%%%%%%%%%%%%%%%%%%%%%%%%%%%%%%%%%%%%%%%%%%%%%%
\section{Tests of Post-Newtonian Gravity}
\label{Sec3_2}

\subsection[Tests of the parameter $\gamma$]{Tests of the parameter
  \boldmath $\gamma$}
\label{gamma}

With the PPN formalism in hand, we are now ready to confront
gravitation theories with the results of solar-system
experiments. In this section we focus on tests of the parameter
$\gamma$, consisting of the deflection of light and the time delay
of light.

%%%%%%%%%%%%%%%%%%%%%%%%%%%%%%%%%%%%%%%%%%%%%%%%%%%%%%%%%%%%%%%%%%%%%%%%%%%%%%%%%%%

\subsubsection{The deflection of light}
\label{deflection}

\begin{figure}[t]
\centering
 % \def\epsfsize#1#2{0.7#1}
  %\centerline{\epsfbox{livingfig4.eps}}
  \includegraphics[width=4in]{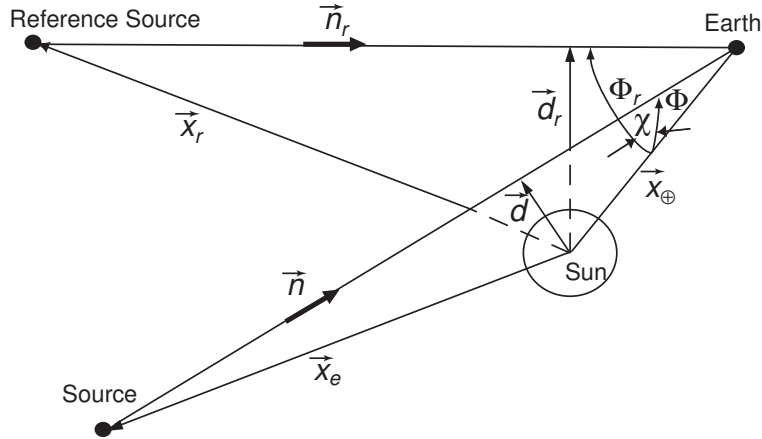}
  \caption{Geometry of light deflection measurements.}
  \label{deflectiongeom}
\end{figure}

A light ray (or photon) which passes the Sun at a distance $d$ is
deflected by an angle
\begin{equation}
  \delta \theta = \frac{1}{2} (1 + \gamma) \frac{4 \, M_\odot}{d}
  \frac{1 + \cos \Phi}{2}
  \label{E28}
\end{equation}
(TEGP~7.1~\cite{tegp}), where $M_\odot$ is the mass of the Sun
and $\Phi$ is the angle between the Earth-Sun line and the
incoming direction of the photon (see Figure~\ref{deflectiongeom}).
For a grazing ray,
$d \approx d_\odot$, $\Phi \approx 0$, and
\begin{equation}
  \delta \theta \approx \frac{1}{2} (1 + \gamma) 1.''7505,
  \label{E29}
\end{equation}
independent of the frequency of light. Another, more useful
expression gives the change in the relative angular separation
between an observed source of light and a nearby reference source
as both rays pass near the Sun:
\begin{equation}
  \delta \theta = \frac{1}{2} (1 + \gamma)
  \left[ -\frac{4 \, M_\odot}{d} \cos \chi + \frac{4 \, M_\odot}{d_\mathrm{r}}
  \left( \frac{1 + \cos \Phi_\mathrm{r}}{2} \right) \right],
  \label{E30}
\end{equation}
where $d$ and $d_\mathrm{r}$ are the distances of closest approach of
the source and reference rays respectively, $\Phi_\mathrm{r}$ is the
angular separation between the Sun and the reference source, and
$\chi$ is the angle between the Sun-source and the Sun-reference
directions, projected on the plane of the sky (see Figure~\ref{deflectiongeom}).
Thus, for example, the relative angular separation between the
two sources may vary if the line of sight of one of them passes
near the Sun ($d \sim R_\odot$, $d_\mathrm{r} \gg d$,
$\chi$ varying with time).

%%%%%%%%%%%%%%%%%%%%%%%%%

%\epubtkImage{livingfig4.png}{
%}

%%%%%%%%%%%%%%%%%

It is interesting to note that the classic derivations of the
deflection of light that use only 
the corpuscular theory of light (Cavendish 1784, von Soldner
1803~\cite{willcavendish}), or
the principle of equivalence (Einstein 1911),
yield only the ``1/2'' part of the coefficient in front of the
expression in Eq.~ (\ref{E28}). But the result of these calculations
is the deflection of light relative to local straight lines, as
established for example by rigid rods; however, because of space
curvature around the Sun, determined by the PPN  parameter
$\gamma$, local straight lines are bent relative to asymptotic
straight lines far from the Sun by just enough to yield the
remaining factor ``$\gamma /2$''. The first factor ``1/2''
holds in any metric theory, the second ``$\gamma /2$'' varies
from theory to theory. Thus, calculations that purport to derive
the full deflection using the equivalence principle alone are
incorrect.

\begin{figure}[t]
\centering
  %\def\epsfsize#1#2{0.5#1}
 %\centerline{\epsfbox{livinggamma.eps}}
 \includegraphics[width=4in]{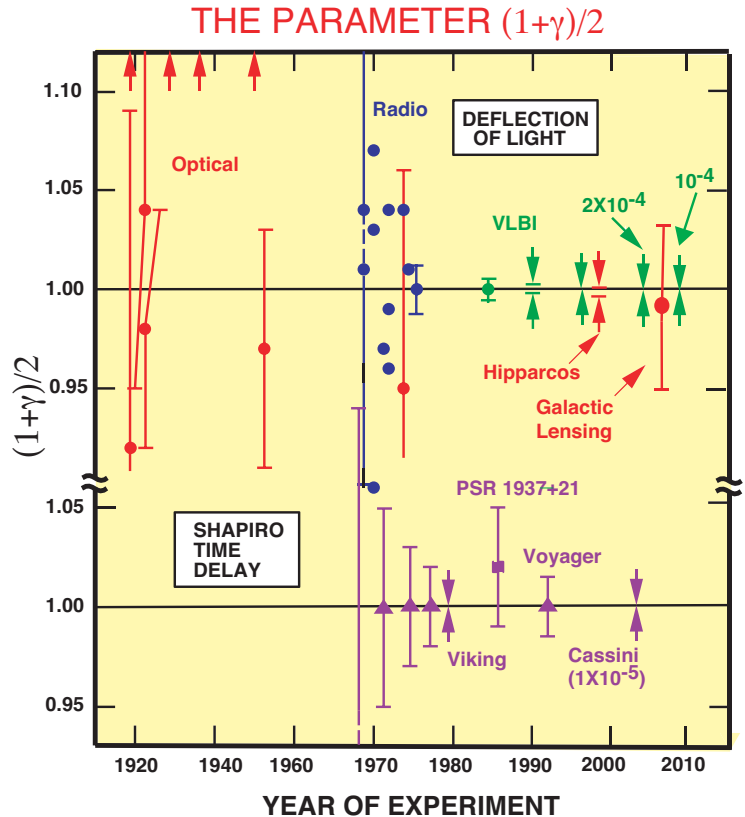}
  \caption{Measurements of the coefficient $(1 + \gamma )/2$ from
    light deflection and time delay measurements. Its GR
    value is unity. The arrows at the top denote anomalously large
    values from early eclipse expeditions. The Shapiro time-delay
    measurements using the Cassini spacecraft yielded an agreement with GR
    to $10^{-3}$~percent, and VLBI light deflection measurements have
    reached 0.02~percent. Hipparcos denotes the optical astrometry
    satellite, which reached 0.1~percent.}
  \label{gammavalues}
\end{figure}

The prediction of the full bending of light by the Sun was one of
the great successes of Einstein's GR.
Eddington's confirmation of the bending of optical starlight
observed during a solar eclipse in the first days following World
War I helped make Einstein famous. However, the experiments of
Eddington and his co-workers had only 30~percent accuracy (for a recent re-evaluation of Eddington's conclusions, see~\cite{2009PhT....62c..37K}).
Succeeding experiments were not much better:  the results were
scattered between one half and twice the Einstein value
(see Figure~\ref{gammavalues}), and the accuracies were low.  For a history of this period see~\cite{Crelinsten}.

%\epubtkImage{livinggamma.png}{
%}

However, the development of radio interferometery, and later
of very-long-baseline radio interfer\-om\-etry (VLBI), produced
greatly improved determinations of
the deflection of light. These techniques now have the capability
of measuring angular separations and changes in angles
to accuracies better than 
100 microarcseconds. Early measurements took advantage
of a series of heavenly coincidences:
Each year, groups of strong quasistellar radio sources pass
very close to the Sun (as seen from the Earth), including the
group 3C273, 3C279, and 3C48, and the group 0111+02, 0119+11,
and 0116+08. As the Earth moves in its orbit, changing the
lines of sight of the quasars relative to the Sun, the angular
separation $\delta \theta$ between pairs of quasars
varies (see Eq.~ (\ref{E30})). The time variation in the
quantities $d$, $d_\mathrm{r}$, $\chi$, and $\Phi_\mathrm{r}$ in Eq.~ (\ref{E30}) is
determined using an accurate ephemeris for the Earth and initial
directions for the quasars, and the resulting prediction for
$\delta \theta$  as a function of time is used as a basis for a
least-squares fit of the measured $\delta \theta$,  with one of
the fitted parameters being the coefficient $\frac{1}{2}(1+ \gamma )$.
A number of measurements of this kind over the period 1969\,--\,1975 yielded
an accurate determination of the coefficient $\frac{1}{2}(1+ \gamma)$, or equivalently $\gamma -1$.
A 1995 VLBI measurement using 3C273 and
3C279 yielded $\gamma -1=(-8 \pm 34) \times 10^{-4}$~\cite{lebach}, while a 2009 measurement using the VLBA targeting the same two quasars plus two other nearby radio sources yielded 
$\gamma -1=(-2 \pm 3)  \times 10^{-4}$~\cite{2009ApJ...699.1395F}.

In recent years, transcontinental and intercontinental VLBI observations 
of quasars and
radio galaxies have been 
made primarily to monitor the Earth's rotation
(``VLBI'' in Figure~\ref{gammavalues}). These measurements are
sensitive to the deflection of light over
almost the entire celestial sphere (at $90 ^\circ$ from the Sun, the
deflection is still 4 milli\-arcseconds).
A 2004 analysis of  almost 2 million VLBI observations
of 541 radio sources, made by 87 VLBI sites
yielded 
$(1+\gamma)/2=0.99992 \pm 0.00023$, or equivalently,
$\gamma-1= (-1.7 \pm 4.5) \times 10^{-4}$~\cite{sshapiro04}.
 A 2009 analysis that incorporated data through 2008 yielded $\gamma-1= (-1.6 \pm 1.5) \times 10^{-4}$~\cite{2009A&A...499..331L}.

Analysis of observations made by the Hipparcos optical astrometry
satellite yielded a test at the level of 0.3
percent~\cite{hipparcos}.
A VLBI
measurement of the deflection of light by Jupiter was
reported in 1991; the predicted deflection of about 300
microarcseconds was seen with about 50 percent accuracy~\cite{treuhaft}.

Finally, a remarkable measurement of $\gamma$ on
galactic scales was reported in 2006~\cite{2006PhRvD..74f1501B}. It used data on gravitational lensing by 15 elliptical
galaxies, collected by the Sloan Digital Sky Survey. The Newtonian
potential $U$ of each lensing galaxy (including the contribution from
dark matter) was derived from
the observed velocity dispersion of stars in the galaxy. Comparing
the observed lensing with the lensing predicted by the models provided a
10 percent bound on $\gamma$, in agreement with general
relativity. Unlike the much tighter bounds described previously, which 
were obtained on the scale of the solar system, this bound was obtained
on a galactic scale.  

The results of light-deflection measurements are summarized in
Figure~\ref{gammavalues}.

%%%%%%%%%%%%%%%%%%%%%%%%%%%%%%%%%%%%%%%%%%%%%%%%%%%%%%%%%%%%%%%%%%%%%%%%%%%%%%%%%%%

\subsubsection{The time delay of light}
\label{timedelay}

A radar signal sent across the solar system past the Sun to a
planet or satellite and returned to the Earth suffers an
additional non-Newtonian delay in its round-trip travel time,
given by (see Figure~\ref{deflectiongeom})
\begin{equation}
  \delta t = 2 (1 + \gamma) M_\odot
  \ln \left( \frac{(r_\oplus + {\bf x}_\oplus \cdot {\bf n})
  (r_\mathrm{e} - {\bf x}_\mathrm{e} \cdot {\bf n})}{d^2} \right),
  \label{E31}
\end{equation}
where $ {\bf x}_\mathrm{e} $ ($ {\bf x}_\oplus $) are the vectors, and
$ r_\mathrm{e} $ ($ r_\oplus $) are the distances from the Sun
to the source (Earth), respectively (TEGP~7.2~\cite{tegp}). For a ray
which passes close to the Sun,
\begin{equation}
  \delta t \approx \frac{1}{2} (1 + \gamma)
  \left [ 240 - 20 \ln \left ( \frac{d^2}{r} \right ) \right ] \mathrm{\ \mu s},
  \label{E32}
\end{equation}
where $d$ is the distance of closest approach of the ray in solar
radii, and $r$ is the distance of the planet or satellite from the
Sun, in astronomical units.

In the two decades following Irwin Shapiro's 1964 discovery of
this effect as a theoretical consequence of GR,
several high-precision measurements were made
using radar ranging to targets passing through superior
conjunction. Since one does not have access to a ``Newtonian''
signal against which to compare the round-trip travel time of the
observed signal, it is necessary to do a differential measurement
of the variations in round-trip travel times as the target passes
through superior conjunction, and to look for the logarithmic
behavior of Eq.~(\ref{E32}). In order to do this accurately however,
one must take into account the variations in round-trip travel
time due to the orbital motion of the target relative to the
Earth. This is done by using radar-ranging (and possibly other)
data on the target taken when it is far from superior conjunction
(i.e.\ when the time-delay term is negligible) to determine
an accurate ephemeris for the target, using the ephemeris to
predict the PPN coordinate trajectory ${\bf x}_\mathrm{e} (t)$ near
superior conjunction, then combining that trajectory with the
trajectory of the Earth ${\bf x}_\oplus (t)$ to determine the
Newtonian round-trip time and the logarithmic term in Eq.~ (\ref{E32}).
The resulting predicted round-trip travel times in terms of the
unknown coefficient $\frac{1}{2}(1+ \gamma)$
are then fit to the measured travel times using the method
of least-squares, and an estimate obtained for
$\frac{1}{2}(1+ \gamma)$.

The targets employed included
planets, such as Mercury or Venus, used as passive reflectors of
the radar signals (``passive radar''), and
artificial satellites, such as Mariners~6 and 7, Voyager~2,
the Viking
Mars landers and orbiters, and the Cassini spacecraft to Saturn,
used as
active retransmitters of the radar signals (``active radar'').

The results for the coefficient $\frac{1}{2}(1+ \gamma)$
of all radar time-delay measurements
performed to date (including a measurement of the one-way time delay
of signals from the millisecond pulsar PSR 1937+21)
are shown in Figure~\ref{gammavalues} (see TEGP~7.2~\cite{tegp} 
for discussion and references). The 1976 Viking experiment resulted in a
0.1 percent measurement~\cite{reasenberg}. 

A significant improvement was reported in 2003 
from Doppler tracking of the Cassini
spacecraft while it was on its way to Saturn~\cite{bertotti03}, with a
result $\gamma -1 = (2.1 \pm 2.3) \times 10^{-5}$. This was made possible
by the ability to do Doppler measurements using both X-band (7175~MHz) and
Ka-band (34316~MHz) radar, thereby significantly reducing the dispersive
effects of the
solar corona. In addition, the 2002 superior conjunction of Cassini was
particularly favorable: with the spacecraft at 8.43 astronomical units from
the Sun, the distance of closest approach of the radar signals to the Sun
was only $1.6 \, R_\odot$.

  From the results of the Cassini experiment, we
can conclude that the coefficient
$\frac{1}{2}(1+ \gamma)$
must be within at most 0.0012~percent of unity.
Scalar-tensor theories must have $\omega > 40,000$ to be compatible with
this constraint.

%%%%%%%%%%%%%%%%%%%%%%%%%%%%%%%%%%%%%%%%%%%%%%%%%%%%%%%%%%%%%%%%%%%%%%%%%%%%%%%%%%%

\subsubsection{Shapiro time delay and the speed of gravity}
\label{gravityspeed}

In 2001, Kopeikin~\cite{kopeikin01}
suggested that a measurement of the time delay of
light from a quasar as the light passed
by the planet Jupiter could be used to measure the speed
of the gravitational interaction. He argued that, since Jupiter is moving
relative to the solar system, and since gravity propagates with a finite
speed, the gravitational field experienced by the light ray should be
affected by gravity's speed, since the field experienced at one time depends on
the location of the source a short time earlier, depending on how fast
gravity propagates. According to his calculations, there should be a
post$^{1/2}$-Newtonian
correction to the normal Shapiro time-delay formula~(\ref{E31})
which depends on the velocity of Jupiter and on the velocity of gravity.
On September~8, 2002, Jupiter passed almost in
front of a quasar, and Kopeikin and Fomalont  made precise
measurements of the Shapiro
delay with picosecond timing accuracy, and claimed to have measured the
correction term
to about 20~percent~\cite{fomalont03, kopeikin02, kopeikin03, kopeikin04}. 

However, several authors pointed out that this 1.5PN effect does \emph{not}
depend on the speed of propagation of gravity, but rather only depends on
the speed of light~\cite{asada02, willspeed03, samuel03, carlip04, samuel04}. 
Intuitively, if one is working to only first order in $v/c$, 
then all that counts is
the uniform motion of the planet, Jupiter (its acceleration about the Sun
contributes a higher-order, unmeasurably small effect). But if that is the
case, then the principle of relativity says that one can view things from the
rest frame of Jupiter. In this frame, Jupiter's gravitational field is
static, and the speed of propagation of gravity is irrelevant.
A detailed post-Newtonian
calculation of the effect was done 
using a variant of the PPN framework, 
in a class of theories
in which the speed of gravity could be different from that
of light~\cite{willspeed03}, 
and found explicitly that, at first order in $v/c$, the effect
depends on the speed of light, not the speed of gravity, in line with
intuition. Effects dependent upon the speed of gravity show up only at
higher order in $v/c$. Kopeikin gave a number of arguments in opposition to
this interpretation~\cite{kopeikin04, kopeikin05a, kopeikin05b}. 
On the other hand, the $v/c$ correction term \emph{does}
show a dependence on the PPN parameter $\alpha_1$, which could be non-zero
in theories of gravity with a differing speed $c_\mathrm{g}$ of gravity (see Eq.~ (7)
of~\cite{willspeed03}). But existing tight bounds on $\alpha_1$ from other
experiments (see Table~\ref{ppnlimits}) already far exceed the capability of the Jupiter VLBI
experiment.

\begin{table}[hptb]
  \caption[Current limits on the PPN parameters.]{Current limits on
    the PPN parameters. }
  \label{ppnlimits}
  \renewcommand{\arraystretch}{1.2}
  \centering
  \begin{tabular}{l|lrl}
    \hline \hline
    Parameter &
    \multicolumn{1}{c}{Effect} &
    \multicolumn{1}{c}{Limit} &
    \multicolumn{1}{c}{Remarks} \\
    \hline \hline
    $\gamma-1$ &
    time delay &
    $2.3 \times 10^{-5 \phantom{0}}$ &
    Cassini tracking \\
    & light deflection &
    $2 \times 10^{-4 \phantom{0}}$ &
    VLBI \\
    $\beta-1$ &
    perihelion shift &
    $8 \times 10^{-5 \phantom{0}}$ &
    $J_{2\odot}=(2.2 \pm 0.1) \times10^{-7}$  \\
    & Nordtvedt effect &
    $2.3 \times 10^{-4 \phantom{0}}$ &
    $\eta_\mathrm{N}=4\beta-\gamma-3$ assumed \\
    $\xi$ &
    spin precession &
    $4 \times  10^{-9 \phantom{0}}$ &
    millisecond pulsars \\
    $\alpha_1$ &
    orbital polarization &
    $ 10^{-4 \phantom{0}}$ &
    Lunar laser ranging \\
    & & $7 \times 10^{-5 \phantom{0}}$ &
    PSR J1738+0333 \\
    $\alpha_2$ &
    spin precession &
    $2 \times 10^{-9 \phantom{0}}$ &
    millisecond pulsars \\
    $\alpha_3$ &
    pulsar acceleration &
    $4 \times 10^{-20}$ &
    pulsar $\dot P$ statistics \\
        $\zeta_1$ &
    \multicolumn{1}{c}{---} &
    $2 \times 10^{-2 \phantom{0}}$ &
    combined PPN bounds \\
    $\zeta_2$ &
    binary acceleration &
    $4 \times 10^{-5 \phantom{0}}$ &
    $\ddot P_\mathrm{p}$ for PSR 1913+16 \\
    $\zeta_3$ &
    Newton's 3rd law &
    $10^{-8 \phantom{0}}$ &
    lunar acceleration \\
    $\zeta_4$ &
    \multicolumn{1}{c}{---} &
    \multicolumn{1}{c}{---} &
    not independent (see Eq.~ (\ref{E37}))\\
    \hline \hline
  \end{tabular}
  \renewcommand{\arraystretch}{1.0}
\end{table}

%%%%%%%%%%%%%%%%%%%%%%%%%%%%%%%%%%%%%%%%%%%%%%%%%%%%%%%%%%%%%%%%%%%%%%%%%%%%%%%%%%%
%%%%%%%%%%%%%%%%%%%%%%%%%%%%%%%%%%%%%%%%%%%%%%%%%%%%%%%%%%%%%%%%%%%%%%%%%%%%%%%%%%%

\subsection{The perihelion shift of Mercury}
\label{perihelion}

The explanation of the anomalous perihelion shift of Mercury's
orbit was another of the triumphs of GR. This had
been an unsolved problem in celestial mechanics for over half a
century, since the announcement by Le Verrier in 1859 that, after
the perturbing effects of the planets on Mercury's orbit had been
accounted for, and after the effect of the precession of the
equinoxes on the astronomical coordinate system had been
subtracted, there remained in the data an unexplained advance
in the perihelion of Mercury. The modern value for this
discrepancy is 43 arcseconds per century. A number of \emph{ad
hoc} proposals were made in an attempt to account for this
excess, including, among others, the existence of a new planet
Vulcan near the Sun, a ring of planetoids, a solar quadrupole
moment and a deviation from the inverse-square law of gravitation,
but none was successful. General relativity accounted
for the anomalous shift in a natural way without disturbing the
agreement with other planetary observations.

The predicted advance per orbit $\Delta \tilde \omega$, including
both relativistic PPN  contributions and the Newtonian
contribution resulting from a
possible solar quadru\-pole moment, is given by
\begin{equation}
  \Delta \tilde \omega = \frac{6 \pi m}{p}
  \left( \frac{1}{3} (2 + 2 \gamma - \beta) + \frac{1}{6}
  (2 \alpha_1 - \alpha_2 + \alpha_3 + 2 \zeta_2) \eta +
  \frac{J_2 R^2}{2 m p} \right),
  \label{E33}
\end{equation}
where $m \equiv m_1 + m_2$ and $\eta  \equiv m_1 m_2 /m^2$
are the total mass and dimensionless reduced mass of the two-body system
respectively; $p \equiv a(1-e^2 )$ is the semi-latus rectum of
the orbit, with the semi-major axis $a$ and the eccentricity $e$; $R$ is
the mean radius of the oblate body; and $J_2$ is a
dimensionless measure of its quadrupole moment, given by
$J_2 = (C-A)/m_1 R^2$, where $C$ and $A$ are the
moments of inertia about the body's rotation and equatorial
axes, respectively (for details of the derivation see TEGP~7.3~\cite{tegp}).
We have ignored preferred-frame and galaxy-induced
contributions to  $\Delta \tilde \omega$; these are discussed in
TEGP~8.3~\cite{tegp}.

The first term in Eq.~ (\ref{E33}) is the classical relativistic
perihelion shift, which depends upon the PPN  parameters $\gamma$
and $\beta$. The second term depends upon the ratio of the masses
of the two bodies; it is zero in any fully conservative theory of
gravity
($ \alpha_1 \equiv \alpha_2 \equiv \alpha_3 \equiv \zeta_2 \equiv 0$);
it is also negligible for Mercury, since
$\eta \approx m_\mathrm{Merc} / M_\odot \approx 2 \times 10^{-7}$.
We shall drop this term henceforth. 

The third term depends upon
the solar quadrupole moment $J_2$. For a Sun that rotates
uniformly with its observed surface angular velocity,  so that
the quadrupole moment is produced by centrifugal flattening, one
may estimate $J_2$ to be
$\sim 1 \times 10^{-7}$. This actually agrees reasonably well with
values inferred from rotating solar models that are in accord with
observations of the normal modes of solar oscillations
(helioseismology); the latest inversions of helioseismology data give $J_2 =
(2.2 \pm 0.1) \times 10^{-7}$~\cite{mecheri04,2008A&A...477..657A}; for a review of measurements of the solar quadrupole moment, see~\cite{2011EPJH...36..407R}.
Substituting standard orbital elements and physical
constants for Mercury and the Sun we obtain the rate of
perihelion shift $\dot {\tilde \omega}$, in seconds of arc per
century,
\begin{equation}
  \dot{\tilde{\omega}} = 42.''98
  \left( \frac{1}{3} (2 + 2 \gamma - \beta) +
  3 \times 10^{-4} \frac{J_2}{10^{-7}} \right).
  \label{E34}
\end{equation}

The most recent fits to planetary data include data from the Messenger spacecraft that orbited Mercury, thereby significantly improving knowledge of its orbit.  Adopting the Cassini bound on $\gamma$ {\em a priori}, these analyses yield a bound on $\beta$  given by $\beta -1 = (-4.1 \pm 7.8) \times 10^{-5}$.   Further analysis could push this bound even lower~\cite{2011CeMDA.111..363F,2014A&A...561A.115V}, although knowledge of $J_2$ would have to improve simultaneously.  A slightly weaker bound $\beta -1 = (0.4 \pm 2.4) \times 10^{-4}$ from the perihelion advance of Mars (again adopting the Cassini bound on $\gamma$) was obtained by exploiting data from the Mars Reconnaissance Orbiter~\cite{2011Icar..211..401K}

Laser tracking of the Earth-orbiting satellite LAGEOS II led to a measurement of its relativistic perigee precession ($3.4$ arcseconds per year) in agreement with GR to $0.2$ percent~\cite{2010PhRvL.105w1103L}.

%%%%%%%%%%%%%%%%%%%%%%%%%%%%%%%%%%%%%%%%%%%%%%%%%%%%%%%%%%%%%%%%%%%%%%%%%%%%%%%%%%%
%%%%%%%%%%%%%%%%%%%%%%%%%%%%%%%%%%%%%%%%%%%%%%%%%%%%%%%%%%%%%%%%%%%%%%%%%%%%%%%%%%%

\subsection{Tests of the strong equivalence principle}
\label{septests}

The next class of solar-system experiments that test relativistic
gravitational effects may be called tests of the strong
equivalence principle (SEP). In Section~\ref{sep} we pointed out that many metric
theories of gravity (perhaps all except GR) can be
expected to violate one or more aspects of SEP. Among the
testable violations of SEP are a
violation of the weak equivalence principle for gravitating bodies
that leads to perturbations in the Earth-Moon
orbit, preferred-location and preferred-frame effects in the
locally measured gravitational constant that could produce
observable geophysical effects, and possible variations in the
gravitational constant over cosmological timescales.

%%%%%%%%%%%%%%%%%%%%%%%%%%%%%%%%%%%%%%%%%%%%%%%%%%%%%%%%%%%%%%%%%%%%%%%%%%%%%%%%%%%

\subsubsection{The Nordtvedt effect and the lunar E\"otv\"os experiment}
\label{Nordtvedteffect}

In a pioneering calculation using his early form of the PPN
formalism, Nord\-tvedt~\cite{nordtvedt1} showed that many metric theories
of gravity predict that massive bodies violate the weak
equivalence principle -- that is, fall with different
accelerations depending on their gravitational self-energy. 
Dicke~\cite{dicke_2} argued that such an effect would occur in theories with
a spatially
varying gravitational constant, such as scalar-tensor
gravity. For a
spherically symmetric body, the acceleration from rest in an
external gravitational potential $U$ has the form
\begin{equation}
  \begin{array}{rcl}
    {\bf a} & = & \displaystyle \frac{m_\mathrm{p}}{m} \nabla U,
    \\ [1.0 em]
    \displaystyle \frac{m_\mathrm{p}}{m} & = & \displaystyle 
    1 - \eta_\mathrm{N} \frac{E_\mathrm{g}}{m},
    \\ [1.0 em]
    \eta_\mathrm{N} & = & \displaystyle
    4 \beta - \gamma - 3 - \frac{10}{3} \xi -
    \alpha_1 + \frac{2}{3} \alpha_2 - \frac{2}{3} \zeta_1 -
    \frac{1}{3} \zeta_2,
  \end{array}
  \label{E35}
\end{equation}
where $E_\mathrm{g}$ is the negative of the gravitational self-energy
of the body ($E_\mathrm{g} >0$). This violation of the massive-body
equivalence principle is known as the ``Nordtvedt effect''. The
effect is absent in GR ($ \eta_\mathrm{N} = 0$) but present
in scalar-tensor theory ($ \eta_\mathrm{N} = 1/(2+ \omega )+4\lambda$). The existence
of the Nordtvedt effect does not violate the results of laboratory
E\"otv\"os experiments, since for laboratory-sized objects
$E_\mathrm{g} /m \le 10^{-27}$, far below the sensitivity of
current or future experiments. However, for astronomical bodies,
$E_\mathrm{g} /m$ may be significant ($3.6 \times 10^{-6}$ for the Sun, $10^{-8}$
for Jupiter,
$4.6 \times 10^{-10}$  for the Earth, $0.2 \times 10^{-10}$
for the Moon). If the Nordtvedt effect is present
($\eta_\mathrm{N} \ne 0$) then the Earth should fall toward the Sun with a
slightly different acceleration than the Moon. This perturbation
in the Earth-Moon orbit leads to a polarization of the orbit that
is directed toward the Sun as it moves around the Earth-Moon
system, as seen from Earth. This polarization represents a
perturbation in the Earth-Moon distance of the form
\begin{equation}
  \delta r = 13.1 \, \eta_\mathrm{N}
  \cos( \omega_0 -\omega_\mathrm{s}) t \mathrm{\ [m]},
  \label{E36}
\end{equation}
where $\omega_0$ and $\omega_\mathrm{s}$ are the angular frequencies
of the orbits of the Moon and Sun around the Earth (see
TEGP~8.1~\cite{tegp} for detailed derivations and references; for
improved
calculations of the numerical coefficient, see~\cite{Nordtvedt95, DamourVokrou96}).

Since August 1969, when the first successful acquisition was made
of a laser signal reflected from the Apollo~11 retroreflector on
the Moon, the LLR experiment has made
regular measurements of the round-trip travel times of laser
pulses between a network of observatories
and the lunar retroreflectors, with accuracies that are
approaching the 5~ps (1~mm) level. These measurements are fit
using the method of least-squares to a theoretical model for the
lunar motion that takes into account perturbations due to the Sun
and the other planets, tidal interactions, and post-Newtonian
gravitational effects. The predicted round-trip travel times
between retroreflector and telescope also take into account the
librations of the Moon, the orientation of the Earth, the
location of the observatories, and atmospheric effects on the
signal propagation. The ``Nordtvedt'' parameter $\eta_\mathrm{N}$ along with
several other important parameters of the model are then
estimated in the least-squares method.  For a review of lunar laser ranging, 
see \cite{2010LRR....13....7M}.

Numerous ongoing analyses of the data find no evidence, within
experimental uncertainty, for the Nordtvedt 
effect~\cite{williams04, williams04ijmp} 
(for earlier results
see~\cite{Dickey, Williams, MullerMG}). 
These results represent a limit on a possible violation of WEP for
massive bodies of about 
1.4 parts in $10^{13}$ (compare Figure~\ref{wepfig}). 

However, 
at this level of precision, one cannot regard the results of LLR
as a ``clean'' test of SEP until one eliminates the
possibility of a compensating violation of WEP for the two bodies,
because the chemical compositions of the Earth
and Moon differ in the relative fractions of iron and silicates. To
this end, the E{\"o}t-Wash group carried out an improved test of WEP
for laboratory bodies whose chemical compositions mimic that of the
Earth and Moon. The resulting bound of 1.4 parts 
in $10^{13}$~\cite{baessler99, adelberger01} from composition effects
reduces the ambiguity in the LLR bound, and establishes the firm SEP test
at the level of about 2 parts in $10^{13}$. These results
can be summarized by the Nordtvedt parameter bound
$| \eta_\mathrm{N} | = (4.4 \pm 4.5) \times 10^{-4}$.

APOLLO, the Apache Point Observatory for Lunar Laser ranging
Operation, a joint effort by researchers from the
Universities of Washington, Seattle, and California, San Diego, has achieved mm ranging precision using enhanced laser and telescope technology, together with a good,
high-altitude site in New Mexico.  However models of the lunar orbit must be improved in parallel in order to achieve an order-of-magnitude improvement in the test of the Nordtvedt effect~\cite{2012CQGra..29r4005M}.
This effort will be aided by the fortuitous 2010 discovery by the Lunar Reconnaissance Orbiter of the precise landing site of the Soviet Lunokhod I rover, which deployed a retroreflector in 1970.  Its uncertain location made it effectively ``lost'' to lunar laser ranging for almost 40 years.  Its location on the lunar surface will make it useful in improving models of the lunar 
libration~\cite{2011Icar..211.1103M}.

In GR, the Nordtvedt effect vanishes; at the level of
several centimeters and below,
a number of non-null general relativistic effects
should be present~\cite{Nordtvedt95}.

Tests of the Nordtvedt effect for neutron stars
have also been carried out using
a class of systems known as wide-orbit binary millisecond pulsars (WBMSP),
which are 
pulsar--white-dwarf binary systems with small orbital eccentricities.
In the gravitational field
of the galaxy, a non-zero Nordtvedt effect can induce an apparent anomalous
eccentricity pointed toward the galactic center~\cite{DamourSchaefer91},
which can be bounded using statistical methods, given enough WBMSPs
(see~\cite{StairsLRR} for a review and references). Using
data from 21 WBMSPs, including recently discovered highly circular systems,
Stairs et al.~\cite{stairs05} obtained the bound $\Delta < 5.6 \times
10^{-3}$, where $\Delta = \eta_\mathrm{N} (E_\mathrm{g}/M)_\mathrm{NS}$. Because 
$(E_\mathrm{g}/M)_\mathrm{NS} \sim 0.1$ for typical neutron stars, this bound does not
compete with the bound on $\eta_\mathrm{N}$ from LLR; on the
other hand, it does test SEP in the strong-field regime because of the
presence of the neutron stars.   The 2013 discovery of a millisecond pulsar in orbit with {\em two} white dwarfs in very circular, coplanar orbits~\cite{2014Natur.505..520R} may lead to a test of the Nordvedt effect in the strong-field regime that surpasses the precision of lunar laser ranging by a substantial factor (see Sec.\ \ref{population}).

%%%%%%%%%%%%%%%%%%%%%%%%%%%%%%%%%%%%%%%%%%%%%%%%%%%%%%%%%%%%%%%%%%%%%%%%%%%%%%%%%%%

\subsubsection{Preferred-frame and preferred-location effects}
\label{preferred}

Some theories of gravity violate SEP by predicting that the
outcomes of local gravitational experiments may depend on the
velocity of the laboratory relative to the mean rest frame of the
universe (preferred-frame effects) or on the location of the
laboratory relative to a nearby gravitating body
(preferred-location effects). In the post-Newtonian limit,
preferred-frame effects are governed by the values of the PPN
parameters $\alpha_1$, $\alpha_2$, and $\alpha_3$, and some
preferred-location effects are governed by $\xi$ (see Table~\ref{ppnmeaning}).

The most important such effects are variations and anisotropies
in the locally-measured value of the gravitational constant which
lead to anomalous Earth tides and variations in the Earth's
rotation rate, anomalous contributions to the
orbital dynamics of planets and the Moon, self-accelerations of
pulsars, and anomalous torques on the Sun that
would cause its spin axis to be randomly oriented relative to the
ecliptic (see TEGP~8.2, 8.3, 9.3, and 14.3~(c)~\cite{tegp}). 

A tight bound on $\alpha_3$ of $4 \times 10^{-20}$ was obtained  from the period derivatives of 21 millisecond pulsars~\cite{Bell,stairs05}.  The best bound on
$\alpha_1$, comes from the orbit of the pulsar--white-dwarf system J1738+0333~\cite{2012CQGra..29u5018S}.  Early bounds on on $\alpha_2$ and $\xi$ came from searches for variations induced by an anisotropy in $G$ on the acceleration of gravity on Earth using gravimeters, and (in the case of $\alpha_2$) from limiting the effects of any anomalous torque on the spinning Sun over the age of the solar system.  Today the best bounds on $\alpha_2$ and $\xi$ come from bounding torques on the solitary millisecond pulsars B1937+21 and J1744--1134~\cite{2012CQGra..29u5018S,2013CQGra..30p5019S,2013CQGra..30p5020S}.  Because these later bounds involve systems with strong internal gravity of the neutron stars, they should strictly speaking be regarded as bounds on ``strong field'' analogues of the PPN parameters.  Here we will treat them as bounds on the standard
PPN  parameters, as shown in Table~\ref{ppnlimits}.

%%%%%%%%%%%%%%%%%%%%%%%%%%%%%%%%%%%%%%%%%%%%%%%%%%%%%%%%%%%%%%%%%%%%%%%%%%%%%%%%%%%

\subsubsection{Constancy of the Newtonian gravitational constant}
\label{bigG}

Most theories of gravity that violate SEP predict that the locally
measured Newtonian gravitational constant may vary with time as
the universe evolves. For the scalar-tensor theories listed in
Table~\ref{ppnvalues},
the predictions for $\dot{G}/G$ can be written in terms of time
derivatives of the asymptotic scalar field.
Where $G$
does change with cosmic evolution, its rate of variation should
be of the order of the expansion rate of the universe,
i.e.\ $\dot{G}/G \sim H_0$, where $H_0$ is the Hubble expansion parameter, given by
$H_0 = 73 \pm 3 \, \mathrm{\ km\ s}^{-1}\mathrm{\ Mpc}^{-1} = 7.4
\times 10^{-11} \,  \mathrm{\ yr}^{-1}$.

Several observational constraints can be placed on $\dot{G}/G$, one kind
coming from bounding the present rate of variation, another from bounding a
difference between the present value and a past value.
The first type of bound typically comes from LLR measurements,
planetary radar-ranging measurements, and pulsar timing data. The
second type comes from studies of the evolution of the Sun, stars and
the Earth, big-bang nucleosynthesis, and analyses of ancient eclipse data.
Recent results are shown in Table~\ref{Gdottable}.

\begin{table}[hptb]
  \caption{Constancy of the gravitational constant. For binary
    pulsar data, the bounds are dependent upon the theory of gravity
    in the strong-field regime and on neutron star equation of
    state. Big-bang nucleosynthesis bounds assume specific form for
    time dependence of $G$.}
  \label{Gdottable}
  \renewcommand{\arraystretch}{1.2}
  \centering
  \begin{tabular}{l|cl}
    \hline \hline
    Method &
    $\rule{0 cm}{1.2 em}\dot{G}/G$ &
    Reference \\
    & ($10^{-13} \mathrm{\ yr}^{-1}$) \\
    \hline \hline
    Mars ephemeris&$0.1 \pm 1.6$&\cite{2011Icar..211..401K}\\
    Lunar laser ranging & $ 4 \pm 9 $ & \cite{williams04} \\
    Binary \& millisecond pulsars& $ -7 \pm 33 $ & \cite{2008ApJ...685L..67D,2009MNRAS.400..805L} \\
    Helioseismology & $ \phantom{0}0 \pm 16 $ & \cite{guenther98} \\
    Big Bang nucleosynthesis & $ 0 \pm 4 $ & \cite{copi04,bambi04} \\
    \hline \hline
  \end{tabular}
  \renewcommand{\arraystretch}{1.0}
\end{table}

The best limits on a current $\dot{G}/G$ come from improvements in the ephemeris of Mars using range and Doppler data from the Mars Global
Surveyor (1998\,--\,2006), Mars Odyssey (2002\,--\,2008), and Mars Reconnaissance Orbiter 
(2006\,--\,2008), together with improved data and modeling of the effects of the asteroid belt~\cite{pitjeva05,2011Icar..211..401K}.  Since the bound is actually on variations of $GM_\odot$, any future improvements in $\dot{G}/G$ beyond a part in $10^{13}$ will have to take into account models of the actual mass loss from the Sun, due to radiation of light and neutrinos ($\sim 0.7 \times 10^{-13} \, {\rm yr}^{-1}$) and due to the solar wind ($\sim 0.2 \times 10^{-13} \, {\rm yr}^{-1}$).
Another bound
comes from LLR measurements (\cite{williams04}; for earlier results
see~\cite{Dickey, Williams, MullerMG}).

Although bounds on $\dot{G}/G$ from solar-system measurements can be
correctly
obtained in a phenomenological manner through the simple expedient of
replacing $G$ by $G_0 + {\dot{G}}_0 (t - t_0 )$ in
Newton's equations of motion, the same does not hold true for pulsar
and binary pulsar timing measurements. The reason is that, in theories
of gravity that violate SEP, such as scalar-tensor theories,
the ``mass'' and moment of inertia of a
gravitationally bound body may vary with $G$. Because
neutron stars are highly relativistic, the fractional variation in
these quantities can be comparable to $\Delta G/G$, the precise
variation depending both on the equation of state of neutron star
matter and on the theory of gravity in the strong-field regime. The
variation in the moment of inertia affects the spin rate of the pulsar,
while the variation in the mass can affect the orbital period in a
manner that can subtract from the direct effect of a variation in $G$,
given by $\dot P_\mathrm{b} /P_\mathrm{b} =-{2} \dot{G}/G$~\cite{nordtvedt3}. Thus,
the bounds quoted in Table~\ref{Gdottable}
for binary and millisecond pulsars are theory-dependent and must be treated as merely
suggestive.

In a similar manner, bounds from helioseismology and big-bang
nucleosynthesis (BBN) assume a model for the evolution of $G$ over the
multi-billion year time spans involved. For example, the concordance of
predictions for 
light elements produced around 3~minutes after the big bang with the
abundances observed indicate that $G$ then was within 20 percent of $G$
today. Assuming a power-law variation of $G \sim t^{-\alpha}$ then yields
a bound on $\dot{G}/G$ today shown in Table~\ref{Gdottable}.

%%%%%%%%%%%%%%%%%%%%%%%%%%%%%%%%%%%%%%%%%%%%%%%%%%%%%%%%%%%%%%%%%%%%%%%%%%%%%%%%%%%
%%%%%%%%%%%%%%%%%%%%%%%%%%%%%%%%%%%%%%%%%%%%%%%%%%%%%%%%%%%%%%%%%%%%%%%%%%%%%%%%%%%

\subsection{Other tests of post-Newtonian gravity}
\label{othertests}

%%%%%%%%%%%%%%%%%%%%%%%%%%%%%%%%%%%%%%%%%%%%%%%%%%%%%%%%%%%%%%%%%%%%%%%%%%%%%%%%%%%

\subsubsection{Search for gravitomagnetism}
\label{gravitomagnetism}

According to GR, moving or rotating matter should
produce a contribution to the gravitational field that is the analogue
of the magnetic field of a moving charge or a magnetic dipole. In
particular, one can view the $g_{0i}$ part of the PPN metric (see
Box~\ref{box2}) as an analogue of the vector potential of
electrodynamics. In a suitable gauge (not the standard PPN gauge), and dropping the preferred-frame
terms, it can be written
\begin{equation}
  g_{0i} = - \frac{1}{2} (4 \gamma + 4 + \alpha_1) V_i \,.
  \label{gravitomag}
\end{equation}
At PN order, this contributes a Lorentz-type acceleration ${\bf v} \times
{\bf B}_\mathrm{g}$ to the equation of motion, where the gravitomagnetic field 
${\bf B}_\mathrm{g}$ is given by ${\bf B}_\mathrm{g} = \nabla \times (g_{0i}{\bf e}^i)$.

Gravitomagnetism plays a role in a variety of measured
relativistic effects involving moving material sources, such as the
Earth-Moon system and binary pulsar systems.
Nordtvedt~\cite{nordtvedt88a, nordtvedt88b} has argued that, if the 
gravitomagnetic potential~(\ref{gravitomag}) were turned off, then there
would be anomalous orbital effects in LLR and binary pulsar data. 

Rotation also produces a gravitomagnetic effect, since for a rotating body,
${\bf V} = -\frac{1}{2} {\bf x} \times {\bf J}/r^3$, where ${\bf J}$ is the
angular momentum of the body. The result is a ``dragging of inertial
frames'' around the body, 
also called the Lense--Thirring effect. A consequence
is a precession of a gyroscope's spin $\bf S$ according to 
\begin{equation}
  \frac{d{\bf S}}{d\tau} =
  {\bf \Omega}_\mathrm{LT} \times {\bf S},
  \qquad
  {\bf \Omega}_\mathrm{LT} =
  - \frac{1}{2} \left( 1 + \gamma + \frac{1}{4} \alpha_1 \right)
  \frac{{\bf J} - 3 {\bf n} ({\bf n} \cdot {\bf J})}{r^3},
  \label{E42}
\end{equation}
where $\bf n$ is a
unit radial vector, and $r$ is the distance from the center of the
body (TEGP~9.1~\cite{tegp}).

In 2011 the Relativity
Gyroscope Experiment (Gravity Probe~B or GPB)
carried out by Stanford University, 
NASA  and Lockheed--Martin Corporation~\cite{gpbwebsite},  finally completed
a space mission to detect this frame-dragging or Lense--Thirring precession,
along with the ``geodetic'' precession 
(see Section~\ref{geodeticprecession}).
Gravity Probe B will very likely go down in the history of science as
one of the most ambitious, difficult, expensive, and controversial
relativity experiments ever performed.\footnote{Full disclosure: The author served as Chair of an external NASA Science Advisory Committee
  for Gravity Probe B from 1998 to 2011.}  It was almost 50 years from inception to completion, although only about half of that time was spent as a full-fledged, approved space program.

The GPB spacecraft was launched on April 20, 2004 into an almost perfectly
circular polar orbit at an altitude of $642$ km, with the orbital plane
parallel to the direction of a guide star known as {\em IM Pegasi} (HR 8703). 
The spacecraft contained four spheres made of fuzed quartz, all
spinning about the same axis (two were spun in the opposite
direction), which was oriented to be in the orbital plane, pointing
toward the guide star.  An onboard telescope pointed continuously at
the guide star, and the direction of each spin was compared with the
direction to the star, which was at a declination of $16^{\rm o}$
relative to the Earth's equatorial plane.  With these conditions, the precessions predicted by 
GR were 
$6630$ milliarcsecond per year for the geodetic effect, and 
 $38$ milliarcsecond per year for frame dragging, the 
former in the orbital plane (in the north-south direction) and
the latter perpendicular to it (in the east-west direction).    

In order to reduce the non-relativistic torques on the
rotors to an acceptable level, the rotors were fabricated to be both
spherical and homogenous to better than a few parts in 10 million.
Each rotor was coated with a thin film of niobium, and the experiment
was conducted at cryogenic temperatures inside a dewar containing 2200 
litres of superfluid liquid helium. As the niobium film becomes a
superconductor, each rotor develops a magnetic moment parallel to its
spin axis.  Variations in the direction of the magnetic moment
relative to the spacecraft were then measured using superconducting current loops
surrounding each rotor.  As the spacecraft orbits the Earth, the
aberration of light 
from the guide star causes an artificial but
predictable change in direction between the rotors and the on-board
telescope; this was an essential tool for calibrating the conversion
between the voltages read by the current loops and the actual angle
between the rotors and the guide star.   The motion of
the guide star relative to distant inertial frames was measured
before, during and after the mission
separately by radio astronomers at Harvard/SAO and elsewhere using
VLBI ({\em IM Pegasi} is a radio star)~\cite{2012ApJS..201....1S}. 

The mission
ended in September 2005, as scheduled, when the last of the
liquid helium boiled off.  Although all subsystems of the spacecraft 
and the apparatus performed extremely well, they were not perfect.
Calibration measurements carried out during the mission, both before
and after the science phase, revealed unexpectedly large torques on
the rotors.  Numerous diagnostic tests worthy of a detective novel showed that these were caused by electrostatic interactions
between surface imperfections (``patch effect'') on the niobium films and the spherical
housings surrounding each rotor.  These effects and other anomalies
greatly contaminated the data and complicated its analysis, but
finally, in October 2010, the Gravity Probe B team announced that the
experiment had successfully measured both the geodetic and
frame-dragging precessions. The outcome was in agreement with general
relativity, with a precision of $0.3$ percent for the geodetic precession, 
and $20$ percent for the frame-dragging effect~\cite{2011PhRvL.106v1101E}.  For a commentary on the GPB result, see~\cite{2011PhyOJ...4...43W}.   The full technical and data analysis details of GPB are expected to be published as a special issue of {\em Classical and Quantum Gravity} in 2015.

Another way to look for frame-dragging is to
measure the precession of orbital planes of bodies circling a rotating
body. One implementation of this idea is to
measure the relative precession, at about 31 milliarcseconds per year,
of the line of nodes of a pair
of laser-ranged geodynamics satellites (LAGEOS), ideally with supplementary
inclination angles; the inclinations must be supplementary in order
to cancel the dominant (126 degrees per year)
nodal precession caused by the Earth's
Newtonian gravitational multipole moments. Unfortunately, the two
existing LAGEOS satellites are not in appropriately inclined orbits. Nevertheless, Ciufolini and collaborators~\cite{ciufolini04,2006NewA...11..527C,2011EPJP..126...72C} 
combined 
nodal precession data
from LAGEOS I and II with improved models for the Earth's multipole moments
provided by 
two orbiting geodesy satellites, Europe's
CHAMP (Challenging Minisatellite Payload)
and NASA's GRACE (Gravity Recovery and Climate Experiment),
and reported a 10 percent confirmation of GR~\cite{2011EPJP..126...72C}. In earlier reports,
Ciufolini et al.\ had reported tests at the the 20\,--\,30 percent
level, without the benefit of the GRACE/CHAMP data~\cite{ciufolini97,
  ciufolini98, ciufolini00}.
Some authors stressed the importance of adequately assessing systematic
errors in the LAGEOS data~\cite{ries, iorio05}.

On February 13, 2012, a third laser-ranged satellite, known as LARES
(Laser Relativity Satellite) was launched by the Italian Space Agency~\cite{2013AcAau..91..313P}.
Its inclination was very close to the required
supplementary angle relative to LAGEOS I, and its eccentricity was
very nearly zero.  However, because its semimajor axis is only $2/3$ that of either LAGEOS I or II, 
and because the Newtonian precession rate is proportional to $a^{-3/2}$, LARES does not provide a cancellation of the Newtonian precession.    Nevertheless, combining data from
all three satellites with continually improving Earth data from GRACE,
the LARES team hopes to achieve a test of
frame-dragging at the one percent level~\cite{2013CQGra..30w5009C}. 

%%%%%%%%%%%%%%%%%%%%%%%%%%%%%%%%%%%%%%%%%%%%%%%%%%%%%%%%%%%%%%%%%%%%%%%%%%%%%%%%%%%

\subsubsection{Geodetic precession}
\label{geodeticprecession}

A gyroscope moving through curved spacetime suffers a precession of
its spin axis given by
\begin{equation}
  \frac{d{\bf S}}{d\tau} =
  {\bf \Omega}_\mathrm{G} \times {\bf S},
  \qquad
  {\bf \Omega}_\mathrm{G} =
  \left( \gamma + \frac{1}{2} \right) {\bf v} \times \nabla U,
  \label{E41}
\end{equation}
where $\bf {v}$ is the velocity of the gyroscope, and $U$ is the
Newtonian gravitational potential of the source (TEGP~9.1~\cite{tegp}).
The Earth-Moon system can be considered as a ``gyroscope'', with its axis
perpendicular to the orbital plane. The predicted precession is about
2 arcseconds per century, an effect first calculated by de Sitter.
This effect has been measured to about 0.6 percent using LLR
data~\cite{Dickey, Williams, williams04}.

For the GPB gyroscopes orbiting the Earth, the precession is 6.63 arcseconds
per year.  GPB 
measured this effect to $3 \times 10^{-3}$; the resulting bound on
 the parameter
$\gamma$ is not competitive with the Cassini bound.

%%%%%%%%%%%%%%%%%%%%%%%%%%%%%%%%%%%%%%%%%%%%%%%%%%%%%%%%%%%%%%%%%%%%%%%%%%%%%%%%%%%

\subsubsection{Tests of post-Newtonian conservation laws}
\label{conservation}

Of the five ``conservation law'' PPN  parameters
$\zeta_1$, $\zeta_2$, $\zeta_3$, $\zeta_4$,
and $\alpha_3$, only three, $\zeta_2$, $\zeta_3$, and $\alpha_3$,
have been constrained directly with any precision;
$\zeta_1$ is constrained
indirectly through its appearance in the Nordtvedt effect parameter
$\eta_\mathrm{N}$, Eq.~ (\ref{E35}). There is strong theoretical
evidence that $\zeta_4$, which is related to the gravity generated by
fluid pressure, is not really an independent parameter -- in any
reasonable theory of gravity there should be a connection between the
gravity produced by kinetic energy ($\rho v^2$), internal energy ($\rho
\Pi$), and pressure ($p$).
From such considerations, there follows~\cite{Will76}
the additional theoretical constraint
\begin{equation}
  6 \zeta_4 = 3 \alpha_3 + 2 \zeta_1 - 3 \zeta_3.
  \label{E37}
\end{equation}

A non-zero value for any of these parameters would result in a
violation of conservation of momentum, or of Newton's third law
in gravitating systems. An alternative statement of Newton's
third law for gravitating systems is that the ``active gravitational mass'',
that is the mass that determines the gravitational potential exhibited
by a body, should equal the ``passive gravitational mass'',
the mass that determines the force on a body in a gravitational field.
Such an equality guarantees the equality of action and reaction and
of conservation of momentum, at least in the Newtonian limit.

A classic test of Newton's third law for gravitating systems was
carried out in 1968 by Kreuzer, in which the gravitational attraction
of fluorine and bromine were compared to a precision of
5\ parts in $10^5$.

A remarkable planetary test was reported by
Bartlett and van Buren~\cite{bartlett}. They noted that current understanding
of the structure of the Moon involves an iron-rich, aluminum-poor
mantle whose center of mass is offset about 10~km from the center of
mass of an aluminum-rich, iron-poor crust. The direction of offset
is toward the Earth, about $14 ^\circ$ to the east of the Earth-Moon line.
Such a model accounts for the basaltic maria which face the Earth,
and the aluminum-rich highlands on the Moon's far side, and for a 2~km
offset between the observed center of mass and center of figure for
the Moon. Because of this asymmetry, a violation of Newton's third
law for aluminum and iron would result in a momentum non-conserving
self-force on the Moon, whose component along the orbital direction
would contribute to the secular acceleration of the lunar orbit.
Improved knowledge of the lunar orbit through LLR, and
a better understanding of tidal effects in the Earth-Moon system
(which also contribute to the secular acceleration) through satellite
data, severely limit any anomalous secular acceleration, with
the resulting limit
\begin{equation}
  \left| \frac{(m_\mathrm{A} / m_\mathrm{P})_\mathrm{Al} -
  (m_\mathrm{A} / m_\mathrm{P})_\mathrm{Fe}}
  {(m_\mathrm{A} / m_\mathrm{P} )_\mathrm{Fe}} \right| <
  4 \times 10^{-12}.
  \label{E38}
\end{equation}
According to the PPN  formalism, in a theory of gravity that violates
conservation of momentum, but that obeys the constraint of Eq.~ (\ref{E37}),
the electrostatic binding energy $E_\mathrm{e}$
of an atomic nucleus could make a contribution to the ratio of active to
passive mass of the form
\begin{equation}
  m_\mathrm{A} = m_\mathrm{P} + \frac{1}{2} \zeta_3 E_\mathrm{e}.
  \label{E39}
\end{equation}
The resulting limit on $\zeta_3$ from the lunar experiment
is $\zeta_3 < 1 \times 10^{-8}$ (TEGP~9.2, 14.3~(d)~\cite{tegp}).
Nordtvedt~\cite{nordtvedt01} has examined whether this bound could be
improved by considering the asymmetric distribution of ocean water on
Earth.

Another consequence of a violation of conservation of momentum is a
self-accel\-er\-at\-ion of the center of mass of a binary stellar system,
given by
\begin{equation}
  {\bf a}_\mathrm{CM} = - \frac{1}{2} (\zeta_2+\alpha_3)
  \frac{m}{a^2} \frac{\mu}{a} \frac{\delta m}{m} \frac{e}{(1 - e^2)^{3/2}}
  {\bf n}_\mathrm{P},
  \label{E40}
\end{equation}
where $\delta m = m_1-m_2$, $a$ is the semi-major axis, and ${\bf n}_\mathrm{P}$
is a unit vector directed from the center of mass to the point of
periastron of $m_1$ (TEGP~9.3~\cite{tegp}).
A consequence of this acceleration would be non-vanishing values for
$d^2 P/dt^2$, where $P$ denotes the period of any intrinsic process in
the system (orbit, spectra,
pulsar periods). The observed upper limit on $d^2 P_\mathrm{p}
/dt^2$ of the binary pulsar PSR 1913+16 places a
strong constraint on such an effect, resulting in the bound $|
\alpha_3 + \zeta_2 |<4 \times 10^{-5}$. Since $\alpha_3$ has already
been constrained to be much less than this (see Table~\ref{ppnlimits}),
we obtain a strong
solitary bound on $\zeta_2 < 4 \times 10^{-5}$~\cite{Will92c}.

%%%%%%%%%%%%%%%%%%%%%%%%%%%%%%%%%%%%%%%%%%%%%%%%%%%%%%%%%%%%%%%%%%%%%%%%%%%%%%%%%%%
%%%%%%%%%%%%%%%%%%%%%%%%%%%%%%%%%%%%%%%%%%%%%%%%%%%%%%%%%%%%%%%%%%%%%%%%%%%%%%%%%%%

\subsection{Prospects for improved PPN parameter values}
\label{improvedPPN}

A number of advanced experiments or space missions are under development or
have been 
proposed which could lead to
significant improvements in values of the PPN  parameters,
of $J_2$ of the Sun, and of $\dot{G}/G$. 

LLR at the  Apache Point Observatory (APOLLO project) could
improve bounds on the Nordvedt parameter to the level $3 \times 10^{-5}$
and on $\dot{G}/G$ to better than 
$10^{-13} \mathrm{\ yr}^{-1}$~\cite{williams04ijmp}.

The  BepiColumbo Mercury orbiter is a joint project of the European and Japanese space agencies, scheduled for launch in 2015~\cite{2010P&SS...58....2B}.  In a two-year experiment, with 6~cm range capability, it
could yield improvements in $\gamma$
to $3 \times 10^{-5}$, in $\beta$ to
$3 \times 10^{-4}$, in $\alpha_1$ to $10^{-5}$, 
in $\dot{G}/G$ to $10^{-13} \mathrm{\ yr}^{-1}$, and in
$J_2$ to $3 \times 10^{-8}$. An eight-year mission could yield further
improvements by factors of 2\,--\,5 in $\beta$, $\alpha_1$, and $J_2$, and a
further factor 15 in $\dot{G}/G$~\cite{milani02,2007PhRvD..75b2001A}. 

GAIA is a high-precision astrometric orbiting telescope launched by ESA in 2013 (a successor to
Hipparcos)~\cite{gaia}.  With astrometric capability ranging from 10 to a few hundred microsarcseconds, plus the ability measure the locations of a billion stars down to 20th magnitude, it could
measure light-deflection and
$\gamma$ to the $10^{-6}$ level~\cite{2002EAS.....2..107M}.  

LATOR (Laser Astrometric Test of Relativity) is a concept for a NASA
mission in which two microsatellites orbit the Sun on Earth-like
orbits near superior conjunction, so that their lines of sight are
close to the Sun. Using optical tracking and an optical interferometer
on the International Space Station, it may be possible to measure the
deflection of light with sufficient accuracy to bound $\gamma$ to a
part in $10^8$ and $J_2$ to a part in $10^{8}$, and to measure the
solar frame-dragging effect to one percent~\cite{turyshev04a,
  turyshev04b}.  

Another concept, proposed for a European Space Agency medium-class mission, is ASTROD I (Astrodynamical Space Test of Relativity using Optical Devices), a variant of LATOR involving a single satellite parked on the far side of the Sun~\cite{2012ExA....34..181B}.   Its goal is to measure $\gamma$ to a few parts in $10^8$, $\beta$ to six parts in $10^6$ and $J_2$ to a part in $10^9$.  A possible follow-on mission, ASTROD-GW, involving three spacecraft, would improve on measurements of those parameters and would also measure the solar frame-dragging effect, as well as look for gravitational waves.

\newpage

%%%%%%%%%%%%%%%%%%%%%%%%%%%%%%%%%%%%%%%%%%%%%%%%%%%%%%%%%%%%%%%%%%%%%%%%%%%%%%%%%%%
%%%%%%%%%%%%%%%%%%%%%%%%%%%%%%%%%%%%%%%%%%%%%%%%%%%%%%%%%%%%%%%%%%%%%%%%%%%%%%%%%%%
%%%%%%%%%%%%%%%%%%%%%%%%%%%%%%%%%%%%%%%%%%%%%%%%%%%%%%%%%%%%%%%%%%%%%%%%%%%%%%%%%%%

\section{Strong Gravity and Gravitational Waves: Tests for the 21st Century}
\label{S4}

%%%%%%%%%%%%%%%%%%%%%%%%%%%%%%%%%%%%%%%%%%%%%%%%%%%%%%%%%%%%%%%%%%%%%%%%%%%%%%%%%%%
%%%%%%%%%%%%%%%%%%%%%%%%%%%%%%%%%%%%%%%%%%%%%%%%%%%%%%%%%%%%%%%%%%%%%%%%%%%%%%%%%%%

\subsection{Strong-field systems in general relativity}
\label{strong}

%%%%%%%%%%%%%%%%%%%%%%%%%%%%%%%%%%%%%%%%%%%%%%%%%%%%%%%%%%%%%%%%%%%%%%%%%%%%%%%%%%%

\subsubsection{Defining weak and strong gravity}
\label{strongvweak}

In the solar system, gravity is weak, in the sense that the  Newtonian
gravitational
potential and related variables ($ U ( {\bf x}, t) \sim v^2 \sim p/\rho \sim
\epsilon$)
are much smaller than unity everywhere.
This is the basis for the post-Newtonian expansion and for the
``parametrized
post-Newtonian'' framework
described in Section~\ref{ppn}.
``Strong-field'' systems are those for which
the simple 1PN approximation of the PPN framework
is no longer
appropriate. This can occur in a number of situations:
\begin{itemize}
\item The system may contain strongly relativistic objects, such as
  neutron stars or black holes, near and inside which
  $\epsilon \sim 1$, and the post-Newtonian approximation breaks
  down. Nevertheless, under some circumstances, the orbital motion may
  be such that the interbody potential and orbital velocities still
  satisfy $\epsilon \ll 1$ so that a kind of post-Newtonian
  approximation for the orbital motion might work; however, the
  strong-field internal gravity of the bodies could (especially in
  alternative theories of gravity) leave imprints on the orbital
  motion.
\item The evolution of the system may be affected by the emission of
  gravitational radiation. The 1PN approximation does not contain the
  effects of gravitational radiation back-reaction. In the expression
  for the metric given in Box~\ref{box2}, radiation back-reaction
  effects in GR do not occur until ${\cal O} (\epsilon^{7/2})$ in $g_{00}$,
  ${\cal O} (\epsilon^{3})$ in $g_{0i}$, and ${\cal O} (\epsilon^{5/2})$
  in $g_{ij}$. Consequently, in order to describe such systems, one
  must carry out a solution of the equations substantially beyond 1PN
  order, sufficient to incorporate the leading radiation damping terms
  at 2.5PN order. In addition, the PPN metric described in
  Section~\ref{ppn} is valid in the \emph{near zone} of the system,
  i.e.\ within one gravitational wavelength of the system's center of
  mass. As such it cannot describe the gravitational waves seen by a
  detector.
\item The system may be highly relativistic in its orbital motion, so
  that $U \sim v^2 \sim 1$ even for the interbody field and orbital
  velocity. Systems like this include the late stage of the inspiral
  of binary systems of neutron stars or black holes, driven by
  gravitational radiation damping, prior to a merger and collapse to a
  final stationary state. Binary inspiral is one of the leading
  candidate sources for detection by the existing LIGO-Virgo network of laser
  interferometric gravitational-wave observatories and by a future space-based interferometer. A proper description of such systems requires not only
  equations for the motion of the binary carried to extraordinarily
  high PN orders (at least 3.5PN), but also requires equations for the
  far-zone gravitational waveform measured at the detector, that are
  equally accurate to high PN orders beyond the leading ``quadrupole''
  approximation.
\end{itemize}

Of course, some systems cannot be properly described by any post-Newt\-onian
approximation because their behavior is fundamentally controlled by
strong gravity. These include the imploding cores of supernovae, the
final merger of two compact objects, the quasinormal-mode vibrations
of neutron stars and black holes, the structure of rapidly rotating
neutron stars, and so on. Phenomena such as these must be analyzed
using different techniques. Chief among these is the full solution of
Einstein's equations via numerical methods. This field of ``numerical
relativity'' has become a mature branch of gravitational
physics, whose description is beyond the scope of this review 
(see~\cite{Lehner01,2012LNP...846.....G,2010nure.book.....B} for reviews).
Another is black-hole perturbation theory 
(see~\cite{msstt97, KokkotasSchmidt99, SasakiTagoshi03,2009CQGra..26p3001B} for  reviews).

%%%%%%%%%%%%%%%%%%%%%%%%%%%%%%%%%%%%%%%%%%%%%%%%%%%%%%%%%%%%%%%%%%%%%%%%%%%%%%%%%%%

\subsubsection{Compact bodies and the strong equivalence principle}
\label{compact-SEP}

When dealing with the motion and gravitational wave generation by
orbiting bodies, one finds a remarkable simplification within GR. 
As long as the bodies are sufficiently well-separated
that one can ignore
tidal interactions and other effects that depend upon the finite extent of
the bodies (such as their quadrupole and higher multipole moments),
then all aspects of their orbital behavior and gravitational wave
generation
can be characterized by just two parameters:
mass and angular momentum.
Whether their internal structure is highly relativistic, as in black
holes or neutron stars, or non-relativistic as in the Earth and Sun,
only the mass and angular momentum are needed. Furthermore, both
quantities are measurable in principle by examining the external
gravitational field of the bodies, and make no reference whatsoever to
their interiors.

Damour~\cite{Damour300} calls this the ``effacement'' of the bodies' internal
structure. It is a consequence of the Strong Equivalence Principle (SEP), described in Section~\ref{sep}.

General relativity satisfies SEP because it contains one and only one
gravitational field, the spacetime metric $g_{\mu\nu}$. Consider the
motion of a body in a binary system, whose size is small compared to
the binary separation. Surround the body by a region that is large
compared to the size of the body, yet small compared to the
separation. Because of the general covariance of the theory, one can
choose a freely-falling coordinate system which comoves with the body,
whose spacetime metric
takes the Minkowski form at its outer boundary (ignoring tidal
effects generated by the companion).
There is thus no evidence of the presence of the companion body,
and the structure of the chosen body can be obtained using the field
equations of GR in this coordinate system. Far from the chosen body,
the metric is characterized by the mass and angular momentum (assuming
that one ignores quadrupole and higher multipole moments of the body) as
measured far from the body using orbiting test particles and gyroscopes.
These asymptotically measured quantities are oblivious to
the body's internal structure. A black hole of mass $m$ and a planet
of mass $m$ would produce identical spacetimes in this outer region.

The geometry of this region surrounding the one body must be matched to
the geometry provided by the companion body. Einstein's equations
provide consistency conditions for this matching that yield
constraints on the motion of the bodies. These are the equations of
motion. As a result, the motion of two planets of mass and angular
momentum $m_1$, $m_2$, ${\bf J}_1$, and ${\bf J}_2$ is
identical to that of two black holes of the same mass and angular
momentum (again, ignoring tidal effects).

This effacement does not occur in an alternative gravitional theory
like scalar-tensor gravity. There, in addition to the spacetime
metric, a scalar field $\phi$ is generated by the masses of
the bodies, and controls the local value of the gravitational coupling
constant (i.e.\ $G_\mathrm{Local}$ is a function of $\phi$). Now, in the local frame
surrounding one of the bodies in our binary system, while the metric
can still be made Minkowskian far away, the scalar field will take on
a value $\phi_0$ determined by the companion body. This can affect
the value of $G_\mathrm{Local}$ inside the chosen body, alter its
internal structure (specifically its gravitational binding energy)
and hence alter its mass.
Effectively, each body can be characterized by several mass 
functions $m_A(\phi)$, which depend on
the value of the scalar field at its location, and several distinct masses
come into play, such as inertial mass, gravitational mass, ``radiation'' mass,
etc. The precise nature of the
functions will depend on the body, specifically on its gravitational
binding energy, and as a result, the motion and
gravitational radiation may depend on the internal structure of each
body. For compact bodies such as neutron stars and black holes these internal
structure effects could be large; for example, the gravitational binding energy
of a neutron star can be 10\,--\,20 percent of its total mass.
At 1PN order, the leading manifestation of this phenomenon is
the Nordtvedt effect.

This is how the study of orbiting systems containing
compact objects provides strong-field tests of GR.
Even though the strong-field nature of the bodies is effaced in GR, it
is not in other theories, thus any result in agreement with the
predictions of GR constitutes a
kind of ``null'' test of strong-field gravity.

%%%%%%%%%%%%%%%%%%%%%%%%%%%%%%%%%%%%%%%%%%%%%%%%%%%%%%%%%%%%%%%%%%%%%%%%%%%%%%%%%%%
%%%%%%%%%%%%%%%%%%%%%%%%%%%%%%%%%%%%%%%%%%%%%%%%%%%%%%%%%%%%%%%%%%%%%%%%%%%%%%%%%%%

\subsection{Motion and gravitational radiation in general relativity: A history}
\label{eomgw}

At the most primitive level, the problem of motion in GR is relatively straightforward, and was an integral part of the theory as proposed by Einstein\footnote{This history is adapted from Ref.~\cite{2011PNAS..108.5938W}.  For a detailed technical and historical review of the problem of motion, see Damour~\cite{Damour300}}.  The first attempts to treat the motion of multiple bodies, each with a finite mass, were made in the period 1916--1917 by Lorentz and Droste and by de Sitter~\cite{LorentzDroste,1916MNRAS..77..155D}.  They derived the metric and   equations of motion for a system of $N$ bodies, in what today would be called the first post-Newtonian approximation of GR (de Sitter's equations turned out to contain some important errors).   In 1916, Einstein took the first crack at a study of gravitational radiation, deriving the energy emitted by a body such as a rotating rod or dumbbell, held together by non-gravitational forces~\cite{1916SPAW.......688E}.  He made some unjustified assumptions as well as a trivial numerical error (later corrected by Eddington~\cite{1922RSPSA.102..268E}), but the underlying conclusion that dynamical systems would radiate gravitational waves was correct. 

The next significant advance in the problem of motion came 20 years later.  In 1938, Einstein, Infeld and Hoffman published the now legendary ``EIH'' paper, a calculation of the $N$-body equations of motion using only the vacuum field equations of GR~\cite{EIH}.  They treated each body in the system as a spherically symmetric object whose nearby vacuum exterior geometry approximated that of the Schwarzschild metric of a static spherical star.   They then solved the vacuum field equations for the metric between each body in the system in a weak field, slow-motion approximation.  Then, using a primitive version of what today would be called ``matched asymptotic expansions'' they showed that, in order for the nearby metric of each body to match smoothly to the interbody metric at each order in the expansion, certain conditions on the motion of each body had to be met.  Together, these conditions turned out to be equivalent to the Droste-Lorentz $N$-body equations of motion.  The internal structure of each body was irrelevant, apart from the requirement that its nearby field be approximately spherically symmetric, a clear illustration of the ``effacement'' principle.

Around the same time, there occurred an unusual detour in the problem of motion.  Using equations of motion based on de Sitter's paper, specialized to two bodies, 
Levi-Civita~\cite{LeviCivita} showed that the center of mass of a binary star system would suffer an acceleration in the direction of the pericenter of the orbit, in an amount proportional to the difference between the two masses, and to the eccentricity of the orbit.  Such an effect would be a violation of the conservation of momentum for isolated systems caused by relativistic gravitational effects.  Levi-Civita even went so far as to suggest looking for this effect in selected nearby close binary star systems.  However,  Eddington and Clark~\cite{1938RSPSA.166..465E} quickly pointed out that Levi-Civita had based his calculations on de Sitter's flawed work; when correct two-body equations of motion were used, the effect vanished, and momentum conservation was upheld.  Robertson confirmed this using the EIH equations of motion~\cite{robertson38}.   Such an effect can only occur in theories of gravity that lack the appropriate conservation laws (Sec.\ \ref{conservation}).

There was ongoing confusion over whether gravitational waves are real or are artifacts of general
covariance.  Although Eddington was credited with making the unfortunate remark that gravitational waves propagate ``with the speed of thought'', he did clearly elucidate the difference between the physical, coordinate independent modes and modes that were purely coordinate artifacts~\cite{1922RSPSA.102..268E}.   But in 1936, in a paper submitted to the {\em Physical Review}, Einstein and Rosen claimed to prove that gravitational waves could not exist; the anonymous referee of their paper found that they had made an error.  Upset that the journal had sent his paper to a referee (a newly instituted practice), Einstein refused to publish there again.  A corrected paper by Einstein and Rosen showing that gravitational waves {\em did} exist -- cylindrical waves in this case -- was published elsewhere~\cite{1937FrInJ.223...43E}.   Fifty years later it was revealed that the anonymous referee was H. P. Robertson~\cite{2005PhT....58i..43K}. 

Roughly 20 more years would pass before another major attack  on the problem of motion.  Fock in the USSR and Chandrasekhar in the US independently developed and systematized the post-Newtonian approximation in a form that laid the foundation for modern post-Newtonian theory~\cite{fock,1965ApJ...142.1488C}.  They developed a full post-Newtonian hydrodynamics, with the ability to treat realistic, self-gravitating bodies of fluid, such as stars and planets.  In the suitable limit of ``point'' particles, or bodies whose size is small enough compared to the interbody separations that finite-size effects such as spin and tidal interactions can be ignored, their equations of motion could be shown to be equivalent to the EIH and the Droste-Lorentz equations of motion.  

The next important period in the history of the problem of motion was 1974 --1979, initiated by the 1974 discovery of the binary pulsar PSR 1913+16 by Hulse and Taylor~\cite{1975ApJ...195L..51H}.  Around the same time there occurred the first serious attempt to calculate the head-on collision of two black holes using purely numerical solutions of Einstein's equations, by Smarr and collaborators~\cite{1976PhRvD..14.2443S}.  

The binary pulsar consists of two neutron stars, one an active pulsar detectable by radio telescopes, the other very likely an old, inactive pulsar (Sec.~\ref{binarypulsars}).  Each neutron star has a mass of around 1.4 solar masses.  The orbit of the system was seen immediately to be quite relativistic, with an orbital period of only eight hours, and a mean orbital speed of 200 km/s, some four times faster than Mercury in its orbit.  Within weeks of its discovery, numerous authors pointed out that PSR 1913+16 would be an important new testing ground for GR.  In particular, it could provide for the first time a test of the effects of the emission of gravitational radiation on the orbit of the system.  

However, the discovery revealed an ugly truth about the ``problem of motion''.  As Ehlers {\em et al.} pointed out in an influential 1976 paper~\cite{1976ApJ...208L..77E}, the general relativistic problem of motion and radiation was full of holes large enough to drive trucks through.  They pointed out that most treatments of the problem used ``delta functions'' as a way to approximate the bodies in the system as point masses.   As a consequence, the ``self-field'', the gravitational field of the body evaluated at its own location, becomes infinite.   While this is not a major issue in Newtonian gravity or classical electrodynamics, the non-linear nature of GR requires that this infinite self-field contribute to gravity.   In the past, such infinities had been simply swept under the rug.   Similarly, because gravitational energy itself produces gravity it thus acts as a source throughout spacetime.  This means that, when calculating radiative fields, integrals for the multipole moments of the source that are so useful in treating radiation begin to diverge.  These divergent integrals had also been routinely swept under the rug.  Ehlers {\em et al.} further pointed out that the true boundary condition for any problem involving radiation by an isolated system should be one of ``no incoming radiation'' from the past.  Connecting this boundary condition with the routine use of retarded solutions of wave equations was not a trivial matter in GR. 
Finally they pointed out that there was no evidence that the post-Newtonian approximation, so central to the problem of motion, was a convergent or even asymptotic sequence.  
Nor had the approximation been carried out to high enough order to make credible error estimates.

During this time, some authors even argued that the ``quadrupole formula'' for the gravitational energy emitted by a system (see below), while correct for a rotating dumbell as calculated by Einstein, was actually {\em wrong} for a binary system moving under its own gravity.  The discovery in 1979 that the rate of decay of the orbit of the binary pulsar was in agreement with the standard quadrupole formula made some of these arguments moot.  Yet the question raised by Ehlers {\em et al.} was still relevant: is the quadrupole formula for binary systems an actual prediction of GR?

Motivated by the Ehlers {\em et al.} critique, numerous workers began to address the holes in the problem of motion, and by the late 1990s most of the criticisms had been answered, particularly those related to divergences.   For a detailed history of the ups and downs of the subject of motion and gravitational waves, see~\cite{2007tste.book.....K}.

The problem of motion and radiation in GR has received renewed interest
since 1990, with proposals for construction of
large-scale laser interferometric
grav\-it\-at\-ional wave observatories.  These proposals culminated in the construction and operation of LIGO in the US,
VIRGO and GEO600 in Europe, and TAMA300 in Japan, the construction of an underground observatory KAGRA in Japan, and the possible construction of a version of LIGO in India.  Advanced versions of LIGO and VIRGO are expected to be online and detecting gravitational waves around 2016.   An interferometer in space has recently been selected by the European Space Agency for a launch in the 2034 time frame.

A leading
candidate source of detectable waves is the inspiral, driven
by gravitational radiation damping,
of a binary system of compact objects (neutron stars
or black holes) (for a review of sources of gravitational waves, see~\cite{2009LRR....12....2S}). The
analysis of signals from such systems
will require theoretical predictions from GR that are
extremely accurate, well beyond the leading-order prediction of
Newtonian or even post-Newtonian gravity for the orbits, and well beyond
the leading-order formulae for gravitational waves.

This presented a major theoretical challenge: to calculate the motion
and radiation of systems of compact objects
to very high PN order, a formidable algebraic task,
while addressing the issues of principle raised by Ehlers {\em et al.}, sufficiently well
to ensure that the results were physically meaningful. This
challenge has been largely met, so that we
may soon see a remarkable convergence between observational
data and accurate predictions of gravitational theory that could
provide new, strong-field tests of GR.

%%%%%%%%%%%%%%%%%%%%%%%%%%%%%%%%%%%%%%%%%%%%%%%%%%%%%%%%%%%%%%%%%%%%%%%%%%%%%%%%%%%
%%%%%%%%%%%%%%%%%%%%%%%%%%%%%%%%%%%%%%%%%%%%%%%%%%%%%%%%%%%%%%%%%%%%%%%%%%%%%%%%%%%

\subsection{Compact binary systems in general relativity}
\label{compactbinaryGR}

\subsubsection{Einstein's equations in ``relaxed'' form}
\label{EErelaxed}

Here we give a brief overview of the modern approach to the problem of motion
and gravitational radiation in GR.  For a full pedagogical treatment, see~\cite{PW2014}.

The Einstein equations $G_{\mu\nu} = 8\pi G T_{\mu\nu}$ are elegant and
deceptively simple, showing geometry (in the form of the Einstein
tensor $G_{\mu\nu}$, which is a function of spacetime curvature)
being generated by matter (in the form of the material stress-energy tensor
$T_{\mu\nu}$). However, this is not the most useful form for actual
calculations. For post-Newtonian calculations, a far more useful form
is the so-called ``relaxed'' Einstein equations, which form the basis of the program of approximating solutions of Einstein's equations known as post-Minkowskian theory and post-Newtonian theory.  The starting point is the so-called ``gothic inverse metric'', defined by $\mathfrak{g}^{\alpha\beta} \equiv \sqrt{-g} g^{\alpha\beta}$, where $g$ is the determinant of $g_{\alpha\beta}$.   One then defines the  gravitational potential $h^{\alpha \beta} \equiv \eta^{\alpha \beta} - \mathfrak{g}^{\alpha
\beta}$.  After imposing the the de~Donder
or harmonic gauge condition
$\partial h^{\alpha \beta} /\partial x^\beta =0$ (summation on
repeated indices is assumed), one can recast the exact Einstein field equations into the form
\begin{equation}
  \Box h^{\alpha\beta} = - 16 \pi G \tau^{\alpha\beta},
  \label{relaxed}
\end{equation}
where $\Box \equiv  -{\partial}^2 / \partial t^2 + {\nabla}^2 $
is the flat-spacetime wave operator.
This form of Einstein's equations bears a striking similarity to Maxwell's
equations for the vector potential $A^\alpha$ in Lorentz gauge: $\Box
A^\alpha = -4\pi J^\alpha$, $\partial A^\alpha /\partial
x^\alpha =0$. There is a key difference, however: The source on the right
hand side of Eq.~ (\ref{relaxed}) is given by the ``effective''
stress-energy pseudotensor
\begin{equation}
  \tau^{\alpha\beta} =
  (-g) \left (T^{\alpha\beta} + t_{\rm LL}^{\alpha\beta} + t_{\rm H}^{\alpha\beta} \right ),
  \label{effective}
\end{equation}
where $t_{\rm LL}^{\alpha\beta}$ and $t_{\rm H}^{\alpha\beta}$ are  the Landau-Lifshitz pseudotensor and a harmonic pseudotensor,
given by terms quadratic (and higher) in $h^{\alpha \beta}$ and its
derivatives (see~\cite{PW2014}, Eqs.~(6.5, 6.52, 6.53) for explicit formulae).
In GR, the gravitational field itself generates
gravity, a reflection of the nonlinearity of Einstein's equations, and
in
contrast to the linearity of Maxwell's equations.

Eq.~ (\ref{relaxed}) is exact, and depends only on the assumption
that the relevant parts of spacetime can be covered by harmonic coordinates. It is called
``relaxed'' because it
can be solved formally as a functional of source variables without
specifying the motion of the source, in the form (with $G =1$)
\begin{equation}
  h^{\alpha \beta} (t, {\bf x}) = 4 \int_{\cal C}
  \frac{\tau^{\alpha \beta} (t -| {\bf x} - {\bf x'} |, {\bf x'})}
  {|{\bf x} - {\bf x'}|} \, d^3x',
  \label{nearintegral}
\end{equation}
where the integration is over the past flat-spacetime null cone $\cal
C$ of the field point $(t,{\bf x})$.
The motion of the source is then determined either by the equation
$\partial {\tau}^{\alpha \beta} /\partial x^\beta =0$ (which follows
from the harmonic gauge condition), or from the usual covariant
equation of motion ${T^{\alpha\beta}}_{;\beta}=0$, where the subscript
$;\beta$ denotes a covariant divergence.
This formal solution can then be iterated in a slow motion ($v<1$)
weak-field ($||h^{\alpha \beta}||<1$) approximation. One begins by
substituting
$h_0^{\alpha \beta} =0$ into the source $\tau^{\alpha \beta}$ in
Eq.~ (\ref{nearintegral}), and
solving for the first iterate $h_1^{\alpha \beta}$, and then repeating the
procedure sufficiently many times to achieve a solution of the desired
accuracy. For example, to obtain the 1PN equations of motion, {\it
two} iterations are needed (i.e.\ $h_2^{\alpha \beta}$ must be
calculated); likewise, to obtain the leading gravitational waveform
for a binary system, two iterations are needed.

At the same time, just as in electromagnetism, the formal
integral~(\ref{nearintegral}) must be handled differently, depending
on whether the field point is in the far zone or the near zone. For
field points in the far zone or radiation zone, $|{\bf x}| >
{\cal R}$, where
$\cal R$ is a distance of the order of a gravitational wavelength, the field can be expanded in inverse powers of
$R=|{\bf x}|$ in a multipole expansion, evaluated at the ``retarded
time'' $t-R$. The leading term in $1/R$ is
the gravitational waveform. For field points in the near zone or
induction zone, $|{\bf x}| \sim |{\bf x}'| <
{\cal R}$, the field is expanded in
powers of $|{\bf x}-{\bf x}'|$ about the local time $t$,
yielding instantaneous potentials that go into the equations of
motion.

However, because the source ${\tau}^{\alpha \beta}$ contains
$h^{\alpha \beta}$ itself, it is not confined to a compact region, but
extends over all spacetime. As a result, there is a danger that the
integrals involved in the various expansions will diverge or be
ill-defined. This consequence of the non-linearity of Einstein's
equations has bedeviled the subject of gravitational radiation for
decades. Numerous approaches have been developed to try to
handle this difficulty. The post-Minkowskian method of Blanchet,
Damour, and Iyer~\cite{bd86, bd88, bd89, di91, bd92, blanchet95}
solves Einstein's equations by two
different techniques, one in the near zone and one in the far zone,
and uses the method of singular asymptotic matching to join the
solutions in an overlap region. The method provides a natural
``regularization'' technique to control potentially divergent
integrals (see~\cite{BlanchetLRR} for a thorough review). 
The ``Direct Integration of the Relaxed Einstein
Equations'' (DIRE) approach of Will, Wiseman, and Pati~\cite{opus, DIRE,2002PhRvD..65j4008P} retains
Eq.~ (\ref{nearintegral}) as the global solution, but splits the
integration into one over the near zone and another over the far zone,
and uses different
integration variables to carry out the explicit integrals over the two
zones. In the DIRE method, all integrals are finite and convergent.
Itoh and Futamase  used an extension of the Einstein--Infeld--Hoffman
matching approach combined with a specific method for taking a
point-particle limit~\cite{ItohFutamase03}, while Damour, Jaranowski, and
Sch\"afer pioneered an ADM Hamiltonian approach that focuses on
the equations of 
motion~\cite{jaraschaefer98, jaranowski, damjaraschaefer, DJSdim, DJSequiv}.

These methods assume from the outset that gravity is
sufficiently weak that $||h^{\alpha\beta}||<1$ and harmonic
coordinates exists everywhere, including inside the bodies. Thus, in
order to apply the results to cases where the bodies may be neutron
stars or black holes, one relies upon the SEP
to argue that, if tidal forces are ignored, and equations are
expressed in terms of masses and spins,  one can simply
extrapolate the results unchanged
to the situation where the bodies are ultrarelativistic.
While no general proof of this exists, it has been shown to be valid
in specific circumstances, such as through 2PN order in the equations of
motion~\cite{1983SvAL....9..230G,2007PhRvD..75l4025M}, and for black holes moving in a Newtonian 
background field~\cite{Damour300}.

Methods such as these
have resolved most of the issues that led to criticism of the
foundations of gravitational radiation theory during the 1970s.

%%%%%%%%%%%%%%%%%%%%%%%%%%%%%%%%%%%%%%%%%%%%%%%%%%%%%%%%%%%%%%%%%%%%%%%%%%%%%%%%%%%
%%%%%%%%%%%%%%%%%%%%%%%%%%%%%%%%%%%%%%%%%%%%%%%%%%%%%%%%%%%%%%%%%%%%%%%%%%%%%%%%%%%

\subsubsection{Equations of motion and gravitational waveform}
\label{eomwaveform}

Among the results of these approaches are formulae for the equations of
motion and gravitational waveform of binary systems of compact
objects, carried out to high orders in a PN expansion.  For a review of the latest results of high-order PN calculations, see~\cite{BlanchetLRR}.
Here we shall
only state the key formulae that will be needed for this review.  
For example,
the relative two-body equation of motion has the form
\begin{equation}
  {\bf a} = \frac{d{\bf v}}{dt} = \frac{m}{r^2}
  \left\{- {\bf \hat{n}} + {\bf A}_{1\mathrm{PN}} +
  {\bf A}_{2\mathrm{PN}} + {\bf A}_{2.5\mathrm{PN}} +
  {\bf A}_{3\mathrm{PN}} + {\bf A}_{3.5\mathrm{PN}} + \dots \right\},
  \label{EOM}
\end{equation}
where $m=m_1+m_2$ is the total mass, $r= |{\bf x}_1 -{\bf x}_2|$,
${\bf v}={\bf v}_1-{\bf v}_2$, and
${\bf \hat{n}} = ({\bf x}_1 -{\bf x}_2)/r$.
The notation ${\bf A}_{n\mathrm{PN}}$ indicates that the term is
${\cal O} (\epsilon^n)$ relative to the Newtonian term $-{\bf \hat{n}}$.
Explicit and unambiguous
formulae for non-spinning bodies through 3.5PN order have been
calculated by
various authors~\cite{BlanchetLRR}.
Here we quote only the 1PN corrections and the
leading radiation-reaction terms at 2.5PN order:
\begin{eqnarray}
  {\bf A}_{1\mathrm{PN}} & = & \left\{ (4 + 2 \eta) \frac{m}{r} -
  (1 + 3 \eta) v^2 + \frac{3}{2} \eta \dot{r}^2 \right\} {\bf \hat{n}} +
  (4 - 2 \eta) \dot{r} {\bf v},
  \label{APN} \\
  {\bf A}_{2.5\mathrm{PN}} & = &  \frac{8}{5} \eta \frac{m}{r}
  \left\{ \left( 3 v^2 + \frac{17}{3} \frac{m}{r} \right) \dot r {\bf \hat{n}} -
  \left(  v^2 + 3 \frac{m}{r} \right) {\bf v} \right\},
  \label{A2.5PN}
\end{eqnarray}%
where $\eta = m_1m_2/(m_1+m_2)^2$.  The radiation-reaction acceleration is expressed in the so-called Damour-Deruelle gauge.
These terms are sufficient to analyze the orbit and evolution of the
binary pulsar (see Section~\ref{binarypulsars}). For example, the 1PN terms are
responsible for
the periastron advance of an eccentric orbit, given by 
\begin{equation}
\dot{\omega} =  \frac{6\pi \alpha m f_b}{a(1-e^2)} \,.
\end{equation}
where $a$ and $e$ are the semi-major axis and
eccentricity of the orbit, respectively, and $f_\mathrm{b}$ is the orbital
frequency, given to the needed order by Kepler's third law
$2 \pi f_\mathrm{b} = (m/a^3)^{1/2}$.

Another product is a formula for the gravitational field
far from the system, written schematically in the form
\begin{equation}
  h^{ij} = \frac{2 m}{R} \left\{ Q^{ij} + Q_{0.5\mathrm{PN}}^{ij} +
  Q_{1\mathrm{PN}}^{ij} + Q_{1.5\mathrm{PN}}^{ij} +
  Q_{2\mathrm{PN}}^{ij} + Q_{2.5\mathrm{PN}}^{ij} + \dots \right\},
  \label{waveform}
\end{equation}
where $R$ is the distance from the source, and the variables
are to be evaluated at retarded time $t-R$. The leading term
is the so-called quadrupole formula
\begin{equation}
  h^{ij} (t, {\bf x}) = \frac{2}{R} \ddot{I}^{ij} (t - R),
  \label{waveformquad}
\end{equation}
where $I^{ij}$ is the quadrupole moment of the source, and overdots
denote time derivatives. For a binary system this leads to
\begin{equation}
  Q^{ij} = 2 \eta \left( v^i v^j - \frac{m \hat{n}^i \hat{n}^j}{r} \right).
  \label{Qij}
\end{equation}
For binary systems, explicit
formulae for the waveform through 3.5PN order have been derived 
(see~\cite{bdiww} for a ready-to-use presentation of the waveform to 2PN order for
circular orbits; see~\cite{BlanchetLRR} for a full review).

Given the gravitational waveform, one can
compute the rate at which energy is carried off by the radiation
(schematically $\int \dot{h} \dot{h} \, d\Omega$,
the gravitational analog of the Poynting
flux).
The lowest-order quadrupole formula leads to the
gravitational wave energy flux
\begin{equation}
  \dot E = \frac{8}{15} \frac{\mu\eta}{r} \left (\frac{m}{r} \right )^3
  \left( 12 v^2 - 11 \dot{r}^2 \right).
  \label{EdotGR}
\end{equation}
This has been extended to 3.5PN order beyond the quadrupole formula~\cite{BlanchetLRR}.
Formulae for fluxes of angular and linear momentum can also be
derived.
The 2.5PN radiation-reaction terms in the equation of motion~(\ref{EOM})
result in
a damping of the orbital energy that precisely balances the energy
flux~(\ref{EdotGR})
determined from the waveform. Averaged over one orbit, this results in
a rate of increase of the binary's orbital frequency given by
\begin{equation}
  \begin{array}{rcl}
    \dot f_\mathrm{b} & = & \displaystyle
    \frac{192 \pi}{5} f_\mathrm{b}^2
    (2 \pi {\cal M} f_\mathrm{b})^{5/3} F (e),
    \\ [0.5 em]
    F (e) & = & \displaystyle (1 - e^2)^{-7/2}
    \left( 1 + \frac{73}{24} e^2 + \frac{37}{96} e^4 \right),
  \end{array}
  \label{fdotGR}
\end{equation}
where ${\cal M}$ is the so-called ``chirp'' mass, given by ${\cal
M}=\eta^{3/5} m$.
Notice that by making precise measurements of the phase $\Phi (t) = 2\pi
\int^t f(t') \, dt'$ of either the orbit or the gravitational waves
(for which $f =2f_\mathrm{b}$ for the dominant component) as a function of
the frequency, one in effect measures the ``chirp'' mass of the
system.

These formalisms have also been generalized to include the leading effects of
spin-orbit and spin-spin coupling between 
the bodies as well as many next-to-leading-order corrections~\cite{BlanchetLRR}.

Another approach to gravitational radiation is applicable to the special
limit in which one mass is much smaller than the other.
This is the method of black hole perturbation theory. One begins with an
exact background spacetime of a black hole, either the non-rotating
Schwarzschild or the rotating Kerr solution, and perturbs it according to
$g_{\mu\nu}=g^{(0)}_{\mu\nu} + h_{\mu\nu}$. The particle moves on a
geodesic of the background spacetime, and a suitably defined source
stress-energy tensor for the particle acts as a source for the gravitational
perturbation and wave field $h_{\mu\nu}$. This method provides
numerical results that are exact in $v$, as well as analytical results
expressed as series in powers of $v$, both for
non-rotating and for rotating black holes. For
non-rotating holes, the analytical expansions have been carried to the impressive level of 
22PN order, or $\epsilon^{22}$ beyond the
quadrupole approximation~\cite{2012PThPh.128..971F}, and for rotating Kerr black holes, to 20PN order~\cite{2014arXiv1403.2697S}. All results of black hole
perturbation agree precisely with the $m_1 \to 0$ limit of the PN results,
up to the highest PN order where they can be compared (for reviews of earlier work
see~\cite{msstt97, KokkotasSchmidt99, SasakiTagoshi03}).

%%%%%%%%%%%%%%%%%%%%%%%%%%%%%%%%%%%%%%%%%%%%%%%%%%%%%%%%%%%%%%%%%%%%%%%%%%%%%%%%%%%
%%%%%%%%%%%%%%%%%%%%%%%%%%%%%%%%%%%%%%%%%%%%%%%%%%%%%%%%%%%%%%%%%%%%%%%%%%%%%%%%%%%

\subsection{Compact binary systems in scalar-tensor theories}
\label{compactbinaryST}

Because of the recent resurgence of interest in scalar-tensor theories of gravity, motivated in part by string theory and $f(R)$ theories, considerable work has been done to analyze the motion and gravitational radiation from systems of compact objects in this class of theories.   In earlier work, Eardley~\cite{1975ApJ...196L..59E} was the first to point out the existence of dipole gravitational radiation from self-gravitating bodies in Brans-Dicke theory, and Will~\cite{1977ApJ...214..826W} worked out the lowest-order monopole, dipole and quadrupole radiation flux in general scalar-tensor theories (as well as in a number of alternative theories) for bodies with weak self-gravity.   Using the approach pioneered by Eardley~\cite{1975ApJ...196L..59E} for incorporating strongly self-gravitating bodies into scalar-tensor calculations, Will and Zaglauer~\cite{zaglauer} calculated the
1PN equations of motion along with the monopole-quadrupole and dipole energy flux for compact binary systems; Alsing {\em et al.}~\cite{2012PhRvD..85f4041A} extended these results to the case of Brans-Dicke theory with a massive scalar field.  However, the expressions for the energy flux in those works were incomplete, because they failed to include some important post-Newtonian corrections in the scalar part of the radiation that actually contribute at the same order as the quadrupole contributions from the tensor part.   Damour and Esposito-Far\`ese~\cite{1996PhRvD..53.5541D} obtained the correct monopole-quadrupole and dipole energy flux, working in the Einstein-frame representation of scalar-tensor theories, and gave partial results for the equations of motion to 2PN order.  Mirshekari and Will~\cite{2013PhRvD..87h4070M} obtained the complete compact-binary equations of motion in general scalar-tensor theories through 2.5PN order, and obtained the energy loss rate in complete agreement with the flux result from Damour and Esposito-Far\`ese. Lang~\cite{2013arXiv1310.3320L} obtained the gravitational-wave signal to 2PN order.   

Notwithstanding the very tight bound on the scalar-tensor coupling parameter $\omega$ from Cassini measurements in the solar system, this effort is motivated by a desire to test this theory in strong-field situations, whether by binary pulsar observations, or by measurements of gravitational radiation from compact binary inspiral.  Here we summarize the key results in a manner that parallels the results for GR.

\subsubsection{Scalar-tensor equations in ``relaxed'' form}
\label{sec:relaxedST}

The field equations of scalar-tensor theory can be cast in a form similar to the ``relaxed'' equations of GR.  Here one works in terms of an auxiliary metric $\tilde{g}_{\alpha\beta} \equiv \varphi {g}_{\alpha\beta}$, where $\varphi \equiv (\phi/\phi_0)$ and $\phi_0$ is the asymptotic value of the scalar field, and defines the auxiliary gothic inverse metric $\tilde{\mathfrak{g}}^{\alpha\beta} \equiv \sqrt{-\tilde{g}} \tilde{g}^{\alpha\beta}$, and the auxiliary tensor gravitational potential $\tilde{h}^{\alpha\beta} \equiv \eta^{\alpha \beta} -\tilde{\mathfrak{g}}^{\alpha\beta}$, along with the harmonic gauge condition $\partial \tilde{h}^{\alpha\beta}/\partial x^\beta =0$.  The field equations then take the form 
\begin{equation}
  \Box \tilde{h}^{\alpha\beta} = - 16 \pi G \tilde{\tau}^{\alpha\beta},
  \label{relaxedST}
\end{equation}
where $\Box \equiv  -{\partial}^2 / \partial t^2 + {\nabla}^2 $
is again the flat-spacetime wave operator, and where
\begin{equation}
 \tilde{\tau}^{\alpha\beta} =
  (-\tilde{g}) \left ( \frac{\varphi}{\phi_0}{T}^{\alpha\beta} + \tilde{t}_\phi^{\alpha\beta} +  \tilde{t}_{\rm LL}^{\alpha\beta} + \tilde{t}_{\rm H}^{\alpha\beta} \right ),
  \label{effectiveST}
\end{equation}
where $ \tilde{t}_\phi^{\alpha\beta} \equiv (3+2\omega) \varphi^{-2} \varphi_{,\mu} \varphi_{,\nu}(\tilde{g}^{\mu\alpha} \tilde{g}^{\nu\beta} - \frac{1}{2}\tilde{g}^{\mu\nu}\tilde{g}^{\alpha\beta})$ is a scalar stress-energy tensor, and
where $\tilde{t}_{\rm LL}^{\alpha\beta}$ and $\tilde{t}_{\rm H}^{\alpha\beta}$ have exactly the same forms, when written in terms of  $\tilde{h}^{\alpha \beta}$, as their counterparts in GR do in terms of ${h}^{\alpha \beta}$.  Note that this is equivalent to formulating the relaxed equations of scalar-tensor theory in the Einstein conformal frame.  The field equation for the scalar field can be written in the form $\Box \varphi = -8\pi G \tilde{\tau}_s$, where $\tilde{\tau}_s$ is a source consisting of a matter term, a scalar energy density term and a term that mixes $\tilde{h}^{\alpha\beta}$ and $\varphi$ (see~\cite{2013PhRvD..87h4070M} for details).

In order to incorporate the internal gravity of compact, self-gravitating bodies, it is common to adopt an 
approach pioneered by Eardley~\cite{1975ApJ...196L..59E}, based in part on general arguments dating back to  Dicke, 
in which one treats the matter energy-momentum tensor as a sum of delta functions located at the position of each body, 
but assumes that the mass of each body is a function $M_A(\phi)$ of the scalar field.  This reflects the fact that the gravitational binding energy of the body is controlled by the value of the gravitational constant, which is directly related to the value of the background scalar field in which the body finds itself.  The underlying assumption is that the timescale for orbital motion is long compared to the internal dynamical timescale of the body, so that the body's structure evolves adiabatically in response to the changing scalar field.  Consequently, the matter action will have an {\em effective} dependence on $\phi$, and as a result the field equations will depend on the ``sensitivity'' of the mass of each body to variations in the scalar field, holding the total number of baryons fixed.  The sensitivity of body $A$ is defined by
\begin{equation}
s_A \equiv \left ( \frac{d \ln M_A(\phi)}{d \ln \phi} \right ) \,,
\label{sensitivity}
\end{equation}
evaluated at a value of the scalar field far from the body.
For neutron stars, the sensitivity depends on the mass and equation of state of the star and is typically of order $0.2$; in the weak-field limit, $s_A$ is proportional to the Newtonian self-gravitational energy per unit mass of the body.  From a theorem of Hawking~\cite{1972CMaPh..25..167H}, for stationary black holes, it is known that $s_{\rm BH} = 1/2$.   This means, among other things, that the source $\tilde{\tau}_s$ for the scalar field will contain an explicit term dependent upon $\partial T/\partial \phi$, because of the dependence on $M_A(\phi)$.

\subsubsection{Equations of motion and gravitational waveform}

By following the methods of post-Minkowskian theory adapted to scalar-tensor theory, it has been possible to derive the equations of motion for binary systems of compact bodies to 2.5PN order~\cite{2013PhRvD..87h4070M} and the gravitational-wave signal and energy flux to 1PN order beyond the quadrupole approximation.   Here we shall quote selected results in parallel with those quoted in Sec.\ \ref{eomwaveform}.   The relative two-body equation of motion has the form
\begin{equation}
  {\bf a} = \frac{d{\bf v}}{dt} = \frac{\alpha m}{r^2}
  \left\{- {\bf \hat{n}} + {\bf A}_{1\mathrm{PN}} + {\bf A}_{1.5\mathrm{PN}} +
  {\bf A}_{2\mathrm{PN}} + {\bf A}_{2.5\mathrm{PN}} +
  {\bf A}_{3\mathrm{PN}} + {\bf A}_{3.5\mathrm{PN}} + \dots \right\} \,.
  \label{EOM_ST}
\end{equation}
The key difference between this PN series and that in GR is the presence of a radiation-reaction term at 1.5PN order, caused by the emission of dipole gravitational radiation.   The key parameters that appear in the two-body equations of motion are given in Table \ref{tab:STparams}.  Notice that $\alpha$ plays the role of a two-body gravitational interaction parameter; $\bar{\gamma}$ and $\bar{\beta}_A$ are the two-body versions of $\gamma-1$ and $\beta-1$ respectively.  In the limit of weakly self-gravitating bodies ($s_A \to 0$), $\alpha \to 1$, $\bar{\gamma} \to \gamma -1 = -2\zeta$ and $\bar{\beta}_A \to \beta -1 = \zeta \lambda$ (compare with Table \ref{ppnvalues}).

\begin{table}
\caption{\label{tab:STparams} Parameters used in the equations of motion}
\centering
\begin{tabular}{cl}
\hline \hline
Parameter&Definition\\
\hline \hline
\multicolumn{2}{l}{\bf Scalar-tensor parameters}\\
$\zeta$&$1/(4+2\omega_0)$\\
$\lambda$&$(d\omega/d\varphi)_0 \zeta^2/(1-\zeta)^2$\\
\multicolumn{2}{l}{\bf Sensitivities}\\
$s_A$&$[d \ln M_A(\phi)/d \ln \phi]_0$\\
$s'_A$&$[d^2 \ln M_A(\phi)/d \ln \phi^2]_0$\\
\multicolumn{2}{l}{\bf Equation of motion parameters}\\
$\alpha $&$1 - \zeta + \zeta (1-2s_1)(1- 2s_2) $
\\
$\bar{\gamma}$ & $-2 \alpha^{-1}\zeta (1-2s_1)(1-2s_2)$
\\
$\bar{\beta}_1 $&$\alpha^{-2} \zeta (1-2s_2)^2 \left ( \lambda (1-2s_1) + 2 \zeta s'_1 \right )$
\\
$\bar{\beta}_2 $&$\alpha^{-2} \zeta (1-2s_1)^2 \left ( \lambda (1-2s_2) + 2 \zeta s'_2 \right )$
\\
\hline\hline
\end{tabular}
\end{table}

Here we quote only the  1PN corrections and the
leading radiation-reaction terms at 1.5PN and 2.5PN order:
\begin{eqnarray}
  {\bf A}_{1\mathrm{PN}} & = & \left\{ (4 + 2 \eta + 2\bar{\gamma} + 2 \bar{\beta}_{+} - 2\psi\bar{\beta}_{-}) \frac{\alpha m}{r} -
  (1 + 3 \eta + \bar{\gamma}) v^2 + \frac{3}{2} \eta \dot{r}^2 \right\} {\bf \hat{n}} 
  \nonumber \\
  && \quad+
  (4 - 2 \eta +2\bar{\gamma}) \dot{r} {\bf v},
  \label{APNST} \\
  {\bf A}_{1.5\mathrm{PN}} & = & \frac{4}{3} \eta \frac{\alpha m}{r} \zeta {\cal S}_{-}^2 (3 \dot{r} {\bf \hat{n}} -{\bf v} ) \,,
  \label{A1.5PNST}
  \\
  {\bf A}_{2.5\mathrm{PN}} & = &  \frac{8}{5} \eta \frac{\alpha m}{r}
  \left\{ \left( a_1 v^2 + a_2 \frac{\alpha m}{r} + a_3 \dot{r}^2 \right) \dot r {\bf \hat{n}} -
  \left( b_1 v^2 + b_2 \frac{m}{r} + b_3 \dot{r}^2  \right) {\bf v} \right\} \,,
  \label{A2.5PNST}
\end{eqnarray}%
where
\begin{eqnarray}
a_1 &=& 3 - \frac{5}{2} \bar{\gamma}  + \frac{15}{2} \bar{\beta}_+   +\frac{5}{8} \zeta {\cal S}_{-}^2 (9 + 4\bar{\gamma} -2\eta)
+ \frac{15}{8} \zeta \psi  {\cal S}_{-} {\cal S}_{+} \,,
\nonumber \\
a_2 &=& \frac{17}{3}  + \frac{35}{6} \bar{\gamma} - \frac{95}{6} \bar{\beta}_+ 
- \frac{5}{24}\zeta {\cal S}_{-}^2  \left [ 135 + 56\bar{\gamma} + 8\eta + 32\bar{\beta}_+  \right ]
 +30 \zeta {\cal S}_{-} \left ( \frac{{\cal S}_{-} \bar{\beta}_+ + {\cal S}_{+} \bar{\beta}_{-}}{\bar{\gamma}} \right )
 \nonumber \\
&&  
-\frac{5}{8} \zeta \psi  {\cal S}_{-} \left ( {\cal S}_{+} - \frac{32}{3} {\cal S}_{-} \bar{\beta}_{-} +16 \frac{{\cal S}_{+} \bar{\beta}_+ + {\cal S}_{-} \bar{\beta}_{-}}{\bar{\gamma}} \right )  -40 \zeta \left (\frac{{\cal S}_{+} \bar{\beta}_+ + {\cal S}_{-} \bar{\beta}_{-}}{\bar{\gamma}} \right )^2 \,,
\nonumber \\
a_3 &=& \frac{25}{8}  \left [ 2\bar{\gamma} - \zeta  {\cal S}_{-}^2 (1-2\eta)
  - 4\bar{\beta}_+ - \zeta \psi {\cal S}_{-} {\cal S}_{+}\right ]  \,, 
\nonumber \\
b_1 &=& 1 - \frac{5}{6}\bar{\gamma} +\frac{5}{2} \bar{\beta}_+ 
-\frac{5}{24} \zeta {\cal S}_{-}^2 (7 + 4\bar{\gamma} -2\eta)
+ \frac{5}{8} \zeta \psi  {\cal S}_{-} {\cal S}_{+} \,,
\nonumber \\
b_2 &=& 3 + \frac{5}{2} \bar{\gamma} -\frac{5}{2} \bar{\beta}_+ 
 -\frac{5}{24} \zeta   {\cal S}_{-}^2  \left [ 23 + 8\bar{\gamma} - 8\eta + 8\bar{\beta}_+  \right ]
 +\frac{10}{3} \zeta {\cal S}_{-}  \left ( \frac{{\cal S}_{-} \bar{\beta}_+ + {\cal S}_{+} \bar{\beta}_{-}}{\bar{\gamma}} \right )
  \nonumber \\
 &&
 -\frac{5}{8} \zeta \psi  {\cal S}_{-} \left ( {\cal S}_{+} - \frac{8}{3} {\cal S}_{-} \bar{\beta}_{-} + \frac{16}{3} \frac{{\cal S}_{+} \bar{\beta}_+ + {\cal S}_{-} \bar{\beta}_{-}}{\bar{\gamma}} \right ) \,,
\nonumber  \\
b_3 &=& \frac{5}{8} \left [ 6\bar{\gamma} + \zeta  {\cal S}_{-}^2 (13 + 8\bar{\gamma}+2\eta)
  - 12\bar{\beta}_+  - 3 \zeta \psi {\cal S}_{-} {\cal S}_{+}  \right ]
  \,.
\label{25PNcoeffs}
\end{eqnarray}
where 
\begin{eqnarray}
\bar{\beta}_\pm &\equiv& \frac{1}{2} (\bar{\beta}_1 \pm \bar{\beta}_2) \,,
\nonumber \\
\psi &\equiv&  (m_1 - m_2)/m \,, 
\nonumber \\
{\cal S}_{-}  &\equiv& - \alpha^{-1/2} (s_1 - s_2) \,,
\nonumber \\
{\cal S}_{+} &\equiv&  \alpha^{-1/2} (1 -s_1 - s_2) \,.
\end{eqnarray}
The periastron advance that results from these equations is given by
\begin{equation}
\dot{\omega} =  \frac{6\pi \alpha m f_b}{a(1-e^2)} \left [ 1 + \frac{2\bar{\gamma} - \bar{\beta}_+ - \psi \bar{\beta}_-}{3} \right ] \,.
\end{equation}
where $2\pi f_b = (\alpha m/a^3)^{1/2}$.

The tensor part of the gravitational waveform has the schematic form
\begin{equation}
  \tilde{h}^{ij} = \frac{2(1-\zeta) m}{R} \left\{ Q^{ij} + Q_{0.5\mathrm{PN}}^{ij} +
  Q_{1\mathrm{PN}}^{ij} + Q_{1.5\mathrm{PN}}^{ij} +
  Q_{2\mathrm{PN}}^{ij}  + \dots \right\},
  \label{STwaveform}
\end{equation}
where
\begin{equation}
  Q^{ij} = 2 \eta \left( v^i v^j - \frac{\alpha m \hat{n}^i \hat{n}^j}{r} \right).
  \label{STQij}
\end{equation}
Contributions to the tensor waveform through 2PN order have been derived by Lang~\cite{2013arXiv1310.3320L}.
The scalar waveform is given by $\phi = \phi_0 (1+ \Psi)$, where,
\begin{equation}
\Psi = \zeta \eta \alpha^{1/2} \frac{m}{R} \left \{ \Psi_{-0.5\mathrm{PN}}+\Psi_{0\mathrm{PN}}+\Psi_{0.5\mathrm{PN}}+\Psi_{1\mathrm{PN}} + \dots \right \} \,,
\label{Psi0}
\end{equation}
where, ignoring terms that are constant in time,
\begin{eqnarray}
\Psi_{-0.5\mathrm{PN}} &=& 4 {\cal S}_{-} ({\bf \hat{N}} \cdot {\bf v} ) \,,
 \nonumber \\
\Psi_{0\mathrm{PN}}&=& 2 \left ( {\cal S}_{+} - \psi {\cal S}_{-} \right ) \left [ ({\bf \hat{N}} \cdot {\bf v} )^2 - \frac{\alpha m}{r} ({\bf \hat{N}} \cdot {\bf x} )^2 \right ]
\nonumber \\
&& \quad 
-2 \frac{\alpha m}{r} \left [ 3 {\cal S}_{+} - \psi {\cal S}_{-} - 8 \left (\frac{{\cal S}_{+} \bar{\beta}_+ + {\cal S}_{-} \bar{\beta}_{-}}{\bar{\gamma}} \right ) \right ] \,,
\nonumber \\
\Psi_{0.5\mathrm{PN}}&=& 
- \frac{\partial}{\partial t} \left \{({\bf \hat{N}} \cdot {\bf x} ) \left [ (3-4\eta) {\cal S}_{-} - \psi {\cal S}_{+}
+ 8 \psi  \left (\frac{{\cal S}_{+} \bar{\beta}_+ + {\cal S}_{-} \bar{\beta}_{-}}{\bar{\gamma}} \right )
- 8 \left (\frac{{\cal S}_{-} \bar{\beta}_+ + {\cal S}_{+} \bar{\beta}_{-}}{\bar{\gamma}} \right ) \right ] \right \}
\nonumber \\
&& 
\quad + \frac{1}{3}  \left [ (1-2\eta){\cal S}_{-} - \psi {\cal S}_{+} \right ] \frac{\partial^3}{\partial t^3} ({\bf \hat{N}} \cdot {\bf x} )^3 \,,
\label{Psiterms}
\end{eqnarray}
where $\bf \hat{N}$ is a unit vector directed toward the observer.

The energy flux is given by
\begin{equation}
dE/dt=  -\frac{4}{3} \zeta  \frac{\mu \eta}{r}  \left ( \frac{\alpha m}{r} \right )^3 {\cal S}_{-}^2- \frac{8}{15} \frac{\mu \eta}{r} \left ( \frac{ \alpha m}{r} \right )^3 \left (\kappa_1 v^2  - \kappa_2 \dot{r}^2 \right ) \,,
\end{equation}
where the first term is the contribution of dipole radiation (formally of -1PN order), and the second term (formally of 0PN order, according to the conventional rules of counting) is a combination of quadrupole radiation, PN corrections to monopole and dipole radiation, and even a cross-term between dipole and octupole radiation.  The coefficients $\kappa_1$ and $\kappa_2$ are given by~\cite{2013PhRvD..87h4070M} 
\begin{eqnarray}
\kappa_1 &=& 12 + 5\bar{\gamma} - 5\zeta {\cal S}_{-}^2 (3 + \bar{\gamma} + 2\bar{\beta}_+ )
+10 \zeta {\cal S}_{-} \left ( \frac{{\cal S}_{-} \bar{\beta}_+ + {\cal S}_{+} \bar{\beta}_{-}}{\bar{\gamma}} \right )
\nonumber \\
&& 
 +10 \zeta \psi   {\cal S}_{-}^2 \bar{\beta}_{-} - 10\zeta \psi  {\cal S}_{-} \left (  \frac{{\cal S}_{+} \bar{\beta}_+ + {\cal S}_{-} \bar{\beta}_{-}}{\bar{\gamma}} \right ) \,,
\nonumber \\
\kappa_2 &=& 11 + \frac{45}{4} \bar{\gamma} - 40 \bar{\beta}_+
- 5\zeta {\cal S}_{-}^2  \left [ 17+ 6\bar{\gamma} + \eta + 8\bar{\beta}_+  \right ]
 +90 \zeta {\cal S}_{-} \left ( \frac{{\cal S}_{-} \bar{\beta}_+ + {\cal S}_{+} \bar{\beta}_{-}}{\bar{\gamma}} \right )
 \nonumber \\
&&  
+ 40 \zeta \psi {\cal S}_{-}^2 \bar{\beta}_{-} - 30\zeta \psi  {\cal S}_{-} \left ( \frac{{\cal S}_{+} \bar{\beta}_+ + {\cal S}_{-} \bar{\beta}_{-}}{\bar{\gamma}} \right )  
-120 \zeta \left (\frac{{\cal S}_{+} \bar{\beta}_+ + {\cal S}_{-} \bar{\beta}_{-}}{\bar{\gamma}} \right )^2 \,.
\label{kappas}
\end{eqnarray}
These results are in complete agreement with the total energy flux to $-1PN$ and $0PN$ orders, as calculated by Damour and Esposito-Far\`ese~\cite{DamourEspo92}.   They disagree with the flux formula of Will and Zaglauer~\cite{zaglauer}, as repeated in earlier versions of this {\em Living Review} as well as in~\cite{2012PhRvD..85f4041A}.  Will and Zaglauer~\cite{zaglauer} failed to take into account PN corrections to the dipole term induced by PN corrections in the equations of motion, and a dipole-octupole cross term in the scalar energy flux, all of which contribute to the flux at the same 0PN order as the quadrupole and monopole contributions.  

In the limit of weakly self-gravitating bodies the equations of motion and energy flux (including the dipole term) reduce to the standard results quoted in TEGP~\cite{tegp}.   

\subsubsection{Binary systems containing black holes}

Roger Penrose was probably the first to conjecture, in a talk at the 1970 Fifth Texas Symposium, that black holes in Brans-Dicke theory are identical to their GR counterparts~\cite{1971ApJ...166L..35T}.  Motivated by this remark, Thorne and Dykla showed that during gravitational collapse to form a black hole, the Brans-Dicke scalar field is radiated away, in accord with Price's theorem, leaving only its constant asymptotic value, and a GR black hole~\cite{1971ApJ...166L..35T}.  Hawking~\cite{1972CMaPh..25..167H} proved on general grounds that stationary, asymptotically flat black holes in vacuum in BD are the black holes of GR.  The basic idea is that black holes in vacuum with non-singular event horizons cannot support scalar ``hair''.   Hawking's theorem was extended to the class of $f(R)$ theories that can be transformed into generalized scalar-tensor theories by Sotiriou and Faraoni~\cite{2012PhRvL.108h1103S}.  

A consequence of these theorems is that, for a stationary black hole, $s = 1/2$.  Another way to see this is to note that, because all information about the matter that formed the black hole has vanished behind the event horizon, the only scale on which the mass of the hole can depend is the Planck scale, and thus $M \propto M_{Planck} \propto G^{-1/2} \propto \phi^{1/2}$.  Hence $s = 1/2$.

If both bodies in the binary system are black holes, then
setting $s_A=1/2$ for each body, all the parameters $\bar{\gamma}$, $\bar{\beta}_A$ and ${\cal S}_\pm$ vanish identically, and $\alpha = 1-\zeta$.  But since $\alpha$ appears only in the combination with $\alpha m$, a simple rescaling of each mass puts all equations into complete agreement with those of GR.  This is also true for the 2PN terms in the equations of motion~\cite{2013PhRvD..87h4070M}.    Thus, in the class of scalar-tensor theories discussed here, binary black holes are observationally indistinguishable from their GR counterparts, at least to high orders in a PN approximation. 

If one of the members of the binary system, say body 2, is a black hole, with $s_2 = 1/2$, then $\alpha = 1-\zeta$, $\bar{\gamma} = \bar{\beta}_A = 0$, and hence, through $1$PN order, the motion is again identical to that in GR.   
At $1.5$PN order, dipole radiation reaction kicks in, since $s_1 < 1/2$.   In this case, ${\cal S}_{-} = {\cal S}_+ = \alpha^{-1/2} (1-2s_1)/2$, and thus the $1.5$PN coefficients in the relative equation of motion (\ref{A1.5PNST}) take the form
\begin{eqnarray}
A_{1.5PN} &=&  \frac{5}{8} Q \,,
\nonumber \\
B_{1.5PN} &=& \frac{5}{24} Q \,,
\end{eqnarray}
where
\begin{equation}
Q \equiv  \frac{\zeta}{1-\zeta} (1-2s_1)^2 = \frac{1}{3+2\omega_0} (1-2s_1)^2 \,,
\label{Qdefinition}
\end{equation}
while the coefficients in the energy loss rate simplify to
\begin{eqnarray}
\kappa_1 &=& 12 - \frac{15}{4} Q \,,
\nonumber \\
\kappa_2 &=& 11 - \frac{5}{4} Q (17 + \eta) \,.
\end{eqnarray}
The result is that the motion of a mixed compact binary system through $2.5$PN order differs from its general relativistic counterpart only by terms that depend on a single parameter $Q$, as defined by Eq.\ (\ref{Qdefinition}).   

It should be pointed out that there {\em are} ways to induce scalar hair on a black hole.  One is to introduce a potential $V(\phi)$, which, depending on its form, can help to support a non-trivial scalar field outside a black hole.   Another is to introduce matter.  A companion neutron star is an obvious choice, and such a binary system  in scalar-tensor theory is clearly different from its general relativistic counterpart.   Another possibility is a distribution of cosmological matter that can support a time-varying scalar field at infinity.  This possibility has been called ``Jacobson's miracle hair-growth formula'' for black holes, based on work by Jacobson~\cite{1999PhRvL..83.2699J,2012JCAP...05..010H}. 

\newpage
%##########

%%%%%%%%%%%%%%%%%%%%%%%%%%%%%%%%%%%%%%%%%%%%%%%%%%%%%%%%%%%%%%%%%%%%%%%%%%%%%%%%%%%
%%%%%%%%%%%%%%%%%%%%%%%%%%%%%%%%%%%%%%%%%%%%%%%%%%%%%%%%%%%%%%%%%%%%%%%%%%%%%%%%%%%
%%%%%%%%%%%%%%%%%%%%%%%%%%%%%%%%%%%%%%%%%%%%%%%%%%%%%%%%%%%%%%%%%%%%%%%%%%%%%%%%%%%

\section{Stellar System Tests of Gravitational Theory}
\label{stellar}

%%%%%%%%%%%%%%%%%%%%%%%%%%%%%%%%%%%%%%%%%%%%%%%%%%%%%%%%%%%%%%%%%%%%%%%%%%%%%%%%%%%
%%%%%%%%%%%%%%%%%%%%%%%%%%%%%%%%%%%%%%%%%%%%%%%%%%%%%%%%%%%%%%%%%%%%%%%%%%%%%%%%%%%

\subsection{The binary pulsar and general relativity}
\label{binarypulsars}

The 1974 discovery of the binary pulsar B1913+16 by Joseph Taylor
and Russell Hulse during a routine search for new pulsars
provided the first possibility of probing new aspects of gravitational
theory:  the effects of strong relativistic internal gravitational fields
on orbital dynamics, and the effects of gravitational radiation reaction.
For reviews of the discovery, see the published
Nobel Prize lectures by Hulse and Taylor~\cite{Hulse, Taylor94}. 
For reviews of the current status of testing general relativity with pulsars, 
including binary and millisecond
pulsars, see~\cite{lrr-2008-8, StairsLRR,2014arXiv1402.5594W}; specific details on every pulsar discovered to date, along with orbit elements of pulsars in binary systems, can be found at the Australia Telescope National Facility (ATNF) online pulsar catalogue~\cite{ATNFpulsarcat}.  Table \ref{bpdata} lists the current values of the key orbital and relativistic parameters for B1913+16, from analysis of data through 2006~\cite{2010ApJ...722.1030W}.

\begin{table}[hptb]
  \caption[Parameters of the binary pulsar B1913+16.]{Parameters of
    the binary pulsar B1913+16. The numbers in parentheses denote
    errors in the last digit. Data taken from~\cite{2010ApJ...722.1030W} and defined as of 11 December 2003 (MJD 52984.0). 
    Note that $\gamma'$ is not the same as the
    PPN parameter $\gamma$ (see Eqs.~ (\ref{pkparameters})).}
  \label{bpdata}
  \renewcommand{\arraystretch}{1.2}
  \centering
  \begin{tabular}{l|ll}
    \hline \hline
    \qquad Parameter &
    \multicolumn{1}{c}{Symbol} &
    \multicolumn{1}{c}{Value} \\
    & \multicolumn{1}{c}{(units)} \\
    \hline \hline
    \phantom{ii}(i) ``Physical'' parameters: \\ [0.3 em]
    \phantom{(iii)~} Right Ascension &
    $\alpha$ &
    $19^\mathrm{h} 15^\mathrm{m}27.^\mathrm{s} 99999(2)$ \\
    \phantom{(iii)~} Declination &
    $\delta$ &
    $16^\circ 06'27.''4034(4)$ \\
    \phantom{(iii)~} Pulsar period &
    $P_\mathrm{p}$ (ms) &
    $59.0299983444181(5)$ \\
    \phantom{(iii)~} Derivative of period &
    $\dot P_\mathrm{p}$ &
    $8.62713(8) \times 10^{-18}$ \\ [1 em]
    \phantom{i}(ii) ``Keplerian'' parameters: \\ [0.3 em]
    \phantom{(iii)~} Projected semimajor axis &
    $a_\mathrm{p} \sin i$ (s) &
    $2.341782(3)$ \\
    \phantom{(iii)~} Eccentricity &
    $e$ &
    $0.6171334(5)$ \\
    \phantom{(iii)~} Orbital period &
    $P_\mathrm{b}$ (day) &
    $0.322997448911(4)$ \\
    \phantom{(iii)~} Longitude of periastron &
    $\omega_0$ (${}^\circ$) &
    $292.54472(4)$ \\
    \phantom{(iii)~} Julian date of periastron &
    $T_0$ (MJD) &
    $52144.90097841(4)$ \\ [1 em]
    (iii) ``Post-Keplerian'' parameters: \\ [0.3 em]
    \phantom{(iii)~} Mean rate of periastron advance &
    $\langle \dot\omega \rangle$ ($ {}^\circ \mathrm{\ yr}^{-1} $) &
    $4.226598(5)$ \\
    \phantom{(iii)~} Redshift/time dilation &
    $\gamma'$ (ms) &
    $4.2992(8)$ \\
    \phantom{(iii)~} Orbital period derivative &
    $\dot P_\mathrm{b}$ ($10^{-12}$) &
    $-2.423(1)$ \\
    \hline \hline
  \end{tabular}
  \renewcommand{\arraystretch}{1.2}
\end{table}

The system consists of a pulsar of nominal period 59~ms in a close binary
orbit with an unseen companion. The orbital period is about
7.75~hours, and the eccentricity is 0.617. From detailed analyses of the
arrival times of pulses (which amounts to an integrated version of the
Doppler-shift methods used in spectroscopic binary systems), extremely
accurate orbital and physical parameters for the system have been obtained
(see Table~\ref{bpdata}). Because the orbit is so close
($\approx 1 R_\odot$)
and because there is no evidence of an eclipse of the pulsar signal or of
mass transfer from the companion, it is generally agreed that the companion
is compact. Evolutionary arguments suggest that it is most likely
a dead pulsar, while B1913+16 is a ``recycled'' pulsar. 
Thus the orbital motion is
very clean, free from tidal or other
complicating effects. Furthermore, the data acquisition is ``clean'' in
the sense that by exploiting the intrinsic stability of the pulsar
clock combined with the ability to maintain and transfer atomic time
accurately using GPS,
the observers can keep track of pulse time-of-arrival with
an accuracy of $13 \mathrm{\ \mu s}$, despite extended gaps between
observing sessions (including a several-year gap in the middle 1990s
for an upgrade of
the Arecibo radio telescope). The pulsar has shown no evidence of ``glitches''
in its pulse period.

Three factors made this system an arena where relativistic celestial
mechanics must be used: the relatively large size of relativistic
effects
[$ v_\mathrm{orbit} \approx (m/r)^{1/2} \approx 10^{-3}$], a factor of 10
larger than the corresponding values for solar-system orbits;
the short orbital period, allowing secular effects to
build up rapidly; and the cleanliness of the system, allowing
accurate determinations of small effects. Because the orbital
separation is large compared to the neutron stars' compact size, tidal
effects can be ignored. Just as Newtonian gravity
is used as a tool for measuring astrophysical parameters of ordinary binary
systems, so GR is used as a tool for measuring
astrophysical parameters in the binary pulsar.

The observational parameters that are obtained from a least-squares
solution of the arrival-time data fall into three groups:
\begin{enumerate}
\item non-orbital parameters, such as the pulsar period and its rate
  of change (defined at a given epoch), and the position of the pulsar
  on the sky;
\item five ``Keplerian'' parameters, most closely related to those
  appropriate for standard Newtonian binary systems, such as the
  eccentricity $e$, the orbital period $P_\mathrm{b}$, and the
  semi-major axis of the pulsar projected along the line of sight,
  $a_\mathrm{p} \sin i$; and
\item five ``post-Keplerian'' parameters.
\end{enumerate}
The five post-Keplerian parameters are: $\langle \dot \omega \rangle$,
the average rate of periastron advance; $\gamma'$, the amplitude of
delays in arrival of pulses caused by the varying effects of the
gravitational redshift and time dilation as the pulsar moves in its
elliptical orbit at varying distances from the companion and with
varying speeds; $\dot P_\mathrm{b}$, the rate of change of orbital
period, caused predominantly by gravitational radiation damping; and
$r$ and $s = \sin i$, respectively the ``range'' and ``shape'' of the
Shapiro time delay of the pulsar signal as it propagates through the
curved spacetime region near the companion, where $i$ is the angle of
inclination of the orbit relative to the plane of the sky. An
additional 14 relativistic parameters are measurable in
principle~\cite{DamourTaylor92}.

In GR, the five post-Keplerian parameters can be related
to the masses of the two bodies and to measured Keplerian parameters
by the equations (TEGP~12.1, 14.6~(a)~\cite{tegp})
\begin{equation}
  \begin{array}{rcl}
    \langle \dot \omega \rangle & = &
    6 \pi f_\mathrm{b} (2\pi m f_\mathrm{b})^{2/3} (1-e^2 )^{-1},
    \\ [0.5 em]
    \gamma' & = & \displaystyle e (2 \pi f_\mathrm{b} )^{-1}
    (2\pi m f_\mathrm{b})^{2/3} \frac{m_2}{m}
    \left( 1 + \frac{m_2}{m} \right),
    \\ [0.5 em]
    \dot P_\mathrm{b} & = & \displaystyle
    - \frac{192 \pi}{5} (2 \pi {\cal M}f_\mathrm{b})^{5/3} F(e),
        \\ [0.3 em]
    r & = & m_2,
    \\ [0.5 em]
    s & = & \sin i,
  \end{array}
  \label{pkparameters}
\end{equation}
where
$m_1$ and $m_2$ denote the pulsar and companion masses, respectively.
The formula for $\langle \dot \omega \rangle$ ignores
possible non-relativistic contributions to the periastron shift,
such as tidally or rotationally induced effects caused by the companion
(for discussion of these effects, see TEGP~12.1~(c)~\cite{tegp}). The formula
for $\dot P_\mathrm{b}$ includes only quadrupole
gravitational radiation;
it ignores other sources of energy loss, such as tidal
dissipation (TEGP~12.1~(f)~\cite{tegp}). Notice that, by virtue of Kepler's 
third
law, $(2\pi f_\mathrm{b})^2 = m/a^3$,  $(2\pi m f_\mathrm{b})^{2/3} = m/a \sim
\epsilon$, thus the first two post-Keplerian parameters can be seen
as ${\cal O} (\epsilon)$, or 1PN corrections to the underlying variable, while the
third is an ${\cal O} (\epsilon^{5/2})$, or 2.5PN correction.
The current observed values for the Keplerian
and post-Keplerian parameters are shown in Table~\ref{bpdata}.
The parameters $r$ and $s$ are not separately
measurable with interesting accuracy for B1913+16 because the
orbit's $47 ^\circ$ inclination does not lead to a substantial Shapiro
delay, however they are measurable in the double pulsar, for example.

Because $f_\mathrm{b}$ and $e$ are separately measured parameters, the
measurement of the three post-Keplerian parameters provides three
constraints on the two unknown masses. The periastron shift measures
the total mass of the system, $\dot P_\mathrm{b}$ measures the chirp mass, and
$\gamma'$ measures a complicated function of the masses.
GR passes the test if it provides a consistent solution to these
constraints, within the measurement errors.

%\epubtkImage{livingbinary1.png}{
\begin{figure}[t]
\centering
 % \def\epsfsize#1#2{0.6#1}
 % \centerline{\epsfbox{binary1.eps}}
 \includegraphics[width=4in]{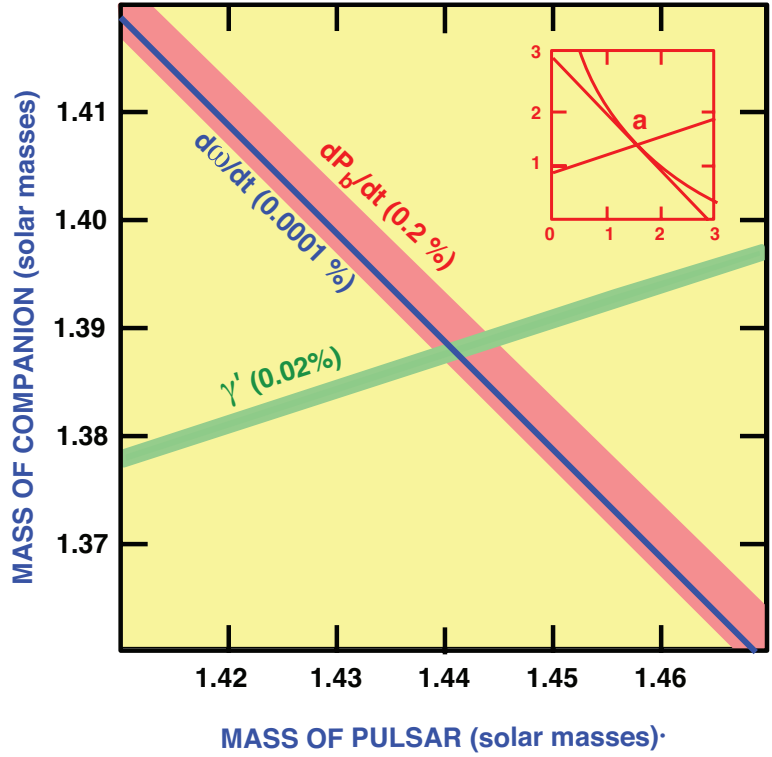}
  \caption{Constraints on masses of the pulsar and its companion
    from data on B1913+16, assuming GR to be valid. The width of
    each strip in the plane reflects observational accuracy, shown as
    a percentage. An inset shows the three constraints on the full
    mass plane;  the intersection region (a) has been magnified 400
    times for the full figure.}
  \label{bpfigure1}
\end{figure}%}

From the intersection of the  $\langle \dot \omega \rangle$
and $\gamma' $ constraints we obtain the values
$m_1 = 1.4398 \pm 0.0002 M_\odot$ and
$m_2 = 1.3886 \pm 0.0002 M_\odot$. The third of Eqs.~(\ref{pkparameters})
then predicts the value
$\dot P_\mathrm{b} = -2.402531 \pm 0.000014 \times 10^{-12}$.
In order to compare the predicted
value for $\dot P_\mathrm{b}$ with the observed value of Table~\ref{bpdata}, it
is necessary to
take into account the small kinematic effect of a relative acceleration between the
binary pulsar system and the solar system caused by the differential
rotation of the galaxy.  Using data
on the location and proper motion of the pulsar, combined with the best
information available on galactic rotation; the current value of this effect
is
 $\dot P_\mathrm{b}^\mathrm{gal} \simeq -(0.027 \pm 0.005) \times 10^{-12}$.
Subtracting this from the observed $\dot P_\mathrm{b}$ (see Table~\ref{bpdata})
gives the corrected 
$\dot P_\mathrm{b}^\mathrm{corr} = -(2.396 \pm 0.005)
\times 10^{-12}$,
which agrees with the prediction within the errors. In other words,
\begin{equation}
  \frac{\dot P_\mathrm{b}^\mathrm{corr}}{\dot P_\mathrm{b}^\mathrm{GR}} =
  0.997 \pm 0.002.
  \label{Pdotcompare}
\end{equation}
The consistency among the measurements is displayed in Figure~\ref{bpfigure1},
in which the regions allowed by the three most precise constraints
have a single common overlap.
Uncertainties in the parameters that go into the galactic correction are now
the limiting factor in the accuracy of the test of gravitational damping.

%\epubtkImage{livingbinary2.png}{
\begin{figure}[ht]
\centering
  %\def\epsfsize#1#2{0.8#1}
  %\centerline{\epsfbox{livingbinary2.eps}}
  \includegraphics[width=4in]{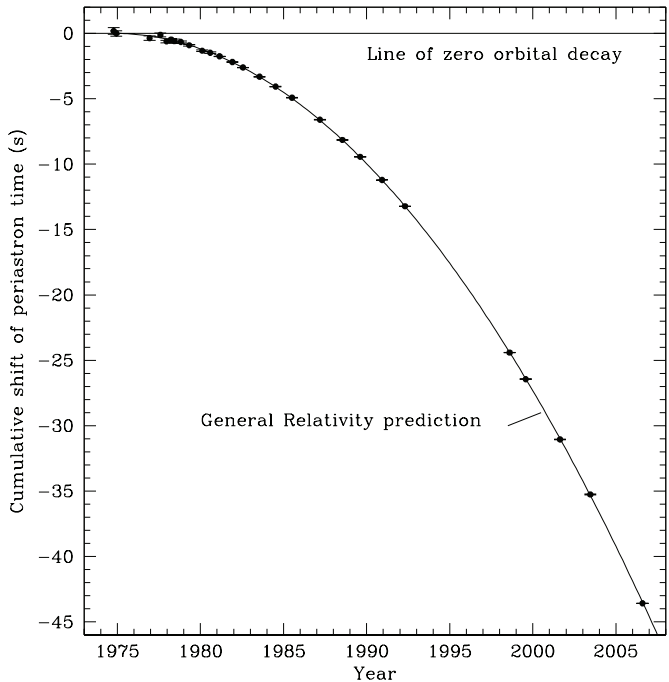}
  \caption{Plot of the cumulative shift of the periastron time
    from 1975\,--\,2005. The points are data, the curve is the GR
    prediction. The gap during the middle 1990s was caused by a
    closure of Arecibo for upgrading~\cite{2010ApJ...722.1030W}.}
  \label{bpfigure2}
\end{figure}%}

A third way to display the agreement with GR is by
comparing the observed phase of the orbit with a theoretical template
phase as a function of time. If $f_\mathrm{b}$ varies slowly in time, then to
first order in a Taylor expansion, the orbital phase is given by
$\Phi_\mathrm{b} (t) = 2\pi f_\mathrm{b0} t + \pi {\dot f}_\mathrm{b0} t^2$. The time of
periastron passage $t_\mathrm{P}$ is given by $\Phi (t_\mathrm{P})=2\pi N$, where $N$ is
an integer, and consequently, the periastron time will not grow
linearly with $N$. Thus the cumulative difference between periastron
time $t_\mathrm{P}$ and $N/f_\mathrm{b0}$, the quantities actually measured in
practice,  should vary according to
$t_\mathrm{P} - N/f_\mathrm{b0} = -{\dot f}_\mathrm{b0} N^2/2 f_\mathrm{b0}^3 \approx - ({\dot
f}_\mathrm{b0}/2f_\mathrm{b0}) t^2$. Figure~\ref{bpfigure2} shows the results: The
dots are the data points, while the curve is the predicted difference
using the measured masses and the quadrupole formula for ${\dot
f}_\mathrm{b0}$~\cite{2010ApJ...722.1030W}.

The consistency among the constraints
provides a test of the assumption that the two bodies
behave as ``point'' masses, without complicated tidal effects, obeying
the general relativistic equations of motion including
gravitational radiation. It is also a test of strong gravity,
in that the highly relativistic internal structure of the
neutron stars does not influence their orbital motion, as predicted by
the SEP of GR.

Observations~\cite{kramer, WeisbergTaylor02} indicate
that the pulse profile is varying with time, which suggests that the pulsar is
undergoing geodetic precession on a 300-year timescale
as it moves through the curved spacetime
generated by its companion (see Section~\ref{geodeticprecession}).
The amount is consistent with GR, assuming that the pulsar's
spin is suitably misaligned with the orbital angular momentum.
Unfortunately, the evidence suggests that the pulsar beam may precess
out of our line of sight by 2025.

%%%%%%%%%%%%%%%%%%%%%%%%%%%%%%%%%%%%%%%%%%%%%%%%%%%%%%%%%%%%%%%%%%%%%%%%%%%%%%%%%%%
%%%%%%%%%%%%%%%%%%%%%%%%%%%%%%%%%%%%%%%%%%%%%%%%%%%%%%%%%%%%%%%%%%%%%%%%%%%%%%%%%%%

\subsection{A zoo of binary pulsars}
\label{population}

More than 70 binary neutron star systems with orbital periods less than
a day are now known. While some are less interesting for
testing relativity, some have yielded interesting tests, and others, notably
the recently discovered ``double pulsar'' are likely to continue to produce significant
results well into the future. Here we describe some of the more interesting
or best studied cases; 

\begin{figure}[h!t]
 \centering
 % \def\epsfsize#1#2{0.8#1}
  %\centerline{\epsfbox{doublepulsarplot.eps}}
  \includegraphics[width=4in]{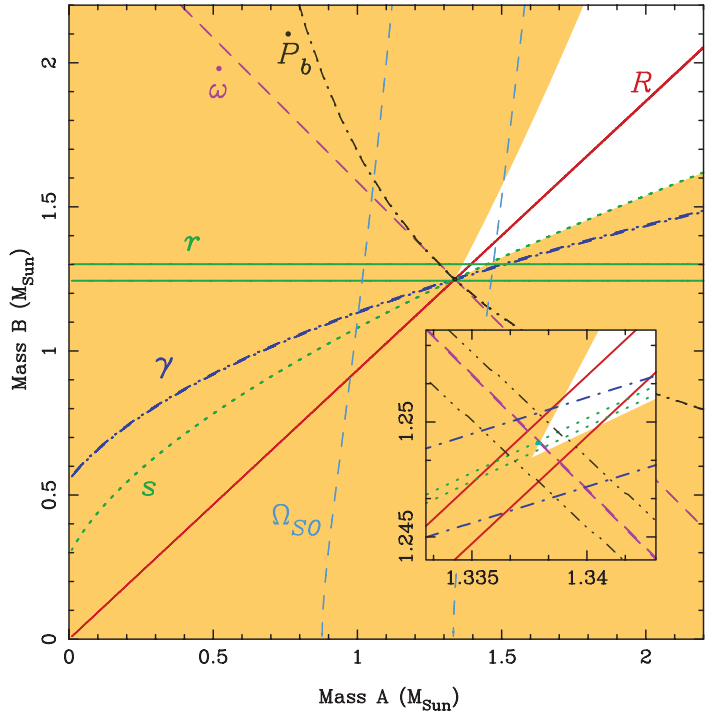}
  \caption{Constraints on masses of the pulsar and its companion
    from data on J0737-3039A,B, assuming GR to be valid.  The inset shows the intersection region magnified by a factor of 80.  (Figure courtesy of M. Kramer).
    }
  \label{bpfigure3}
\end{figure}

\medskip
\noindent
{\bf The ``double'' pulsar: J0737-3039A, B.}  This binary pulsar system, discovered in 2003~\cite{burgay03}, was
  already remarkable for its extraordinarily short orbital period (0.1
  days) and large periastron advance
  ($16.88^\circ \mathrm{\ yr}^{-1}$), but then the companion was also
  discovered to be a pulsar~\cite{lyne04}.  Because two projected
  semi-major axes could be measured,  the mass ratio was obtained
  directly from the ratio of the two values of $a_\mathrm{p} \sin i$,
  and thereby the two masses could be obtained by combining that ratio with the
  periastron advance, assuming GR. The results are
  $m_A = 1.337 \pm 0.005 \, M_\odot$ and
  $m_B = 1.250 \pm 0.005 \, M_\odot$, where $A$ denotes the primary
  (first) pulsar. From these values, one finds that the orbit is
  nearly edge-on, with $\sin i = 0.9991$, a value which is
  completely consistent with that inferred from the Shapiro delay
  parameter. In fact, the five measured
  post-Keplerian parameters plus the ratio of the projected semi-major
  axes give six constraints on the masses (assuming GR): as seen in Fig.\ \ref{bpfigure3}, all six
  overlap within their measurement errors~\cite{2006Sci...314...97K}.  Because of the location of the system, galactic proper-motion
  effects play a significantly smaller role in the interpretation
  of $\dot P_\mathrm{b}$ measurements than they did in B1913+16; this and the reduced effect of interstellar dispersion means that the accuracy of measuring the gravitational-wave damping may soon beat that from the Hulse-Taylor system.   The geodetic precession of pulsar B's spin axis has also been measured by monitoring changes in the patterns of eclipses of the signal from pulsar A, with a result in agreement with GR to about 13 percent~\cite{2008Sci...321..104B}; the constraint on the masses from that effect (assuming GR to be correct) is also shown in Fig.\  \ref{bpfigure3}.

%
%\epubtkImage{doublepulsarplot.png}{
%}

\medskip
\noindent
{\bf J1738+0333: A white-dwarf companion.}
This is a low-eccentricity, $8.5$-hour period system in which the white-dwarf companion is bright enough to permit detailed spectroscopy, allowing the companion mass to be determined directly to be $0.181 \, M_\odot$.  The mass ratio is determined from Doppler shifts of the spectral lines of the companion and of the pulsar period, giving the pulsar mass $1.46 \, M_\odot$.  Ten years of observation of the system yielded both a measurement of the apparent orbital period decay, and enough information about its parallax and proper motion to account for the substantial galactic kinematic effect to give a value of the intrinsic period decay of $\dot{P}_\mathrm{b} =  (-25.9 \pm 3.2) \times 10^{-15} \mathrm{ s \, s^{-1}}$ in agreement with the predicted effect~ \cite{2012MNRAS.423.3328F}.   But because of the asymmetry of the system, the result also places a significant bound on the existence of dipole radiation, predicted by many alternative theories of gravity (see Sec. \ref{binarypulsarsalt} below for discussion).  Data from this system were also used to place the tight bound on the PPN parameter $\alpha_1$ shown in Table \ref{ppnlimits}.

\medskip
\noindent
{\bf J1141-6545: A white-dwarf companion.}
This system is similar in some ways to the Hulse-Taylor binary: short orbital period ($0.20$ days), significant orbital eccentricity ($0.172$), rapid periastron advance ($5.3$ degrees per year) and massive components ($M_p = 1.27 \pm 0.01 M_\odot$, $M_c = 1.02 \pm 0.01 M_\odot$).  The key difference is that the companion is again a white dwarf.  The intrinsic orbit period decay has been measured in agreement with GR to about six percent, again placing limits on dipole gravitational radiation~\cite{2008PhRvD..77l4017B}.

\medskip
\noindent
{\bf J0348+0432: The most massive neutron star.}
Discovered in 2011, this is another neutron-star white-dwarf system, in a very short period ($0.1$ day), low eccentricity ($2 \times 10^{-6}$) orbit.  Timing of the neutron star and spectroscopy of the white dwarf have led to mass values of $0.172 \, M_\odot$ for the white dwarf and $2.01 \pm 0.04 \, M_\odot$ for the pulsar, making it the most massive accurately measured neutron star yet.  This result ruled out a number of heretofore viable soft equations of state for nuclear matter.   The orbit period decay agrees with the GR prediction within 20 percent and is expected to improve steadily with time. 

\medskip
\noindent
{\bf J0337+1715: Two white-dwarf companions.}
This remarkable system was reported in 2014 ~\cite{2014Natur.505..520R}.  It consists of a 2.73 millisecond pulsar ($M=1.44 \, M_\odot$) with extremely good timing precision, accompanied by {\em two} white dwarfs in coplanar circular orbits.
The inner white dwarf ($M = 0.1975 \, M_\odot$) has an orbital period of $1.629$ days, with $e = 6.918 \times 10^{-4}$, and the outer white dwarf ($M = 0.41 \, M_\odot$) has a period of $327.26$ days, with $e = 3.536 \times 10^{-2}$.   This is an ideal system for testing the Nordtvedt effect in the strong-field regime.  Here the inner system is the analogue of the Earth-Moon system, and the outer white dwarf plays the role of the Sun.  Because the outer semi-major axis is about 1/3 of an astronomical unit, the basic driving perturbation is comparable to that provided by the Sun.  However, the self-gravitational binding energy per unit mass of the neutron star is almost a billion times larger than that of the Earth, greatly amplifying the size of the Nordtvedt effect.  Depending on the details, this system could exceed lunar laser ranging in testing the Nordtvedt effect by several orders of magnitude.

\medskip
\noindent
{\bf Other binary pulsars.}
Two of the earliest binary pulsars, B1534+12 and B2127+11C, discovered in 1990, failed to live up to their early promise despite being similar to the Hulse-Taylor system in most respects (both were believed to be double neutron-star systems).  The main reason was the significant uncertainty in the kinematic effect on $\dot{P}_\mathrm{b}$  of local accelerations, galactic in the case of B1534+12, and those arising from the globular cluster that was home to B2127+11C.

\begin{table}[hptb]
  \caption[Parameters of other binary pulsars.]{Parameters of other
    binary pulsars. References may be found in the text.  Values for orbit period 
    derivatives include corrections for galactic kinematic effects}
  \label{bpdata2}
  \renewcommand{\arraystretch}{1.2}
  \centering
  \begin{tabular}{l|lll}
    \hline \hline
    Parameter &
    \multicolumn{1}{c}{J0737--3039(A, B)} &
    \multicolumn{1}{c}{J1738+0333} &
    \multicolumn{1}{c}{J1141--6545} 
   \\
    \hline \hline
    \phantom{i}(i) Keplerian: \\ [0.3 em]
    \phantom{(ii)} $a_\mathrm{p} \sin i$ (s) &
    $ \phantom{-} 1.41504(2) / 1.513(3) $ &
    $ \phantom{-} 0.34342913(2) $ &
    $ \phantom{-} 1.858922(6) $  \\
    \phantom{(ii)} $e$ &
    $ \phantom{-} 0.087779(5) $ &
    $ \phantom{-} (3.4\pm 1.1)\times 10^{-7} $ &
    $ \phantom{-} 0.171884(2) $ 
    \\
    \phantom{(ii)} $P_\mathrm{b}$ (day) &
    $ \phantom{-} 0.102251563(1)  $ &
    $ \phantom{-} 0.354790739872(1) $ &
    $ \phantom{-} 0.1976509593(1) $ 
    \\ [1 em]
    (ii) Post-Keplerian: \\ [0.3 em]
    \phantom{(ii)} $\langle \dot \omega \rangle$ (${}^\circ \mathrm{\ yr}^{-1}$) &
    $ \phantom{-} 16.90(1) $ &
     &
    $ \phantom{-} 5.3096(4) $ 
    \\
    \phantom{(ii)} $\gamma'$ (ms) &
    $ \phantom{-}  0.382(5) $ &
   &
    $ \phantom{-} 0.77(1) $ 
     \\
    \phantom{(ii)} ${\dot P}_\mathrm{b}$ ($10^{-12}$) &
    $ -1.21(6) $ &
    $ -0.026(3) $ &
    $ -0.401(25) $ 
   \\
    \phantom{(ii)} $r$ ($\mathrm{\mu s}$) &
    $ \phantom{-} 6.2(5) $ &  & 
         \\
    \phantom{(ii)} $s=\sin i$ &
      $\phantom{-} 0.9995(4) $&  &
     \\
    \hline \hline
  \end{tabular}
  \renewcommand{\arraystretch}{1.0}
\end{table}

%%%%%%%%%%%%%%%%%%%%%%%%%%%%%%%%%%%%%%%%%%%%%%%%%%%%%%%%%%%%%%%%%%%%%%%%%%%%%%%%%%%
%%%%%%%%%%%%%%%%%%%%%%%%%%%%%%%%%%%%%%%%%%%%%%%%%%%%%%%%%%%%%%%%%%%%%%%%%%%%%%%%%%%

\subsection{Binary pulsars and alternative theories}
\label{binarypulsarsalt}

Soon after the discovery of the binary pulsar it was widely hailed as
a new testing ground for relativistic gravitational effects.
As we have seen in the case of GR, in most respects,
the system has lived up to, indeed exceeded, the early expectations.

In another respect, however, the system has only partially lived up to its
promise, namely as a direct testing ground for alternative theories of
gravity. The origin of this promise was the discovery~\cite{1975ApJ...196L..59E,1977ApJ...214..826W}
that alternative theories of gravity generically predict the emission
of dipole gravitational radiation from binary star systems.
In GR, there is no dipole radiation because the
``dipole moment'' (center of mass) of isolated systems is
uniform in time (conservation
of momentum), and because the ``inertial mass'' that determines the
dipole moment is the same as the mass that generates gravitational
waves (SEP). In other theories, while the
inertial dipole moment may remain uniform, the ``gravity wave'' dipole
moment need not, because the mass that generates gravitational waves
depends differently on the internal
gravitational binding energy of each body than does the inertial mass
(violation of SEP).
Schematically, in a coordinate system in which the center of inertial
mass is at the origin, so that $m_\mathrm{I,1} {\bf x}_1 + m_\mathrm{I,2} {\bf x}_2 =0$,
the dipole part of the retarded gravitational field would be given by
\begin{equation}
  h \sim \frac{1}{R} \frac{d}{dt}
  (m_\mathrm{GW,1} {\bf x}_1 + m_\mathrm{GW,2} {\bf x}_2) \sim
  \frac{\eta m}{R} {\bf v}
  \left( \frac{m_\mathrm{GW,1}}{m_\mathrm{I,1}} -
  \frac{m_\mathrm{GW,2}}{m_\mathrm{I,2}} \right),
  \label{hdipole}
\end{equation}
where ${\bf v} = {\bf v}_1 -{\bf v}_2$ and $\eta$ and $m$ are defined
using inertial masses. In theories that violate SEP,
the difference between gravitational wave mass and inertial mass is a
function of the internal gravitational binding energy of the bodies.
This additional form of gravitational radiation damping could,
at least in principle, be significantly stronger than the usual quadrupole
damping, because it depends on fewer powers of the orbital velocity $v$,
and it depends on the gravitational binding energy per unit mass of
the bodies, which, for neutron stars, could be as large as 20~percent
(see TEGP~10~\cite{tegp} for further details).
As one fulfillment of this promise, Will and Eardley worked out in
detail the effects of dipole gravitational radiation in the bimetric theory
of Rosen, and, when the first observation of the decrease of
the orbital period was announced in 1979, the Rosen theory
suffered a terminal blow. A wide
class of alternative theories also fails the binary pulsar test because
of dipole gravitational radiation (TEGP~12.3~\cite{tegp}).

On the other hand, the early observations of PSR 1913+16
already indicated that, in
GR, the masses of the two bodies were nearly equal, so
that, in theories of gravity that are in some sense ``close'' to
GR, dipole gravitational radiation would not be a
strong effect, because of the apparent symmetry of the system.
The Rosen theory, and others like it, are not ``close'' to GR,
except in their predictions for the weak-field, slow-motion
regime of the solar system. When relativistic neutron stars are present,
theories like these can predict strong effects on the motion of the bodies
resulting from their internal highly relativistic gravitational structure
(violations of SEP). As a consequence,
the masses inferred from observations of the periastron shift
and $\gamma'$ may
be significantly different from those inferred using GR,
and may be different from each other, leading to strong
dipole gravitational radiation damping. By contrast, the Brans--Dicke
theory is ``close'' to GR, roughly
speaking within $1/ \omega_\mathrm{BD}$ of the predictions of the latter,
for large values of the coupling constant $\omega_\mathrm{BD}$. 
Thus, despite the presence of dipole gravitational radiation,
the Hulse-Taylor binary pulsar provides at present only a weak test of pure Brans--Dicke
theory, not competitive with solar-system tests.   

However, the discovery of binary pulsar systems with a white dwarf companion, such as 
J1738+0333, J1141-6545 and J0348+0432 has made it possible to perform strong tests of the existence of dipole radiation.   This is because such systems are necessarily asymmetrical, since the gravitational binding energy per unit mass of white dwarfs is of order $10^{-4}$, much less than that of the neutron star.   Already, significant bounds have been placed on dipole radiation using J1738+0333 and J1141-6545~\cite{2012MNRAS.423.3328F,2008PhRvD..77l4017B}.  

Because the gravitational-radiation and strong-field properties of alternative theories of gravity can be dramatically different from those of GR and each other, it is difficult to parametrize these aspects of the theories in the manner of the PPN framework.  In addition, because of the generic violation of the Strong Equivalence Principle in these theories, the results can be very sensitive to the equation of state and mass of the neutron star(s) in the system.  In the end, there is no way around having to analyze every theory in turn.  
On the other hand, because of their relative simplicity, scalar-tensor theories provide an illustration of the essential effects, and so we shall discuss binary pulsars within this class of theories.

%%%%%%%%%%%%%%%%%%%%%%%%%%%%%%%%%%%%%%%%%%%%%%%%%%%%%%%%%%%%%%%%%%%%%%%%%%%%%%%%%%%
%%%%%%%%%%%%%%%%%%%%%%%%%%%%%%%%%%%%%%%%%%%%%%%%%%%%%%%%%%%%%%%%%%%%%%%%%%%%%%%%%%%

\subsection{Binary pulsars and scalar-tensor gravity}
\label{binarypulsarsscalar}

Making the usual assumption that both members of the system are neutron
stars, and using the methods summarized in TEGP~10\,--\,12~\cite{tegp} (see also~\cite{2013PhRvD..87h4070M})
one can obtain
formulas for the periastron shift, the gravitational redshift/second-order
Doppler shift parameter, the Shapiro delay coefficients, and the rate of change of orbital period,
analogous to Eqs.~ (\ref{pkparameters}). These formulas depend on the
masses of the two neutron stars, on their sensitivities $s_A$, and on
the scalar-tensor parameters, as defined in Table~\ref{tab:STparams} (and on a new sensitivity $\kappa^*$, defined below). First, there is
a modification of Kepler's third law, given by
\begin{equation}
  2 \pi f_\mathrm{b} = \left( \frac{\alpha m}{a^3} \right)^{1/2}\!\!\!\!\!\!\!.
  \label{KeplerBD}
\end{equation}
Then the predictions for $\langle \dot \omega \rangle$, $\gamma^\prime$, $\dot P_\mathrm{b}$, $r$ and $s$ are
\begin{eqnarray}
  \langle \dot \omega \rangle & = &
  6 \pi f_\mathrm{b} (2\pi \alpha m f_\mathrm{b})^{2/3} (1-e^2 )^{-1}
  {\cal P} \,,
  \label{periastronBD}
  \\
  \gamma^\prime & = &
  e (2 \pi f_\mathrm{b} )^{-1} (2\pi \alpha m f_\mathrm{b})^{2/3}
  \frac{m_2}{m} \alpha^{-1}
  \left [1-2\zeta s_2 + \alpha \frac{m_2}{m} + \zeta \kappa_1^* (1-2s_2) \right ],
  \label{gammaBD}
  \\
  \dot P_\mathrm{b} & = &
  - \frac{192 \pi}{5} (2 \pi \alpha {\cal M} f_\mathrm{b})^{5/3} F^\prime (e) -
  8 \pi \zeta (2 \pi \mu f_\mathrm{b} ) {\cal S}^2 G(e) \,,
  \label{PdotBD}
  \\
  r &=& m_2 (1-\zeta) \,,
  \\
  s &=& \sin i \,,
\end{eqnarray}%
where 
\begin{eqnarray}
{\cal P} & = & 1 + \frac{1}{3} \left ( 2\bar{\gamma} - \bar{\beta}_{+} +\psi \bar{\beta}_{-} \right )\,,
\\
  F^\prime (e) & = & \frac{1}{12}  (1-e^2 )^{-7/2}
  \left [ \kappa_1 \left ( 1 + \frac{7}{2} e^2 + \frac{1}{2} e^4 \right ) - \frac{1}{2} \kappa_2 e^2 \left ( 1+ \frac{1}{2} e^2 \right ) \right ] \,,
  \\
  G(e) & = &
  (1-e^2 )^{-5/2} \left( 1+ \frac{1}{2} e^2 \right) \,,
  \label{BDcoefficients}
\end{eqnarray}%
where $\kappa_1$ and $\kappa_2$ are defined in Eq.~ (\ref{kappas}).
The quantity $\kappa_\mathrm{A}^*$ is defined by
\begin{equation}
  \kappa_\mathrm{A}^* =
   \left( \frac{\partial (\ln I_\mathrm{A} )}{\partial (\ln \phi)} \right) \,, 
  \label{sensitivities}
\end{equation}
and measures the ``sensitivity'' of the  moment
of inertia $I_\mathrm{A}$ of each body to changes in the scalar field
 for a fixed baryon number $N$ (see
TEGP~11, 12 and 14.6~(c)~\cite{tegp} for further details). The sensitivities $s_A$ and $\kappa_A^*$ will
depend on the neutron-star equation of state. Notice how the violation of
SEP in scalar-tensor theory introduces complex structure-dependent
effects in everything from the Newtonian limit (modification of the
effective coupling constant in Kepler's third law) to
gravitational radiation. In the limit $\zeta \to 0$, we recover GR, and
all structure dependence disappears.
The first term in $\dot P_\mathrm{b}$ (see Eq.~ (\ref{PdotBD}))
is the combined effect of quadrupole and
monopole gravitational radiation, post-Newtonian corrections to dipole radiation, and a dipole-octupole coupling term, all contributing at $0$PN order, while the second term is the
effect of dipole radiation, contributing at the dominant $-1$PN order.

\begin{figure}[h!t]
\centering
 % \def\epsfsize#1#2{0.6#1}
  %\centerline{\epsfbox{STbounds.eps}}
  \includegraphics[width=4in]{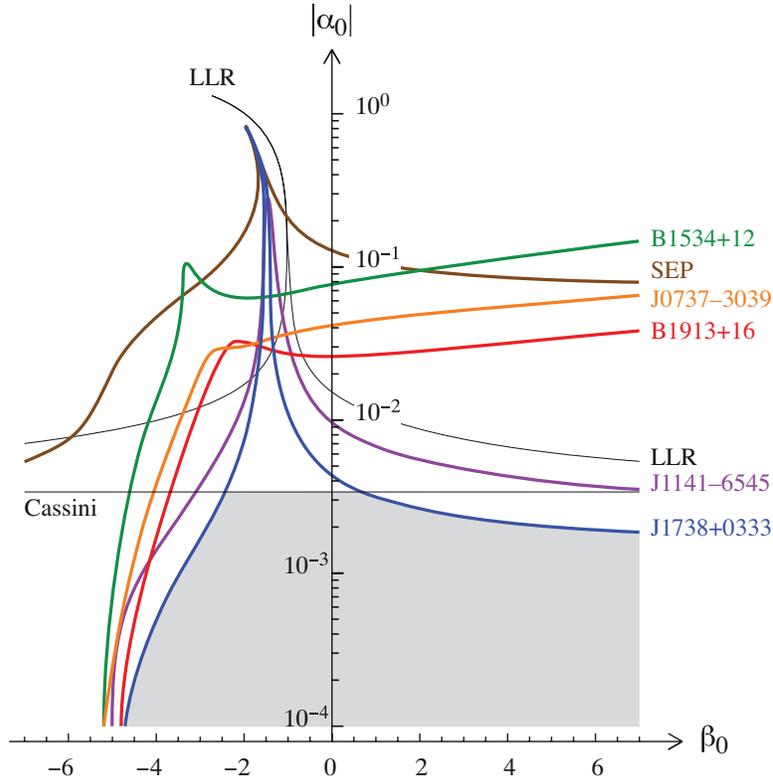}
  \caption{Bounds on the scalar-tensor parameters $\alpha_0$ and $\beta_0$ from solar-system and binary pulsar measurements.  Bounds from tests of the Nordtvedt effect using lunar laser ranging and circular pulsar--white-dwarf binary systems are denoted LLR and SEP, respectively.  Credit Ref.~\cite{2012MNRAS.423.3328F}} \label{STbounds}
\end{figure}

Unfortunately, because of the near equality of  neutron star
masses in typical double neutron star binary pulsars, dipole radiation is somewhat suppressed, and the
bounds obtained are typically not competitive with the 
Cassini bound on $\gamma$, 
except for those generalized scalar-tensor theories,
with $\beta_0 < 0$ where the strong gravity of the neutron stars induces spontaneous scalarization effects~\cite{DamourEspo98}. 
Figure \ref{STbounds} illustrates this:  the bounds on $\alpha_0$ and $\beta_0$ from the three binary neutron star systems B1913+16, J0737-3039, and B1534+12 are not close to being competitive with the Cassini bound on $\alpha_0$, except for very negative values of $\beta_0$ (recall that $\alpha_0 = (3 + 2\omega_0)^{-1/2}$).

On the other hand, a binary pulsar system with dissimilar objects, such as
a white dwarf or black hole companion, provides potentially more
promising tests of dipole radiation.  As a result, the
neutron-star--white-dwarf systems J1141-6545 and J1738+0333 yield much more stringent bounds.  Indeed, the latter system surpasses the Cassini bound for $\beta_0 > 1$ and $\beta_0 < -2$, and is close to that bound for the pure Brans-Dicke case $\beta_0 =0$~\cite{2012MNRAS.423.3328F}.  

It is worth pointing out that the bounds displayed in Fig.~\ref{STbounds} have been calculated using a specific choice of scalar-tensor theory, in which the function $A(\varphi)$ is given by 
\begin{equation}
A(\varphi) = \exp \left [\alpha_0  (\varphi-\varphi_0) + \frac{1}{2} \beta_0 (\varphi -\varphi_0)^2 \right ] \,,
\end{equation}
where $\alpha_0$, and $\beta_0$, are arbitrary parameters, and $\varphi_0$ is the asymptotic value of the scalar field.  In the notation for scalar-tensor theories used here, this theory corresponds to the choice
\begin{equation}
\omega (\phi) = -\frac{1}{2}  \left (3 -  \frac{1}{\alpha_0^2 - \beta_0 \ln \phi} \right )\,,
\end{equation}
where $\phi_0 = A(\varphi_0)^{-2} =1$.  The 
parameters $\zeta$ and $\lambda$ are given by
\begin{eqnarray}
\zeta &=& \frac{\alpha_0^2}{1+\alpha_0^2} \,,
\nonumber \\
\lambda &=& \frac{1}{2} \frac{\beta_0}{1+\alpha_0^2} \,.
\end{eqnarray}
The bounds were also calculated using a polytropic equation of state, which tends to give lower maximum masses for neutron stars than do more realistic equations of state.
%

%\epubtkImage{livingbinary1.png}{
%}

Bounds on various versions of TeVeS theories have also been established, with the tightest constraints again coming from neutron-star--white-dwarf binaries~\cite{2012MNRAS.423.3328F}; in the case of TeVeS, the theory naturally predicts $\gamma =1$ in the post-Newtonian limit, so the Cassini measurements are irrelevant here.

\newpage

%%%%%%%%%%%%%%%%%%%%%%%%%%%%%%%%%%%%%%%%%%%%%%%%%%%%%%%%%%%%%%%%%%%%%%%%%%%%%%%%%%%
%%%%%%%%%%%%%%%%%%%%%%%%%%%%%%%%%%%%%%%%%%%%%%%%%%%%%%%%%%%%%%%%%%%%%%%%%%%%%%%%%%%
%%%%%%%%%%%%%%%%%%%%%%%%%%%%%%%%%%%%%%%%%%%%%%%%%%%%%%%%%%%%%%%%%%%%%%%%%%%%%%%%%%%

\section{Gravitational-Wave Tests of Gravitational Theory}
\label{gwaves}

%%%%%%%%%%%%%%%%%%%%%%%%%%%%%%%%%%%%%%%%%%%%%%%%%%%%%%%%%%%%%%%%%%%%%%%%%%%%%%%%%%%
%%%%%%%%%%%%%%%%%%%%%%%%%%%%%%%%%%%%%%%%%%%%%%%%%%%%%%%%%%%%%%%%%%%%%%%%%%%%%%%%%%%

\subsection{Gravitational-wave observatories}
\label{gwobservatories}

Soon after the publication of this update, a new method of testing
relativistic gravity will be realized, 
when a worldwide network of upgraded laser interferometric
gravitational wave observatories in the U.S.\ (LIGO Hanford and LIGO Livingston) and Europe
(VIRGO and GEO600) begins regular
detection and analysis of gravitational wave signals from astrophysical
sources.  Within a few years, they may be joined by an underground cryogenic interferometer (KAGRA) in Japan, and around 2022, by a LIGO-type interferometer in India.   These
broad-band antennas will have the capability of detecting and
measuring the gravitational waveforms from astronomical sources in a
frequency band between about 10~Hz (the seismic noise cutoff) and
500~Hz (the photon counting noise cutoff), with a maximum
sensitivity to strain at around 100~Hz of $h \sim \Delta l/l \sim 10^{-22}$
(rms), for the kilometer-scale LIGO/VIRGO projects. 
The most promising source for detection and study of
the gravitational wave signal is the ``inspiralling compact binary''
-- a binary system of neutron stars or black holes (or one of each) in
the final minutes of a death spiral leading to a violent merger.
Such is the fate, for example, of the Hulse--Taylor binary pulsar 
B1913+16 in about 300~Myr, or the ``double pulsar'' J0737-3039
in about 85~Myr. Given the expected sensitivity of the
advanced LIGO-Virgo detectors, which could see such sources out to
many hundreds of megaparsecs, it has been estimated that from 40 to
several hundred annual inspiral events could be detectable.
Other sources, such as supernova core collapse events, instabilities
in rapidly rotating newborn neutron stars, signals from
non-axisymmetric pulsars, and a stochastic background of waves, may be
detectable (see~\cite{2009LRR....12....2S} for a review).  

In addition, plans are being developed for orbiting laser
interferometer space antennae, such as DECIGO in Japan and eLISA in Europe.  The eLISA system
would consist of three spacecraft orbiting the sun in a triangular
formation separated from each other by a million kilometers, and
would be sensitive primarily in the very low-frequency band between
$10^{-4}$ and $10^{-1} \mathrm{\ Hz}$, with peak strain sensitivity of order $h \sim
10^{-23}$.  

A third approach that focuses on the ultra low-frequency band (nanohertz) is that of Pulsar Timing Arrays (PTA), whereby a network of highly stable millisecond pulsars is monitored in a coherent way using radio telescopes, in hopes of detecting the fluctuations in arrival times induced by passing gravitational waves.

For recent reviews of the status of all these approaches to gravitational-wave detection, see the Proceedings of the 8th Edoardo Amaldi Conference on Gravitational Waves~\cite{0264-9381-27-8-080301}.

In addition to opening a new astronomical window, the
detailed observation of gravitational waves by such observatories may
provide the means to test general relativistic predictions for the
polarization and speed of the waves, for gravitational radiation
damping and for strong-field gravity.   These topics have been thoroughly covered in two recent {\em Living Reviews} by Gair {\em et al.}~\cite{2013LRR....16....7G} and by Yunes and Siemens~\cite{2013LRR....16....9Y}.  Here we present a brief overview.

\subsection{Gravitational-wave amplitude and polarization}
\label{gwdetection}

\subsubsection{General relativity}

A generic gravitational wave detector can be modelled as a body of mass $M$ at
a
distance $L$ from a fiducial laboratory point, connected to the point
by a spring of resonant frequency $\omega_0$ and quality factor $Q$.
From the equation of geodesic deviation, the infinitesimal
displacement $\xi$ of the mass along the line of separation from its
equilibrium position satisfies the equation of motion
\begin{equation}
  \ddot \xi + \frac{2 \omega_0}{Q} \dot \xi + \omega_0^2 \xi =
  \frac{L}{2} \left( F_+ (\theta, \phi, \psi) {\ddot h}_+ (t) +
  F_\times (\theta, \phi, \psi) {\ddot h}_\times (t) \right),
  \label{detector}
\end{equation}
where $F_+ (\theta,\phi,\psi)$ and $F_\times
(\theta,\phi,\psi)$ are ``beam-pattern'' factors that depend on the
direction of the source $(\theta,\phi)$ and on a polarization  angle
$\psi$, and $h_+(t)$ and $h_\times (t)$ are gravitational waveforms
corresponding to the two polarizations of
the gravitational wave (for pedagogical reviews, see~\cite{Thorne300,PW2014}). In
a source coordinate system in which the $x\mbox{\,--\,}y$
plane is the plane of the sky and the $z$-direction points toward the
detector, these two modes are given by
\begin{equation}
  h_+ (t) = \frac{1}{2}
  \left( h^{xx}_\mathrm{TT} (t) - h^{yy}_\mathrm{TT} (t) \right),
  \qquad
  h_\times (t) = h^{xy}_\mathrm{TT} (t),
  \label{modes}
\end{equation}
where $h^{ij}_\mathrm{TT}$ represent transverse-traceless (TT) projections of the
calculated waveform of Eq.~ (\ref{waveform}), given by
\begin{equation}
  h^{ij}_\mathrm{TT} =
  h^{kl} \left[ \left( \delta^{ik} - \hat{N}^i \hat{N}^k \right)
  \left( \delta^{jl} - \hat{N}^j \hat{N}^l \right) -
  \frac{1}{2} \left( \delta^{ij} - \hat{N}^i \hat{N}^j \right)
  \left( \delta^{kl} - \hat{N}^k \hat{N}^l \right) \right],
  \label{TTprojection}
\end{equation}
where $\hat{N}^j$ is a unit vector pointing toward the detector.
The beam pattern factors depend on the orientation and nature of the
detector. For a wave approaching along the laboratory
$z$-direction, and for a mass whose location on the $x\mbox{\,--\,}y$ plane
makes an angle $\phi$
with the $x$-axis, the beam pattern factors are given by $F_+ =
\cos 2 \phi$ and $F_\times = \sin 2 \phi$.
For a laser interferometer with one arm along
the laboratory $x$-axis, the other along the $y$-axis, with $\xi$
defined as the \emph{differential} displacement along the two arms, the beam pattern functions are 
\begin{eqnarray}
F_+ &=& \frac{1}{2} (1+\cos^2 \theta
)\cos 2 \phi \, \cos 2 \psi - \cos \theta \, \sin 2 \phi \, \sin 2 \psi \,,
\nonumber \\
F_\times &=& \frac{1}{2} (1+\cos^2 \theta ) \cos 2 \phi \,
\sin 2 \psi + \cos \theta \, \sin 2 \phi \, \cos 2 \psi \,.
\end{eqnarray}
Here, we assume that $\omega_0 \approx 0$ in Eq.~(\ref{detector}), corresponding to the essentially free motion of the suspended mirrors in the horizontal direction.  
The waveforms $h_+ (t)$ and $h_\times (t)$ depend on the nature and
evolution of the source. For example, for a binary system in a
circular orbit, with an inclination $i$ relative to the plane of the
sky, and the $x$-axis oriented along the major axis of the projected
orbit, the quadrupole approximation of Eq.~ (\ref{Qij}) gives
\begin{eqnarray}
  h_+ (t) & = &
  - \frac{2{\cal M}}{R} (2\pi {\cal M} f_\mathrm{b})^{2/3}
  (1 + \cos^2 i) \, \cos 2 \Phi_\mathrm{b} (t),
  \\
  h_\times (t) & = & - \frac{2{\cal M}}{R}
  (2\pi {\cal M} f_\mathrm{b})^{2/3} \, (2 \cos i )\, \cos 2 \Phi_\mathrm{b} (t),
\end{eqnarray}%
where $\Phi_\mathrm{b} (t) = 2\pi \int^t f_\mathrm{b} (t') \, dt'$ is the orbital
phase.

%%%%%%%%%%%%%%%%%%%%%%%%%%%%%%%%%%%%%%%%%%%%%%%%%%%%%%%%%%%%%%%%%%%%%%%%%%%%%%%%%%%
%%%%%%%%%%%%%%%%%%%%%%%%%%%%%%%%%%%%%%%%%%%%%%%%%%%%%%%%%%%%%%%%%%%%%%%%%%%%%%%%%%%

\subsubsection{Alternative theories of gravity}

\begin{figure}[h!t]
\centering
 % \def\epsfsize#1#2{0.5#1}
 % \centerline{\epsfbox{livingmodes.eps}}
 \includegraphics[width=4in]{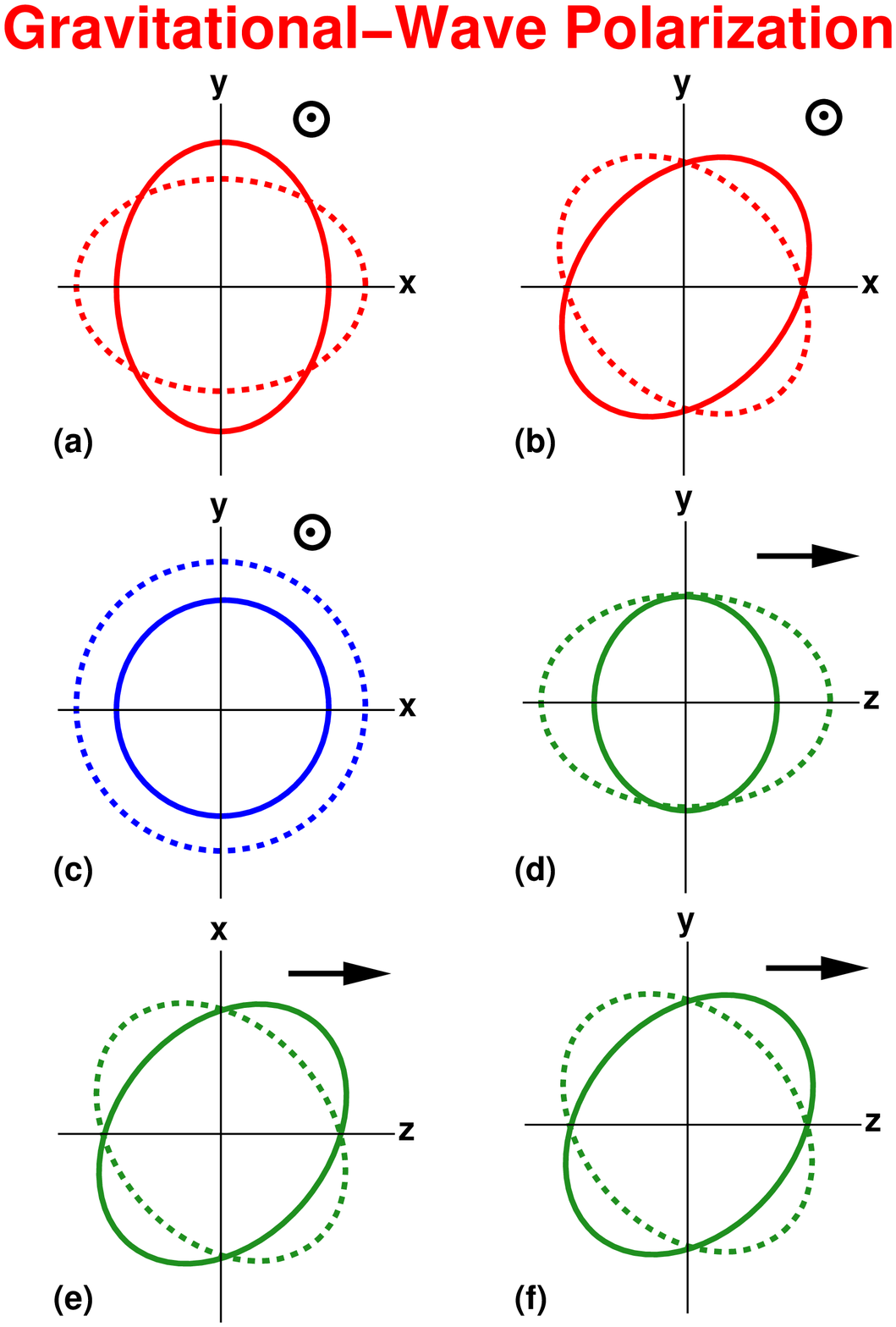}
  \caption{The six polarization modes for gravitational waves
    permitted in any metric theory of gravity. Shown is the
    displacement that each mode induces on a ring of test particles.
    The wave propagates in the $+z$ direction. There is no
    displacement out of the plane of the picture. In (a), (b), and
    (c), the wave propagates out of the plane; in (d), (e), and (f),
    the wave propagates in the plane. In GR, only (a)
    and (b) are present; in massless scalar-tensor gravity, (c) may also be
    present.}
  \label{wavemodes}
\end{figure}

A generic gravitational wave detector
whose scale is small compared to the gravitational wavelength
measures the local components of a symmetric $3\times3$ tensor which
is composed of the ``electric'' components of the Riemann curvature tensor,
$R_{0i0j}$, via the equation of geodesic deviation, given, for a pair
of freely falling particles by
$ {\ddot x}^i =  - R_{0i0j} x^j $,
where $x^i$ denotes the spatial separation.
In general there are six independent components, which can be expressed in
terms of polarizations (modes with specific transformation properties
under rotations and boosts); for a wave propagating in the $z$-direction, they can be displayed by the matrix 
\begin{equation}
S^{jk} = \left( 
\begin{array}{ccc} 
A_{\rm S} + A_+  & A_\times & A_{{\rm V} 1} \\ 
A_\times & A_{\rm S} -A_+ & A_{{\rm V} 2} \\ 
A_{{\rm V}1} & A_{{\rm V}2} & A_{\rm L} 
\end{array} \right) \,. 
\label{eq14:smatrix}
\end{equation}
Three modes ($A_{+}$, $A_\times$, and $A_{\rm S}$) are transverse to the direction of
propagation, with two representing quadrupolar deformations and one
representing a monopolar ``breathing'' deformation. Three modes are
longitudinal, with one ($A_{\rm L}$) an axially symmetric
stretching mode in the propagation direction,
and one quadrupolar mode in each of the two orthogonal planes containing the
propagation direction ($A_{{\rm V} 1}$ and $A_{{\rm V} 2}$). Figure~\ref{wavemodes} shows the displacements
induced on a ring of freely falling test particles by each of these
modes.
General relativity predicts only the first two
transverse quadrupolar modes (a) and (b) independently of the source; these
correspond to the waveforms $h_+$
and $h_\times$
discussed earlier (note the $\cos 2 \phi $ and $\sin 2 \phi $
dependences of the displacements).

Massless scalar-tensor gravitational waves
can in addition contain the transverse breathing mode (c).  This can be obtained from the physical waveform $h^{\alpha\beta}$, which is related to $\tilde{h}^{\alpha\beta}$ and $\varphi$ to the required order by
\begin{equation}
h^{\alpha\beta} = \tilde{h}^{\alpha\beta} + \Psi \eta^{\alpha\beta} \,,
\end{equation}
where $\Psi = \varphi-1$.   In this case, $A_{+(-)} \propto \tilde{h}_{+(-)}$, while $A_{\rm S} \propto \Psi$ (see Eqs.~(\ref{STwaveform}), (\ref{STQij}), (\ref{Psi0}) and (\ref{Psiterms}) for the leading contributions to these fields).
In massive
scalar-tensor theories, the longitudinal mode (d) can also be present, but
is
suppressed relative to (c) by a factor $(\lambda/\lambda_\mathrm{C})^2$, where $\lambda$
is the wavelength of the radiation, and $\lambda_\mathrm{C}$ is the Compton
wavelength of the massive scalar. 

More general
metric theories predict additional longitudinal modes,
up to the full complement of
six (TEGP~10.2~\cite{tegp}).

A suitable array of gravitational antennas could delineate or limit
the number of modes present in a given wave. The strategy depends on
whether or not the source direction is known.
In general there are eight unknowns (six polarizations and two direction
cosines), but only six measurables ($R_{0i0j}$). If the direction can
be established by either association of the waves with optical or
other observations, or by time-of-flight measurements between
separated detectors, then six suitably oriented detectors suffice to
determine all six components. If the direction cannot be established,
then the system is underdetermined, and no unique solution can be
found. However, if one assumes that only transverse waves are
present, then there are only three unknowns if the source direction is
known, or five unknowns otherwise. Then the corresponding number
(three or five) of detectors can determine the polarization. If
distinct evidence were found of any mode other than the two
transverse quadrupolar modes of GR, the result would be disastrous for
GR. On the other hand, the absence of a breathing mode would not
necessarily rule out scalar-tensor gravity, because the strength
of that mode depends on the nature of the source.

%\epubtkImage{livingmodes.png}{
%}

For laser interferometers, the signal controlling the laser phase output can be written in the form
\begin{equation}
S(t) = \frac{1}{2} \bigl( {e}_1^j {e}_1^k - {e}_2^j {e}_2^k \bigr) 
S^{jk}  \,, 
\end{equation}
where ${\bf e}_1$ and ${\bf e_2}$ are unit vectors directed along the two arms of the interferometer. The final result is
\begin{equation}
S(t) = F_{\rm S} A_{\rm S} + F_{\rm L} A_{\rm L} 
+ F_{{\rm V}1} A_{{\rm V}1} + F_{{\rm V}2} A_{{\rm V}2}
+ F_+ A_+ + F_\times A_\times\,,
\label{patternfunctions1}
\end{equation}
where the angular pattern functions $F_A (\theta, \phi, \psi)$ are
given by  
\begin{eqnarray}
F_{\rm S} &=& -\frac{1}{2} \sin^2\theta \cos2\phi \,, 
\nonumber \\
F_{\rm L} &=& \frac{1}{2} \sin^2\theta \cos2\phi \,, 
\nonumber \\ 
F_{{\rm V}1} &=& -\sin\theta (\cos\theta \cos2\phi \cos\psi 
- \sin2\phi \sin\psi ) \,, 
\nonumber \\ 
F_{{\rm V}2} &=& -\sin\theta (\cos\theta \cos2\phi \sin\psi 
+ \sin2\phi \cos\psi ) \,, 
\nonumber \\ 
F_+ &=& \frac{1}{2} (1+ \cos^2\theta ) \cos2\phi \cos2\psi
- \cos\theta \sin2\phi \sin2\psi \,, 
\nonumber \\ 
F_\times &=& \frac{1}{2} (1+ \cos^2\theta ) \cos2\phi \sin2\psi 
+ \cos\theta \sin2\phi \cos2\psi \,,
\label{patternfunctions2}
\end{eqnarray}
(see~\cite{PW2014} for detailed derivations and definitions).
Note that the scalar and longitudinal pattern functions are degenerate and thus no array of laser interferometers can measure these two modes separately.  

Some
of the details of implementing such polarization observations have
been worked out for arrays of resonant cylindrical, disk-shaped,
spherical, and truncated icosahedral
detectors (TEGP~10.2 \cite{tegp}, for recent reviews see~\cite{lobo, wagoner}).
Early work to assess whether
the ground-based or space-based
laser interferometers (or combinations of the two types) could perform
interesting polarization 
measurements was carried out in~\cite{wagoner2, brunetti, maggiore, gasperini, WenSchutz05}; for a recent detailed analysis see \cite{2009PhRvD..79h2002N}.
Unfortunately for this purpose, the two LIGO observatories (in Washington
and Louisiana states, respectively) have been constructed to have their
respective arms as parallel as possible, apart from the curvature of the
Earth; while this maximizes the joint sensitivity of the two detectors to
gravitational waves, it minimizes their ability to detect two modes of
polarization. In this regard the addition of Virgo, and the future KAGRA and LIGO-India systems will be crucial to polarization measurements.   The capability of space-based interferometers to measure the polarization modes was assessed in detail in~\cite{2010PhRvD..82l2003T,2010PhRvD..81j4043N}.  For pulsar timing arrays, see~\cite{2008ApJ...685.1304L,2011PhRvD..83l3529A,2012PhRvD..85h2001C}.

%%%%%%%%%%%%%%%%%%%%%%%%%%%%%%%%%%%%%%%%%%%%%%%%%%%%%%%%%%%%%%%%%%%%%%%%%%%%%%%%%%%
%%%%%%%%%%%%%%%%%%%%%%%%%%%%%%%%%%%%%%%%%%%%%%%%%%%%%%%%%%%%%%%%%%%%%%%%%%%%%%%%%%%

\subsection{Gravitational-wave phase evolution}
\label{backreaction}

\subsubsection{General relativity}

In the binary pulsar, a test of GR was made possible by measuring at
least three relativistic effects that depended upon only two unknown
masses. The evolution of the orbital phase under the damping effect
of gravitational radiation played a crucial role. Another situation
in which measurement of orbital phase can lead to tests of GR is that
of the inspiralling compact binary system. The key differences are
that here gravitational radiation itself is the detected signal,
rather than radio pulses, and the phase evolution alone carries all
the information. In the binary pulsar, the first derivative of the
binary frequency $\dot f_\mathrm{b}$ was measured; here the full nonlinear
variation of $f_\mathrm{b}$ as a function of time is measured.

Broad-band laser interferometers
are especially sensitive to the phase evolution of the gravitational
waves, which carry the information about the orbital phase evolution.
The analysis of gravitational wave data from such sources will involve
some form of matched filtering of the noisy detector output against an
ensemble of theoretical ``template'' waveforms which depend on the
intrinsic parameters of the inspiralling binary, such as the component
masses, spins, and so on, and on its inspiral evolution.
How accurate must a template be in order to
``match'' the waveform from a given source (where by a match we mean
maximizing the cross-correlation or the
signal-to-noise ratio)?  In the total accumulated phase
of the wave detected in the sensitive bandwidth, the template must
match the signal to a fraction of a cycle. For two inspiralling
neutron stars, around 16,000~cycles should be detected during the final few
minutes of inspiral; this implies a
phasing accuracy of $10^{-5}$ or better. Since $v \sim 1/10$ during
the late inspiral, this means that correction terms in the phasing
at the level of
$v^5$ or higher are needed. More formal analyses confirm this
intuition~\cite{3min, finnchern, cutlerflan, poissonwill}.

Because it is a slow-motion system ($v \sim 10^{-3}$), the binary
pulsar is sensitive only to the lowest-order effects of gravitational
radiation as predicted by the quadrupole formula. Nevertheless, the
first correction terms of order $v$ and $v^2$ to the quadrupole formula
were
calculated as early as 1976~\cite{wagwill} (see TEGP~10.3~\cite{tegp}).

But for laser interferometric observations of gravitational waves,
the bottom line is that, in order to measure the astrophysical
parameters of the source and to test the properties of the
gravitational waves, it is necessary to derive the
gravitational waveform and the resulting radiation back-reaction on the orbit
phasing at least to 3PN
order beyond the quadrupole
approximation.

For the special case of
non-spinning bodies moving on quasi-circular orbits
(i.e.\ circular apart from a slow inspiral), the evolution of the
gravitational wave frequency $f = 2f_\mathrm{b}$ through 2PN
order has the form
\begin{eqnarray}
  \dot f = \frac{96 \pi}{5} f^2 (\pi {\cal M} f)^{5/3}
  \biggl [ \!\!\! & & 1 - \left( \frac{743}{336} + \frac{11}{4} \eta \right)
  (\pi mf)^{2/3} + 4\pi(\pi mf)
  \nonumber
  \\
  & & + \left( \frac{34103}{18144} + \frac{13661}{2016} \eta +
  \frac{59}{18} \eta^2 \right) (\pi mf)^{4/3} +
  {\cal O} [(\pi mf)^{5/3}] \biggr ],
  \label{fdot2PN}
\end{eqnarray}%
where $\eta= m_1m_2/m^2$. The first term is the quadrupole
contribution (compare Eq.~ (\ref{fdotGR})),
the second term is the 1PN contribution, the
third term, with the coefficient $4\pi$,
is the ``tail'' contribution,
and the fourth term is the 2PN
contribution.
Two decades of intensive work by many groups have led to the development of waveforms in GR that are accurate to 3.5PN order, and for some specific effects, such as those related to spin, to 4.5PN order
(see~\cite{BlanchetLRR} for a thorough review).

Similar expressions can be derived for the loss of angular momentum
and linear momentum. Expressions for non-circular
orbits have also been derived~\cite{gopu, damgopuiyer04}. These losses react
back on the orbit to circularize it and cause it to inspiral. The
result is that the orbital phase (and consequently the
gravitational wave
phase) evolves non-linearly with time. It is the sensitivity of the
broad-band laser interferometric detectors to phase that makes the
higher-order contributions to $df/dt$ so observationally relevant.

If the coefficients of each of the powers of $f$ in Eq.~ (\ref{fdot2PN})
can be measured, then one again obtains more than two constraints on
the two unknowns $m_1$ and $m_2$, leading to the possibility to test
GR. For example,
Blanchet and Sathyaprakash~\cite{lucsathya2, lucsathya} have shown that, by
observing a source with a sufficiently strong signal, an interesting
test of the $4\pi$ coefficient of the ``tail'' term could be
performed.

Another possibility involves gravitational waves from a small mass
orbiting and inspiralling into a (possibly supermassive) spinning black hole.
A general non-circular, non-equatorial orbit will precess around the
hole, both in periastron and in orbital plane,
leading to a complex gravitational waveform that carries
information about the non-spherical, strong-field spacetime around the hole.
According to GR, this spacetime must be the Kerr spacetime of a
rotating black hole, uniquely specified by its mass and angular
momentum, and consequently, observation of the waves could test this
fundamental hypothesis of GR~\cite{ryan, poissonBH}.

\subsubsection{Alternative theories of gravity}

In general, alternative theories of gravity will predict rather different phase evolution from that of GR, notably via the addition of
dipole gravitational radiation.  
For example, the dipole gravitational radiation predicted by scalar-tensor
theories modifies the gravitational radiation
back-reaction, and thereby the phase evolution.
Including only the leading 0PN and $-1$PN (dipole) contributions, one
obtains,
\begin{equation}
  \dot f = \frac{96 \pi}{5} f^2 (\pi \alpha {\cal M} f)^{5/3} \frac{\kappa_1}{12}
  \left [ 1 + b (\pi mf)^{-2/3} \right],
  \label{fdotBD}
\end{equation}
where ${\cal M} = \eta^{3/5} m$,
and $b$ is the coefficient of the dipole term, given to first order in $\zeta$ by
$b= (5/24) \zeta \alpha^{-5/3} {\cal S}^2$, where
$\kappa_1$ is given by Eq.~ (\ref{kappas}), ${\cal S} = \alpha^{-1/2} (s_1-s_2)$
and $\zeta = 1/(4+2\omega_0)$.
Double neutron star systems are not
promising because the small range of masses available near
$1.4 \, M_\odot$ results in suppression of dipole radiation by symmetry.
For black holes, $s=0.5$ identically, consequently
double black hole systems turn out to be observationally
identical in the two theories.
Thus mixed systems involving a neutron star and a
black hole are preferred. However, a number of analyses of the
capabilities of both ground-based
and space-based (LISA) observatories have shown that observing waves from
neutron-star--black-hole inspirals is not likely to bound scalar-tensor
gravity at a level competitive with the Cassini bound, with future
solar-system improvements, or with binary pulsar observations~\cite{willbd, krolak95, DamourEspo98, scharrewill, willyunes, bbw1, bbw2}. 

These considerations suggest that it might be fruitful to attempt to parametrize the phasing formulae in a manner reminiscent of the PPN framework for post-Newtonian gravity.  A number of approaches along this line have been developed~\cite{2009PhRvD..80l2003Y,2010PhRvD..82f4010M}.

%%%%%%%%%%%%%%%%%%%%%%%%%%%%%%%%%%%%%%%%%%%%%%%%%%%%%%%%%%%%%%%%%%%%%%%%%%%%%%%%%%%
%%%%%%%%%%%%%%%%%%%%%%%%%%%%%%%%%%%%%%%%%%%%%%%%%%%%%%%%%%%%%%%%%%%%%%%%%%%%%%%%%%%

\subsection{Speed of gravitational waves}

According to GR, in the limit in which the wavelength of gravitational
waves is small compared to the radius of curvature of the background
spacetime, the waves propagate along null geodesics of the background
spacetime, i.e.\ they have the same speed $c$ as light (in this
section, we do not set $c=1$). In other
theories, the speed could differ from $c$ because of coupling of
gravitation to ``background'' gravitational fields. For example, in
the Rosen bimetric theory with a flat background metric
{\boldmath $\eta$},
gravitational waves follow null geodesics of {\boldmath $\eta$},
while light follows null geodesics of ${\bf {g}}$ (TEGP~10.1~\cite{tegp}).

Another way in which the speed of gravitational waves could differ
from $c$ is if gravitation were propagated by a massive field (a
massive graviton), in which case $v_\mathrm{g}$ would
be given by, in a local inertial frame,
\begin{equation}
  \frac{v_\mathrm{g}^2}{c^2} = 1 - \frac{m_\mathrm{g}^2 c^4}{E^2},
  \label{eq1}
\end{equation}
where $m_\mathrm{g}$ and $E$ are the graviton rest mass and energy,
respectively. 

The most obvious way to test this is to
compare the arrival times of a gravitational wave and an
electromagnetic
wave from the same event, e.g., a supernova or a prompt gamma-ray burst.
For a source at a distance $D$, the
resulting value of the difference $1-v_\mathrm{g}/c$ is
\begin{equation}
  1 - \frac{v_\mathrm{g}}{c} = 5 \times 10^{-17}
  \left( \frac{200 \mathrm{\ Mpc}}{D} \right)
  \left( \frac{\Delta t}{1 \mathrm{\ s}} \right),
  \label{eq2}
\end{equation}
where $\Delta t \equiv \Delta t_\mathrm{a} - (1+Z) \Delta t_\mathrm{e} $
is the ``time difference'',
where $\Delta t_\mathrm{a}$ and $\Delta t_\mathrm{e}$ are the differences
in arrival time and emission time of the
two signals, respectively, and $Z$ is the redshift of the source.
In many cases, $\Delta t_\mathrm{e}$ is unknown,
so that the best one can do is employ an upper bound on
$\Delta t_\mathrm{e}$ based on observation or modelling.
The result will then be a bound on $1-v_\mathrm{g}/c$.

For a massive graviton, if
the frequency of the gravitational waves is such that $hf \gg
m_\mathrm{g}c^2$,
where $h$ is Planck's constant, then $v_\mathrm{g}/c \approx 1 - \frac{1}{2}
(c/\lambda_\mathrm{g}
f)^{2}$, where $\lambda_\mathrm{g}= h/m_\mathrm{g}c$ is the graviton Compton wavelength,
and
the bound on $1-v_\mathrm{g}/c$ can be converted to a bound on $\lambda_\mathrm{g}$,
given by
\begin{equation}
  \lambda_\mathrm{g} > 3 \times 10^{12} \mathrm{\ km}
  \left( \frac{D}{200 \mathrm{\ Mpc}}
  \frac{100 \mathrm{\ Hz}}{f} \right)^{1/2}
  \left( \frac{1}{f \Delta t} \right)^{1/2}\!\!\!\!\!\!\!.
  \label{eq4}
\end{equation}

The foregoing discussion assumes that the source emits \emph{both}
gravitational and electromagnetic radiation in detectable amounts, and
that the relative time of emission can be established
to sufficient accuracy, or can be shown to be sufficiently
small.

However, there is a situation in which a bound on the graviton mass
can be set using gravitational radiation alone~\cite{graviton}.
That is the case of
the inspiralling compact binary. Because the frequency of the
gravitational radiation sweeps from low frequency at the initial
moment of observation to higher frequency at the final moment, the
speed of the gravitons emitted will vary, from lower speeds initially
to higher speeds (closer to $c$) at the end. This will cause a
distortion of the observed phasing of the waves and result in a
shorter than expected
overall time $\Delta t_\mathrm{a}$ of passage of a given number of cycles.
Furthermore, through the technique of matched filtering, the
parameters of the compact binary can be measured accurately
(assuming that GR is a good approximation to the orbital evolution, even in
the presence of a massive graviton), and
thereby the emission time $\Delta t_\mathrm{e}$ can be determined accurately.
Roughly speaking, the ``phase interval'' $f\Delta t$ in
Eq.~ (\ref{eq4}) can be measured to an accuracy $1/\rho$, where $\rho$ is the
signal-\-to-\-noise
ratio.

Thus one can estimate the bounds on $\lambda_\mathrm{g}$ achievable for various
compact inspiral systems, and for various detectors. For
stellar-mass inspiral (neutron stars or black holes) observed by the
LIGO/VIRGO class of ground-based interferometers,
$D \approx 200 \mathrm{\ Mpc}$, $f \approx 100 \mathrm{\ Hz}$, and
$f\Delta t \sim \rho^{-1} \approx 1/10$.
The result is $\lambda_\mathrm{g} > 10^{13} \mathrm{\ km}$. For supermassive binary
black holes ($10^4$ to $10^7 \, M_\odot$) observed by the proposed laser
interferometer space antenna (LISA), $D \approx 3 \mathrm{\ Gpc}$, $f \approx 10^{-3} \mathrm{\ Hz}$,
and $f\Delta t \sim
\rho^{-1} \approx 1/1000$. The result is $\lambda_\mathrm{g} >
10^{17} \mathrm{\ km}$.

A full noise analysis using proposed noise curves for the advanced
LIGO and for LISA weakens these crude bounds by factors between two and 
10~\cite{graviton, willyunes, bbw1, bbw2,2009CQGra..26o5002A,2009PhRvD..80d4002S}. For example, for the inspiral of
two $10^6 \, M_\odot$ black holes with aligned spins at a distance of
$3 \mathrm{\ Gpc}$ observed by LISA,
a bound of $2 \times 10^{16} \mathrm{\ km}$ could be placed~\cite{bbw1}. Other
possibilities include using binary pulsar data to bound modifications of
gravitational radiation damping by a massive graviton~\cite{FinnSutton02},
using LISA observations of the phasing of waves from compact white-dwarf
binaries, eccentric galactic binaries, and eccentric inspiral 
binaries~\cite{CutlerLarHis03, Jones05}, and using pulsar timing arrays~\cite{2010ApJ...722.1589L}.

Bounds obtainable from gravitational radiation effects
should be compared 
with the solid bound 
$\lambda_\mathrm{g} > 2.8 \times 10^{12} \mathrm{\ km}$~\cite{talmadge}
derived from solar system dynamics, which limit
the presence of a Yukawa modification of Newtonian gravity of the form
\begin{equation}
  V(r) = \frac{GM}{r} \exp(-r/\lambda_\mathrm{g}),
\end{equation}
and with the model-dependent bound
$\lambda_\mathrm{g} > 6\times 10^{19} \mathrm{\ km}$ from consideration of
galactic and cluster dynamics~\cite{visser}.

Mirshekari {\em et al.}~\cite{2012PhRvD..85b4041M} studied bounds that could be placed on more general graviton dispersion relations that could emerge from alternative theories with Lorentz violation, in which the effective propagation speed is given by
\begin{equation}
\label{dispersionMYW}
\frac{v_{g}^{2}}{c^2}  = 1 - \frac{m_{\mathrm g}^2 c^4}{E^{2}} - \mathbb{A} E^{\alpha-2}
\,,
\end{equation}
where $\mathbb{A}$ and $\alpha$ are parameters that depend on the theory.

%%%%%%%%%%%%%%%%%%%%%%%%%%%%%%%%%%%%%%%%%%%%%%%%%%%%%%%%%%%%%%%%%%%%%%%%%%%%%%%%%%%
%%%%%%%%%%%%%%%%%%%%%%%%%%%%%%%%%%%%%%%%%%%%%%%%%%%%%%%%%%%%%%%%%%%%%%%%%%%%%%%%%%%

\section{Astrophysical and cosmological tests}
\label{other}

One of the central difficulties of testing GR in the
strong-field regime is the possibility of contamination by uncertain
or complex physics. In the solar system, weak-field gravitational effects
can
in most cases be measured cleanly and separately from
non-gravitational effects. The remarkable cleanliness of many binary pulsars
permits precise measurements of gravitational phenomena
in a strong-field context.

Unfortunately, nature is rarely so kind.
Still, under suitable conditions, qualitative and even quantitative
strong-field tests of GR could be carried out.

One example is the exploration of the spacetime near black holes
and neutron stars. 
Studies of certain kinds of accretion known as advection-dominated
accretion flow (ADAF) in low-luminosity binary X-ray sources
may yield the signature of the black hole event
horizon~\cite{2008NewAR..51..733N}.
The spectrum of frequencies of quasi-periodic oscillations (QPO) from galactic
black hole binaries may permit measurement of the spins of the black
holes~\cite{Psaltis04}.
Aspects of strong-field gravity and frame-dragging may be revealed in
spectral shapes of iron fluorescence lines from the inner regions of
accretion disks~\cite{2013SSRv..tmp...81R,2013CQGra..30x4004R}.  Using submm VLBI, a collaboration dubbed the Event Horizon Telescope could image our galactic center black hole SgrA* and the black hole in M87 with horizon-scale angular resolution; observation of accretion phenomena at these angular resolutions could provide tests of the spacetime geometry very close to the black hole~\cite{2009astro2010S..68D}.
Tracking of hypothetical stars whose orbits are within a fraction of a milliparsec of SgrA* could test the black hole ``ho-hair'' theorem, via a direct measurement of both the angular momentum $J$ and quadrupole moment $Q$ of the black hole, and a test of the requirement that $Q=-J^2/M$~\cite{2008ApJ...674L..25W}.

Because of uncertainties in the detailed models, the results to date of 
studies like these are
suggestive at best, but the combination of future higher-resolution
observations and better modelling could lead to striking tests of
strong-field predictions of GR.

For a detailed review of strong-field tests of GR using electromagnetic observations, see~\cite{2008LRR....11....9P}.

Another example is in cosmology. From a few seconds after the big bang
until the present, the underlying physics of the universe is well
understood, in terms of a Standard Model of a nearly
spatially flat universe, 13.6~Gyr old,
dominated by cold dark matter and dark energy ($\Lambda$CDM).
Some alternative theories of gravity that
are qualitatively different from GR fail to produce cosmologies that
meet even the minimum requirements of agreeing qualitatively with big-bang
nucleosynthesis (BBN) or the properties of the cosmic microwave background
(TEGP~13.2~\cite{tegp}).
Others, such as Brans--Dicke theory, are sufficiently close to GR (for
large enough $\omega_\mathrm{BD}$) that they conform to all cosmological
observations, given the underlying uncertainties. The generalized
scalar-tensor theories and $f(R)$ theories, however, could have small effective $\omega$ at
early times, while evolving through the attractor mechanism to large
$\omega$ today. 

One way to test such theories is through
big-bang nucleosynthesis, since the abundances of the light elements
produced when the temperature of the universe was about 1~MeV are
sensitive to the rate of expansion at that epoch, which in turn
depends on the strength of interaction between geometry and the
scalar field. Because the universe is radiation-dominated at that
epoch, uncertainties in the amount of cold dark matter or of the
cosmological constant are unimportant. The nuclear reaction rates are
reasonably well understood from laboratory experiments and theory, and
the number of light neutrino families (3) conforms to evidence from
particle accelerators. Thus, within modest uncertainties, one can
assess the quantitative difference between the BBN predictions of GR and
scalar-tensor gravity under strong-field conditions and compare with
observations.
For recent analyses, see~\cite{Santiago97, damourpichon, Clifton05, coc06}.

In addition, many alternative theories, such as $f(R)$ theories have been developed in order to provide an alternative to the dark energy of the standard $\Lambda$CDM model, in particular by modifying gravity on large, cosmological scales, while preserving the conventional solar and stellar system phenomenology of GR.   Since we are now in a period of what may be called ``precision cosmology'', one can begin to envision trying to test alternative theories using the accumulation of data on many fronts, including the growth of large scale structure, cosmic background fluctuations, galactic rotation curves, BBN, weak lensing, baryon acoustic oscillations, etc.  The ``parametrized post-Friedmann'' framework is one initial foray into this arena~\cite{2013PhRvD..87b4015B}.  Other approaches can be found in~\cite{2010PhRvD..81l3508D,2011PhRvD..84l3001D,2012PhRvD..86j3008D,2012JCAP...06..032Z,2012PhRvD..85d3508H}.

%%%%%%%%%%%%%%%%%%%%%%%%%%%%%%%%%%%%%%%%%%%%%%%%%%%%%%%%%%%%%%%%%%%%%%%%%%%%%%%%%%%
%%%%%%%%%%%%%%%%%%%%%%%%%%%%%%%%%%%%%%%%%%%%%%%%%%%%%%%%%%%%%%%%%%%%%%%%%%%%%%%%%%%
%%%%%%%%%%%%%%%%%%%%%%%%%%%%%%%%%%%%%%%%%%%%%%%%%%%%%%%%%%%%%%%%%%%%%%%%%%%%%%%%%%%

\section{Conclusions}
\label{S5}

General relativity has held up under extensive experimental scrutiny.
The question then arises, why bother to continue to test it?  One
reason is that gravity is a fundamental interaction of nature, and as
such requires the most solid empirical underpinning we can provide.
Another is that all attempts to quantize gravity and to unify it with
the other forces suggest that the standard general relativity of Einstein may not be the last word.
Furthermore, the predictions of general relativity are fixed;
the pure theory contains
no adjustable constants so nothing can be changed. Thus every test
of the theory is either a potentially deadly test or a possible probe for
new physics. Although it is remarkable
that this theory, born 100 years ago out of almost pure thought,
has managed to survive every test, the possibility of finding
a discrepancy will continue to drive experiments for years to come.
These experiments will search for new physics beyond Einstein at many different scales: the large distance scales of the astrophysical, galactic, and cosmological realms; scales of very short distances or high energy; and scales related to strong or dynamical gravity.

\newpage

%%%%%%%%%%%%%%%%%%%%%%%%%%%%%%%%%%%%%%%%%%%%%%%%%%%%%%%%%%%%%%%%%%%%%%%%%%%%%%%%%%%
%%%%%%%%%%%%%%%%%%%%%%%%%%%%%%%%%%%%%%%%%%%%%%%%%%%%%%%%%%%%%%%%%%%%%%%%%%%%%%%%%%%
%%%%%%%%%%%%%%%%%%%%%%%%%%%%%%%%%%%%%%%%%%%%%%%%%%%%%%%%%%%%%%%%%%%%%%%%%%%%%%%%%%%

\section{Acknowledgments}

This work has been supported since the initial version 
in part by the National Science Foundation,
Grant Numbers PHY 96-00049, 00-96522, 03-53180, 06-52448, 09Ð65133, 12-60995 and 13-06069, and by the National
Aeronautics and Space Administration, Grant Numbers NAG5-10186 and NNG-
06GI60G. We also
gratefully acknowledge the continuing hospitality of the Institut d'Astrophysique de Paris,
where portions of this update were completed.   We thank Luc Blanchet for helpful comments, and Michael Kramer and Norbert Wex for providing useful figures.
\newpage

%%%%%%%%%%%%%%%%%%%%%%%%%%%%%%%%%%%%%%%%%%%%%%%%%%%%%%%%%%%%%%%%%%%%%%%%%%%%%%%%%%%
%%%%%%%%%%%%%%%%%%%%%%%%%%%%%%%%%%%%%%%%%%%%%%%%%%%%%%%%%%%%%%%%%%%%%%%%%%%%%%%%%%%
%%%%%%%%%%%%%%%%%%%%%%%%%%%%%%%%%%%%%%%%%%%%%%%%%%%%%%%%%%%%%%%%%%%%%%%%%%%%%%%%%%%

\bibliography{refs}

\begin{thebibliography}{100}

\bibitem{adelberger01}
Adelberger, E.G., ``New tests of Einstein's equivalence principle and Newton's
  inverse-square law'', {\em Class. Quantum Grav.}, {\bf 18}, 2397--2405
  (2001). {\small[\href{http://dx.doi.org/10.1088/0264-9381/18/13/302}{DOI}]}.

\bibitem{adelberger03}
Adelberger, E.G., Heckel, B.R.  and Nelson, A.E., ``Tests of the Gravitational
  Inverse-Square Law'', {\em Annu. Rev. Nucl. Part. Sci.}, {\bf 53}, 77--121
  (2003).
  {\small[\href{http://dx.doi.org/10.1146/annurev.nucl.53.041002.110503}{DOI}]},
  {\small[\href{http://arxiv.org/abs/hep-ph/0307284}{{arXiv:hep-ph/0307284
  {\small[hep-ph]}}}]}.

\bibitem{Adelberger91}
Adelberger, E.G., Heckel, B.R., Stubbs, C.W.  and Rogers, W.F., ``Searches for
  new macroscopic forces'', {\em Annu. Rev. Nucl. Sci.}, {\bf 41}, 269--320
  (1991).
  {\small[\href{http://dx.doi.org/10.1146/annurev.ns.41.120191.001413}{DOI}]}.

\bibitem{2007PhRvL..98m1104A}
{Adelberger}, E.~G., {Heckel}, B.~R., {Hoedl}, S., {Hoyle}, C.~D., {Kapner},
  D.~J.  and {Upadhye}, A., ``{Particle-Physics Implications of a Recent Test
  of the Gravitational Inverse-Square Law}'', {\em Phys. Rev. Lett.}, {\bf 98},
  131104 (2007).
  {\small[\href{http://dx.doi.org/10.1103/PhysRevLett.98.131104}{DOI}]},
  {\small[\href{http://adsabs.harvard.edu/abs/2007PhRvL..98m1104A}{ADS}]},
  {\small[\href{http://arxiv.org/abs/hep-ph/0611223}{{hep-ph/0611223}}]}.

\bibitem{2009PhR...480....1A}
{Alexander}, S.  and {Yunes}, N., ``{Chern-Simons modified general
  relativity}'', {\em Phys. Rep.}, {\bf 480}, 1--55 (2009).
  {\small[\href{http://dx.doi.org/10.1016/j.physrep.2009.07.002}{DOI}]},
  {\small[\href{http://adsabs.harvard.edu/abs/2009PhR...480....1A}{ADS}]},
  {\small[\href{http://arxiv.org/abs/0907.2562}{{arXiv:0907.2562
  {\small[hep-th]}}}]}.

\bibitem{2012PhRvD..85f4041A}
{Alsing}, J., {Berti}, E., {Will}, C.~M.  and {Zaglauer}, H., ``{Gravitational
  radiation from compact binary systems in the massive Brans-Dicke theory of
  gravity}'', {\em \prd}, {\bf 85}, 064041 (2012).
  {\small[\href{http://dx.doi.org/10.1103/PhysRevD.85.064041}{DOI}]},
  {\small[\href{http://adsabs.harvard.edu/abs/2012PhRvD..85f4041A}{ADS}]},
  {\small[\href{http://arxiv.org/abs/1112.4903}{{arXiv:1112.4903
  {\small[gr-qc]}}}]}.

\bibitem{alvager}
Alv{\"{a}}ger, T., Farley, F.J.M., Kjellman, J.  and Wallin, I., ``Test of the
  second postulate of special relativity in the GeV region'', {\em Phys.
  Lett.}, {\bf 12}, 260--262 (1977).

\bibitem{AlvarezMann96b}
Alvarez, C.  and Mann, R.B., ``The equivalence principle and anomalous magnetic
  moment experiments'', {\em Phys. Rev. D}, {\bf 54}, 7097--7107 (1996).
  {\small[\href{http://dx.doi.org/10.1103/PhysRevD.54.7097}{DOI}]},
  {\small[\href{http://arxiv.org/abs/gr-qc/9511028}{{gr-qc/9511028}}]}.

\bibitem{AlvarezMann96a}
Alvarez, C.  and Mann, R.B., ``Testing the equivalence principle by Lamb shift
  energies'', {\em Phys. Rev. D}, {\bf 54}, 5954--5974 (1996).
  {\small[\href{http://dx.doi.org/10.1103/PhysRevD.54.5954}{DOI}]},
  {\small[\href{http://arxiv.org/abs/gr-qc/9507040}{{gr-qc/9507040}}]}.

\bibitem{AlvarezMann97a}
Alvarez, C.  and Mann, R.B., ``The equivalence principle and g-2 experiments'',
  {\em Phys. Lett. B}, {\bf 409}, 83--87 (1997).
  {\small[\href{http://dx.doi.org/10.1016/S0370-2693(97)00801-0}{DOI}]},
  {\small[\href{http://arxiv.org/abs/gr-qc/9510070}{{gr-qc/9510070}}]}.

\bibitem{AlvarezMann97b}
Alvarez, C.  and Mann, R.B., ``The equivalence principle in the non-baryonic
  regime'', {\em Phys. Rev. D}, {\bf 55}, 1732--1740 (1997).
  {\small[\href{http://dx.doi.org/10.1103/PhysRevD.55.1732}{DOI}]},
  {\small[\href{http://arxiv.org/abs/gr-qc/9609039}{{gr-qc/9609039}}]}.

\bibitem{AlvarezMann97c}
Alvarez, C.  and Mann, R.B., ``Testing the equivalence principle using atomic
  vacuum energy shifts'', {\em Mod. Phys. Lett. A}, {\bf 11}, 1757--1763
  (1997). {\small[\href{http://arxiv.org/abs/gr-qc/9612031}{{gr-qc/9612031}}]}.

\bibitem{2011PhRvD..83l3529A}
{Alves}, M.~E.~S.  and {Tinto}, M., ``{Pulsar timing sensitivities to
  gravitational waves from relativistic metric theories of gravity}'', {\em
  \prd}, {\bf 83}, 123529 (2011).
  {\small[\href{http://dx.doi.org/10.1103/PhysRevD.83.123529}{DOI}]},
  {\small[\href{http://adsabs.harvard.edu/abs/2011PhRvD..83l3529A}{ADS}]},
  {\small[\href{http://arxiv.org/abs/1102.4824}{{arXiv:1102.4824
  {\small[gr-qc]}}}]}.

\bibitem{1998PhRvL..81.2858A}
{Anderson}, J.~D., {Laing}, P.~A., {Lau}, E.~L., {Liu}, A.~S., {Nieto}, M.~M.
  and {Turyshev}, S.~G., ``{Indication, from Pioneer 10/11, Galileo, and
  Ulysses Data, of an Apparent Anomalous, Weak, Long-Range Acceleration}'',
  {\em Phys. Rev. Lett.}, {\bf 81}, 2858--2861 (1998).
  {\small[\href{http://dx.doi.org/10.1103/PhysRevLett.81.2858}{DOI}]},
  {\small[\href{http://adsabs.harvard.edu/abs/1998PhRvL..81.2858A}{ADS}]},
  {\small[\href{http://arxiv.org/abs/gr-qc/9808081}{{gr-qc/9808081}}]}.

\bibitem{2008A&A...477..657A}
{Antia}, H.~M., {Chitre}, S.~M.  and {Gough}, D.~O., ``{Temporal variations in
  the Sun's rotational kinetic energy}'', {\em Astron. Astrophys.}, {\bf 477},
  657--663 (2008).
  {\small[\href{http://dx.doi.org/10.1051/0004-6361:20078209}{DOI}]},
  {\small[\href{http://adsabs.harvard.edu/abs/2008A%26A...477..657A}{ADS}]},
  {\small[\href{http://arxiv.org/abs/0711.0799}{{arXiv:0711.0799}}]}.

\bibitem{antoniadis98}
Antoniadis, I., Arkani-Hamed, N., Dimopoulos, S.  and Dvali, G., ``New
  dimensions at a millimeter to a fermi and superstrings at a TeV'', {\em Phys.
  Lett. B}, {\bf 436}, 257--263 (1998).
  {\small[\href{http://dx.doi.org/10.1016/S0370-2693(98)00860-0}{DOI}]},
  {\small[\href{http://arxiv.org/abs/hep-ph/9804398}{{hep-ph/9804398}}]}.

\bibitem{antonini05}
Antonini, P., Okhapkin, M., G{\"{o}}kl{\"{u}}, E.  and Schiller, S., ``Test of
  constancy of speed of light with rotating cryogenic optical resonators'',
  {\em Phys. Rev. A}, {\bf 71}, 050101 (2005).
  {\small[\href{http://arxiv.org/abs/gr-qc/0504109}{{gr-qc/0504109}}]}.

\bibitem{add98}
Arkani-Hamed, N., Dimopoulos, S.  and Dvali, G.R., ``The hierarchy problem and
  new dimensions at a millimeter'', {\em Phys. Lett. B}, {\bf 429}, 263--272
  (1998).
  {\small[\href{http://dx.doi.org/10.1016/S0370-2693(98)00466-3}{DOI}]},
  {\small[\href{http://arxiv.org/abs/hep-ph/9803315}{{arXiv:hep-ph/9803315}}]}.

\bibitem{2009CQGra..26o5002A}
{Arun}, K.~G.  and {Will}, C.~M., ``{Bounding the mass of the graviton with
  gravitational waves: effect of higher harmonics in gravitational waveform
  templates}'', {\em Class. Quantum Grav.}, {\bf 26}, 155002 (2009).
  {\small[\href{http://dx.doi.org/10.1088/0264-9381/26/15/155002}{DOI}]},
  {\small[\href{http://adsabs.harvard.edu/abs/2009CQGra..26o5002A}{ADS}]},
  {\small[\href{http://arxiv.org/abs/0904.1190}{{arXiv:0904.1190
  {\small[gr-qc]}}}]}.

\bibitem{asada02}
Asada, H., ``The light cone effect on the Shapiro time delay'', {\em Astrophys.
  J. Lett.}, {\bf 574}, L69--L70 (2002).
  {\small[\href{http://dx.doi.org/10.1086/342369}{DOI}]},
  {\small[\href{http://arxiv.org/abs/astro-ph/0206266}{{astro-ph/0206266}}]}.

\bibitem{ashby1}
Ashby, N., ``Relativistic effects in the Global Positioning System'', in
  Dadhich, N.  and Narlikar, J.V., eds., {\em Gravitation and Relativity: At
  the Turn of the Millenium}, Proceedings of the 15th International Conference
  on General Relativity and Gravitation, pp. 231--258, (Inter-University Center
  for Astronomy and Astrophysics, Pune, India, 1998).

\bibitem{ashby2}
Ashby, N., ``Relativity in the Global Positioning System'', {\em Living Rev.
  Relativity}, {\bf 6}, lrr-2003-1 (2003).
  {\small[\href{http://dx.doi.org/10.12942/lrr-2003-1}{DOI}]}. URL (accessed 1
  June 2014): \newline\url{http://www.livingreviews.org/lrr-2003-1}.

\bibitem{2007PhRvD..75b2001A}
{Ashby}, N., {Bender}, P.~L.  and {Wahr}, J.~M., ``{Future gravitational
  physics tests from ranging to the BepiColombo Mercury planetary orbiter}'',
  {\em \prd}, {\bf 75}, 022001 (2007).
  {\small[\href{http://dx.doi.org/10.1103/PhysRevD.75.022001}{DOI}]},
  {\small[\href{http://adsabs.harvard.edu/abs/2007PhRvD..75b2001A}{ADS}]}.

\bibitem{ATNFpulsarcat}
``ATNF Pulsar Catalogue'', web interface to database, Australia Telescope
  National Facility. URL (accessed 1 June 2014):
  \newline\url{http://www.atnf.csiro.au/research/pulsar/psrcat/}.

\bibitem{baessler99}
Bae{\ss}ler, S., Heckel, B.R., Adelberger, E.G., Gundlach, J.H., Schmidt, U.
  and Swanson, H.E., ``Improved Test of the Equivalence Principle for
  Gravitational Self-Energy'', {\em Phys. Rev. Lett.}, {\bf 83}, 3585--3588
  (1999). {\small[\href{http://dx.doi.org/10.1103/PhysRevLett.83.3585}{DOI}]}.

\bibitem{2013PhRvD..87b4015B}
{Baker}, T., {Ferreira}, P.~G.  and {Skordis}, C., ``{The parameterized
  post-Friedmann framework for theories of modified gravity: Concepts,
  formalism, and examples}'', {\em \prd}, {\bf 87}, 024015 (2013).
  {\small[\href{http://dx.doi.org/10.1103/PhysRevD.87.024015}{DOI}]},
  {\small[\href{http://adsabs.harvard.edu/abs/2013PhRvD..87b4015B}{ADS}]},
  {\small[\href{http://arxiv.org/abs/1209.2117}{{arXiv:1209.2117
  {\small[astro-ph.CO]}}}]}.

\bibitem{bambi04}
Bambi, C., Giannotti, M.  and Villante, F.L., ``Response of primordial
  abundances to a general modification of $G_{N}$ and/or of the early universe
  expansion rate'', {\em Phys. Rev. D}, {\bf 71}, 123524 (2005).
  {\small[\href{http://dx.doi.org/10.1103/PhysRevD.71.123524}{DOI}]},
  {\small[\href{http://arxiv.org/abs/astro-ph/0503502}{{astro-ph/0503502}}]}.

\bibitem{bartlett}
Bartlett, D.F.  and Van~Buren, D., ``Equivalence of Active and Passive
  Gravitational Mass Using the Moon'', {\em Phys. Rev. Lett.}, {\bf 57}, 21--24
  (1986). {\small[\href{http://dx.doi.org/10.1103/PhysRevLett.57.21}{DOI}]}.

\bibitem{bauch02}
Bauch, A.  and Weyers, S., ``New experimental limit on the validity of local
  position invariance'', {\em Phys. Rev. D}, {\bf 65}, 081101R (2002).
  {\small[\href{http://dx.doi.org/10.1103/PhysRevD.65.081101}{DOI}]}.

\bibitem{2010nure.book.....B}
{Baumgarte}, T.~W.  and {Shapiro}, S.~L., {\em {Numerical Relativity: Solving
  Einstein's Equations on the Computer}}, (Cambridge University Press,
  Cambridge, 2010).
  {\small[\href{http://adsabs.harvard.edu/abs/2010nure.book.....B}{ADS}]}.

\bibitem{PhysRevD.70.083509}
Bekenstein, J.~D., ``Relativistic gravitation theory for the modified Newtonian
  dynamics paradigm'', {\em Phys. Rev. D}, {\bf 70}, 083509 (2004).
  {\small[\href{http://dx.doi.org/10.1103/PhysRevD.70.083509}{DOI}]}.

\bibitem{Bell}
Bell, J.F.  and Damour, T., ``A new test of conservation laws and Lorentz
  invariance in relativistic gravity'', {\em Class. Quantum Grav.}, {\bf 13},
  3121--3127 (1996).
  {\small[\href{http://dx.doi.org/10.1088/0264-9381/13/12/003}{DOI}]},
  {\small[\href{http://arxiv.org/abs/gr-qc/9606062}{{gr-qc/9606062}}]}.

\bibitem{2010P&SS...58....2B}
{Benkhoff}, J. {et~al.}, ``{BepiColombo -- Comprehensive exploration of
  Mercury: Mission overview and science goals}'', {\em Planetary Sp. Sci.},
  {\bf 58}, 2--20 (2010).
  {\small[\href{http://dx.doi.org/10.1016/j.pss.2009.09.020}{DOI}]},
  {\small[\href{http://adsabs.harvard.edu/abs/2010P26SS...58....2B}{ADS}]}.

\bibitem{PhysRevD.86.010001}
Beringer, J. {et~al.} (Particle Data Group), ``Review of Particle Physics'',
  {\em Phys. Rev. D}, {\bf 86}, 010001 (2012).
  {\small[\href{http://dx.doi.org/10.1103/PhysRevD.86.010001}{DOI}]}.

\bibitem{bbw1}
Berti, E., Buonanno, A.  and Will, C.M., ``Estimating spinning binary
  parameters and testing alternative theories of gravity with LISA'', {\em
  Phys. Rev. D}, {\bf 71}, 084025 (2005).
  {\small[\href{http://dx.doi.org/10.1103/PhysRevD.71.084025}{DOI}]},
  {\small[\href{http://arxiv.org/abs/gr-qc/0411129}{{gr-qc/0411129}}]}.

\bibitem{bbw2}
Berti, E., Buonanno, A.  and Will, C.M., ``Testing general relativity and
  probing the merger history of massive black holes with LISA'', {\em Class.
  Quantum Grav.}, {\bf 22}, S943--S954 (2005).
  {\small[\href{http://dx.doi.org/10.1088/0264-9381/22/18/S08}{DOI}]},
  {\small[\href{http://arxiv.org/abs/gr-qc/0504017}{{gr-qc/0504017}}]}.

\bibitem{2009CQGra..26p3001B}
{Berti}, E., {Cardoso}, V.  and {Starinets}, A.~O., ``{TOPICAL REVIEW:
  Quasinormal modes of black holes and black branes}'', {\em Class. Quantum
  Grav.}, {\bf 26}, 163001 (2009).
  {\small[\href{http://dx.doi.org/10.1088/0264-9381/26/16/163001}{DOI}]},
  {\small[\href{http://adsabs.harvard.edu/abs/2009CQGra..26p3001B}{ADS}]},
  {\small[\href{http://arxiv.org/abs/0905.2975}{{arXiv:0905.2975
  {\small[gr-qc]}}}]}.

\bibitem{bertotti03}
Bertotti, B., Iess, L.  and Tortora, P., ``A test of general relativity using
  radio links with the Cassini spacecraft'', {\em Nature}, {\bf 425}, 374--376
  (2003). {\small[\href{http://dx.doi.org/10.1038/nature01997}{DOI}]}.

\bibitem{2011PhRvD..83g5004B}
{Bezerra}, V.~B., {Klimchitskaya}, G.~L., {Mostepanenko}, V.~M.  and {Romero},
  C., ``{Constraints on non-Newtonian gravity from measuring the Casimir force
  in a configuration with nanoscale rectangular corrugations}'', {\em \prd},
  {\bf 83}, 075004 (2011).
  {\small[\href{http://dx.doi.org/10.1103/PhysRevD.83.075004}{DOI}]},
  {\small[\href{http://adsabs.harvard.edu/abs/2011PhRvD..83g5004B}{ADS}]},
  {\small[\href{http://arxiv.org/abs/1103.0993}{{arXiv:1103.0993
  {\small[hep-ph]}}}]}.

\bibitem{2008PhRvD..77l4017B}
{Bhat}, N.~D.~R., {Bailes}, M.  and {Verbiest}, J.~P.~W.,
  ``{Gravitational-radiation losses from the pulsar white-dwarf binary PSR
  J1141 6545}'', {\em \prd}, {\bf 77}, 124017 (2008).
  {\small[\href{http://dx.doi.org/10.1103/PhysRevD.77.124017}{DOI}]},
  {\small[\href{http://adsabs.harvard.edu/abs/2008PhRvD..77l4017B}{ADS}]},
  {\small[\href{http://arxiv.org/abs/0804.0956}{{arXiv:0804.0956}}]}.

\bibitem{2009PhRvD..79h3015B}
{Bi}, X.-J., {Cao}, Z., {Li}, Y.  and {Yuan}, Q., ``{Testing Lorentz invariance
  with the ultrahigh energy cosmic ray spectrum}'', {\em \prd}, {\bf 79},
  083015 (2009).
  {\small[\href{http://dx.doi.org/10.1103/PhysRevD.79.083015}{DOI}]},
  {\small[\href{http://adsabs.harvard.edu/abs/2009PhRvD..79h3015B}{ADS}]},
  {\small[\href{http://arxiv.org/abs/0812.0121}{{arXiv:0812.0121}}]}.

\bibitem{biller}
Biller, S.D. {et~al.}, ``Limits to Quantum Gravity Effects on Energy Dependence
  of the Speed of Light from Observations of TeV Flares in Active Galaxies'',
  {\em Phys. Rev. Lett.}, {\bf 83}, 2108--2111 (1999).
  {\small[\href{http://dx.doi.org/10.1103/PhysRevLett.83.2108}{DOI}]},
  {\small[\href{http://arxiv.org/abs/gr-qc/9810044}{{arXiv:gr-qc/9810044}}]}.

\bibitem{bize03}
Bize, S. {et~al.}, ``Testing the Stability of Fundamental Constants with
  $^{199}$Hg$^+$ Single-Ion Optical Clock'', {\em Phys. Rev. Lett.}, {\bf 90},
  150802 (2003).
  {\small[\href{http://dx.doi.org/10.1103/PhysRevLett.90.150802}{DOI}]},
  {\small[\href{http://arxiv.org/abs/physics/0212109}{{physics/0212109}}]}.

\bibitem{blanchet95}
Blanchet, L., ``Second-post-Newtonian generation of gravitational radiation'',
  {\em Phys. Rev. D}, {\bf 51}, 2559--2583 (1995).
  {\small[\href{http://dx.doi.org/10.1103/PhysRevD.51.2559}{DOI}]},
  {\small[\href{http://arxiv.org/abs/gr-qc/9501030}{{gr-qc/9501030}}]}.

\bibitem{BlanchetLRR}
{Blanchet}, L., ``{Gravitational Radiation from Post-Newtonian Sources and
  Inspiralling Compact Binaries}'', {\em Living Rev. Relativity}, {\bf 17},
  lrr-2014-2 (2014).
  {\small[\href{http://dx.doi.org/10.12942/lrr-2014-2}{DOI}]},
  {\small[\href{http://adsabs.harvard.edu/abs/2014LRR....17....2B}{ADS}]},
  {\small[\href{http://arxiv.org/abs/1310.1528}{{arXiv:1310.1528
  {\small[gr-qc]}}}]}. URL (accessed 1 June 2014):
  \newline\url{http://www.livingreviews.org/lrr-2014-2}.

\bibitem{bd86}
Blanchet, L.  and Damour, T., ``Radiative gravitational fields in general
  relativity I. General structure of the field outside the source'', {\em Phil.
  Trans. R. Soc. London, Ser. A}, {\bf 320}, 379--430 (1986).

\bibitem{bd88}
Blanchet, L.  and Damour, T., ``Tail-transported temporal correlations in the
  dynamics of a gravitating system'', {\em Phys. Rev. D}, {\bf 37}, 1410--1435
  (1988). {\small[\href{http://dx.doi.org/10.1103/PhysRevD.37.1410}{DOI}]}.

\bibitem{bd89}
Blanchet, L.  and Damour, T., ``Post-Newtonian generation of gravitational
  waves'', {\em Ann. Inst. Henri Poincar\'e A}, {\bf 50}, 377--408 (1989).

\bibitem{bd92}
Blanchet, L.  and Damour, T., ``Hereditary effects in gravitational
  radiation'', {\em Phys. Rev. D}, {\bf 46}, 4304--4319 (1992).
  {\small[\href{http://dx.doi.org/10.1103/PhysRevD.46.4304}{DOI}]}.

\bibitem{bdiww}
Blanchet, L., Damour, T., Iyer, B.R., Will, C.M.  and Wiseman, A.G.,
  ``Gravitational-Radiation Damping of Compact Binary Systems to Second
  Post-Newtonian Order'', {\em Phys. Rev. Lett.}, {\bf 74}, 3515--3518 (1995).
  {\small[\href{http://dx.doi.org/10.1103/PhysRevLett.74.3515}{DOI}]},
  {\small[\href{http://arxiv.org/abs/gr-qc/9501027}{{gr-qc/9501027}}]}.

\bibitem{2011MNRAS.412.2530B}
{Blanchet}, L.  and {Novak}, J., ``{External field effect of modified Newtonian
  dynamics in the Solar system}'', {\em \mnras}, {\bf 412}, 2530--2542 (2011).
  {\small[\href{http://dx.doi.org/10.1111/j.1365-2966.2010.18076.x}{DOI}]},
  {\small[\href{http://adsabs.harvard.edu/abs/2011MNRAS.412.2530B}{ADS}]},
  {\small[\href{http://arxiv.org/abs/1010.1349}{{arXiv:1010.1349
  {\small[astro-ph.CO]}}}]}.

\bibitem{2011arXiv1105.5815B}
{Blanchet}, L.  and {Novak}, J., ``{Testing MOND in the Solar System}'', {\em
  ArXiv e-prints} (2011).
  {\small[\href{http://adsabs.harvard.edu/abs/2011arXiv1105.5815B}{ADS}]},
  {\small[\href{http://arxiv.org/abs/1105.5815}{{arXiv:1105.5815
  {\small[astro-ph.CO]}}}]}.

\bibitem{lucsathya2}
Blanchet, L.  and Sathyaprakash, B.S., ``Signal analysis of gravitational wave
  tails'', {\em Class. Quantum Grav.}, {\bf 11}, 2807--2831 (1994).
  {\small[\href{http://dx.doi.org/10.1088/0264-9381/11/11/020}{DOI}]}.

\bibitem{lucsathya}
Blanchet, L.  and Sathyaprakash, B.S., ``Detecting the tail effect in
  gravitational wave experiments'', {\em Phys. Rev. Lett.}, {\bf 74},
  1067--1070 (1995).
  {\small[\href{http://dx.doi.org/10.1103/PhysRevLett.74.1067}{DOI}]}.

\bibitem{2008PhRvL.100n0801B}
{Blatt}, S. {et~al.}, ``{New Limits on Coupling of Fundamental Constants to
  Gravity Using Sr87 Optical Lattice Clocks}'', {\em Phys. Rev. Lett.}, {\bf
  100}, 140801 (2008).
  {\small[\href{http://dx.doi.org/10.1103/PhysRevLett.100.140801}{DOI}]},
  {\small[\href{http://adsabs.harvard.edu/abs/2008PhRvL.100n0801B}{ADS}]},
  {\small[\href{http://arxiv.org/abs/0801.1874}{{arXiv:0801.1874
  {\small[physics.atom-ph]}}}]}.

\bibitem{2006PhRvD..74f1501B}
{Bolton}, A.~S., {Rappaport}, S.  and {Burles}, S., ``{Constraint on the
  post-Newtonian parameter {$\gamma$} on galactic size scales}'', {\em \prd},
  {\bf 74}, 061501 (2006).
  {\small[\href{http://dx.doi.org/10.1103/PhysRevD.74.061501}{DOI}]},
  {\small[\href{http://adsabs.harvard.edu/abs/2006PhRvD..74f1501B}{ADS}]},
  {\small[\href{http://arxiv.org/abs/arXiv:astro-ph/0607657}{{arXiv:astro-ph/0607657}}]}.

\bibitem{braginsky}
Braginsky, V.B.  and Panov, V.I., ``Verification of the equivalence of inertial
  and gravitational mass'', {\em Sov. Phys. JETP}, {\bf 34}, 463--466 (1972).

\bibitem{2012ExA....34..181B}
{Braxmaier}, C. {et~al.}, ``{Astrodynamical Space Test of Relativity using
  Optical Devices I (ASTROD I) -- a class-M fundamental physics mission
  proposal for cosmic vision 2015-2025: 2010 Update}'', {\em Experimental
  Astronomy}, {\bf 34}, 181--201 (2012).
  {\small[\href{http://dx.doi.org/10.1007/s10686-011-9281-y}{DOI}]},
  {\small[\href{http://adsabs.harvard.edu/abs/2012ExA....34..181B}{ADS}]},
  {\small[\href{http://arxiv.org/abs/1104.0060}{{arXiv:1104.0060
  {\small[gr-qc]}}}]}.

\bibitem{brecher}
Brecher, K., ``Is the speed of light independent of the velocity of the
  source?'', {\em Phys. Rev. Lett.}, {\bf 39}, 1051--1054 (1977).

\bibitem{2008Sci...321..104B}
{Breton}, R.~P. {et~al.}, ``{Relativistic Spin Precession in the Double
  Pulsar}'', {\em Science}, {\bf 321}, 104 (2008).
  {\small[\href{http://dx.doi.org/10.1126/science.1159295}{DOI}]},
  {\small[\href{http://adsabs.harvard.edu/abs/2008Sci...321..104B}{ADS}]},
  {\small[\href{http://arxiv.org/abs/0807.2644}{{arXiv:0807.2644}}]}.

\bibitem{brillethall}
Brillet, A.  and Hall, J.L., ``Improved laser test of the isotropy of space'',
  {\em Phys. Rev. Lett.}, {\bf 42}, 549--552 (1979).
  {\small[\href{http://dx.doi.org/10.1103/PhysRevLett.42.549}{DOI}]}.

\bibitem{brunetti}
Brunetti, M., Coccia, E., Fafone, V.  and Fucito, F., ``Gravitational-wave
  radiation from compact binary systems in the Jordan--Brans--Dicke theory'',
  {\em Phys. Rev. D}, {\bf 59}, 044027 (1999).
  {\small[\href{http://dx.doi.org/10.1103/PhysRevD.59.044027}{DOI}]},
  {\small[\href{http://arxiv.org/abs/gr-qc/9805056}{{gr-qc/9805056}}]}.

\bibitem{burgay03}
Burgay, M. {et~al.}, ``An increased estimate of the merger rate of double
  neutron stars from observations of a highly relativistic system'', {\em
  Nature}, {\bf 426}, 531--533 (2003).
  {\small[\href{http://dx.doi.org/10.1038/nature02124}{DOI}]},
  {\small[\href{http://adsabs.harvard.edu/abs/2003Natur.426..531B}{ADS}]},
  {\small[\href{http://arxiv.org/abs/astro-ph/0312071}{{arXiv:astro-ph/0312071
  {\small[astro-ph]}}}]}.

\bibitem{carlip04}
Carlip, S., ``Model-dependence of Shapiro time delay and the `speed of
  gravity/speed of light' controversy'', {\em Class. Quantum Grav.}, {\bf 21},
  3803--3812 (2004).
  {\small[\href{http://dx.doi.org/10.1088/0264-9381/21/15/011}{DOI}]},
  {\small[\href{http://arxiv.org/abs/gr-qc/0403060}{{gr-qc/0403060}}]}.

\bibitem{2012PhRvD..85h2001C}
{Chamberlin}, S.~J.  and {Siemens}, X., ``{Stochastic backgrounds in
  alternative theories of gravity: Overlap reduction functions for pulsar
  timing arrays}'', {\em \prd}, {\bf 85}, 082001 (2012).
  {\small[\href{http://dx.doi.org/10.1103/PhysRevD.85.082001}{DOI}]},
  {\small[\href{http://adsabs.harvard.edu/abs/2012PhRvD..85h2001C}{ADS}]},
  {\small[\href{http://arxiv.org/abs/1111.5661}{{arXiv:1111.5661
  {\small[astro-ph.HE]}}}]}.

\bibitem{Champeney}
Champeney, D.C., Isaak, G.R.  and Khan, A.M., ``An `aether drift' experiment
  based on the M\"ossbauer effect'', {\em Phys. Lett.}, {\bf 7}, 241--243
  (1963). {\small[\href{http://dx.doi.org/10.1016/0031-9163(63)90312-3}{DOI}]}.

\bibitem{petitjean2}
Chand, H., Petitjean, P., Srianand, R.  and Aracil, B., ``Probing the
  time-variation of the fine-structure constant: Results based on Si IV
  doublets from a UVES sample'', {\em Astron. Astrophys.}, {\bf 430}, 47--58
  (2005). {\small[\href{http://dx.doi.org/10.1051/0004-6361:20041186}{DOI}]},
  {\small[\href{http://arxiv.org/abs/astro-ph/0408200}{{astro-ph/0408200}}]}.

\bibitem{1965ApJ...142.1488C}
{Chandrasekhar}, S., ``{The Post-Newtonian Equations of Hydrodynamics in
  General Relativity.}'', {\em \apj}, {\bf 142}, 1488 (November 1965).
  {\small[\href{http://dx.doi.org/10.1086/148432}{DOI}]},
  {\small[\href{http://adsabs.harvard.edu/abs/1965ApJ...142.1488C}{ADS}]}.

\bibitem{kapitulnik}
Chiaverini, J., Smullin, S.J., Geraci, A.A., Weld, D.M.  and Kapitulnik, A.,
  ``New experimental constraints on non-Newtonian forces below 100 {$\mu$}m'',
  {\em Phys. Rev. Lett.}, {\bf 90}, 151101 (2003).
  {\small[\href{http://dx.doi.org/10.1103/PhysRevLett.90.151101}{DOI}]},
  {\small[\href{http://arxiv.org/abs/hep-ph/0209325}{{hep-ph/0209325}}]}.

\bibitem{2010Sci...329.1630C}
{Chou}, C.~W., {Hume}, D.~B., {Rosenband}, T.  and {Wineland}, D.~J.,
  ``{Optical Clocks and Relativity}'', {\em Science}, {\bf 329}, 1630--1633
  (2010). {\small[\href{http://dx.doi.org/10.1126/science.1192720}{DOI}]},
  {\small[\href{http://adsabs.harvard.edu/abs/2010Sci...329.1630C}{ADS}]}.

\bibitem{chupp}
Chupp, T.E., Hoare, R.J., Loveman, R.A., Oteiza, E.R., Richardson, J.M.,
  Wagshul, M.E.  and Thompson, A.K., ``Results of a new test of local Lorentz
  invariance: A search for mass anisotropy in $^{21}$Ne'', {\em Phys. Rev.
  Lett.}, {\bf 63}, 1541--1545 (1989).
  {\small[\href{http://dx.doi.org/10.1103/PhysRevLett.63.1541}{DOI}]}.

\bibitem{ciufolini00}
Ciufolini, I., ``The 1995--99 measurements of the Lense--Thirring effect using
  laser-ranged satellites'', {\em Class. Quantum Grav.}, {\bf 17}, 2369--2380
  (2000). {\small[\href{http://dx.doi.org/10.1088/0264-9381/17/12/309}{DOI}]}.

\bibitem{ciufolini97}
Ciufolini, I., Chieppa, F., Lucchesi, D.  and Vespe, F., ``Test of Lense -
  Thirring orbital shift due to spin'', {\em Class. Quantum Grav.}, {\bf 14},
  2701--2726 (1997).
  {\small[\href{http://dx.doi.org/10.1088/0264-9381/14/10/003}{DOI}]}.

\bibitem{2013CQGra..30w5009C}
{Ciufolini}, I., {Moreno Monge}, B., {Paolozzi}, A., {Koenig}, R., {Sindoni},
  G., {Michalak}, G.  and {Pavlis}, E.~C., ``{Monte Carlo simulations of the
  LARES space experiment to test General Relativity and fundamental physics}'',
  {\em Class. Quantum Grav.}, {\bf 30}, 235009 (2013).
  {\small[\href{http://dx.doi.org/10.1088/0264-9381/30/23/235009}{DOI}]},
  {\small[\href{http://adsabs.harvard.edu/abs/2013CQGra..30w5009C}{ADS}]},
  {\small[\href{http://arxiv.org/abs/1310.2601}{{arXiv:1310.2601
  {\small[gr-qc]}}}]}.

\bibitem{2011EPJP..126...72C}
{Ciufolini}, I., {Paolozzi}, A., {Pavlis}, E.~C., {Ries}, J., {Koenig}, R.,
  {Matzner}, R., {Sindoni}, G.  and {Neumeyer}, H., ``{Testing gravitational
  physics with satellite laser ranging}'', {\em Eur. Phys. J. Plus}, {\bf 126},
  72 (2011).
  {\small[\href{http://dx.doi.org/10.1140/epjp/i2011-11072-2}{DOI}]},
  {\small[\href{http://adsabs.harvard.edu/abs/2011EPJP..126...72C}{ADS}]}.

\bibitem{ciufolini04}
Ciufolini, I.  and Pavlis, E.C., ``A confirmation of the general relativistic
  prediction of the Lense--Thirring effect'', {\em Nature}, {\bf 431}, 958--960
  (2004). {\small[\href{http://dx.doi.org/10.1038/nature03007}{DOI}]}.

\bibitem{ciufolini98}
Ciufolini, I., Pavlis, E.C., Chieppa, F., Fernandes-Vieira, E.  and
  P{\'{e}}rez-Mercader, J., ``Test of general relativity and measurement of the
  Lense--Thirring effect with two Earth satellites'', {\em Science}, {\bf 279},
  2100--2103 (1998).
  {\small[\href{http://dx.doi.org/10.1126/science.279.5359.2100}{DOI}]}.

\bibitem{2006NewA...11..527C}
{Ciufolini}, I., {Pavlis}, E.~C.  and {Peron}, R., ``{Determination of
  frame-dragging using Earth gravity models from CHAMP and GRACE}'', {\em \na},
  {\bf 11}, 527--550 (2006).
  {\small[\href{http://dx.doi.org/10.1016/j.newast.2006.02.001}{DOI}]},
  {\small[\href{http://adsabs.harvard.edu/abs/2006NewA...11..527C}{ADS}]}.

\bibitem{Clifton05}
Clifton, T., Barrow, J.D.  and Scherrer, R.J., ``Constraints on the variation
  of $G$ from primordial nucleosynthesis'', {\em Phys. Rev. D}, {\bf 71},
  123526 (2005).
  {\small[\href{http://dx.doi.org/10.1103/PhysRevD.71.123526}{DOI}]},
  {\small[\href{http://arxiv.org/abs/astro-ph/0504418}{{astro-ph/0504418}}]}.

\bibitem{coc06}
Coc, A., Olive, K.A., Uzan, J.-P.  and Vangioni, E., ``Big bang nucleosynthesis
  constraints on scalar-tensor theories of gravity'', {\em Phys. Rev. D}, {\bf
  73}, 083525 (2006).
  {\small[\href{http://dx.doi.org/10.1103/PhysRevD.73.083525}{DOI}]},
  {\small[\href{http://arxiv.org/abs/astro-ph/0601299}{{astro-ph/0601299}}]}.

\bibitem{Coley82}
Coley, A., ``Schiff's Conjecture on Gravitation'', {\em Phys. Rev. Lett.}, {\bf
  49}, 853--855 (1982).
  {\small[\href{http://dx.doi.org/10.1103/PhysRevLett.49.853}{DOI}]}.

\bibitem{colladay97}
Colladay, D.  and Kosteleck{\'{y}}, V.A., ``CPT violation and the standard
  model'', {\em Phys. Rev. D}, {\bf 55}, 6760--6774 (1997).
  {\small[\href{http://dx.doi.org/10.1103/PhysRevD.55.6760}{DOI}]},
  {\small[\href{http://arxiv.org/abs/hep-ph/9703464}{{arXiv:hep-ph/9703464}}]}.

\bibitem{colladay98}
Colladay, D.  and Kosteleck{\'{y}}, V.A., ``Lorentz-violating extension of the
  standard model'', {\em Phys. Rev. D}, {\bf 58}, 116002 (1998).
  {\small[\href{http://dx.doi.org/10.1103/PhysRevD.58.116002}{DOI}]},
  {\small[\href{http://arxiv.org/abs/hep-ph/9809521}{{arXiv:hep-ph/9809521}}]}.

\bibitem{copi04}
Copi, C.J., Davis, A.N.  and Krauss, L.M., ``New Nucleosynthesis Constraint on
  the Variation of $G$'', {\em Phys. Rev. Lett.}, {\bf 92}, 171301 (2004).
  {\small[\href{http://dx.doi.org/10.1103/PhysRevLett.92.171301}{DOI}]},
  {\small[\href{http://arxiv.org/abs/astro-ph/0311334}{{astro-ph/0311334}}]}.

\bibitem{Crelinsten}
Crelinsten, J., {\em Einstein's Jury: The Race to Test Relativity}, (Princeton
  University Press, Princeton, 2006).

\bibitem{creminelli05}
Creminelli, P., Nicolis, A., Papucci, M.  and Trincherini, E., ``Ghosts in
  massive gravity'', {\em J. High En. Phys.}, {\bf 2005}, 003 (2005).
  {\small[\href{http://dx.doi.org/10.1088/1126-6708/2005/09/003}{DOI}]},
  {\small[\href{http://arxiv.org/abs/hep-th/0505147}{{hep-th/0505147}}]}.

\bibitem{cutlerflan}
Cutler, C.  and Flanagan, {\'{E}}.{\'{E}}., ``Gravitational waves from merging
  compact binaries: How accurately can one extract the binary's parameters from
  the inspiral wave form?'', {\em Phys. Rev. D}, {\bf 49}, 2658--2697 (1994).
  {\small[\href{http://dx.doi.org/10.1103/PhysRevD.49.2658}{DOI}]},
  {\small[\href{http://arxiv.org/abs/gr-qc/9402014}{{gr-qc/9402014}}]}.

\bibitem{CutlerLarHis03}
Cutler, C., Hiscock, W.A.  and Larson, S.L., ``LISA, binary stars, and the mass
  of the graviton'', {\em Phys. Rev. D}, {\bf 67}, 024015 (2003).
  {\small[\href{http://dx.doi.org/10.1103/PhysRevD.67.024015}{DOI}]},
  {\small[\href{http://arxiv.org/abs/gr-qc/0209101}{{gr-qc/0209101}}]}.

\bibitem{3min}
Cutler, C. {et~al.}, ``The Last Three Minutes: Issues in Gravitational-Wave
  Measurements of Coalescing Compact Binaries'', {\em Phys. Rev. Lett.}, {\bf
  70}, 2984--2987 (1993).
  {\small[\href{http://dx.doi.org/10.1103/PhysRevLett.70.2984}{DOI}]},
  {\small[\href{http://arxiv.org/abs/astro-ph/9208005}{{astro-ph/9208005}}]}.

\bibitem{Damour300}
Damour, T., ``The problem of motion in Newtonian and Einsteinian gravity'', in
  Hawking, S.W.  and Israel, W., eds., {\em Three Hundred Years of
  Gravitation}, pp. 128--198, (Cambridge University Press, Cambridge; New York,
  1987).
  {\small[\href{http://adsabs.harvard.edu/abs/1987thyg.book..128D}{ADS}]}.

\bibitem{damourdyson}
Damour, T.  and Dyson, F.J., ``The Oklo bound on the time variation of the
  fine-structure constant revisited'', {\em Nucl. Phys. B}, {\bf 480}, 37--54
  (1996).
  {\small[\href{http://dx.doi.org/10.1016/S0550-3213(96)00467-1}{DOI}]},
  {\small[\href{http://arxiv.org/abs/hep-ph/9606486}{{hep-ph/9606486}}]}.

\bibitem{DamourEspo92}
Damour, T.  and Esposito-Far{\`{e}}se, G., ``Tensor-multi-scalar theories of
  gravitation'', {\em Class. Quantum Grav.}, {\bf 9}, 2093--2176 (1992).
  {\small[\href{http://dx.doi.org/10.1088/0264-9381/9/9/015}{DOI}]}.

\bibitem{1996PhRvD..53.5541D}
{Damour}, T.  and {Esposito-Far{\`e}se}, G., ``{Testing gravity to second
  post-Newtonian order: A field-theory approach}'', {\em \prd}, {\bf 53},
  5541--5578 (1996).
  {\small[\href{http://dx.doi.org/10.1103/PhysRevD.53.5541}{DOI}]},
  {\small[\href{http://adsabs.harvard.edu/abs/1996PhRvD..53.5541D}{ADS}]},
  {\small[\href{http://arxiv.org/abs/gr-qc/9506063}{{gr-qc/9506063}}]}.

\bibitem{DamourEspo98}
Damour, T.  and Esposito-Far{\`{e}}se, G., ``Gravitational-wave versus
  binary-pulsar tests of strong-field gravity'', {\em Phys. Rev. D}, {\bf 58},
  042001 (1998).
  {\small[\href{http://dx.doi.org/10.1103/PhysRevD.58.042001}{DOI}]},
  {\small[\href{http://arxiv.org/abs/gr-qc/9803031}{{gr-qc/9803031}}]}.

\bibitem{damgopuiyer04}
Damour, T., Gopakumar, A.  and Iyer, B.R., ``Phasing of gravitational waves
  from inspiralling eccentric binaries'', {\em Phys. Rev. D}, {\bf 70}, 064028
  (2004). {\small[\href{http://dx.doi.org/10.1103/PhysRevD.70.064028}{DOI}]},
  {\small[\href{http://arxiv.org/abs/gr-qc/0404128}{{gr-qc/0404128}}]}.

\bibitem{di91}
Damour, T.  and Iyer, B.R., ``Post-Newtonian generation of gravitational waves.
  II. The spin moments'', {\em Ann. Inst. Henri Poincar\'e A}, {\bf 54},
  115--164 (1991).

\bibitem{damjaraschaefer}
Damour, T., Jaranowski, P.  and Sch{\"{a}}fer, G., ``Poincar\'e invariance in
  the ADM Hamiltonian approach to the general relativistic two-body problem'',
  {\em Phys. Rev. D}, {\bf 62}, 021501 (2000).
  {\small[\href{http://arxiv.org/abs/gr-qc/0003051}{{gr-qc/0003051}}]}.
  Erratum: Phys.Rev. D 63 (2001) 029903.

\bibitem{DJSdim}
Damour, T., Jaranowski, P.  and Sch{\"{a}}fer, G., ``Dimensional regularization
  of the gravitational interaction of point masses'', {\em Phys. Lett. B}, {\bf
  513}, 147--155 (2001).
  {\small[\href{http://dx.doi.org/10.1016/S0370-2693(01)00642-6}{DOI}]},
  {\small[\href{http://arxiv.org/abs/gr-qc/0105038}{{gr-qc/0105038}}]}.

\bibitem{DJSequiv}
Damour, T., Jaranowski, P.  and Sch{\"{a}}fer, G., ``Equivalence between the
  ADM-Hamiltonian and the harmonic-coordinates approaches to the third
  post-Newtonian dynamics of compact binaries'', {\em Phys. Rev. D}, {\bf 63},
  044021 (2001).
  {\small[\href{http://dx.doi.org/10.1103/PhysRevD.63.044021}{DOI}]},
  {\small[\href{http://arxiv.org/abs/gr-qc/0010040}{{gr-qc/0010040}}]}. Erratum
  Phys. Rev. D 66 (2002) 029901(E).

\bibitem{DamourNord93a}
Damour, T.  and Nordtvedt~Jr., K., ``General relativity as a cosmological
  attractor of tensor-scalar theories'', {\em Phys. Rev. Lett.}, {\bf 70},
  2217--2219 (1993).
  {\small[\href{http://dx.doi.org/10.1103/PhysRevLett.70.2217}{DOI}]}.

\bibitem{DamourNord93b}
Damour, T.  and Nordtvedt~Jr., K., ``Tensor-scalar cosmological models and
  their relaxation toward general relativity'', {\em Phys. Rev. D}, {\bf 48},
  3436--3450 (1993).
  {\small[\href{http://dx.doi.org/10.1103/PhysRevD.48.3436}{DOI}]}.

\bibitem{damourpiazza02a}
Damour, T., Piazza, F.  and Veneziano, G., ``Runaway dilaton and equivalence
  principle violations'', {\em Phys. Rev. Lett.}, {\bf 89}, 081601 (2002).
  {\small[\href{http://dx.doi.org/10.1103/PhysRevLett.89.081601}{DOI}]},
  {\small[\href{http://arxiv.org/abs/gr-qc/0204094}{{arXiv:gr-qc/0204094}}]}.

\bibitem{damourpiazza02b}
Damour, T., Piazza, F.  and Veneziano, G., ``Violations of the equivalence
  principle in a dilaton-runaway scenario'', {\em Phys. Rev. D}, {\bf 66},
  046007 (2002).
  {\small[\href{http://dx.doi.org/10.1103/PhysRevD.66.046007}{DOI}]},
  {\small[\href{http://arxiv.org/abs/hep-th/0205111}{{arXiv:hep-th/0205111}}]}.

\bibitem{damourpichon}
Damour, T.  and Pichon, B., ``Big bang nucleosynthesis and tensor-scalar
  gravity'', {\em Phys. Rev. D}, {\bf 59}, 123502 (1999).
  {\small[\href{http://dx.doi.org/10.1103/PhysRevD.59.123502}{DOI}]},
  {\small[\href{http://arxiv.org/abs/astro-ph/9807176}{{astro-ph/9807176}}]}.

\bibitem{DamourPolyakov}
Damour, T.  and Polyakov, A.M., ``The string dilaton and a least coupling
  principle'', {\em Nucl. Phys. B}, {\bf 423}, 532--558 (1994).
  {\small[\href{http://dx.doi.org/10.1016/0550-3213(94)90143-0}{DOI}]},
  {\small[\href{http://arxiv.org/abs/hep-th/9401069}{{arXiv:hep-th/9401069}}]}.

\bibitem{DamourSchaefer91}
Damour, T.  and Sch{\"{a}}fer, G., ``New tests of the strong equivalence
  principle using binary-pulsar data'', {\em Phys. Rev. Lett.}, {\bf 66},
  2549--2552 (1991).
  {\small[\href{http://dx.doi.org/10.1103/PhysRevLett.66.2549}{DOI}]}.

\bibitem{DamourTaylor92}
Damour, T.  and Taylor, J.H., ``Strong-field tests of relativistic gravity and
  binary pulsars'', {\em Phys. Rev. D}, {\bf 45}, 1840--1868 (1992).
  {\small[\href{http://dx.doi.org/10.1103/PhysRevD.45.1840}{DOI}]}.

\bibitem{DamourVokrou96}
Damour, T.  and Vokrouhlick{\'{y}}, D., ``Equivalence principle and the Moon'',
  {\em Phys. Rev. D}, {\bf 53}, 4177--4201 (1996).
  {\small[\href{http://dx.doi.org/10.1103/PhysRevD.53.4177}{DOI}]},
  {\small[\href{http://arxiv.org/abs/gr-qc/9507016}{{gr-qc/9507016}}]}.

\bibitem{2010PhRvD..81l3508D}
{Daniel}, S.~F., {Linder}, E.~V., {Smith}, T.~L., {Caldwell}, R.~R., {Cooray},
  A., {Leauthaud}, A.  and {Lombriser}, L., ``{Testing general relativity with
  current cosmological data}'', {\em \prd}, {\bf 81}, 123508 (2010).
  {\small[\href{http://dx.doi.org/10.1103/PhysRevD.81.123508}{DOI}]},
  {\small[\href{http://adsabs.harvard.edu/abs/2010PhRvD..81l3508D}{ADS}]},
  {\small[\href{http://arxiv.org/abs/1002.1962}{{arXiv:1002.1962
  {\small[astro-ph.CO]}}}]}.

\bibitem{lrr-2010-3}
De~Felice, A.  and Tsujikawa, S., ``$f(R)$ Theories'', {\em Living Rev.
  Relativity}, {\bf 13}, lrr-2010-3 (2010).
  {\small[\href{http://dx.doi.org/10.12942/lrr-2010-3}{DOI}]}. URL (accessed 1
  June 2014): \newline\url{http://www.livingreviews.org/lrr-2010-3}.

\bibitem{2014arXiv1401.4173D}
{de Rham}, C., ``{Massive Gravity}'', {\em ArXiv e-prints} (January 2014).
  {\small[\href{http://adsabs.harvard.edu/abs/2014arXiv1401.4173D}{ADS}]},
  {\small[\href{http://arxiv.org/abs/1401.4173}{{arXiv:1401.4173
  {\small[hep-th]}}}]}.

\bibitem{1916MNRAS..77..155D}
{de Sitter}, W., ``{On Einstein's theory of gravitation and its astronomical
  consequences. Second paper}'', {\em \mnras}, {\bf 77}, 155--184 (December
  1916).
  {\small[\href{http://adsabs.harvard.edu/abs/1916MNRAS..77..155D}{ADS}]}.

\bibitem{deffayet02}
Deffayet, C., Dvali, G., Gabadadze, G.  and Vainshtein, A., ``Nonperturbative
  continuity in graviton mass versus perturbative discontinuity'', {\em Phys.
  Rev. D}, {\bf 65}, 044026 (2002).
  {\small[\href{http://dx.doi.org/10.1103/PhysRevD.65.044026}{DOI}]},
  {\small[\href{http://arxiv.org/abs/hep-th/0106001}{{hep-th/0106001}}]}.

\bibitem{2008ApJ...685L..67D}
{Deller}, A.~T., {Verbiest}, J.~P.~W., {Tingay}, S.~J.  and {Bailes}, M.,
  ``{Extremely High Precision VLBI Astrometry of PSR J0437-4715 and
  Implications for Theories of Gravity}'', {\em \apjl}, {\bf 685}, L67--L70
  (2008). {\small[\href{http://dx.doi.org/10.1086/592401}{DOI}]},
  {\small[\href{http://adsabs.harvard.edu/abs/2008ApJ...685L..67D}{ADS}]},
  {\small[\href{http://arxiv.org/abs/0808.1594}{{arXiv:0808.1594}}]}.

\bibitem{2013arXiv1310.7426D}
{Di Casola}, E., {Liberati}, S.  and {Sonego}, S., ``{Nonequivalence of
  equivalence principles}'', {\em ArXiv e-prints} (2013).
  {\small[\href{http://adsabs.harvard.edu/abs/2013arXiv1310.7426D}{ADS}]},
  {\small[\href{http://arxiv.org/abs/1310.7426}{{arXiv:1310.7426
  {\small[gr-qc]}}}]}.

\bibitem{dick98}
Dick, R., ``Inequivalence of Jordan and Einstein frame: What is the low energy
  gravity in string theory?'', {\em Gen. Relativ. Gravit.}, {\bf 30}, 435--444
  (1998). {\small[\href{http://dx.doi.org/10.1023/A:1018810926163}{DOI}]}.

\bibitem{dicke64}
Dicke, R.H., ``Experimental relativity'', in DeWitt, C.M.  and DeWitt, B.S.,
  eds., {\em Relativity, Groups and Topology}, Les Houches Summer School 1963,
  pp. 165--313, (Gordon and Breach, New York; London, 1964).

\bibitem{dicke1}
Dicke, R.H., {\em Gravitation and the Universe}, Jayne Lecture for 1969,
  (American Philosophical Society, Philadelphia, 1969).

\bibitem{Dickey}
Dickey, J.O. {et~al.}, ``Lunar Laser Ranging: A Continuing Legacy of the Apollo
  Program'', {\em Science}, {\bf 265}, 482--490 (1994).
  {\small[\href{http://dx.doi.org/10.1126/science.265.5171.482}{DOI}]}.

\bibitem{2009astro2010S..68D}
{Doeleman}, S. {et~al.}, ``{Imaging an Event Horizon: submm-VLBI of a Super
  Massive Black Hole}'', in {\em Astro2010: The Astronomy and Astrophysics
  Decadal Survey}, p.~68, (2009).
  {\small[\href{http://adsabs.harvard.edu/abs/2009astro2010S..68D}{ADS}]},
  {\small[\href{http://arxiv.org/abs/0906.3899}{{arXiv:0906.3899
  {\small[astro-ph.CO]}}}]}.

\bibitem{2012PhRvD..86j3008D}
{Dossett}, J.~N.  and {Ishak}, M., ``{Spatial curvature and cosmological tests
  of general relativity}'', {\em \prd}, {\bf 86}, 103008 (2012).
  {\small[\href{http://dx.doi.org/10.1103/PhysRevD.86.103008}{DOI}]},
  {\small[\href{http://adsabs.harvard.edu/abs/2012PhRvD..86j3008D}{ADS}]},
  {\small[\href{http://arxiv.org/abs/1205.2422}{{arXiv:1205.2422
  {\small[astro-ph.CO]}}}]}.

\bibitem{2011PhRvD..84l3001D}
{Dossett}, J.~N., {Ishak}, M.  and {Moldenhauer}, J., ``{Testing general
  relativity at cosmological scales: Implementation and parameter
  correlations}'', {\em \prd}, {\bf 84}, 123001 (2011).
  {\small[\href{http://dx.doi.org/10.1103/PhysRevD.84.123001}{DOI}]},
  {\small[\href{http://adsabs.harvard.edu/abs/2011PhRvD..84l3001D}{ADS}]},
  {\small[\href{http://arxiv.org/abs/1109.4583}{{arXiv:1109.4583
  {\small[astro-ph.CO]}}}]}.

\bibitem{drever}
Drever, R.W.P., ``A search for anisotropy of inertial mass using a free
  precession technique'', {\em Philos. Mag.}, {\bf 6}, 683--687 (1961).
  {\small[\href{http://dx.doi.org/10.1080/14786436108244418}{DOI}]}.

\bibitem{dyson72}
Dyson, F.J., ``The Fundamental Constants and Their Time Variation'', in Salam,
  A.  and Wigner, E.P., eds., {\em Aspects of Quantum Theory}, pp. 213--236,
  (Cambridge University Press, Cambridge; New York, 1972).
  {\small[\href{http://books.google.com/books?id=CpJiqUFkHGoC&pg=PA213}{Google
  Books}]}.

\bibitem{1975ApJ...196L..59E}
{Eardley}, D.~M., ``{Observable effects of a scalar gravitational field in a
  binary pulsar}'', {\em \apjl}, {\bf 196}, L59--L62 (1975).
  {\small[\href{http://dx.doi.org/10.1086/181744}{DOI}]},
  {\small[\href{http://adsabs.harvard.edu/abs/1975ApJ...196L..59E}{ADS}]}.

\bibitem{1938RSPSA.166..465E}
{Eddington}, A.  and {Clark}, G.~L., ``{The Problem of n Bodies in General
  Relativity Theory}'', {\em Proc. Roy. Soc. London A}, {\bf 166}, 465--475
  (1938). {\small[\href{http://dx.doi.org/10.1098/rspa.1938.0104}{DOI}]},
  {\small[\href{http://adsabs.harvard.edu/abs/1938RSPSA.166..465E}{ADS}]}.

\bibitem{1922RSPSA.102..268E}
{Eddington}, A.~S., ``{The Propagation of Gravitational Waves}'', {\em Proc.
  Roy. Soc. London A}, {\bf 102}, 268--282 (1922).
  {\small[\href{http://dx.doi.org/10.1098/rspa.1922.0085}{DOI}]},
  {\small[\href{http://adsabs.harvard.edu/abs/1922RSPSA.102..268E}{ADS}]}.

\bibitem{1976ApJ...208L..77E}
{Ehlers}, J., {Rosenblum}, A., {Goldberg}, J.~N.  and {Havas}, P., ``{Comments
  on gravitational radiation damping and energy loss in binary systems}'', {\em
  Astrophys. J. Lett.}, {\bf 208}, L77--L81 (1976).
  {\small[\href{http://dx.doi.org/10.1086/182236}{DOI}]},
  {\small[\href{http://adsabs.harvard.edu/abs/1976ApJ...208L..77E}{ADS}]}.

\bibitem{1916SPAW.......688E}
{Einstein}, A., ``{N{\"a}herungsweise Integration der Feldgleichungen der
  Gravitation}'', {\em Sitzungsberichte der K{\"o}niglich Preu{\ss}ischen
  Akademie der Wissenschaften (Berlin)}, 688--696 (1916).
  {\small[\href{http://adsabs.harvard.edu/abs/1916SPAW.......688E}{ADS}]}.

\bibitem{EIH}
{Einstein}, A., {Infeld}, L.  and {Hoffmann}, B., ``{The Gravitational
  Equations and the Problem of Motion}'', {\em Ann. Math.}, {\bf 39}, 65--100
  (1938).
  {\small[\href{http://adsabs.harvard.edu/abs/1938AnMat..39...65E}{ADS}]}.

\bibitem{1937FrInJ.223...43E}
{Einstein}, A.  and {Rosen}, N., ``{On Gravitational Waves}'', {\em J. Franklin
  Institute}, {\bf 223}, 43--54 (1937).
  {\small[\href{http://dx.doi.org/10.1016/S0016-0032(37)90583-0}{DOI}]},
  {\small[\href{http://adsabs.harvard.edu/abs/1937FrInJ.223...43E}{ADS}]}.

\bibitem{eling04}
Eling, C.  and Jacobson, T., ``Static post-Newtonian equivalence of general
  relativity and gravity with a dynamical preferred frame'', {\em Phys. Rev.
  D}, {\bf 69}, 064005 (2004).
  {\small[\href{http://dx.doi.org/10.1103/PhysRevD.69.064005}{DOI}]},
  {\small[\href{http://arxiv.org/abs/gr-qc/0310044}{{gr-qc/0310044}}]}.

\bibitem{eotvos}
E{\"{o}}tv{\"{o}}s, R.V., Pek{\'{a}}r, V.  and Fekete, E., ``Beitrage zum
  Gesetze der Proportionalit\"at von Tr\"agheit und Gravit\"at'', {\em Ann.
  Phys. (Leipzig)}, {\bf 68}, 11--66 (1922).

\bibitem{2011PhRvL.106v1101E}
{Everitt}, C.~W.~F. {et~al.}, ``{Gravity Probe B: Final Results of a Space
  Experiment to Test General Relativity}'', {\em Phys. Rev. Lett.}, {\bf 106},
  221101 (2011).
  {\small[\href{http://dx.doi.org/10.1103/PhysRevLett.106.221101}{DOI}]},
  {\small[\href{http://adsabs.harvard.edu/abs/2011PhRvL.106v1101E}{ADS}]},
  {\small[\href{http://arxiv.org/abs/1105.3456}{{arXiv:1105.3456
  {\small[gr-qc]}}}]}.

\bibitem{2012LRR....15...10F}
{Famaey}, B.  and {McGaugh}, S.~S., ``{Modified Newtonian Dynamics (MOND):
  Observational Phenomenology and Relativistic Extensions}'', {\em Living Rev.
  Relativity}, {\bf 15}, lrr-2012-10 (2012).
  {\small[\href{http://dx.doi.org/10.12942/lrr-2012-10}{DOI}]},
  {\small[\href{http://adsabs.harvard.edu/abs/2012LRR....15...10F}{ADS}]},
  {\small[\href{http://arxiv.org/abs/1112.3960}{{arXiv:1112.3960
  {\small[astro-ph.CO]}}}]}. URL (accessed 1 June 2014):
  \newline\url{http://www.livingreviews.org/lrr-2012-10}.

\bibitem{farley}
Farley, F.J.M., Bailey, J., Brown, R.C.A., Giesch, M., J{\"{o}}stlein, H.,
  van~der Meer, S., Picasso, E.  and Tannenbaum, M., ``The Anomalous Magnetic
  Moment of the Negative Muon'', {\em Nuovo Cimento}, {\bf 45}, 281--286
  (1966).

\bibitem{2011CeMDA.111..363F}
{Fienga}, A., {Laskar}, J., {Kuchynka}, P., {Manche}, H., {Desvignes}, G.,
  {Gastineau}, M., {Cognard}, I.  and {Theureau}, G., ``{The INPOP10a planetary
  ephemeris and its applications in fundamental physics}'', {\em Celestial
  Mechanics and Dynamical Astronomy}, {\bf 111}, 363--385 (2011).
  {\small[\href{http://dx.doi.org/10.1007/s10569-011-9377-8}{DOI}]},
  {\small[\href{http://adsabs.harvard.edu/abs/2011CeMDA.111..363F}{ADS}]},
  {\small[\href{http://arxiv.org/abs/1108.5546}{{arXiv:1108.5546
  {\small[astro-ph.EP]}}}]}.

\bibitem{finnchern}
Finn, L.S.  and Chernoff, D.F., ``Observing binary inspiral in gravitational
  radiation: One interferometer'', {\em Phys. Rev. D}, {\bf 47}, 2198--2219
  (1993). {\small[\href{http://dx.doi.org/10.1103/PhysRevD.47.2198}{DOI}]},
  {\small[\href{http://arxiv.org/abs/gr-qc/9301003}{{gr-qc/9301003}}]}.

\bibitem{FinnSutton02}
Finn, L.S.  and Sutton, P.J., ``Bounding the mass of the graviton using binary
  pulsar observations'', {\em Phys. Rev. D}, {\bf 65}, 044022 (2002).
  {\small[\href{http://dx.doi.org/10.1103/PhysRevD.65.044022}{DOI}]},
  {\small[\href{http://arxiv.org/abs/gr-qc/0109049}{{gr-qc/0109049}}]}.

\bibitem{fischbach92}
Fischbach, E., Gillies, G.T., Krause, D.E., Schwan, J.G.  and Talmadge, C.L.,
  ``Non-Newtonian gravity and new weak forces: An index of measurements and
  theory'', {\em Metrologia}, {\bf 29}, 213--260 (1992).
  {\small[\href{http://dx.doi.org/10.1088/0026-1394/29/3/001}{DOI}]}.

\bibitem{fischbach5}
Fischbach, E., Sudarsky, D., Szafer, A., Talmadge, C.L.  and Aronson, S.H.,
  ``Reanalysis of the E\"otv\"os experiment'', {\em Phys. Rev. Lett.}, {\bf
  56}, 3--6 (1986).
  {\small[\href{http://dx.doi.org/10.1103/PhysRevLett.56.1427}{DOI}]}. Erratum:
  Phys. Rev. Lett. 56 (1986) 1427.

\bibitem{FischbachTalmadge}
Fischbach, E.  and Talmadge, C.L., ``Six years of the fifth force'', {\em
  Nature}, {\bf 356}, 207--215 (1992).
  {\small[\href{http://dx.doi.org/10.1038/356207a0}{DOI}]}.

\bibitem{FischbachTalmadge2}
Fischbach, E.  and Talmadge, C.L., {\em The Search for Non-Newtonian Gravity},
  (Springer, New York, 1998).
  {\small[\href{http://books.google.com/books?id=G7wZtGHPiCEC}{Google Books}]}.

\bibitem{fischer04}
Fischer, M. {et~al.}, ``New limits on the drift of fundamental constants from
  laboratory measurements'', {\em Phys. Rev. Lett.}, {\bf 92}, 230802 (2004).
  {\small[\href{http://dx.doi.org/10.1103/PhysRevLett.92.230802}{DOI}]},
  {\small[\href{http://arxiv.org/abs/physics/0312086}{{physics/0312086}}]}.

\bibitem{fock}
Fock, V.A., {\em The Theory of Space, Time and Gravitation}, (MacMillan, New
  York, 1964).

\bibitem{fomalont03}
Fomalont, E.B.  and Kopeikin, S.M., ``The measurement of the light deflection
  from Jupiter: experimental results'', {\em Astrophys. J.}, {\bf 598},
  704--711 (2003). {\small[\href{http://dx.doi.org/10.1086/378785}{DOI}]},
  {\small[\href{http://arxiv.org/abs/astro-ph/0302294}{{astro-ph/0302294}}]}.

\bibitem{2009ApJ...699.1395F}
{Fomalont}, E., {Kopeikin}, S., {Lanyi}, G.  and {Benson}, J., ``{Progress in
  Measurements of the Gravitational Bending of Radio Waves Using the VLBA}'',
  {\em \apj}, {\bf 699}, 1395--1402 (2009).
  {\small[\href{http://dx.doi.org/10.1088/0004-637X/699/2/1395}{DOI}]},
  {\small[\href{http://adsabs.harvard.edu/abs/2009ApJ...699.1395F}{ADS}]},
  {\small[\href{http://arxiv.org/abs/0904.3992}{{arXiv:0904.3992
  {\small[astro-ph.CO]}}}]}.

\bibitem{foster05}
Foster, B.Z.  and Jacobson, T., ``Post-Newtonian parameters and constraints on
  Einstein-{\AE}ther theory'', {\em Phys. Rev. D}, {\bf 73}, 064015 (2006).
  {\small[\href{http://dx.doi.org/10.1103/PhysRevD.73.064015}{DOI}]},
  {\small[\href{http://arxiv.org/abs/gr-qc/0509083}{{arXiv:gr-qc/0509083
  {\small[gr-qc]}}}]}.

\bibitem{2012MNRAS.423.3328F}
{Freire}, P.~C.~C. {et~al.}, ``{The relativistic pulsar-white dwarf binary PSR
  J1738+0333 - II. The most stringent test of scalar-tensor gravity}'', {\em
  \mnras}, {\bf 423}, 3328--3343 (2012).
  {\small[\href{http://dx.doi.org/10.1111/j.1365-2966.2012.21253.x}{DOI}]},
  {\small[\href{http://adsabs.harvard.edu/abs/2012MNRAS.423.3328F}{ADS}]},
  {\small[\href{http://arxiv.org/abs/1205.1450}{{arXiv:1205.1450
  {\small[astro-ph.GA]}}}]}.

\bibitem{hipparcos}
Froeschl{\'{e}}, M., Mignard, F.  and Arenou, F., ``Determination of the PPN
  parameter $\gamma$ with the Hipparcos data'', in {\em Proceedings of the
  Hipparcos Venice Symposium}, (ESA, Noordwijk, Netherlands, 1997). URL
  (accessed 1 June 2014):
  \newline\url{http://astro.estec.esa.nl/Hipparcos/venice.html}.

\bibitem{fujii04}
Fujii, Y., ``Oklo Constraint on the Time-Variability of the Fine-Structure
  Constant'', in Karshenboim, S.G.  and Peik, E., eds., {\em Astrophysics,
  Clocks and Fundamental Constants}, 302nd WE-Heraeus-Seminar, June 2003, Bad
  Honnef, Germany, Lecture Notes in Physics, 648, pp. 167--185, (Springer,
  Berlin; New York, 2004).
  {\small[\href{http://arxiv.org/abs/hep-ph/0311026}{{hep-ph/0311026}}]}.

\bibitem{2007sttg.book.....F}
{Fujii}, Y.  and {Maeda}, K.-I., {\em {The Scalar-Tensor Theory of
  Gravitation}}, (Cambridge University Press, Cambridge, 2007).
  {\small[\href{http://adsabs.harvard.edu/abs/2007sttg.book.....F}{ADS}]}.

\bibitem{2012PThPh.128..971F}
{Fujita}, R., ``{Gravitational Waves from a Particle in Circular Orbits around
  a Schwarzschild Black Hole to the 22nd Post-Newtonian Order}'', {\em Prog.
  Theor. Phys.}, {\bf 128}, 971--992 (2012).
  {\small[\href{http://adsabs.harvard.edu/abs/2012PThPh.128..971F}{ADS}]},
  {\small[\href{http://arxiv.org/abs/1211.5535}{{arXiv:1211.5535
  {\small[gr-qc]}}}]}.

\bibitem{gaia}
``Gaia - Taking The Galactic Census'', project homepage, ESA. URL (accessed 1
  June 2014): \newline\url{http://www.cosmos.esa.int/web/gaia/}.

\bibitem{2013LRR....16....7G}
{Gair}, J.~R., {Vallisneri}, M., {Larson}, S.~L.  and {Baker}, J.~G.,
  ``{Testing General Relativity with Low-Frequency, Space-Based
  Gravitational-Wave Detectors}'', {\em Living Rev. Relativity}, {\bf 16},
  lrr-2013-7 (2013).
  {\small[\href{http://adsabs.harvard.edu/abs/2013LRR....16....7G}{ADS}]},
  {\small[\href{http://arxiv.org/abs/1212.5575}{{arXiv:1212.5575
  {\small[gr-qc]}}}]}. URL (accessed 1 June 2014):
  \newline\url{http://www.livingreviews.org/lrr-2013-7}.

\bibitem{gasperini}
Gasperini, M., ``On the response of gravitational antennas to dilatonic
  waves'', {\em Phys. Lett. B}, {\bf 470}, 67--72 (1999).
  {\small[\href{http://dx.doi.org/10.1016/S0370-2693(99)01309-X}{DOI}]},
  {\small[\href{http://arxiv.org/abs/gr-qc/9910019}{{gr-qc/9910019}}]}.

\bibitem{2008PhRvD..78b2002G}
{Geraci}, A.~A., {Smullin}, S.~J., {Weld}, D.~M., {Chiaverini}, J.  and
  {Kapitulnik}, A., ``{Improved constraints on non-Newtonian forces at
  10microns}'', {\em \prd}, {\bf 78}, 022002 (2008).
  {\small[\href{http://dx.doi.org/10.1103/PhysRevD.78.022002}{DOI}]},
  {\small[\href{http://adsabs.harvard.edu/abs/2008PhRvD..78b2002G}{ADS}]},
  {\small[\href{http://arxiv.org/abs/0802.2350}{{arXiv:0802.2350
  {\small[hep-ex]}}}]}.

\bibitem{gleiser}
Gleiser, R.J.  and Kozameh, C.N., ``Astrophysical limits on quantum gravity
  motivated birefringence'', {\em Phys. Rev. D}, {\bf 64}, 083007 (2001).
  {\small[\href{http://dx.doi.org/10.1103/PhysRevD.64.083007}{DOI}]},
  {\small[\href{http://arxiv.org/abs/gr-qc/0102093}{{arXiv:gr-qc/0102093}}]}.

\bibitem{Godone}
Godone, A., Novero, C.  and Tavella, P., ``Null gravitational redshift
  experiment with nonidentical atomic clocks'', {\em Phys. Rev. D}, {\bf 51},
  319--323 (1995).
  {\small[\href{http://dx.doi.org/10.1103/PhysRevD.51.319}{DOI}]}.

\bibitem{gopu}
Gopakumar, A.  and Iyer, B.R., ``Gravitational waves from inspiraling compact
  binaries: Angular momentum flux, evolution of the orbital elements and the
  waveform to the second post-Newtonian order'', {\em Phys. Rev. D}, {\bf 56},
  7708--7731 (1997).
  {\small[\href{http://dx.doi.org/10.1103/PhysRevD.56.7708}{DOI}]},
  {\small[\href{http://arxiv.org/abs/gr-qc/9710075}{{gr-qc/9710075}}]}.

\bibitem{2012LNP...846.....G}
Gourgoulhon, E., {\em {3+1 Formalism in General Relativity: Bases of Numerical
  Relativity}}, Lecture Notes in Physics, 846, (Springer-Verlag, Berlin, 2012).
  {\small[\href{http://dx.doi.org/10.1007/978-3-642-24525-1}{DOI}]},
  {\small[\href{http://adsabs.harvard.edu/abs/2012LNP...846.....G}{ADS}]}.

\bibitem{gpbwebsite}
``Gravity Probe B: Testing Einstein's Universe'', project homepage, Stanford
  University. URL (accessed 1 June 2014):
  \newline\url{http://einstein.stanford.edu/}.

\bibitem{1983SvAL....9..230G}
{Grishchuk}, L.~P.  and {Kopeikin}, S.~M., ``{The Motion of a Pair of
  Gravitating Bodies Including the Radiation Reaction Force}'', {\em Soviet
  Astronomy Letters}, {\bf 9}, 230--232 (April 1983).
  {\small[\href{http://adsabs.harvard.edu/abs/1983SvAL....9..230G}{ADS}]}.

\bibitem{2012PhRvL.109h0801G}
{Gu{\'e}na}, J., {Abgrall}, M., {Rovera}, D., {Rosenbusch}, P., {Tobar}, M.~E.,
  {Laurent}, P., {Clairon}, A.  and {Bize}, S., ``{Improved Tests of Local
  Position Invariance Using Rb87 and Cs133 Fountains}'', {\em Phys. Rev.
  Lett.}, {\bf 109}, 080801 (2012).
  {\small[\href{http://dx.doi.org/10.1103/PhysRevLett.109.080801}{DOI}]},
  {\small[\href{http://adsabs.harvard.edu/abs/2012PhRvL.109h0801G}{ADS}]},
  {\small[\href{http://arxiv.org/abs/1205.4235}{{arXiv:1205.4235
  {\small[physics.atom-ph]}}}]}.

\bibitem{guenther98}
Guenther, D.B., Krauss, L.M.  and Demarque, P., ``Testing the Constancy of the
  Gravitational Constant Using Helioseismology'', {\em Astrophys. J.}, {\bf
  498}, 871--876 (1998).
  {\small[\href{http://dx.doi.org/10.1086/305567}{DOI}]}.

\bibitem{haugan}
Haugan, M.P., ``Energy conservation and the principle of equivalence'', {\em
  Ann. Phys. (N.Y.)}, {\bf 118}, 156--186 (1979).
  {\small[\href{http://dx.doi.org/10.1016/0003-4916(79)90238-0}{DOI}]}.

\bibitem{hauganlammer1}
Haugan, M.P.  and L{\"{a}}mmerzahl, C., ``On the interpretation of
  Michelson--Morley experiments'', {\em Phys. Lett. A}, {\bf 282}, 223--229
  (2001). {\small[\href{http://arxiv.org/abs/gr-qc/0103052}{{gr-qc/0103052}}]}.

\bibitem{hauganlammer2}
Haugan, M.P.  and L{\"{a}}mmerzahl, C., ``Principles of equivalence: Their role
  in gravitation physics and experiments that test them'', in L{\"{a}}mmerzahl,
  C., Everitt, C.W.F.  and Hehl, F.W., eds., {\em Gyros, Clocks, and
  Interferometers...: Testing Relativistic Gravity in Space}, Lecture Notes in
  Physics, 562, pp. 195--212, (Springer, Berlin; New York, 2001).
  {\small[\href{http://arxiv.org/abs/gr-qc/0103067}{{gr-qc/0103067}}]}.

\bibitem{hauganwill}
Haugan, M.P.  and Will, C.M., ``Modern tests of special relativity'', {\em
  Phys. Today}, {\bf 40}(5), 69--76 (1987).
  {\small[\href{http://dx.doi.org/10.1063/1.881074}{DOI}]}.

\bibitem{1972CMaPh..25..167H}
{Hawking}, S.~W., ``{Black holes in the Brans-Dicke Theory of gravitation}'',
  {\em Commun. Math. Phys.}, {\bf 25}, 167--171 (1972).
  {\small[\href{http://dx.doi.org/10.1007/BF01877518}{DOI}]},
  {\small[\href{http://adsabs.harvard.edu/abs/1972CMaPh..25..167H}{ADS}]}.

\bibitem{2014arXiv1402.6950H}
{Hees}, A., {Folkner}, W.~M., {Jacobson}, R.~A.  and {Park}, R.~S.,
  ``{Constraints on MOND theory from radio tracking data of the Cassini
  spacecraft}'', {\em ArXiv e-prints} (February 2014).
  {\small[\href{http://adsabs.harvard.edu/abs/2014arXiv1402.6950H}{ADS}]},
  {\small[\href{http://arxiv.org/abs/1402.6950}{{arXiv:1402.6950
  {\small[gr-qc]}}}]}.

\bibitem{hellings73}
Hellings, R.W.  and Nordtvedt~Jr., K., ``Vector-Metric Theory of Gravity'',
  {\em Phys. Rev. D}, {\bf 7}, 3593--3602 (1973).
  {\small[\href{http://dx.doi.org/10.1103/PhysRevD.7.3593}{DOI}]}.

\bibitem{2012RvMP...84..671H}
{Hinterbichler}, K., ``{Theoretical aspects of massive gravity}'', {\em Rev.
  Mod. Phys.}, {\bf 84}, 671--710 (2012).
  {\small[\href{http://dx.doi.org/10.1103/RevModPhys.84.671}{DOI}]},
  {\small[\href{http://adsabs.harvard.edu/abs/2012RvMP...84..671H}{ADS}]},
  {\small[\href{http://arxiv.org/abs/1105.3735}{{arXiv:1105.3735
  {\small[hep-th]}}}]}.

\bibitem{2012PhRvD..85d3508H}
{Hojjati}, A., {Zhao}, G.-B., {Pogosian}, L., {Silvestri}, A., {Crittenden}, R.
   and {Koyama}, K., ``{Cosmological tests of general relativity: A principal
  component analysis}'', {\em \prd}, {\bf 85}, 043508 (2012).
  {\small[\href{http://dx.doi.org/10.1103/PhysRevD.85.043508}{DOI}]},
  {\small[\href{http://adsabs.harvard.edu/abs/2012PhRvD..85d3508H}{ADS}]},
  {\small[\href{http://arxiv.org/abs/1111.3960}{{arXiv:1111.3960
  {\small[astro-ph.CO]}}}]}.

\bibitem{2012JCAP...05..010H}
{Horbatsch}, M.~W.  and {Burgess}, C.~P., ``{Cosmic black-hole hair growth and
  quasar OJ287}'', {\em J. Cosm. Astropart. Phys.}, {\bf 5}, 010 (2012).
  {\small[\href{http://dx.doi.org/10.1088/1475-7516/2012/05/010}{DOI}]},
  {\small[\href{http://adsabs.harvard.edu/abs/2012JCAP...05..010H}{ADS}]},
  {\small[\href{http://arxiv.org/abs/1111.4009}{{arXiv:1111.4009
  {\small[gr-qc]}}}]}.

\bibitem{hoyle04}
Hoyle, C.D., Kapner, D.J., Heckel, B.R., Adelberger, E.G., Gundlach, J.H.,
  Schmidt, U.  and Swanson, H.E., ``Submillimeter tests of the gravitational
  inverse-square law'', {\em Phys. Rev. D}, {\bf 70}, 042004 (2004).
  {\small[\href{http://dx.doi.org/10.1103/PhysRevD.70.042004}{DOI}]},
  {\small[\href{http://arxiv.org/abs/hep-ph/0405262}{{hep-ph/0405262}}]}.

\bibitem{hoyle01}
Hoyle, C.D., Schmidt, U., Heckel, B.R., Adelberger, E.G., Gundlach, J.H.,
  Kapner, D.J.  and Swanson, H.E., ``Submillimeter Test of the Gravitational
  Inverse-Square Law: A Search for `large' Extra Dimensions'', {\em Phys. Rev.
  Lett.}, {\bf 86}, 1418--1421 (2001).
  {\small[\href{http://dx.doi.org/10.1103/PhysRevLett.86.1418}{DOI}]},
  {\small[\href{http://arxiv.org/abs/hep-ph/0011014}{{arXiv:hep-ph/0011014}}]}.

\bibitem{hughes}
Hughes, V.W., Robinson, H.G.  and Beltran-Lopez, V., ``Upper limit for the
  anisotropy of inertial mass from nuclear resonance experiments'', {\em Phys.
  Rev. Lett.}, {\bf 4}, 342--344 (1960).
  {\small[\href{http://dx.doi.org/10.1103/PhysRevLett.4.342}{DOI}]}.

\bibitem{Hulse}
Hulse, R.A., ``Nobel Lecture: The discovery of the binary pulsar'', {\em Rev.
  Mod. Phys.}, {\bf 66}, 699--710 (1994).
  {\small[\href{http://dx.doi.org/10.1103/RevModPhys.66.699}{DOI}]}.

\bibitem{1975ApJ...195L..51H}
{Hulse}, R.~A.  and {Taylor}, J.~H., ``{Discovery of a pulsar in a binary
  system}'', {\em \apjl}, {\bf 195}, L51--L53 (1975).
  {\small[\href{http://dx.doi.org/10.1086/181708}{DOI}]},
  {\small[\href{http://adsabs.harvard.edu/abs/1975ApJ...195L..51H}{ADS}]}.

\bibitem{iorio05}
Iorio, L., ``On the reliability of the so-far performed tests for measuring the
  Lense--Thirring effect with the LAGEOS satellites'', {\em New Astronomy},
  {\bf 10}, 603--615 (2005).
  {\small[\href{http://dx.doi.org/10.1016/j.newast.2005.01.001}{DOI}]},
  {\small[\href{http://arxiv.org/abs/gr-qc/0411024}{{gr-qc/0411024}}]}.

\bibitem{ItohFutamase03}
Itoh, Y.  and Futamase, T., ``New derivation of a third post-Newtonian equation
  of motion for relativistic compact binaries without ambiguity'', {\em Phys.
  Rev. D}, {\bf 68}, 121501(R) (2003).
  {\small[\href{http://dx.doi.org/10.1103/PhysRevD.68.121501}{DOI}]}.

\bibitem{ivanchik05}
Ivanchik, A., Petitjean, P., Varshalovich, D., Aracil, B., Srianand, R., Chand,
  H., Ledoux, C.  and Boiss{\'{e}}, P., ``A new constraint on the time
  dependence of the proton-to-electron mass ratio: Analysis of the Q 0347-383
  and Q 0405-443 spectra'', {\em Astron. Astrophys.}, {\bf 440}, 45--52 (2005).
  {\small[\href{http://dx.doi.org/10.1051/0004-6361:20052648}{DOI}]},
  {\small[\href{http://arxiv.org/abs/astro-ph/0507174}{{astro-ph/0507174}}]}.

\bibitem{ives}
Ives, H.E.  and Stilwell, G.R., ``An experimental study of the rate of a moving
  atomic clock'', {\em J. Opt. Soc. Am.}, {\bf 28}, 215--226 (1938).
  {\small[\href{http://dx.doi.org/10.1364/JOSA.28.000215}{DOI}]}.

\bibitem{1999PhRvL..83.2699J}
{Jacobson}, T., ``{Primordial Black Hole Evolution in Tensor-Scalar
  Cosmology}'', {\em Phys. Rev. Lett.}, {\bf 83}, 2699--2702 (1999).
  {\small[\href{http://dx.doi.org/10.1103/PhysRevLett.83.2699}{DOI}]},
  {\small[\href{http://adsabs.harvard.edu/abs/1999PhRvL..83.2699J}{ADS}]},
  {\small[\href{http://arxiv.org/abs/astro-ph/9905303}{{astro-ph/9905303}}]}.

\bibitem{jacobson01}
Jacobson, T.  and Mattingly, D., ``Gravity with a dynamical preferred frame'',
  {\em Phys. Rev. D}, {\bf 64}, 024028 (2001).
  {\small[\href{http://dx.doi.org/10.1103/PhysRevD.64.024028}{DOI}]},
  {\small[\href{http://adsabs.harvard.edu/abs/2001PhRvD..64b4028J}{ADS}]},
  {\small[\href{http://arxiv.org/abs/gr-qc/0007031}{{arXiv:gr-qc/0007031}}]}.

\bibitem{jacobson04}
Jacobson, T.  and Mattingly, D., ``Einstein-aether waves'', {\em Phys. Rev. D},
  {\bf 70}, 024003 (2004).
  {\small[\href{http://dx.doi.org/10.1103/PhysRevD.70.024003}{DOI}]},
  {\small[\href{http://arxiv.org/abs/gr-qc/0402005}{{arXiv:gr-qc/0402005
  {\small[gr-qc]}}}]}.

\bibitem{jaraschaefer98}
Jaranowski, P.  and Sch{\"{a}}fer, G., ``3rd post-Newtonian higher order
  Hamilton dynamics for two-body point-mass systems'', {\em Phys. Rev. D}, {\bf
  57}, 7274--7291 (1998).
  {\small[\href{http://dx.doi.org/10.1103/PhysRevD.57.7274}{DOI}]},
  {\small[\href{http://arxiv.org/abs/gr-qc/9712075}{{gr-qc/9712075}}]}.
  Erratum: Phys. Rev. D 63 (2001) 029902.

\bibitem{jaranowski}
Jaranowski, P.  and Sch{\"{a}}fer, G., ``Binary black-hole problem at the third
  post-Newtonian approximation in the orbital motion: Static part'', {\em Phys.
  Rev. D}, {\bf 60}, 124003 (1999).
  {\small[\href{http://dx.doi.org/10.1103/PhysRevD.60.124003}{DOI}]},
  {\small[\href{http://arxiv.org/abs/gr-qc/9906092}{{gr-qc/9906092}}]}.

\bibitem{jaseja}
Jaseja, T.S., Javan, A., Murray, J.  and Townes, C.H., ``Test of special
  relativity or of the isotropy of space by use of infrared masers'', {\em
  Phys. Rev.}, {\bf 133}, A1221--A1225 (1964).
  {\small[\href{http://dx.doi.org/10.1103/PhysRev.133.A1221}{DOI}]}.

\bibitem{Jones05}
Jones, D.I., ``Bounding the mass of the graviton using eccentric binaries'',
  {\em Astrophys. J. Lett.}, {\bf 618}, L115--L118 (2005).
  {\small[\href{http://dx.doi.org/10.1086/427773}{DOI}]},
  {\small[\href{http://arxiv.org/abs/gr-qc/0411123}{{gr-qc/0411123}}]}.

\bibitem{2012ApJ...746L..16K}
{Kanekar}, N., {Langston}, G.~I., {Stocke}, J.~T., {Carilli}, C.~L.  and
  {Menten}, K.~M., ``{Constraining Fundamental Constant Evolution with H I and
  OH Lines}'', {\em \apjl}, {\bf 746}, L16 (2012).
  {\small[\href{http://dx.doi.org/10.1088/2041-8205/746/2/L16}{DOI}]},
  {\small[\href{http://adsabs.harvard.edu/abs/2012ApJ...746L..16K}{ADS}]},
  {\small[\href{http://arxiv.org/abs/1201.3372}{{arXiv:1201.3372
  {\small[astro-ph.CO]}}}]}.

\bibitem{2007PhRvL..98b1101K}
{Kapner}, D.~J., {Cook}, T.~S., {Adelberger}, E.~G., {Gundlach}, J.~H.,
  {Heckel}, B.~R., {Hoyle}, C.~D.  and {Swanson}, H.~E., ``{Tests of the
  Gravitational Inverse-Square Law below the Dark-Energy Length Scale}'', {\em
  Phys. Rev. Lett.}, {\bf 98}, 021101 (2007).
  {\small[\href{http://dx.doi.org/10.1103/PhysRevLett.98.021101}{DOI}]},
  {\small[\href{http://adsabs.harvard.edu/abs/2007PhRvL..98b1101K}{ADS}]},
  {\small[\href{http://arxiv.org/abs/hep-ph/0611184}{{hep-ph/0611184}}]}.

\bibitem{1999PhRvL..83.1892K}
{Katz}, J.~I., ``{Comment on ``Indication, from Pioneer 10/11, Galileo, and
  Ulysses Data, of an Apparent Anomalous, Weak, Long-Range Acceleration''}'',
  {\em Phys. Rev. Lett.}, {\bf 83}, 1892 (1999).
  {\small[\href{http://dx.doi.org/10.1103/PhysRevLett.83.1892}{DOI}]},
  {\small[\href{http://adsabs.harvard.edu/abs/1999PhRvL..83.1892K}{ADS}]},
  {\small[\href{http://arxiv.org/abs/gr-qc/9809070}{{gr-qc/9809070}}]}.

\bibitem{2005PhT....58i..43K}
{Kennefick}, D., ``{Einstein Versus the Physical Review}'', {\em Phys. Today},
  {\bf 58}(9), 43 (2005).
  {\small[\href{http://dx.doi.org/10.1063/1.2117822}{DOI}]},
  {\small[\href{http://adsabs.harvard.edu/abs/2005PhT....58i..43K}{ADS}]}.

\bibitem{2007tste.book.....K}
{Kennefick}, D., {\em {Traveling at the Speed of Thought: Einstein and the
  Quest for Gravitational Waves}}, (Princeton University Press, Princeton,
  2007).
  {\small[\href{http://adsabs.harvard.edu/abs/2007tste.book.....K}{ADS}]}.

\bibitem{2009PhT....62c..37K}
{Kennefick}, D., ``{Testing relativity from the 1919 eclipse -- a question of
  bias}'', {\em Phys. Today}, {\bf 62}(3), 37 (2009).
  {\small[\href{http://dx.doi.org/10.1063/1.3099578}{DOI}]},
  {\small[\href{http://adsabs.harvard.edu/abs/2009PhT....62c..37K}{ADS}]}.

\bibitem{2004PhRvL..93q1104K}
{Khoury}, J.  and {Weltman}, A., ``{Chameleon Fields: Awaiting Surprises for
  Tests of Gravity in Space}'', {\em Phys. Rev. Lett.}, {\bf 93}, 171104
  (2004).
  {\small[\href{http://dx.doi.org/10.1103/PhysRevLett.93.171104}{DOI}]},
  {\small[\href{http://adsabs.harvard.edu/abs/2004PhRvL..93q1104K}{ADS}]},
  {\small[\href{http://arxiv.org/abs/astro-ph/0309300}{{astro-ph/0309300}}]}.

\bibitem{2012MNRAS.422.3370K}
{King}, J.~A., {Webb}, J.~K., {Murphy}, M.~T., {Flambaum}, V.~V., {Carswell},
  R.~F., {Bainbridge}, M.~B., {Wilczynska}, M.~R.  and {Koch}, F.~E.,
  ``{Spatial variation in the fine-structure constant - new results from
  VLT/UVES}'', {\em \mnras}, {\bf 422}, 3370--3414 (2012).
  {\small[\href{http://dx.doi.org/10.1111/j.1365-2966.2012.20852.x}{DOI}]},
  {\small[\href{http://adsabs.harvard.edu/abs/2012MNRAS.422.3370K}{ADS}]},
  {\small[\href{http://arxiv.org/abs/1202.4758}{{arXiv:1202.4758
  {\small[astro-ph.CO]}}}]}.

\bibitem{2013PhRvD..87l5031K}
{Klimchitskaya}, G.~L., {Mohideen}, U.  and {Mostepanenko}, V.~M.,
  ``{Constraints on corrections to Newtonian gravity from two recent
  measurements of the Casimir interaction between metallic surfaces}'', {\em
  \prd}, {\bf 87}, 125031 (2013).
  {\small[\href{http://dx.doi.org/10.1103/PhysRevD.87.125031}{DOI}]},
  {\small[\href{http://adsabs.harvard.edu/abs/2013PhRvD..87l5031K}{ADS}]},
  {\small[\href{http://arxiv.org/abs/1306.4979}{{arXiv:1306.4979
  {\small[gr-qc]}}}]}.

\bibitem{KokkotasSchmidt99}
Kokkotas, K.D.  and Schmidt, B.G., ``Quasi-Normal Modes of Stars and Black
  Holes'', {\em Living Rev. Relativity}, {\bf 2}, lrr-1999-2 (1999).
  {\small[\href{http://dx.doi.org/10.12942/lrr-1999-2}{DOI}]}. URL (accessed 1
  June 2014): \newline\url{http://www.livingreviews.org/lrr-1999-2}.

\bibitem{2011Icar..211..401K}
{Konopliv}, A.~S., {Asmar}, S.~W., {Folkner}, W.~M., {Karatekin}, {\"O}.,
  {Nunes}, D.~C., {Smrekar}, S.~E., {Yoder}, C.~F.  and {Zuber}, M.~T., ``{Mars
  high resolution gravity fields from MRO, Mars seasonal gravity, and other
  dynamical parameters}'', {\em Icarus}, {\bf 211}, 401--428 (2011).
  {\small[\href{http://dx.doi.org/10.1016/j.icarus.2010.10.004}{DOI}]},
  {\small[\href{http://adsabs.harvard.edu/abs/2011Icar..211..401K}{ADS}]}.

\bibitem{kopeikin01}
Kopeikin, S.M., ``Testing the relativistic effect of the propagation of gravity
  by very long baseline interferometry'', {\em Astrophys. J. Lett.}, {\bf 556},
  L1--L5 (2001). {\small[\href{http://dx.doi.org/10.1086/322872}{DOI}]},
  {\small[\href{http://arxiv.org/abs/gr-qc/0105060}{{gr-qc/0105060}}]}.

\bibitem{kopeikin03}
Kopeikin, S.M., ``The post-Newtonian treatment of the VLBI experiment on
  September 8, 2002'', {\em Phys. Lett. A}, {\bf 312}, 147--157 (2003).
  {\small[\href{http://dx.doi.org/10.1016/S0375-9601(03)00613-3}{DOI}]},
  {\small[\href{http://arxiv.org/abs/gr-qc/0212121}{{gr-qc/0212121}}]}.

\bibitem{kopeikin04}
Kopeikin, S.M., ``The speed of gravity in general relativity and theoretical
  interpretation of the Jovian deflection experiment'', {\em Class. Quantum
  Grav.}, {\bf 21}, 3251--3286 (2004).
  {\small[\href{http://dx.doi.org/10.1088/0264-9381/21/13/010}{DOI}]},
  {\small[\href{http://arxiv.org/abs/gr-qc/0310059}{{gr-qc/0310059}}]}.

\bibitem{kopeikin05b}
Kopeikin, S.M., ``Comment on {`Model-dependence of Shapiro time delay and the
  ``speed of gravity/speed of light'' controversy'}'', {\em Class. Quantum
  Grav.}, {\bf 22}, 5181 (2005).
  {\small[\href{http://dx.doi.org/10.1088/0264-9381/22/23/N01}{DOI}]},
  {\small[\href{http://arxiv.org/abs/gr-qc/0501048}{{gr-qc/0501048}}]}.

\bibitem{kopeikin02}
Kopeikin, S.M.  and Fomalont, E.B., ``General relativistic model for
  experimental measurement of the speed of propagation of gravity by VLBI'', in
  Ros, E., Porcas, R.W., Lobanov, A.P.  and Zensus, J.A., eds., {\em
  Proceedings of the 6th European VLBI Network Symposium}, June 25--28 2002,
  Bonn, Germany, pp. 49--52, (Max-Planck-Institut f\"ur Radioastronomie, Bonn,
  2002). {\small[\href{http://arxiv.org/abs/gr-qc/0206022}{{gr-qc/0206022}}]}.

\bibitem{kopeikin05a}
{Kopeikin}, S.~M., ``{Comments on `On the Speed of Gravity and the
  Jupiter/quasar Measurement' by S. Samuel}'', {\em Int. J. Mod. Phys. D}, {\bf
  15}, 273--288 (2006).
  {\small[\href{http://dx.doi.org/10.1142/S021827180600853X}{DOI}]},
  {\small[\href{http://adsabs.harvard.edu/abs/2006IJMPD..15..273K}{ADS}]},
  {\small[\href{http://arxiv.org/abs/gr-qc/0501001}{{gr-qc/0501001}}]}.

\bibitem{kosteleckylane99}
Kosteleck{\'{y}}, V.A.  and Lane, C.D., ``Constraints on Lorentz violation from
  clock-comparison experiments'', {\em Phys. Rev. D}, {\bf 60}, 116010 (1999).
  {\small[\href{http://dx.doi.org/10.1103/PhysRevD.60.116010}{DOI}]},
  {\small[\href{http://arxiv.org/abs/hep-ph/9908504}{{arXiv:hep-ph/9908504}}]}.

\bibitem{kosteleckymewes02}
Kosteleck{\'{y}}, V.A.  and Mewes, M., ``Signals for Lorentz violation in
  electrodynamics'', {\em Phys. Rev. D}, {\bf 66}, 056005 (2002).
  {\small[\href{http://dx.doi.org/10.1103/PhysRevD.66.056005}{DOI}]},
  {\small[\href{http://arxiv.org/abs/hep-ph/0205211}{{arXiv:hep-ph/0205211}}]}.

\bibitem{RevModPhys.83.11}
Kosteleck\'y, V.A.  and Russell, N., ``Data tables for Lorentz and $CPT$
  violation'', {\em Rev. Mod. Phys.}, {\bf 83}, 11--31 (2011).
  {\small[\href{http://dx.doi.org/10.1103/RevModPhys.83.11}{DOI}]}.

\bibitem{kosteleckysamuel}
Kosteleck{\'{y}}, V.A.  and Samuel, S., ``Gravitational phenomenology in
  higher-dimensional theories and strings'', {\em Phys. Rev. D}, {\bf 40},
  1886--1903 (1989).
  {\small[\href{http://dx.doi.org/10.1103/PhysRevD.40.1886}{DOI}]}.

\bibitem{kramer}
Kramer, M., ``Determination of the geometry of the PSR B1913+16 system by
  geodetic precession'', {\em Astrophys. J.}, {\bf 509}, 856--860 (1998).
  {\small[\href{http://dx.doi.org/10.1086/306535}{DOI}]},
  {\small[\href{http://arxiv.org/abs/astro-ph/9808127}{{astro-ph/9808127}}]}.

\bibitem{2006Sci...314...97K}
{Kramer}, M. {et~al.}, ``{Tests of General Relativity from Timing the Double
  Pulsar}'', {\em Science}, {\bf 314}, 97--102 (2006).
  {\small[\href{http://dx.doi.org/10.1126/science.1132305}{DOI}]},
  {\small[\href{http://adsabs.harvard.edu/abs/2006Sci...314...97K}{ADS}]},
  {\small[\href{http://arxiv.org/abs/arXiv:astro-ph/0609417}{{arXiv:astro-ph/0609417}}]}.

\bibitem{krisher90a}
Krisher, T.P., Anderson, J.D.  and Campbell, J.K., ``Test of the gravitational
  redshift effect at Saturn'', {\em Phys. Rev. Lett.}, {\bf 64}, 1322--1325
  (1990). {\small[\href{http://dx.doi.org/10.1103/PhysRevLett.64.1322}{DOI}]}.

\bibitem{krisher90}
Krisher, T.P., Maleki, L., Lutes, G.F., Primas, L.E., Logan, R.T., Anderson,
  J.D.  and Will, C.M., ``Test of the isotropy of the one-way speed of light
  using hydrogen-maser frequency standards'', {\em Phys. Rev. D}, {\bf 42},
  731--734 (1990).
  {\small[\href{http://dx.doi.org/10.1103/PhysRevD.42.731}{DOI}]}.

\bibitem{krisher93}
Krisher, T.P., Morabito, D.D.  and Anderson, J.D., ``The Galileo solar redshift
  experiment'', {\em Phys. Rev. Lett.}, {\bf 70}, 2213--2216 (1993).

\bibitem{krolak95}
Kr{\'{o}}lak, A., Kokkotas, K.D.  and Sch{\"{a}}fer, G., ``Estimation of the
  post-Newtonian parameters in the gravitational-wave emission of a coalescing
  binary'', {\em Phys. Rev. D}, {\bf 52}, 2089--2111 (1995).
  {\small[\href{http://arxiv.org/abs/gr-qc/9503013}{{gr-qc/9503013}}]}.

\bibitem{2009A&A...499..331L}
{Lambert}, S.~B.  and {Le Poncin-Lafitte}, C., ``{Determining the relativistic
  parameter {$\gamma$} using very long baseline interferometry}'', {\em \aap},
  {\bf 499}, 331--335 (2009).
  {\small[\href{http://dx.doi.org/10.1051/0004-6361/200911714}{DOI}]},
  {\small[\href{http://adsabs.harvard.edu/abs/2009A%26A...499..331L}{ADS}]},
  {\small[\href{http://arxiv.org/abs/0903.1615}{{arXiv:0903.1615
  {\small[gr-qc]}}}]}.

\bibitem{lammer03}
L{\"{a}}mmerzahl, C., ``The Einstein equivalence principle and the search for
  new physics'', in Giulini, D.J.W., Kiefer, C.  and L{\"{a}}mmerzahl, C.,
  eds., {\em Quantum Gravity: From Theory to Experimental Search}, Lecture
  Notes in Physics, 631, pp. 367--394, (Springer, Berlin; New York, 2003).
  {\small[\href{http://books.google.com/books?id=SBUWhufgFgkC}{Google Books}]}.

\bibitem{lamoreaux86}
Lamoreaux, S.K., Jacobs, J.P., Heckel, B.R., Raab, F.J.  and Fortson, E.N.,
  ``New limits on spatial anisotropy from optically-pumped $^{201}$Hg and
  $^{199}$Hg'', {\em Phys. Rev. Lett.}, {\bf 57}, 3125--3128 (1986).
  {\small[\href{http://dx.doi.org/10.1103/PhysRevLett.57.3125}{DOI}]}.

\bibitem{2013arXiv1310.3320L}
{Lang}, R.~N., ``{Compact binary systems in scalar-tensor gravity. II. Tensor
  gravitational waves to second post-Newtonian order}'', {\em ArXiv e-prints}
  (October 2013).
  {\small[\href{http://adsabs.harvard.edu/abs/2013arXiv1310.3320L}{ADS}]},
  {\small[\href{http://arxiv.org/abs/1310.3320}{{arXiv:1310.3320
  {\small[gr-qc]}}}]}.

\bibitem{2009MNRAS.400..805L}
{Lazaridis}, K. {et~al.}, ``{Generic tests of the existence of the
  gravitational dipole radiation and the variation of the gravitational
  constant}'', {\em \mnras}, {\bf 400}, 805--814 (2009).
  {\small[\href{http://dx.doi.org/10.1111/j.1365-2966.2009.15481.x}{DOI}]},
  {\small[\href{http://adsabs.harvard.edu/abs/2009MNRAS.400..805L}{ADS}]},
  {\small[\href{http://arxiv.org/abs/0908.0285}{{arXiv:0908.0285
  {\small[astro-ph.GA]}}}]}.

\bibitem{lebach}
Lebach, D.E., Corey, B.E., Shapiro, I.I., Ratner, M.I., Webber, J.C., Rogers,
  A.E.E., Davis, J.L.  and Herring, T.A., ``Measurement of the Solar
  Gravitational Deflection of Radio Waves Using Very-Long-Baseline
  Interferometry'', {\em Phys. Rev. Lett.}, {\bf 75}, 1439--1442 (1995).
  {\small[\href{http://dx.doi.org/10.1103/PhysRevLett.75.1439}{DOI}]}.

\bibitem{2010ApJ...722.1589L}
{Lee}, K., {Jenet}, F.~A., {Price}, R.~H., {Wex}, N.  and {Kramer}, M.,
  ``{Detecting Massive Gravitons Using Pulsar Timing Arrays}'', {\em \apj},
  {\bf 722}, 1589--1597 (2010).
  {\small[\href{http://dx.doi.org/10.1088/0004-637X/722/2/1589}{DOI}]},
  {\small[\href{http://adsabs.harvard.edu/abs/2010ApJ...722.1589L}{ADS}]},
  {\small[\href{http://arxiv.org/abs/1008.2561}{{arXiv:1008.2561
  {\small[astro-ph.HE]}}}]}.

\bibitem{2008ApJ...685.1304L}
{Lee}, K.~J., {Jenet}, F.~A.  and {Price}, R.~H., ``{Pulsar Timing as a Probe
  of Non-Einsteinian Polarizations of Gravitational Waves}'', {\em \apj}, {\bf
  685}, 1304--1319 (2008).
  {\small[\href{http://dx.doi.org/10.1086/591080}{DOI}]},
  {\small[\href{http://adsabs.harvard.edu/abs/2008ApJ...685.1304L}{ADS}]}.

\bibitem{2013PhRvL.111f0801L}
{Leefer}, N., {Weber}, C.~T.~M., {Cing{\"o}z}, A., {Torgerson}, J.~R.  and
  {Budker}, D., ``{New Limits on Variation of the Fine-Structure Constant Using
  Atomic Dysprosium}'', {\em Phys. Rev. Lett.}, {\bf 111}, 060801 (2013).
  {\small[\href{http://dx.doi.org/10.1103/PhysRevLett.111.060801}{DOI}]},
  {\small[\href{http://adsabs.harvard.edu/abs/2013PhRvL.111f0801L}{ADS}]},
  {\small[\href{http://arxiv.org/abs/1304.6940}{{arXiv:1304.6940
  {\small[physics.atom-ph]}}}]}.

\bibitem{Lehner01}
Lehner, L., ``Numerical relativity: a review'', {\em Class. Quantum Grav.},
  {\bf 18}, R25--R86 (2001).
  {\small[\href{http://dx.doi.org/10.1088/0264-9381/18/17/202}{DOI}]},
  {\small[\href{http://arxiv.org/abs/gr-qc/0106072}{{gr-qc/0106072}}]}.

\bibitem{2013MNRAS.430.2454L}
{Lentati}, L. {et~al.}, ``{Variations in the fundamental constants in the QSO
  host J1148+5251 at z = 6.4 and the BR1202-0725 system at z = 4.7}'', {\em
  \mnras}, {\bf 430}, 2454--2463 (2013).
  {\small[\href{http://dx.doi.org/10.1093/mnras/stt070}{DOI}]},
  {\small[\href{http://adsabs.harvard.edu/abs/2013MNRAS.430.2454L}{ADS}]},
  {\small[\href{http://arxiv.org/abs/1211.3316}{{arXiv:1211.3316
  {\small[astro-ph.CO]}}}]}.

\bibitem{LeviCivita}
Levi-Civita, T., ``Astronomical consequences of the relativistic two-body
  problem'', {\em Am. J. Math.}, {\bf 59}, 225--334 (1937).

\bibitem{Liberati2013}
{Liberati}, S., ``{Tests of Lorentz invariance: a 2013 update}'', {\em
  Classical and Quantum Gravity}, {\bf 30}, 133001 (2013).
  {\small[\href{http://dx.doi.org/10.1088/0264-9381/30/13/133001}{DOI}]},
  {\small[\href{http://adsabs.harvard.edu/abs/2013CQGra..30m3001L}{ADS}]},
  {\small[\href{http://arxiv.org/abs/1304.5795}{{arXiv:1304.5795
  {\small[gr-qc]}}}]}.

\bibitem{lightmanlee}
Lightman, A.P.  and Lee, D.L., ``Restricted proof that the weak equivalence
  principle implies the Einstein equivalence principle'', {\em Phys. Rev. D},
  {\bf 8}, 364--376 (1973).
  {\small[\href{http://dx.doi.org/10.1103/PhysRevD.8.364}{DOI}]}.

\bibitem{lineweaver96}
Lineweaver, C.H., Tenorio, L., Smoot, G.F., Keegstra, P., Banday, A.J.  and
  Lubin, P., ``The dipole observed in the COBE DMR 4 year data'', {\em
  Astrophys. J.}, {\bf 470}, 38--42 (1996).
  {\small[\href{http://dx.doi.org/10.1086/177846}{DOI}]}.

\bibitem{lipa03}
Lipa, J.A., Nissen, J.A., Wang, S., Stricker, D.A.  and Avaloff, D., ``New
  limit on signals of Lorentz violation in electrodynamics'', {\em Phys. Rev.
  Lett.}, {\bf 90}, 060403 (2003).
  {\small[\href{http://dx.doi.org/10.1103/PhysRevLett.90.060403}{DOI}]},
  {\small[\href{http://arxiv.org/abs/physics/0302093}{{physics/0302093}}]}.

\bibitem{lobo}
Lobo, J.A., ``Spherical GW detectors and geometry'', in Coccia, E., Veneziano,
  G.  and Pizzella, G., eds., {\em Second Edoardo Amaldi Conference on
  Gravitational Waves}, Edoardo Amaldi Foundation Series, pp. 168--179, (World
  Scientific, Singapore, 1998).

\bibitem{long03}
Long, J.C., Chan, H.W., Churnside, A.B., Gulbis, E.A., Varney, M.C.M.  and
  Price, J.C., ``Upper limits to submillimetre-range forces from extra
  space-time dimensions'', {\em Nature}, {\bf 421}, 922--925 (2003).
  {\small[\href{http://dx.doi.org/10.1038/nature01432}{DOI}]},
  {\small[\href{http://arxiv.org/abs/hep-ph/0210004}{{hep-ph/0210004}}]}.

\bibitem{long99}
Long, J.C., Chan, H.W.  and Price, J.C., ``Experimental status of
  gravitational-strength forces in the sub-centimeter regime'', {\em Nucl.
  Phys. B}, {\bf 539}, 23--34 (1999).
  {\small[\href{http://dx.doi.org/10.1016/S0550-3213(98)00711-1}{DOI}]},
  {\small[\href{http://arxiv.org/abs/hep-ph/9805217}{{hep-ph/9805217}}]}.

\bibitem{lopresto91}
LoPresto, J.C., Schrader, C.  and Pierce, A.K., ``Solar gravitational redshift
  from the infrared oxygen triplet'', {\em Astrophys. J.}, {\bf 376}, 757--760
  (1991). {\small[\href{http://dx.doi.org/10.1086/170323}{DOI}]}.

\bibitem{LorentzDroste}
Lorentz, H.~A.  and Droste, J., ``{The motion of a system of bodies under the
  influence of their mutual attraction, according to Einstein's theory.}'',
  {\em Versl.\ K.\ Akad.\ Wetensch.\ Amsterdam}, {\bf 26}, 392 (1917).

\bibitem{lrr-2008-8}
{Lorimer}, D.~R., ``{Binary and Millisecond Pulsars}'', {\em Living Rev.
  Relativity}, {\bf 11}, lrr-2008-8 (2008).
  {\small[\href{http://dx.doi.org/10.12942/lrr-2008-8}{DOI}]},
  {\small[\href{http://adsabs.harvard.edu/abs/2008LRR....11....8L}{ADS}]},
  {\small[\href{http://arxiv.org/abs/0811.0762}{{arXiv:0811.0762}}]}. URL
  (accessed 1 June 2014):
  \newline\url{http://www.livingreviews.org/lrr-2008-8}.

\bibitem{2010PhRvL.105w1103L}
{Lucchesi}, D.~M.  and {Peron}, R., ``{Accurate Measurement in the Field of the
  Earth of the General-Relativistic Precession of the LAGEOS II Pericenter and
  New Constraints on Non-Newtonian Gravity}'', {\em Phys. Rev. Lett.}, {\bf
  105}, 231103 (2010).
  {\small[\href{http://dx.doi.org/10.1103/PhysRevLett.105.231103}{DOI}]},
  {\small[\href{http://adsabs.harvard.edu/abs/2010PhRvL.105w1103L}{ADS}]},
  {\small[\href{http://arxiv.org/abs/1106.2905}{{arXiv:1106.2905
  {\small[gr-qc]}}}]}.

\bibitem{lyne04}
Lyne, A.G. {et~al.}, ``A Double-Pulsar System: A Rare Laboratory for
  Relativistic Gravity and Plasma Physics'', {\em Science}, {\bf 303},
  1153--1157 (2004).
  {\small[\href{http://dx.doi.org/10.1126/science.1094645}{DOI}]},
  {\small[\href{http://arxiv.org/abs/astro-ph/0401086}{{astro-ph/0401086}}]}.

\bibitem{maeda88}
Maeda, K.-I., ``On time variation of fundamental constants in superstring
  theories'', {\em Mod. Phys. Lett. A}, {\bf 3}, 243--249 (1988).
  {\small[\href{http://dx.doi.org/10.1142/S0217732388000295}{DOI}]}.

\bibitem{maggiore}
Maggiore, M.  and Nicolis, A., ``Detection strategies for scalar gravitational
  waves with interferometers and resonant spheres'', {\em Phys. Rev. D}, {\bf
  62}, 024004 (1999).
  {\small[\href{http://arxiv.org/abs/gr-qc/9907055}{{gr-qc/9907055}}]}.

\bibitem{magueijo03}
Magueijo, J., ``New varying speed of light theories'', {\em Rep. Prog. Phys.},
  {\bf 66}, 2025--2068 (2003).
  {\small[\href{http://dx.doi.org/10.1088/0034-4885/66/11/R04}{DOI}]},
  {\small[\href{http://arxiv.org/abs/astro-ph/0305457}{{arXiv:astro-ph/0305457}}]}.

\bibitem{malaney}
Malaney, R.A.  and Mathews, G.J., ``Probing the early universe: A review of
  primordial nucleosynthesis beyond the standard big bang'', {\em Phys. Rep.},
  {\bf 229}, 147--219 (1993).
  {\small[\href{http://dx.doi.org/10.1016/0370-1573(93)90134-Y}{DOI}]}.

\bibitem{maleki01}
Maleki, L.  and Prestage, J.D., ``SpaceTime Mission: Clock test of relativity
  at four solar radii'', in L{\"{a}}mmerzahl, C., Everitt, C.W.F.  and Hehl,
  F.W., eds., {\em Gyros, Clocks, and Interferometers...: Testing Relativistic
  Gravity in Space}, Proceedings of a meeting held in Bad Honnef, Germany,
  August 21--27, 1999, Lecture Notes in Physics, 562, p. 369, (Springer,
  Berlin; New York, 2001).

\bibitem{salomon03}
Marion, H. {et~al.}, ``A search for variations of fundamental constants using
  atomic fountain clock'', {\em Phys. Rev. Lett.}, {\bf 90}, 150801 (2003).
  {\small[\href{http://dx.doi.org/10.1103/PhysRevLett.90.150801}{DOI}]},
  {\small[\href{http://arxiv.org/abs/physics/0212112}{{physics/0212112}}]}.

\bibitem{0264-9381-27-8-080301}
Marka, Z.  and Marka, S., ``Selected articles from `The 8th Edoardo Amaldi
  Conference on Gravitational Waves (Amaldi 8)', Columbia University, New York,
  22 -- 26 June 2009'', {\em Class. Quantum Grav.}, {\bf 27}(8), 080301 (2010).

\bibitem{mattingly}
Mattingly, D., ``Modern Tests of Lorentz Invariance'', {\em Living Rev.
  Relativity}, {\bf 8}, lrr-2005-5 (2005).
  {\small[\href{http://dx.doi.org/10.12942/lrr-2005-5}{DOI}]},
  {\small[\href{http://arxiv.org/abs/gr-qc/0502097}{{arXiv:gr-qc/0502097}}]}.
  URL (accessed 1 June 2014):
  \newline\url{http://www.livingreviews.org/lrr-2005-5}.

\bibitem{mattingly02}
Mattingly, D.  and Jacobson, T.A., ``Relativistic Gravity with a Dynamical
  Preferred Frame'', in Kosteleck{\'{y}}, V.A., ed., {\em CPT and Lorentz
  Symmetry II}, Proceedings of the Second Meeting, held at Indiana University,
  Bloomington, August 15--18, 2001, pp. 331--335, (World Scientific, Singapore;
  River Edge, 2002).
  {\small[\href{http://arxiv.org/abs/gr-qc/0112012}{{gr-qc/0112012}}]}.

\bibitem{mecheri04}
Mecheri, R., Abdelatif, T., Irbah, A., Provost, J.  and Berthomieu, G., ``New
  values of gravitational moments $J_{2}$ and $J_{4}$ deduced from
  helioseismology'', {\em Solar Phys.}, {\bf 222}, 191--197 (2004).
  {\small[\href{http://dx.doi.org/10.1023/B:SOLA.0000043563.96766.21}{DOI}]},
  {\small[\href{http://adsabs.harvard.edu/abs/2004SoPh..222..191M}{ADS}]}.

\bibitem{2010LRR....13....7M}
{Merkowitz}, S.~M., ``{Tests of Gravity Using Lunar Laser Ranging}'', {\em
  Living Rev. Relativity}, {\bf 13}, lrr-2010-7 (2010).
  {\small[\href{http://dx.doi.org/10.12942/lrr-2010-7}{DOI}]},
  {\small[\href{http://adsabs.harvard.edu/abs/2010LRR....13....7M}{ADS}]}. URL
  (accessed 1 June 2014):
  \newline\url{http://www.livingreviews.org/lrr-2010-7}.

\bibitem{2010Metro..47L...9M}
{Merlet}, S., {Bodart}, Q., {Malossi}, N., {Landragin}, A., {Pereira Dos
  Santos}, F., {Gitlein}, O.  and {Timmen}, L., ``{Comparison between two
  mobile absolute gravimeters: optical versus atomic interferometers}'', {\em
  Metrologia}, {\bf 47}, L9--L11 (2010).
  {\small[\href{http://dx.doi.org/10.1088/0026-1394/47/4/L01}{DOI}]},
  {\small[\href{http://adsabs.harvard.edu/abs/2010Metro..47L...9M}{ADS}]},
  {\small[\href{http://arxiv.org/abs/1005.0357}{{arXiv:1005.0357
  {\small[physics.atom-ph]}}}]}.

\bibitem{mm}
Michelson, A.A.  and Morley, E.W., ``On the Relative Motion of the Earth and
  the Luminiferous Ether'', {\em Am. J. Sci.}, {\bf 34}, 333--345 (1887).
  Online version (accessed 1 June 2014):
  \newline\url{http://www.aip.org/history/gap/Michelson/Michelson.html}.

\bibitem{microscope}
``MICROSCOPE'', project homepage, CNES. URL (accessed 1 June 2014):
  \newline\url{http://smsc.cnes.fr/MICROSCOPE/}.

\bibitem{2002EAS.....2..107M}
{Mignard}, F., ``{Fundamental Physics with GAIA}'', in {Bienayme}, O.  and
  {Turon}, C., eds., {\em EAS Publications Series}, 2, pp. 107--121, (2002).
  {\small[\href{http://dx.doi.org/10.1051/eas:2002009}{DOI}]},
  {\small[\href{http://adsabs.harvard.edu/abs/2002EAS.....2..107M}{ADS}]}.

\bibitem{milani02}
Milani, A., Vokrouhlick{\'{y}}, D., Villani, D., Bonanno, C.  and Rossi, A.,
  ``Testing general relativity with the BepiColombo radio science experiment'',
  {\em Phys. Rev. D}, {\bf 66}, 082001 (2002).
  {\small[\href{http://dx.doi.org/10.1103/PhysRevD.66.082001}{DOI}]}.

\bibitem{milgrom83}
Milgrom, M., ``A modification of the Newtonian dynamics as a possible
  alternative to the hidden mass hypothesis'', {\em Astrophys. J.}, {\bf 270},
  365--370 (1983). {\small[\href{http://dx.doi.org/10.1086/161130}{DOI}]},
  {\small[\href{http://adsabs.harvard.edu/abs/1983ApJ...270..365M}{ADS}]}.

\bibitem{2009MNRAS.399..474M}
{Milgrom}, M., ``{MOND effects in the inner Solar system}'', {\em \mnras}, {\bf
  399}, 474--486 (2009).
  {\small[\href{http://dx.doi.org/10.1111/j.1365-2966.2009.15302.x}{DOI}]},
  {\small[\href{http://adsabs.harvard.edu/abs/2009MNRAS.399..474M}{ADS}]},
  {\small[\href{http://arxiv.org/abs/0906.4817}{{arXiv:0906.4817
  {\small[astro-ph.CO]}}}]}.

\bibitem{msstt97}
Mino, Y., Sasaki, M., Shibata, M., Tagoshi, H.  and Tanaka, T., ``Black Hole
  Perturbation'', {\em Prog. Theor. Phys. Suppl.}, {\bf 128}, 1--121 (1997).
  {\small[\href{http://dx.doi.org/10.1143/PTPS.128.1}{DOI}]},
  {\small[\href{http://arxiv.org/abs/gr-qc/9712057}{{gr-qc/9712057}}]}.

\bibitem{2013PhRvD..87h4070M}
{Mirshekari}, S.  and {Will}, C.~M., ``{Compact binary systems in scalar-tensor
  gravity: Equations of motion to 2.5 post-Newtonian order}'', {\em \prd}, {\bf
  87}, 084070 (2013).
  {\small[\href{http://dx.doi.org/10.1103/PhysRevD.87.084070}{DOI}]},
  {\small[\href{http://adsabs.harvard.edu/abs/2013PhRvD..87h4070M}{ADS}]},
  {\small[\href{http://arxiv.org/abs/1301.4680}{{arXiv:1301.4680
  {\small[gr-qc]}}}]}.

\bibitem{2012PhRvD..85b4041M}
{Mirshekari}, S., {Yunes}, N.  and {Will}, C.~M., ``{Constraining
  Lorentz-violating, modified dispersion relations with gravitational waves}'',
  {\em \prd}, {\bf 85}, 024041 (2012).
  {\small[\href{http://dx.doi.org/10.1103/PhysRevD.85.024041}{DOI}]},
  {\small[\href{http://adsabs.harvard.edu/abs/2012PhRvD..85b4041M}{ADS}]},
  {\small[\href{http://arxiv.org/abs/1110.2720}{{arXiv:1110.2720
  {\small[gr-qc]}}}]}.

\bibitem{2010PhRvD..82f4010M}
{Mishra}, C.~K., {Arun}, K.~G., {Iyer}, B.~R.  and {Sathyaprakash}, B.~S.,
  ``{Parametrized tests of post-Newtonian theory using Advanced LIGO and
  Einstein Telescope}'', {\em \prd}, {\bf 82}, 064010 (2010).
  {\small[\href{http://dx.doi.org/10.1103/PhysRevD.82.064010}{DOI}]},
  {\small[\href{http://adsabs.harvard.edu/abs/2010PhRvD..82f4010M}{ADS}]},
  {\small[\href{http://arxiv.org/abs/1005.0304}{{arXiv:1005.0304
  {\small[gr-qc]}}}]}.

\bibitem{MTW}
Misner, C.W., Thorne, K.S.  and Wheeler, J.A., {\em Gravitation}, (W.H.
  Freeman, San Francisco, 1973).
  {\small[\href{http://adsabs.harvard.edu/abs/1973grav.book.....M}{ADS}]}.

\bibitem{2007PhRvD..75l4025M}
{Mitchell}, T.  and {Will}, C.~M., ``{Post-Newtonian gravitational radiation
  and equations of motion via direct integration of the relaxed Einstein
  equations. V. Evidence for the strong equivalence principle to second
  post-Newtonian order}'', {\em \prd}, {\bf 75}, 124025 (2007).
  {\small[\href{http://dx.doi.org/10.1103/PhysRevD.75.124025}{DOI}]},
  {\small[\href{http://adsabs.harvard.edu/abs/2007PhRvD..75l4025M}{ADS}]},
  {\small[\href{http://arxiv.org/abs/0704.2243}{{arXiv:0704.2243
  {\small[gr-qc]}}}]}.

\bibitem{2013arXiv1311.4978M}
{Modenini}, D.  and {Tortora}, P., ``{Pioneer 10 and 11 orbit determination
  analysis shows no discrepancy with Newton-Einstein's laws of gravity}'', {\em
  ArXiv e-prints} (November 2013).
  {\small[\href{http://adsabs.harvard.edu/abs/2013arXiv1311.4978M}{ADS}]},
  {\small[\href{http://arxiv.org/abs/1311.4978}{{arXiv:1311.4978
  {\small[gr-qc]}}}]}.

\bibitem{muller03}
M{\"{u}}ller, H., Herrmann, S., Braxmaier, C., Schiller, S.  and Peters, A.,
  ``Modern Michelson--Morley experiment using cryogenic optical resonators'',
  {\em Phys. Rev. Lett.}, {\bf 91}, 020401 (2003).
  {\small[\href{http://dx.doi.org/10.1103/PhysRevLett.91.020401}{DOI}]},
  {\small[\href{http://arxiv.org/abs/physics/0305117}{{physics/0305117}}]}.

\bibitem{2010Natur.463..926M}
{M{\"u}ller}, H., {Peters}, A.  and {Chu}, S., ``{A precision measurement of
  the gravitational redshift by the interference of matter waves}'', {\em
  Nature}, {\bf 463}, 926--929 (2010).
  {\small[\href{http://dx.doi.org/10.1038/nature08776}{DOI}]},
  {\small[\href{http://adsabs.harvard.edu/abs/2010Natur.463..926M}{ADS}]}.

\bibitem{MullerMG}
M{\"{u}}ller, J., Schneider, M., Nordtvedt~Jr., K.  and Vokrouhlick{\'{y}}, D.,
  ``What can LLR provide to relativity?'', in Piran, T., ed., {\em The Eighth
  Marcel Grossmann Meeting on Recent Developments in Theoretical and
  Experimental General Relativity, Gravitation and Relativistic Field
  Theories}, Proceedings of the meeting held at the Hebrew University of
  Jerusalem, June 22--27, 1997, pp. 1151--1153, (World Scientific, Singapore,
  1999).

\bibitem{murphy01}
Murphy, M.T., Webb, J.K., Flambaum, V.V., Dzuba, V.A., Churchill, C.W.,
  Prochaska, J.X., Barrow, J.D.  and Wolfe, A.M., ``Possible evidence for a
  variable fine-structure constant from QSO absorption lines: motivations,
  analysis and results'', {\em Mon. Not. R. Astron. Soc.}, {\bf 327},
  1208--1222 (2001).
  {\small[\href{http://dx.doi.org/10.1046/j.1365-8711.2001.04840.x}{DOI}]},
  {\small[\href{http://arxiv.org/abs/astro-ph/0012419}{{astro-ph/0012419}}]}.

\bibitem{2011Icar..211.1103M}
{Murphy}, T.~W. {et~al.}, ``{Laser ranging to the lost Lunokhod 1 reflector}'',
  {\em Icarus}, {\bf 211}, 1103--1108 (2011).
  {\small[\href{http://dx.doi.org/10.1016/j.icarus.2010.11.010}{DOI}]},
  {\small[\href{http://adsabs.harvard.edu/abs/2011Icar..211.1103M}{ADS}]},
  {\small[\href{http://arxiv.org/abs/1009.5720}{{arXiv:1009.5720
  {\small[astro-ph.EP]}}}]}.

\bibitem{2012CQGra..29r4005M}
{Murphy}, Jr., T.~W., {Adelberger}, E.~G., {Battat}, J.~B.~R., {Hoyle}, C.~D.,
  {Johnson}, N.~H., {McMillan}, R.~J., {Stubbs}, C.~W.  and {Swanson}, H.~E.,
  ``{APOLLO: millimeter lunar laser ranging}'', {\em Class. Quantum Grav.},
  {\bf 29}, 184005 (2012).
  {\small[\href{http://dx.doi.org/10.1088/0264-9381/29/18/184005}{DOI}]},
  {\small[\href{http://adsabs.harvard.edu/abs/2012CQGra..29r4005M}{ADS}]}.

\bibitem{2008NewAR..51..733N}
{Narayan}, R.  and {McClintock}, J.~E., ``{Advection-dominated accretion and
  the black hole event horizon}'', {\em New Astron. Rev.}, {\bf 51}, 733--751
  (2008). {\small[\href{http://dx.doi.org/10.1016/j.newar.2008.03.002}{DOI}]},
  {\small[\href{http://adsabs.harvard.edu/abs/2008NewAR..51..733N}{ADS}]},
  {\small[\href{http://arxiv.org/abs/0803.0322}{{arXiv:0803.0322}}]}.

\bibitem{Ni77}
Ni, W.-T., ``Equivalence principles and electromagnetism'', {\em Phys. Rev.
  Lett.}, {\bf 38}, 301--304 (1977).
  {\small[\href{http://dx.doi.org/10.1103/PhysRevLett.38.301}{DOI}]}.

\bibitem{2009PhRvD..79h2002N}
{Nishizawa}, A., {Taruya}, A., {Hayama}, K., {Kawamura}, S.  and {Sakagami},
  M.-A., ``{Probing nontensorial polarizations of stochastic gravitational-wave
  backgrounds with ground-based laser interferometers}'', {\em \prd}, {\bf 79},
  082002 (2009).
  {\small[\href{http://dx.doi.org/10.1103/PhysRevD.79.082002}{DOI}]},
  {\small[\href{http://adsabs.harvard.edu/abs/2009PhRvD..79h2002N}{ADS}]},
  {\small[\href{http://arxiv.org/abs/0903.0528}{{arXiv:0903.0528
  {\small[astro-ph.CO]}}}]}.

\bibitem{2010PhRvD..81j4043N}
{Nishizawa}, A., {Taruya}, A.  and {Kawamura}, S., ``{Cosmological test of
  gravity with polarizations of stochastic gravitational waves around 0.1-1
  Hz}'', {\em \prd}, {\bf 81}, 104043 (2010).
  {\small[\href{http://dx.doi.org/10.1103/PhysRevD.81.104043}{DOI}]},
  {\small[\href{http://adsabs.harvard.edu/abs/2010PhRvD..81j4043N}{ADS}]},
  {\small[\href{http://arxiv.org/abs/0911.0525}{{arXiv:0911.0525
  {\small[gr-qc]}}}]}.

\bibitem{nordstrom13}
Nordstr{\"{o}}m, G., ``Zur Theorie der Gravitation vom Standpunkt des
  Relativit\"atsprinzips'', {\em Ann. Phys. (Leipzig)}, {\bf 42}, 533--554
  (1913). {\small[\href{http://dx.doi.org/10.1002/andp.19133471303}{DOI}]}.

\bibitem{nordtvedt1}
Nordtvedt~Jr., K., ``Equivalence Principle for Massive Bodies. I.
  Phenomenology'', {\em Phys. Rev.}, {\bf 169}, 1014--1016 (1968).
  {\small[\href{http://dx.doi.org/10.1103/PhysRev.169.1014}{DOI}]},
  {\small[\href{http://adsabs.harvard.edu/abs/1968PhRv..169.1014N}{ADS}]}.

\bibitem{nordtvedt2}
Nordtvedt~Jr., K., ``Equivalence Principle for Massive Bodies. II. Theory'',
  {\em Phys. Rev.}, {\bf 169}, 1017--1025 (1968).
  {\small[\href{http://dx.doi.org/10.1103/PhysRev.169.1017}{DOI}]}.

\bibitem{nordtvedt88b}
Nordtvedt~Jr., K., ``Existence of the gravitomagnetic interaction'', {\em Int.
  J. Theor. Phys.}, {\bf 27}, 1395--1404 (1988).
  {\small[\href{http://dx.doi.org/10.1007/BF00671317}{DOI}]}.

\bibitem{nordtvedt88a}
Nordtvedt~Jr., K., ``Gravitomagnetic interaction and laser ranging to Earth
  satellites'', {\em Phys. Rev. Lett.}, {\bf 61}, 2647--2649 (1988).
  {\small[\href{http://dx.doi.org/10.1103/PhysRevLett.61.2647}{DOI}]}.

\bibitem{nordtvedt3}
Nordtvedt~Jr., K., ``$\dot{G}/G$ and a cosmological acceleration of
  gravitationally compact bodies'', {\em Phys. Rev. Lett.}, {\bf 65}, 953--956
  (1990). {\small[\href{http://dx.doi.org/10.1103/PhysRevLett.65.953}{DOI}]}.

\bibitem{Nordtvedt95}
Nordtvedt~Jr, K.L., ``The Relativistic Orbit Observables in Lunar Laser
  Ranging'', {\em Icarus}, {\bf 114}, 51--62 (1995).
  {\small[\href{http://dx.doi.org/10.1006/icar.1995.1042}{DOI}]}.

\bibitem{nordtvedt01}
Nordtvedt~Jr., K., ``Testing Newton's third law using lunar laser ranging'',
  {\em Class. Quantum Grav.}, {\bf 18}, L133--L137 (2001).
  {\small[\href{http://dx.doi.org/10.1088/0264-9381/18/20/101}{DOI}]}.

\bibitem{Ohanian74}
Ohanian, H.C., ``Comment on the Schiff Conjecture'', {\em Phys. Rev. D}, {\bf
  10}, 2041--2042 (1974).
  {\small[\href{http://dx.doi.org/10.1103/PhysRevD.10.2041}{DOI}]}.

\bibitem{olive04}
Olive, K.A., Pospelov, M., Qian, Y.-Z., Manh{\`{e}}s, G., Vangioni-Flam, E.,
  Coc, A.  and Cass{\'{e}}, M., ``Reexamination of the $^{187}$Re bound on the
  variation of fundamental couplings'', {\em Phys. Rev. D}, {\bf 69}, 027701
  (2004). {\small[\href{http://dx.doi.org/10.1103/PhysRevD.69.027701}{DOI}]},
  {\small[\href{http://arxiv.org/abs/astro-ph/0309252}{{astro-ph/0309252}}]}.

\bibitem{2013AcAau..91..313P}
{Paolozzi}, A.  and {Ciufolini}, I., ``{LARES successfully launched in orbit:
  Satellite and mission description}'', {\em Acta Astronautica}, {\bf 91},
  313--321 (2013).
  {\small[\href{http://dx.doi.org/10.1016/j.actaastro.2013.05.011}{DOI}]},
  {\small[\href{http://adsabs.harvard.edu/abs/2013AcAau..91..313P}{ADS}]},
  {\small[\href{http://arxiv.org/abs/1305.6823}{{arXiv:1305.6823
  {\small[astro-ph.IM]}}}]}.

\bibitem{DIRE}
Pati, M.E.  and Will, C.M., ``Post-Newtonian gravitational radiation and
  equations of motion via direct integration of the relaxed Einstein equations:
  Foundations'', {\em Phys. Rev. D}, {\bf 62}, 124015 (2000).
  {\small[\href{http://dx.doi.org/10.1103/PhysRevD.62.124015}{DOI}]},
  {\small[\href{http://arxiv.org/abs/gr-qc/0007087}{{gr-qc/0007087}}]}.

\bibitem{2002PhRvD..65j4008P}
{Pati}, M.~E.  and {Will}, C.~M., ``{Post-Newtonian gravitational radiation and
  equations of motion via direct integration of the relaxed Einstein equations.
  II. Two-body equations of motion to second post-Newtonian order, and
  radiation reaction to 3.5 post-Newtonian order}'', {\em \prd}, {\bf 65},
  104008 (2002).
  {\small[\href{http://dx.doi.org/10.1103/PhysRevD.65.104008}{DOI}]},
  {\small[\href{http://adsabs.harvard.edu/abs/2002PhRvD..65j4008P}{ADS}]},
  {\small[\href{http://arxiv.org/abs/gr-qc/0201001}{{gr-qc/0201001}}]}.

\bibitem{peik04}
Peik, E., Lipphardt, B., Schnatz, H., Schneider, T.  and Tamm, C., ``Limit on
  the Present Temporal Variation of the Fine Structure Constant'', {\em Phys.
  Rev. Lett.}, {\bf 93}, 170801 (2004).
  {\small[\href{http://dx.doi.org/10.1103/PhysRevLett.93.170801}{DOI}]},
  {\small[\href{http://arxiv.org/abs/physics/0402132}{{physics/0402132}}]}.

\bibitem{2013PhRvA..87a0102P}
{Peil}, S., {Crane}, S., {Hanssen}, J.~L., {Swanson}, T.~B.  and {Ekstrom},
  C.~R., ``{Tests of local position invariance using continuously running
  atomic clocks}'', {\em Phys. Rev. A}, {\bf 87}, 010102 (2013).
  {\small[\href{http://dx.doi.org/10.1103/PhysRevA.87.010102}{DOI}]},
  {\small[\href{http://adsabs.harvard.edu/abs/2013PhRvA..87a0102P}{ADS}]},
  {\small[\href{http://arxiv.org/abs/1301.6145}{{arXiv:1301.6145
  {\small[physics.atom-ph]}}}]}.

\bibitem{2006PhRvC..74f4610P}
{Petrov}, Y.~V., {Nazarov}, A.~I., {Onegin}, M.~S., {Petrov}, V.~Y.  and
  {Sakhnovsky}, E.~G., ``{Natural nuclear reactor at Oklo and variation of
  fundamental constants: Computation of neutronics of a fresh core}'', {\em
  \prc}, {\bf 74}, 064610 (2006).
  {\small[\href{http://dx.doi.org/10.1103/PhysRevC.74.064610}{DOI}]},
  {\small[\href{http://adsabs.harvard.edu/abs/2006PhRvC..74f4610P}{ADS}]},
  {\small[\href{http://arxiv.org/abs/hep-ph/0506186}{{hep-ph/0506186}}]}.

\bibitem{pitjeva05}
Pitjeva, E.V., ``Relativistic effects and solar oblateness from radar
  observations of planets and spacecraft'', {\em Astron. Lett.}, {\bf 31},
  340--349 (2005). {\small[\href{http://dx.doi.org/10.1134/1.1922533}{DOI}]}.

\bibitem{poissonBH}
Poisson, E., ``Measuring black-hole parameters and testing general relativity
  using gravitational-wave data from space-based interferometers'', {\em Phys.
  Rev. D}, {\bf 54}, 5939--5953 (1996).
  {\small[\href{http://dx.doi.org/10.1103/PhysRevD.54.5939}{DOI}]},
  {\small[\href{http://arxiv.org/abs/gr-qc/9606024}{{gr-qc/9606024}}]}.

\bibitem{poissonwill}
Poisson, E.  and Will, C.M., ``Gravitational waves from inspiraling compact
  binaries: Parameter estimation using second-post-Newtonian wave forms'', {\em
  Phys. Rev. D}, {\bf 52}, 848--855 (1995).
  {\small[\href{http://dx.doi.org/10.1103/PhysRevD.52.848}{DOI}]},
  {\small[\href{http://arxiv.org/abs/gr-qc/9502040}{{gr-qc/9502040}}]}.

\bibitem{PW2014}
Poisson, E.  and Will, C.M., {\em Gravity: Newtonian, Post-Newtonian,
  Relativistic}, (Cambridge University Press, Cambridge, 2014).

\bibitem{prestage85}
Prestage, J.D., Bollinger, J.J., Itano, W.M.  and Wineland, D.J., ``Limits for
  Spatial Anisotropy by Use of Nuclear-Spin--Polarized $^{9}$Be$^{+}$ Ions'',
  {\em Phys. Rev. Lett.}, {\bf 54}, 2387--2390 (1985).
  {\small[\href{http://dx.doi.org/10.1103/PhysRevLett.54.2387}{DOI}]}.

\bibitem{prestage95}
Prestage, J.D., Tjoelker, R.L.  and Maleki, L., ``Atomic clocks and variation
  of the fine structure constant'', {\em Phys. Rev. Lett.}, {\bf 74},
  3511--3514 (1995).
  {\small[\href{http://dx.doi.org/10.1103/PhysRevLett.74.3511}{DOI}]}.

\bibitem{Psaltis04}
Psaltis, D., ``Measurements of black hole spins and tests of strong-field
  general relativity'', in Kaaret, P., Lamb, F.K.  and Swank, J.H., eds., {\em
  X-Ray Timing 2003: Rossi and Beyond}, Proceedings of the conference held 3--5
  November 2003 in Cambridge, MA, AIP Conference Proceedings, 714, pp. 29--35,
  (American Institute of Physics, Melville, 2004).
  {\small[\href{http://arxiv.org/abs/astro-ph/0402213}{{astro-ph/0402213}}]}.

\bibitem{2008LRR....11....9P}
{Psaltis}, D., ``{Probes and Tests of Strong-Field Gravity with Observations in
  the Electromagnetic Spectrum}'', {\em Living Rev. Relativity}, {\bf 11},
  lrr-2008-9 (2008).
  {\small[\href{http://dx.doi.org/10.12942/lrr-2008-9}{DOI}]},
  {\small[\href{http://adsabs.harvard.edu/abs/2008LRR....11....9P}{ADS}]},
  {\small[\href{http://arxiv.org/abs/0806.1531}{{arXiv:0806.1531}}]}. URL
  (accessed 1 June 2014):
  \newline\url{http://www.\-living\-reviews.org/lrr-2008-9}.

\bibitem{quast04}
Quast, R., Reimers, D.  and Levshakov, S.A., ``Probing the variability of the
  fine-structure constant with the VLT/UVES'', {\em Astron. Astrophys.}, {\bf
  415}, L7--L11 (2004).
  {\small[\href{http://dx.doi.org/10.1051/0004-6361:20040013}{DOI}]},
  {\small[\href{http://arxiv.org/abs/astro-ph/0311280}{{astro-ph/0311280}}]}.

\bibitem{randall2}
Randall, L.  and Sundrum, R., ``An alternative to compactification'', {\em
  Phys. Rev. Lett.}, {\bf 83}, 4690--4693 (1999).
  {\small[\href{http://dx.doi.org/10.1103/PhysRevLett.83.4690}{DOI}]},
  {\small[\href{http://arxiv.org/abs/hep-ph/9906064}{{hep-ph/9906064}}]}.

\bibitem{randall1}
Randall, L.  and Sundrum, R., ``Large Mass Hierarchy from a Small Extra
  Dimension'', {\em Phys. Rev. Lett.}, {\bf 83}, 3370--3373 (1999).
  {\small[\href{http://dx.doi.org/10.1103/PhysRevLett.83.3370}{DOI}]},
  {\small[\href{http://arxiv.org/abs/hep-ph/9905021}{{hep-ph/9905021}}]}.

\bibitem{2014Natur.505..520R}
{Ransom}, S.~M. {et~al.}, ``{A millisecond pulsar in a stellar triple
  system}'', {\em \nat}, {\bf 505}, 520--524 (2014).
  {\small[\href{http://dx.doi.org/10.1038/nature12917}{DOI}]},
  {\small[\href{http://adsabs.harvard.edu/abs/2014Natur.505..520R}{ADS}]},
  {\small[\href{http://arxiv.org/abs/1401.0535}{{arXiv:1401.0535
  {\small[astro-ph.SR]}}}]}.

\bibitem{reasenberg}
Reasenberg, R.D. {et~al.}, ``Viking relativity experiment: Verification of
  signal retardation by solar gravity'', {\em Astrophys. J. Lett.}, {\bf 234},
  L219--L221 (1979). {\small[\href{http://dx.doi.org/10.1086/183144}{DOI}]},
  {\small[\href{http://adsabs.harvard.edu/abs/1979ApJ...234L.219R}{ADS}]}.

\bibitem{reeves}
Reeves, H., ``On the origin of the light elements ($Z < 6$)'', {\em Rev. Mod.
  Phys.}, {\bf 66}, 193--216 (1994).
  {\small[\href{http://dx.doi.org/10.1103/RevModPhys.66.193}{DOI}]}.

\bibitem{2009SSRv..148..233R}
{Reynaud}, S., {Salomon}, C.  and {Wolf}, P., ``{Testing General Relativity
  with Atomic Clocks}'', {\em Space Sci. Rev.}, {\bf 148}, 233--247 (2009).
  {\small[\href{http://dx.doi.org/10.1007/s11214-009-9539-0}{DOI}]},
  {\small[\href{http://adsabs.harvard.edu/abs/2009SSRv..148..233R}{ADS}]},
  {\small[\href{http://arxiv.org/abs/0903.1166}{{arXiv:0903.1166
  {\small[quant-ph]}}}]}.

\bibitem{2013SSRv..tmp...81R}
{Reynolds}, C.~S., ``{Measuring Black Hole Spin Using X-Ray Reflection
  Spectroscopy}'', {\em Space Sci. Rev. On Line}, 1--18 (2013).
  {\small[\href{http://dx.doi.org/10.1007/s11214-013-0006-6}{DOI}]},
  {\small[\href{http://arxiv.org/abs/1302.3260}{{arXiv:1302.3260
  {\small[astro-ph.HE]}}}]}. URL (accessed 1 June, 2014):
  \newline\url{http://adsabs.harvard.edu/abs/2013SSRv..tmp...81R}.

\bibitem{2013CQGra..30x4004R}
{Reynolds}, C.~S., ``{The spin of supermassive black holes}'', {\em Class.
  Quantum Grav.}, {\bf 30}, 244004 (2013).
  {\small[\href{http://dx.doi.org/10.1088/0264-9381/30/24/244004}{DOI}]},
  {\small[\href{http://adsabs.harvard.edu/abs/2013CQGra..30x4004R}{ADS}]},
  {\small[\href{http://arxiv.org/abs/1307.3246}{{arXiv:1307.3246
  {\small[astro-ph.HE]}}}]}.

\bibitem{ries}
Ries, J.C., Eanes, R.J., Tapley, B.D.  and Peterson, G.E., ``Prospects for an
  Improved Lense--Thirring Test with SLR and the GRACE Gravity Mission'', in
  Noomen, R., Klosko, S., Noll, C.  and Pearlman, M., eds., {\em Proceedings of
  the 13th International Workshop on Laser Ranging: Science Session and Full
  Proceedings CD-ROM}, `Toward Millimeter Accuracy' Workshop held in
  Washington, DC, October 07--11, 2002, NASA Conference Proceedings, pp.
  211--248. NASA, (2003). URL (accessed 1 June 2014):
  \newline\url{http://cddis.gsfc.nasa.gov/lw13/lw_proceedings.html}.

\bibitem{2011AnP...523..439R}
{Rievers}, B.  and {L{\"a}mmerzahl}, C., ``{High precision thermal modeling of
  complex systems with application to the flyby and Pioneer anomaly}'', {\em
  Annalen der Physik}, {\bf 523}, 439--449 (2011).
  {\small[\href{http://dx.doi.org/10.1002/andp.201100081}{DOI}]},
  {\small[\href{http://adsabs.harvard.edu/abs/2011AnP...523..439R}{ADS}]},
  {\small[\href{http://arxiv.org/abs/1104.3985}{{arXiv:1104.3985
  {\small[gr-qc]}}}]}.

\bibitem{riis}
Riis, E., Anderson, L.-U.A., Bjerre, N., Poulson, O., Lee, S.A.  and Hall,
  J.L., ``Test of the Isotropy of the Speed of Light Using Fast-Beam Laser
  Spectroscopy'', {\em Phys. Rev. Lett.}, {\bf 60}, 81--84 (1988).
  {\small[\href{http://dx.doi.org/10.1103/PhysRevLett.60.81}{DOI}]}.

\bibitem{robertson38}
Robertson, H.P., ``The two-body problem in general relativity.'', {\em Ann.
  Math.}, {\bf 39}, 101--104 (1938).

\bibitem{dicke_2}
Roll, P.G., Krotkov, R.  and Dicke, R.H., ``The equivalence of inertial and
  passive gravitational mass'', {\em Ann. Phys. (N.Y.)}, {\bf 26}, 442--517
  (1964). {\small[\href{http://dx.doi.org/10.1016/0003-4916(64)90259-3}{DOI}]}.

\bibitem{rossi}
Rossi, B.  and Hall, D.B., ``Variation of the rate of decay of mesotrons with
  momentum'', {\em Phys. Rev.}, {\bf 59}, 223--228 (1941).
  {\small[\href{http://dx.doi.org/10.1103/PhysRev.59.223}{DOI}]}.

\bibitem{2011EPJH...36..407R}
{Rozelot}, J.-P.  and {Damiani}, C., ``{History of solar oblateness
  measurements and interpretation}'', {\em Eur. Phys. J. H}, {\bf 36}, 407--436
  (2011). {\small[\href{http://dx.doi.org/10.1140/epjh/e2011-20017-4}{DOI}]},
  {\small[\href{http://adsabs.harvard.edu/abs/2011EPJH...36..407R}{ADS}]}.

\bibitem{ryan}
Ryan, F.D., ``Gravitational waves from the inspiral of a compact object into a
  massive, axisymmetric body with arbitrary multipole moments'', {\em Phys.
  Rev. D}, {\bf 52}, 5707--5718 (1995).
  {\small[\href{http://dx.doi.org/10.1103/PhysRevD.52.5707}{DOI}]}.

\bibitem{PhysRevD.80.044032}
Sagi, Eva, ``Preferred frame parameters in the tensor-vector-scalar theory of
  gravity and its generalization'', {\em Phys. Rev. D}, {\bf 80}, 044032
  (2009). {\small[\href{http://dx.doi.org/10.1103/PhysRevD.80.044032}{DOI}]}.

\bibitem{samuel03}
Samuel, S., ``On the speed of gravity and the v/c corrections to the Shapiro
  time delay'', {\em Phys. Rev. Lett.}, {\bf 90}, 231101 (2003).
  {\small[\href{http://dx.doi.org/10.1103/PhysRevLett.90.231101}{DOI}]},
  {\small[\href{http://arxiv.org/abs/astro-ph/0304006}{{astro-ph/0304006}}]}.

\bibitem{samuel04}
Samuel, S., ``On the Speed of Gravity and the Jupiter/quasar Measurement'',
  {\em Int. J. Mod. Phys. D}, {\bf 13}, 1753--1770 (2004).
  {\small[\href{http://dx.doi.org/10.1142/S0218271804005900}{DOI}]},
  {\small[\href{http://arxiv.org/abs/astro-ph/0412401}{{astro-ph/0412401}}]}.

\bibitem{Santiago97}
Santiago, D.I., Kalligas, D.  and Wagoner, R.V., ``Nucleosynthesis constraints
  on scalar-tensor theories of gravity'', {\em Phys. Rev. D}, {\bf 56},
  7627--7637 (1997).
  {\small[\href{http://dx.doi.org/10.1103/PhysRevD.56.7627}{DOI}]}.

\bibitem{SasakiTagoshi03}
Sasaki, M.  and Tagoshi, H., ``Analytic Black Hole Perturbation Approach to
  Gravitational Radiation'', {\em Living Rev. Relativity}, {\bf 6}, lrr-2003-6
  (2003). {\small[\href{http://dx.doi.org/10.12942/lrr-2003-6}{DOI}]},
  {\small[\href{http://adsabs.harvard.edu/abs/2003LRR.....6....6S}{ADS}]}. URL
  (accessed 1 June 2014):
  \newline\url{http://www.livingreviews.org/lrr-2003-6}.

\bibitem{2009LRR....12....2S}
{Sathyaprakash}, B.~S.  and {Schutz}, B.~F., ``{Physics, Astrophysics and
  Cosmology with Gravitational Waves}'', {\em Living Rev. Relativity}, {\bf
  12}, lrr-2009-2 (2009).
  {\small[\href{http://dx.doi.org/10.12942/lrr-2009-2}{DOI}]},
  {\small[\href{http://adsabs.harvard.edu/abs/2009LRR....12....2S}{ADS}]},
  {\small[\href{http://arxiv.org/abs/0903.0338}{{arXiv:0903.0338
  {\small[gr-qc]}}}]}. URL (accessed 1 June 2014):
  \newline\url{http://www.livingreviews.org/lrr-2009-2}.

\bibitem{scharrewill}
Scharre, P.D.  and Will, C.M., ``Testing scalar-tensor gravity using space
  gravitational-wave interferometers'', {\em Phys. Rev. D}, {\bf 65}, 042002
  (2002). {\small[\href{http://dx.doi.org/10.1103/PhysRevD.65.042002}{DOI}]},
  {\small[\href{http://arxiv.org/abs/gr-qc/0109044}{{gr-qc/0109044}}]}.

\bibitem{2008PhRvL.100d1101S}
{Schlamminger}, S., {Choi}, K.-Y., {Wagner}, T.~A., {Gundlach}, J.~H.  and
  {Adelberger}, E.~G., ``{Test of the Equivalence Principle Using a Rotating
  Torsion Balance}'', {\em Phys. Rev. Lett.}, {\bf 100}, 041101 (2008).
  {\small[\href{http://dx.doi.org/10.1103/PhysRevLett.100.041101}{DOI}]},
  {\small[\href{http://adsabs.harvard.edu/abs/2008PhRvL.100d1101S}{ADS}]},
  {\small[\href{http://arxiv.org/abs/0712.0607}{{arXiv:0712.0607
  {\small[gr-qc]}}}]}.

\bibitem{2009fcgr.book.....S}
{Schutz}, B.F., {\em A First Course in General Relativity}, (Cambridge
  University Press, Cambridge, 2009).
  {\small[\href{http://adsabs.harvard.edu/abs/2009fcgr.book.....S}{ADS}]}.

\bibitem{2014arXiv1403.2697S}
{Shah}, A.~G, ``{Gravitational-wave flux for a particle orbiting a Kerr black
  hole to 20th post-Newtonian order: a numerical approach}'', {\em ArXiv
  e-prints} (March 2014).
  {\small[\href{http://adsabs.harvard.edu/abs/2014arXiv1403.2697S}{ADS}]},
  {\small[\href{http://arxiv.org/abs/1403.2697}{{arXiv:1403.2697
  {\small[gr-qc]}}}]}.

\bibitem{shankland}
Shankland, R.S., McCuskey, S.W., Leone, F.C.  and Kuerti, G., ``New analysis of
  the interferometer observations of Dayton C. Miller'', {\em Rev. Mod. Phys.},
  {\bf 27}, 167--178 (1955).
  {\small[\href{http://dx.doi.org/10.1103/RevModPhys.27.167}{DOI}]}.

\bibitem{2013CQGra..30p5019S}
{Shao}, L., {Caballero}, R.~N., {Kramer}, M., {Wex}, N., {Champion}, D.~J.  and
  {Jessner}, A., ``{A new limit on local Lorentz invariance violation of
  gravity from solitary pulsars}'', {\em Class. Quantum Grav.}, {\bf 30},
  165019 (2013).
  {\small[\href{http://dx.doi.org/10.1088/0264-9381/30/16/165019}{DOI}]},
  {\small[\href{http://adsabs.harvard.edu/abs/2013CQGra..30p5019S}{ADS}]},
  {\small[\href{http://arxiv.org/abs/1307.2552}{{arXiv:1307.2552
  {\small[gr-qc]}}}]}.

\bibitem{2012CQGra..29u5018S}
{Shao}, L.  and {Wex}, N., ``{New tests of local Lorentz invariance of gravity
  with small-eccentricity binary pulsars}'', {\em Class. Quantum Grav.}, {\bf
  29}, 215018 (2012).
  {\small[\href{http://dx.doi.org/10.1088/0264-9381/29/21/215018}{DOI}]},
  {\small[\href{http://adsabs.harvard.edu/abs/2012CQGra..29u5018S}{ADS}]},
  {\small[\href{http://arxiv.org/abs/1209.4503}{{arXiv:1209.4503
  {\small[gr-qc]}}}]}.

\bibitem{2013CQGra..30p5020S}
{Shao}, L.  and {Wex}, N., ``{New limits on the violation of local position
  invariance of gravity}'', {\em Class. Quantum Grav.}, {\bf 30}, 165020
  (2013).
  {\small[\href{http://dx.doi.org/10.1088/0264-9381/30/16/165020}{DOI}]},
  {\small[\href{http://adsabs.harvard.edu/abs/2013CQGra..30p5020S}{ADS}]},
  {\small[\href{http://arxiv.org/abs/1307.2637}{{arXiv:1307.2637
  {\small[gr-qc]}}}]}.

\bibitem{shapiro}
Shapiro, I.I., ``A century of relativity'', {\em Rev. Mod. Phys.}, {\bf 71},
  S41--S53 (1999).
  {\small[\href{http://dx.doi.org/10.1103/RevModPhys.71.S41}{DOI}]}.

\bibitem{2012ApJS..201....1S}
{Shapiro}, I.~I., {Bartel}, N., {Bietenholz}, M.~F., {Lebach}, D.~E.,
  {Lestrade}, J.-F., {Ransom}, R.~R.  and {Ratner}, M.~I., ``{VLBI for Gravity
  Probe B. I. Overview}'', {\em Astrophys. J. Suppl.}, {\bf 201}, 1 (2012).
  {\small[\href{http://dx.doi.org/10.1088/0067-0049/201/1/1}{DOI}]},
  {\small[\href{http://adsabs.harvard.edu/abs/2012ApJS..201....1S}{ADS}]},
  {\small[\href{http://arxiv.org/abs/1204.4630}{{arXiv:1204.4630
  {\small[astro-ph.IM]}}}]}.

\bibitem{sshapiro04}
Shapiro, S.S., Davis, J.L., Lebach, D.E.  and Gregory, J.S., ``Measurement of
  the Solar Gravitational Deflection of Radio Waves using Geodetic
  Very-Long-Baseline Interferometry Data, 1979--1999'', {\em Phys. Rev. Lett.},
  {\bf 92}, 121101 (2004).
  {\small[\href{http://dx.doi.org/10.1103/PhysRevLett.92.121101}{DOI}]}.

\bibitem{shlyakter}
Shlyakter, A.I., ``Direct test of the constancy of fundamental nuclear
  constants'', {\em Nature}, {\bf 264}, 340 (1976).
  {\small[\href{http://dx.doi.org/10.1038/264340a0}{DOI}]}.

\bibitem{PhysRevD.77.123502}
Skordis, C., ``Generalizing tensor-vector-scalar cosmology'', {\em Phys. Rev.
  D}, {\bf 77}, 123502 (2008).
  {\small[\href{http://dx.doi.org/10.1103/PhysRevD.77.123502}{DOI}]}.

\bibitem{2009CQGra..26n3001S}
{Skordis}, C., ``{The tensor-vector-scalar theory and its cosmology}'', {\em
  Class. Quantum Grav.}, {\bf 26}, 143001 (2009).
  {\small[\href{http://dx.doi.org/10.1088/0264-9381/26/14/143001}{DOI}]},
  {\small[\href{http://adsabs.harvard.edu/abs/2009CQGra..26n3001S}{ADS}]},
  {\small[\href{http://arxiv.org/abs/0903.3602}{{arXiv:0903.3602
  {\small[astro-ph.CO]}}}]}.

\bibitem{1976PhRvD..14.2443S}
{Smarr}, L., {{\v C}ade{\v z}}, A., {Dewitt}, B.  and {Eppley}, K.,
  ``{Collision of two black holes: Theoretical framework}'', {\em \prd}, {\bf
  14}, 2443--2452 (1976).
  {\small[\href{http://dx.doi.org/10.1103/PhysRevD.14.2443}{DOI}]},
  {\small[\href{http://adsabs.harvard.edu/abs/1976PhRvD..14.2443S}{ADS}]}.

\bibitem{2011PhRvL.107q1604S}
{Smiciklas}, M., {Brown}, J.~M., {Cheuk}, L.~W., {Smullin}, S.~J.  and
  {Romalis}, M.~V., ``{New Test of Local Lorentz Invariance Using a
  ${}^{21}$Ne-Rb-K Comagnetometer}'', {\em Phys. Rev. Lett.}, {\bf 107}, 171604
  (2011).
  {\small[\href{http://dx.doi.org/10.1103/PhysRevLett.107.171604}{DOI}]},
  {\small[\href{http://adsabs.harvard.edu/abs/2011PhRvL.107q1604S}{ADS}]},
  {\small[\href{http://arxiv.org/abs/1106.0738}{{arXiv:1106.0738
  {\small[physics.atom-ph]}}}]}.

\bibitem{2010RvMP...82..451S}
{Sotiriou}, T.~P.  and {Faraoni}, V., ``{$f(R)$ theories of gravity}'', {\em
  Rev. Mod. Phys.}, {\bf 82}, 451--497 (2010).
  {\small[\href{http://dx.doi.org/10.1103/RevModPhys.82.451}{DOI}]},
  {\small[\href{http://adsabs.harvard.edu/abs/2010RvMP...82..451S}{ADS}]},
  {\small[\href{http://arxiv.org/abs/0805.1726}{{arXiv:0805.1726
  {\small[gr-qc]}}}]}.

\bibitem{2012PhRvL.108h1103S}
{Sotiriou}, T.~P.  and {Faraoni}, V., ``{Black Holes in Scalar-Tensor
  Gravity}'', {\em Phys. Rev. Lett.}, {\bf 108}, 081103 (2012).
  {\small[\href{http://dx.doi.org/10.1103/PhysRevLett.108.081103}{DOI}]},
  {\small[\href{http://adsabs.harvard.edu/abs/2012PhRvL.108h1103S}{ADS}]},
  {\small[\href{http://arxiv.org/abs/1109.6324}{{arXiv:1109.6324
  {\small[gr-qc]}}}]}.

\bibitem{0264-9381-29-18-180301}
Speake, C.C.  and Will, C.M., ``Tests of the weak equivalence principle'', {\em
  Class. Quantum Grav.}, {\bf 29}, 180301 (2012).

\bibitem{petitjean1}
Srianand, R., Chand, H., Petitjean, P.  and Aracil, B., ``Limits on the time
  variation of the electromagnetic fine-structure constant in the low energy
  limit from absorption lines in the spectra of distant quasars'', {\em Phys.
  Rev. Lett.}, {\bf 92}, 121302 (2004).
  {\small[\href{http://dx.doi.org/10.1103/PhysRevLett.92.121302}{DOI}]},
  {\small[\href{http://arxiv.org/abs/astro-ph/0402177}{{astro-ph/0402177}}]}.

\bibitem{StairsLRR}
Stairs, I.H., ``Testing General Relativity with Pulsar Timing'', {\em Living
  Rev. Relativity}, {\bf 6}, lrr-2003-5 (2003).
  {\small[\href{http://dx.doi.org/10.12942/lrr-2003-5}{DOI}]}. URL (accessed 1
  June 2014): \newline\url{http://www.livingreviews.org/lrr-2003-5}.

\bibitem{stairs05}
Stairs, I.H. {et~al.}, ``Discovery of three wide-orbit binary pulsars:
  Implications for Binary Evolution and Equivalence Principles'', {\em
  Astrophys. J.}, {\bf 632}, 1060--1068 (2005).
  {\small[\href{http://dx.doi.org/10.1086/432526}{DOI}]},
  {\small[\href{http://arxiv.org/abs/astro-ph/0506188}{{astro-ph/0506188}}]}.

\bibitem{stanwix05}
Stanwix, P.L., Tobar, M.E., Wolf, P., Susli, M., Locke, C.R., Ivanov, E.N.,
  Winterflood, J.  and van Kann, F., ``Test of Lorentz Invariance in
  Electrodynamics Using Rotating Cryogenic Sapphire Microwave Oscillators'',
  {\em Phys. Rev. Lett.}, {\bf 95}, 040404 (2005).
  {\small[\href{http://dx.doi.org/10.1103/PhysRevLett.95.040404}{DOI}]},
  {\small[\href{http://arxiv.org/abs/hep-ph/0506074}{{hep-ph/0506074}}]}.

\bibitem{2009PhRvD..80d4002S}
{Stavridis}, A.  and {Will}, C.~M., ``{Bounding the mass of the graviton with
  gravitational waves: Effect of spin precessions in massive black hole
  binaries}'', {\em \prd}, {\bf 80}, 044002 (2009).
  {\small[\href{http://dx.doi.org/10.1103/PhysRevD.80.044002}{DOI}]},
  {\small[\href{http://adsabs.harvard.edu/abs/2009PhRvD..80d4002S}{ADS}]},
  {\small[\href{http://arxiv.org/abs/0906.3602}{{arXiv:0906.3602
  {\small[gr-qc]}}}]}.

\bibitem{2009NJPh...11h5003S}
{Stecker}, F.~W.  and {Scully}, S.~T., ``{Searching for new physics with
  ultrahigh energy cosmic rays}'', {\em New Journal of Physics}, {\bf 11},
  085003 (2009).
  {\small[\href{http://dx.doi.org/10.1088/1367-2630/11/8/085003}{DOI}]},
  {\small[\href{http://adsabs.harvard.edu/abs/2009NJPh...11h5003S}{ADS}]},
  {\small[\href{http://arxiv.org/abs/0906.1735}{{arXiv:0906.1735
  {\small[astro-ph.HE]}}}]}.

\bibitem{Su94}
Su, Y., Heckel, B.R., Adelberger, E.G., Gundlach, J.H., Harris, M., Smith, G.L.
   and Swanson, H.E., ``New tests of the universality of free fall'', {\em
  Phys. Rev. D}, {\bf 50}, 3614--3636 (1994).
  {\small[\href{http://dx.doi.org/10.1103/PhysRevD.50.3614}{DOI}]}.

\bibitem{2011PhRvL.107q1101S}
{Sushkov}, A.~O., {Kim}, W.~J., {Dalvit}, D.~A.~R.  and {Lamoreaux}, S.~K.,
  ``{New Experimental Limits on Non-Newtonian Forces in the Micrometer
  Range}'', {\em Phys. Rev. Lett.}, {\bf 107}, 171101 (2011).
  {\small[\href{http://dx.doi.org/10.1103/PhysRevLett.107.171101}{DOI}]},
  {\small[\href{http://adsabs.harvard.edu/abs/2011PhRvL.107q1101S}{ADS}]},
  {\small[\href{http://arxiv.org/abs/1108.2547}{{arXiv:1108.2547
  {\small[quant-ph]}}}]}.

\bibitem{talmadge}
Talmadge, C.L., Berthias, J.-P., Hellings, R.W.  and Standish, E.M.,
  ``Model-Independent Constraints on Possible Modifications of Newtonian
  Gravity'', {\em Phys. Rev. Lett.}, {\bf 61}, 1159--1162 (1988).
  {\small[\href{http://dx.doi.org/10.1103/PhysRevLett.61.1159}{DOI}]}.

\bibitem{Taylor87}
Taylor, J.H., ``Astronomical and Space Experiments to Test Relativity'', in
  MacCallum, M.A.H., ed., {\em General Relativity and Gravitation}, p. 209,
  (Cambridge University Press, Cambridge; New York, 1987).
  {\small[\href{http://books.google.com/books?id=ejw9AAAAIAAJ}{Google Books}]}.

\bibitem{TaylorVeneziano}
Taylor, T.R.  and Veneziano, G., ``Dilaton Couplings at Large Distances'', {\em
  Phys. Lett. B}, {\bf 213}, 450--454 (1988).
  {\small[\href{http://dx.doi.org/10.1016/0370-2693(88)91290-7}{DOI}]}.

\bibitem{Taylor94}
Taylor~Jr, J.H., ``Nobel Lecture: Binary pulsars and relativistic gravity'',
  {\em Rev. Mod. Phys.}, {\bf 66}, 711--719 (1994).
  {\small[\href{http://dx.doi.org/10.1103/RevModPhys.66.711}{DOI}]}.

\bibitem{Thorne300}
Thorne, K.S., ``Gravitational radiation'', in Hawking, S.W.  and Israel, W.,
  eds., {\em Three Hundred Years of Gravitation}, pp. 330--458, (Cambridge
  University Press, Cambridge; New York, 1987).
  {\small[\href{http://books.google.com/books?id=Vq787qC5PWQC&pg=PA330}{Google
  Books}]}.

\bibitem{1971ApJ...166L..35T}
{Thorne}, K.~S.  and {Dykla}, J.~J., ``{Black Holes in the Dicke-Brans Theory
  of Gravity}'', {\em \apjl}, {\bf 166}, L35--L38 (1971).
  {\small[\href{http://dx.doi.org/10.1086/180734}{DOI}]},
  {\small[\href{http://adsabs.harvard.edu/abs/1971ApJ...166L..35T}{ADS}]}.

\bibitem{2010PhRvD..82l2003T}
{Tinto}, M.  and {Alves}, M.~E.~D.~S., ``{LISA sensitivities to gravitational
  waves from relativistic metric theories of gravity}'', {\em \prd}, {\bf 82},
  122003 (2010).
  {\small[\href{http://dx.doi.org/10.1103/PhysRevD.82.122003}{DOI}]},
  {\small[\href{http://adsabs.harvard.edu/abs/2010PhRvD..82l2003T}{ADS}]},
  {\small[\href{http://arxiv.org/abs/1010.1302}{{arXiv:1010.1302
  {\small[gr-qc]}}}]}.

\bibitem{treuhaft}
Treuhaft, R.N.  and Lowe, S.T., ``A measurement of planetary relativistic
  deflection'', {\em Astron. J.}, {\bf 102}, 1879--1888 (1991).
  {\small[\href{http://dx.doi.org/10.1086/116010}{DOI}]}.

\bibitem{2007PhRvL..98t1101T}
{Tu}, L.-C., {Guan}, S.-G., {Luo}, J., {Shao}, C.-G.  and {Liu}, L.-X., ``{Null
  Test of Newtonian Inverse-Square Law at Submillimeter Range with a
  Dual-Modulation Torsion Pendulum}'', {\em Phys. Rev. Lett.}, {\bf 98}, 201101
  (2007).
  {\small[\href{http://dx.doi.org/10.1103/PhysRevLett.98.201101}{DOI}]},
  {\small[\href{http://adsabs.harvard.edu/abs/2007PhRvL..98t1101T}{ADS}]}.

\bibitem{turneaure}
Turneaure, J.P., Will, C.M., Farrell, B.F., Mattison, E.M.  and Vessot, R.F.C.,
  ``Test of the principle of equivalence by a null gravitational redshift
  experiment'', {\em Phys. Rev. D}, {\bf 27}, 1705--1714 (1983).
  {\small[\href{http://dx.doi.org/10.1103/PhysRevD.27.1705}{DOI}]}.

\bibitem{turyshev04a}
Turyshev, S.G., Shao, M.  and Nordtvedt~Jr., K., ``Experimental design for the
  LATOR mission'', {\em Int. J. Mod. Phys. D}, {\bf 13}, 2035--2063 (2004).
  {\small[\href{http://dx.doi.org/10.1142/S0218271804006528}{DOI}]},
  {\small[\href{http://arxiv.org/abs/gr-qc/0410044}{{gr-qc/0410044}}]}.

\bibitem{turyshev04b}
Turyshev, S.G., Shao, M.  and Nordtvedt~Jr., K., ``The laser astrometric test
  of relativity mission'', {\em Class. Quantum Grav.}, {\bf 21}, 2773--2799
  (2004). {\small[\href{http://dx.doi.org/10.1088/0264-9381/21/12/001}{DOI}]},
  {\small[\href{http://arxiv.org/abs/gr-qc/0311020}{{gr-qc/0311020}}]}.

\bibitem{2008ARNPS..58..207T}
{Turyshev}, S.~G., ``{Experimental Tests of General Relativity}'', {\em Ann.
  Rev. Nucl. Part. Sci.}, {\bf 58}, 207--248 (November 2008).
  {\small[\href{http://dx.doi.org/10.1146/annurev.nucl.58.020807.111839}{DOI}]},
  {\small[\href{http://adsabs.harvard.edu/abs/2008ARNPS..58..207T}{ADS}]},
  {\small[\href{http://arxiv.org/abs/0806.1731}{{arXiv:0806.1731
  {\small[gr-qc]}}}]}.

\bibitem{2010LRR....13....4T}
{Turyshev}, S.~G.  and {Toth}, V.~T., ``{The Pioneer Anomaly}'', {\em Living
  Rev. Relativity}, {\bf 13}, lrr-2010-4 (2010).
  {\small[\href{http://dx.doi.org/10.12942/lrr-2010-4}{DOI}]},
  {\small[\href{http://adsabs.harvard.edu/abs/2010LRR....13....4T}{ADS}]},
  {\small[\href{http://arxiv.org/abs/1001.3686}{{arXiv:1001.3686
  {\small[gr-qc]}}}]}. URL (accessed 1 June 2014):
  \newline\url{http://www.livingreviews.org/lrr-2010-4}.

\bibitem{2012PhRvL.108x1101T}
{Turyshev}, S.~G., {Toth}, V.~T., {Kinsella}, G., {Lee}, S.-C., {Lok}, S.~M.
  and {Ellis}, J., ``{Support for the Thermal Origin of the Pioneer Anomaly}'',
  {\em Phys. Rev. Lett.}, {\bf 108}, 241101 (2012).
  {\small[\href{http://dx.doi.org/10.1103/PhysRevLett.108.241101}{DOI}]},
  {\small[\href{http://adsabs.harvard.edu/abs/2012PhRvL.108x1101T}{ADS}]},
  {\small[\href{http://arxiv.org/abs/1204.2507}{{arXiv:1204.2507
  {\small[gr-qc]}}}]}.

\bibitem{2011LRR....14....2U}
{Uzan}, J.-P., ``{Varying Constants, Gravitation and Cosmology}'', {\em
  http://www.livingreviews.org/lrr-2010-4. Relativity}, {\bf 14}, lrr-2011-2
  (2011). {\small[\href{http://dx.doi.org/10.12942/lrr-2011-2}{DOI}]},
  {\small[\href{http://adsabs.harvard.edu/abs/2011LRR....14....2U}{ADS}]},
  {\small[\href{http://arxiv.org/abs/1009.5514}{{arXiv:1009.5514
  {\small[astro-ph.CO]}}}]}. URL (accessed 1 June 2014):
  \newline\url{http://www.livingreviews.org/lrr-2011-2}.

\bibitem{vdv70}
van Dam, H.  and Veltman, M.J.G., ``Massive and massless Yang-Mills and
  gravitational fields'', {\em Nucl. Phys. B}, {\bf 22}, 397--411 (1970).
  {\small[\href{http://dx.doi.org/10.1016/0550-3213(70)90416-5}{DOI}]}.

\bibitem{2014A&A...561A.115V}
{Verma}, A.~K., {Fienga}, A., {Laskar}, J., {Manche}, H.  and {Gastineau}, M.,
  ``{Use of MESSENGER radioscience data to improve planetary ephemeris and to
  test general relativity}'', {\em \aap}, {\bf 561}, A115 (2014).
  {\small[\href{http://dx.doi.org/10.1051/0004-6361/201322124}{DOI}]},
  {\small[\href{http://adsabs.harvard.edu/abs/2014A%26A...561A.115V}{ADS}]},
  {\small[\href{http://arxiv.org/abs/1306.5569}{{arXiv:1306.5569
  {\small[astro-ph.EP]}}}]}.

\bibitem{vessot}
Vessot, R.F.C. {et~al.}, ``Test of Relativistic Gravitation with a Space-Borne
  Hydrogen Maser'', {\em Phys. Rev. Lett.}, {\bf 45}, 2081--2084 (1980).
  {\small[\href{http://dx.doi.org/10.1103/PhysRevLett.45.2081}{DOI}]},
  {\small[\href{http://adsabs.harvard.edu/abs/1980PhRvL..45.2081V}{ADS}]}.

\bibitem{visser}
Visser, M., ``Mass for the graviton'', {\em Gen. Relativ. Gravit.}, {\bf 30},
  1717--1728 (1998).
  {\small[\href{http://dx.doi.org/10.1023/A:1026611026766}{DOI}]},
  {\small[\href{http://arxiv.org/abs/gr-qc/9705051}{{gr-qc/9705051}}]}.

\bibitem{2012CQGra..29r4002W}
{Wagner}, T.~A., {Schlamminger}, S., {Gundlach}, J.~H.  and {Adelberger},
  E.~G., ``{Torsion-balance tests of the weak equivalence principle}'', {\em
  Class. Quantum Grav.}, {\bf 29}, 184002 (2012).
  {\small[\href{http://dx.doi.org/10.1088/0264-9381/29/18/184002}{DOI}]},
  {\small[\href{http://adsabs.harvard.edu/abs/2012CQGra..29r4002W}{ADS}]},
  {\small[\href{http://arxiv.org/abs/1207.2442}{{arXiv:1207.2442
  {\small[gr-qc]}}}]}.

\bibitem{wagoner}
Wagoner, R.V., ``Resonant-mass detection of tensor and scalar waves'', in
  Marck, J.-A.  and Lasota, J.-P., eds., {\em Relativistic Gravitation and
  Gravitational Radiation}, Proceedings of the Les Houches School of Physics,
  held in Les Houches, Haute Savoie, 26 September -- 6 October, 1995, pp.
  419--432, (Cambridge University Press, Cambridge, 1997).

\bibitem{wagoner2}
Wagoner, R.V.  and Kalligas, D., ``Scalar-tensor theories and gravitational
  radiation'', in Marck, J.-A.  and Lasota, J.-P., eds., {\em Relativistic
  Gravitation and Gravitational Radiation}, Proceedings of the Les Houches
  School of Physics, held in Les Houches, Haute Savoie, 26 September -- 6
  October, 1995, pp. 433--446, (Cambridge University Press, Cambridge, 1997).

\bibitem{wagwill}
Wagoner, R.V.  and Will, C.M., ``Post-Newtonian gravitational radiation from
  orbiting point masses'', {\em Astrophys. J.}, {\bf 210}, 764--775 (1976).
  {\small[\href{http://dx.doi.org/10.1086/154886}{DOI}]}.

\bibitem{webb99}
Webb, J.K., Flambaum, V.V., Churchill, C.W., Drinkwater, M.J.  and Barrow,
  J.D., ``Search for time variation of the fine structure constant'', {\em
  Phys. Rev. Lett.}, {\bf 82}, 884--887 (1999).
  {\small[\href{http://dx.doi.org/10.1103/PhysRevLett.82.884}{DOI}]},
  {\small[\href{http://arxiv.org/abs/astro-ph/9803165}{{astro-ph/9803165}}]}.

\bibitem{Weinberg}
Weinberg, S., {\em Gravitation and Cosmology: Principles and Applications of
  the General Theory of Relativity}, (Wiley, New York, 1972).

\bibitem{WeisbergTaylor02}
Weisberg, J.M.  and Taylor, J.H., ``General Relativistic Geodetic Spin
  Precession in Binary Pulsar B1913+16: Mapping the Emission Beam in Two
  Dimensions'', {\em Astrophys. J.}, {\bf 576}, 942--949 (2002).
  {\small[\href{http://dx.doi.org/10.1086/341803}{DOI}]},
  {\small[\href{http://arxiv.org/abs/astro-ph/0205280}{{astro-ph/0205280}}]}.

\bibitem{2010ApJ...722.1030W}
{Weisberg}, J.~M., {Nice}, D.~J.  and {Taylor}, J.~H., ``{Timing Measurements
  of the Relativistic Binary Pulsar PSR B1913+16}'', {\em Astrophys. J.}, {\bf
  722}, 1030--1034 (2010).
  {\small[\href{http://dx.doi.org/10.1088/0004-637X/722/2/1030}{DOI}]},
  {\small[\href{http://adsabs.harvard.edu/abs/2010ApJ...722.1030W}{ADS}]},
  {\small[\href{http://arxiv.org/abs/1011.0718}{{arXiv:1011.0718
  {\small[astro-ph.GA]}}}]}.

\bibitem{WenSchutz05}
Wen, L.  and Schutz, B.F., ``Coherent network detection of gravitational waves:
  the redundancy veto'', {\em Class. Quantum Grav.}, {\bf 22}, S1321--S1336
  (2005). {\small[\href{http://dx.doi.org/10.1088/0264-9381/22/18/S46}{DOI}]},
  {\small[\href{http://arxiv.org/abs/gr-qc/0508042}{{gr-qc/0508042}}]}.

\bibitem{2014arXiv1402.5594W}
{Wex}, N., ``{Testing Relativistic Gravity with Radio Pulsars}'', {\em ArXiv
  e-prints} (February 2014).
  {\small[\href{http://adsabs.harvard.edu/abs/2014arXiv1402.5594W}{ADS}]},
  {\small[\href{http://arxiv.org/abs/1402.5594}{{arXiv:1402.5594
  {\small[gr-qc]}}}]}.

\bibitem{Will71a}
Will, C.M., ``Theoretical Frameworks for Testing Relativistic Gravity. II.
  Parametrized Post-Newtonian Hydrodynamics, and the Nordtvedt Effect'', {\em
  Astrophys. J.}, {\bf 163}, 611--628 (1971).
  {\small[\href{http://dx.doi.org/10.1086/150804}{DOI}]},
  {\small[\href{http://adsabs.harvard.edu/abs/1971ApJ...163..611W}{ADS}]}.

\bibitem{Will76}
Will, C.M., ``Active mass in relativistic gravity: Theoretical interpretation
  of the Kreuzer experiment'', {\em Astrophys. J.}, {\bf 204}, 224--234 (1976).
  {\small[\href{http://dx.doi.org/10.1086/154164}{DOI}]}.

\bibitem{willcavendish}
Will, C.M., ``Henry Cavendish, Johann von Soldner, and the deflection of
  light'', {\em Am. J. Phys.}, {\bf 56}, 413--415 (1988).
  {\small[\href{http://dx.doi.org/10.1119/1.15622}{DOI}]}.

\bibitem{WillSky}
Will, C.M., ``Twilight time for the fifth force?'', {\em Sky and Telescope},
  {\bf 80}, 472--479 (1990).

\bibitem{Will92b}
Will, C.M., ``Clock synchronization and isotropy of the one-way speed of
  light'', {\em Phys. Rev. D}, {\bf 45}, 403--411 (1992).
  {\small[\href{http://dx.doi.org/10.1103/PhysRevD.45.403}{DOI}]}.

\bibitem{Will92c}
Will, C.M., ``Is momentum conserved? A test in the binary system PSR 1913+16'',
  {\em Astrophys. J. Lett.}, {\bf 393}, L59--L61 (1992).
  {\small[\href{http://dx.doi.org/10.1086/186451}{DOI}]}.

\bibitem{tegp}
Will, C.M., {\em Theory and Experiment in Gravitational Physics}, (Cambridge
  University Press, Cambridge; New York, 1993), 2nd edition.
  {\small[\href{http://books.google.com/books?id=BhnUITA7sDIC}{Google Books}]}.

\bibitem{WER}
Will, C.M., {\em Was Einstein Right?: Putting General Relativity to the Test},
  (Basic Books, New York, 1993), 2nd edition.

\bibitem{willbd}
Will, C.M., ``Testing scalar-tensor gravity with gravitational-wave
  observations of inspiralling compact binaries'', {\em Phys. Rev. D}, {\bf
  50}, 6058--6067 (1994).
  {\small[\href{http://dx.doi.org/10.1103/PhysRevD.50.6058}{DOI}]},
  {\small[\href{http://arxiv.org/abs/gr-qc/9406022}{{gr-qc/9406022}}]}.

\bibitem{graviton}
Will, C.M., ``Bounding the mass of the graviton using gravitional-wave
  observations of inspiralling compact binaries'', {\em Phys. Rev. D}, {\bf
  57}, 2061 (1998).
  {\small[\href{http://dx.doi.org/10.1103/PhysRevD.57.2061}{DOI}]},
  {\small[\href{http://arxiv.org/abs/gr-qc/9709011}{{gr-qc/9709011}}]}.

\bibitem{physicscentral}
Will, C.M., ``Einstein's relativity and everyday life'', online article,
  American Physical Society, (2000). URL (accessed 1 June 2014):
  \newline\url{http://www.physicscentral.com/writers/writers-00-2.html}.

\bibitem{willspeed03}
Will, C.M., ``Propagation speed of gravity and the relativistic time delay'',
  {\em Astrophys. J.}, {\bf 590}, 683--690 (2003).
  {\small[\href{http://dx.doi.org/10.1086/375164}{DOI}]},
  {\small[\href{http://arxiv.org/abs/astro-ph/0301145}{{astro-ph/0301145}}]}.

\bibitem{2006eins.book...33W}
{Will}, C.M., ``{Special Relativity: A Centenary Perspective}'', in {Damour},
  T., {Darrigol}, O., {Duplantier}, B.  and {Rivasseau}, V., eds., {\em
  Einstein, 1905-2005: Poincar{\'e} Seminar 2005}, p.~33, (Birk\"auser Verlag,
  Basel, 2006).
  {\small[\href{http://dx.doi.org/10.1007/3-7643-7436-5_2}{DOI}]},
  {\small[\href{http://adsabs.harvard.edu/abs/2006eins.book...33W}{ADS}]}.

\bibitem{2008ApJ...674L..25W}
{Will}, C.M., ``{Testing the General Relativistic ``No-Hair'' Theorems Using
  the Galactic Center Black Hole Sagittarius A*}'', {\em Astrophys. J. Lett.},
  {\bf 674}, L25--L28 (2008).
  {\small[\href{http://dx.doi.org/10.1086/528847}{DOI}]},
  {\small[\href{http://adsabs.harvard.edu/abs/2008ApJ...674L..25W}{ADS}]},
  {\small[\href{http://arxiv.org/abs/0711.1677}{{arXiv:0711.1677}}]}.

\bibitem{2010AmJPh..78.1240W}
{Will}, C.M., ``{Resource Letter PTG-1: Precision Tests of Gravity}'', {\em Am.
  J. Phys.}, {\bf 78}, 1240--1247 (2010).
  {\small[\href{http://dx.doi.org/10.1119/1.3481700}{DOI}]},
  {\small[\href{http://adsabs.harvard.edu/abs/2010AmJPh..78.1240W}{ADS}]},
  {\small[\href{http://arxiv.org/abs/1008.0296}{{arXiv:1008.0296
  {\small[gr-qc]}}}]}.

\bibitem{willnordtvedt72}
Will, C.M.  and Nordtvedt~Jr., K.L., ``Conservation Laws and Preferred Frames
  in Relativistic Gravity. I. Preferred-Frame Theories and an Extended PPN
  Formalism'', {\em Astrophys. J.}, {\bf 177}, 757--774 (1972).
  {\small[\href{http://dx.doi.org/10.1086/151754}{DOI}]},
  {\small[\href{http://adsabs.harvard.edu/abs/1972ApJ...177..757W}{ADS}]}.

\bibitem{opus}
Will, C.M.  and Wiseman, A.G., ``Gravitational radiation from compact binary
  systems: Gravitational waveforms and energy loss to second post-Newtonian
  order'', {\em Phys. Rev. D}, {\bf 54}, 4813--4848 (1996).
  {\small[\href{http://dx.doi.org/10.1103/PhysRevD.54.4813}{DOI}]},
  {\small[\href{http://arxiv.org/abs/gr-qc/9608012}{{gr-qc/9608012}}]}.

\bibitem{willyunes}
Will, C.M.  and Yunes, N., ``Testing alternative theories of gravity using
  LISA'', {\em Class. Quantum Grav.}, {\bf 21}, 4367--4381 (2004).
  {\small[\href{http://dx.doi.org/10.1088/0264-9381/21/18/006}{DOI}]},
  {\small[\href{http://arxiv.org/abs/gr-qc/0403100}{{gr-qc/0403100}}]}.

\bibitem{zaglauer}
Will, C.M.  and Zaglauer, H.W., ``Gravitational radiation, close binary
  systems, and the Brans--Dicke theory of gravity'', {\em Astrophys. J.}, {\bf
  346}, 366--377 (1989).
  {\small[\href{http://dx.doi.org/10.1086/168016}{DOI}]},
  {\small[\href{http://adsabs.harvard.edu/abs/1989ApJ...346..366W}{ADS}]}.

\bibitem{1977ApJ...214..826W}
{Will}, C.~M., ``{Gravitational radiation from binary systems in alternative
  metric theories of gravity - Dipole radiation and the binary pulsar}'', {\em
  \apj}, {\bf 214}, 826--839 (June 1977).
  {\small[\href{http://dx.doi.org/10.1086/155313}{DOI}]},
  {\small[\href{http://adsabs.harvard.edu/abs/1977ApJ...214..826W}{ADS}]}.

\bibitem{2011PhyOJ...4...43W}
{Will}, C.~M., ``{Finally, results from Gravity Probe B}'', {\em Physics Online
  Journal}, {\bf 4}, 43 (2011).
  {\small[\href{http://dx.doi.org/10.1103/Physics.4.43}{DOI}]},
  {\small[\href{http://adsabs.harvard.edu/abs/2011PhyOJ...4...43W}{ADS}]},
  {\small[\href{http://arxiv.org/abs/1106.1198}{{arXiv:1106.1198
  {\small[gr-qc]}}}]}. URL (accessed 1 June 2014):
  \newline\url{http://physics.aps.org/articles/v4/43}.

\bibitem{2011PNAS..108.5938W}
{Will}, C.~M., ``{On the unreasonable effectiveness of the post-Newtonian
  approximation in gravitational physics}'', {\em Proc. Nat. Acad. Sci. (US)},
  {\bf 108}, 5938--5945 (April 2011).
  {\small[\href{http://dx.doi.org/10.1073/pnas.1103127108}{DOI}]},
  {\small[\href{http://adsabs.harvard.edu/abs/2011PNAS..108.5938W}{ADS}]},
  {\small[\href{http://arxiv.org/abs/1102.5192}{{arXiv:1102.5192
  {\small[gr-qc]}}}]}.

\bibitem{Williams}
Williams, J.G., Newhall, X.X.  and Dickey, J.O., ``Relativity parameters
  determined from lunar laser ranging'', {\em Phys. Rev. D}, {\bf 53},
  6730--6739 (1996).
  {\small[\href{http://dx.doi.org/10.1103/PhysRevD.53.6730}{DOI}]}.

\bibitem{williams04}
Williams, J.G., Turyshev, S.G.  and Boggs, D.H., ``Progress in Lunar Laser
  Ranging Tests of Relativistic Gravity'', {\em Phys. Rev. Lett.}, {\bf 93},
  261101 (2004).
  {\small[\href{http://dx.doi.org/10.1103/PhysRevLett.93.261101}{DOI}]},
  {\small[\href{http://arxiv.org/abs/gr-qc/0411113}{{gr-qc/0411113}}]}.

\bibitem{williams04ijmp}
Williams, J.G., Turyshev, S.G.  and Murphy~Jr, T.W., ``Improving LLR Tests Of
  Gravitational Theory'', {\em Int. J. Mod. Phys. D}, {\bf 13}, 567--582
  (2004). {\small[\href{http://dx.doi.org/10.1142/S0218271804004682}{DOI}]},
  {\small[\href{http://arxiv.org/abs/gr-qc/0311021}{{gr-qc/0311021}}]}.

\bibitem{wolf03}
Wolf, P., Bize, S., Clairon, A., Luiten, A.N., Santarelli, G.  and Tobar, M.E.,
  ``Tests of Lorentz invariance using a microwave resonator'', {\em Phys. Rev.
  Lett.}, {\bf 90}, 060402 (2003).
  {\small[\href{http://dx.doi.org/10.1103/PhysRevLett.90.060402}{DOI}]},
  {\small[\href{http://arxiv.org/abs/gr-qc/0210049}{{gr-qc/0210049}}]}.

\bibitem{2011CQGra..28n5017W}
{Wolf}, P., {Blanchet}, L., {Bord{\'e}}, C.~J., {Reynaud}, S., {Salomon}, C.
  and {Cohen-Tannoudji}, C., ``{Does an atom interferometer test the
  gravitational redshift at the Compton frequency?}'', {\em Class. Quantum
  Grav.}, {\bf 28}, 145017 (2011).
  {\small[\href{http://dx.doi.org/10.1088/0264-9381/28/14/145017}{DOI}]},
  {\small[\href{http://adsabs.harvard.edu/abs/2011CQGra..28n5017W}{ADS}]},
  {\small[\href{http://arxiv.org/abs/1012.1194}{{arXiv:1012.1194
  {\small[gr-qc]}}}]}.

\bibitem{2006PhRvL..96f0801W}
{Wolf}, P., {Chapelet}, F., {Bize}, S.  and {Clairon}, A., ``{Cold Atom Clock
  Test of Lorentz Invariance in the Matter Sector}'', {\em Phys. Rev. Lett.},
  {\bf 96}, 060801 (2006).
  {\small[\href{http://dx.doi.org/10.1103/PhysRevLett.96.060801}{DOI}]},
  {\small[\href{http://adsabs.harvard.edu/abs/2006PhRvL..96f0801W}{ADS}]},
  {\small[\href{http://arxiv.org/abs/hep-ph/0601024}{{hep-ph/0601024}}]}.

\bibitem{wolfe76}
Wolfe, A.M., Brown, R.L.  and Roberts, M.S., ``Limits on the Variation of
  Fundamental Atomic Quantities over Cosmic Time Scales'', {\em Phys. Rev.
  Lett.}, {\bf 37}, 179--181 (1976).
  {\small[\href{http://dx.doi.org/10.1103/PhysRevLett.37.179}{DOI}]}.

\bibitem{2012PhRvL.108h1101Y}
{Yang}, S.-Q., {Zhan}, B.-F., {Wang}, Q.-L., {Shao}, C.-G., {Tu}, L.-C., {Tan},
  W.-H.  and {Luo}, J., ``{Test of the Gravitational Inverse Square Law at
  Millimeter Ranges}'', {\em Phys. Rev. Lett.}, {\bf 108}, 081101 (2012).
  {\small[\href{http://dx.doi.org/10.1103/PhysRevLett.108.081101}{DOI}]},
  {\small[\href{http://adsabs.harvard.edu/abs/2012PhRvL.108h1101Y}{ADS}]}.

\bibitem{2009PhRvD..80l2003Y}
{Yunes}, N.  and {Pretorius}, F., ``{Fundamental theoretical bias in
  gravitational wave astrophysics and the parametrized post-Einsteinian
  framework}'', {\em \prd}, {\bf 80}, 122003 (2009).
  {\small[\href{http://dx.doi.org/10.1103/PhysRevD.80.122003}{DOI}]},
  {\small[\href{http://adsabs.harvard.edu/abs/2009PhRvD..80l2003Y}{ADS}]},
  {\small[\href{http://arxiv.org/abs/0909.3328}{{arXiv:0909.3328
  {\small[gr-qc]}}}]}.

\bibitem{2013LRR....16....9Y}
{Yunes}, N.  and {Siemens}, X., ``{Gravitational-Wave Tests of General
  Relativity with Ground-Based Detectors and Pulsar-Timing Arrays}'', {\em
  Living Rev. Relativity}, {\bf 16}, lrr-2013-9 (2013).
  {\small[\href{http://dx.doi.org/10.12942/lrr-2013-9}{DOI}]},
  {\small[\href{http://adsabs.harvard.edu/abs/2013LRR....16....9Y}{ADS}]},
  {\small[\href{http://arxiv.org/abs/1304.3473}{{arXiv:1304.3473
  {\small[gr-qc]}}}]}. URL (accessed 1 June 2014):
  \newline\url{http://www.livingreviews.org/lrr-2013-9}.

\bibitem{zakharov70}
Zakharov, V.I., ``Linearized gravitation theory and the graviton mass'', {\em
  JETP Lett.}, {\bf 12}, 312 (1970).

\bibitem{2012JCAP...06..032Z}
{Zuntz}, J., {Baker}, T., {Ferreira}, P.~G.  and {Skordis}, C., ``{Ambiguous
  tests of general relativity on cosmological scales}'', {\em J. Cosm.
  Astropart. Phys.}, {\bf 6}, 032 (2012).
  {\small[\href{http://dx.doi.org/10.1088/1475-7516/2012/06/032}{DOI}]},
  {\small[\href{http://adsabs.harvard.edu/abs/2012JCAP...06..032Z}{ADS}]},
  {\small[\href{http://arxiv.org/abs/1110.3830}{{arXiv:1110.3830
  {\small[astro-ph.CO]}}}]}.

\end{thebibliography}

\end{document}